\documentclass[letterpaper,11pt]{article}
\pdfoutput=1 

\usepackage{jheppub} 

\usepackage{graphicx}
\usepackage{epstopdf}
\usepackage{amsmath, amssymb}
\usepackage{comment}
\allowdisplaybreaks

\usepackage{subcaption}
\usepackage{float}
\usepackage{cellspace}
\usepackage{longtable,makecell}
\usepackage{tabularx}
\usepackage{ytableau}
\usepackage{tablefootnote}
\usepackage{hyphenat}
\ytableausetup{smalltableaux, aligntableaux = center}
\def\fund{{{\ydiagram{1}}}}
\def\sym{{{\ydiagram{2}}}}
\def\antisym{{{\ydiagram{1,1}}}}

\usepackage{multirow}

\newcommand{\be}{\begin{eqnarray}}
\newcommand{\ee}{\end{eqnarray}}

\newcommand{\bn}{\begin{enumerate}}
\newcommand{\en}{\end{enumerate}}

\def\IZ{\mathbb{Z}}

\def\CA{{\cal A}}

\def\CI{{\cal I}}

\def\CN{{\cal N}}
\def\CO{{\cal O}}

\def\CQ{{\cal Q}}

\def\a{\alpha}
\def\b{\beta}

\def\d{\delta}
\def\e{\epsilon}

\def\m{\mu}
\def\n{\nu}

\def\r{\rho}

\def\s{\sigma}

\def\D{\Delta}

\def\O{\Omega}

\def\half{\frac{1}{2}}

\def\goto{\rightarrow}

\def\vev#1{\langle #1 \rangle}

\def\Tr{{\rm Tr}}

\begin{document}

\title{Large $N$ Universality of 4d $\CN=1$ SCFTs \\with Simple Gauge Groups }
 
\author[a]{Minseok Cho,}
\author[a, b]{Ki-Hong Lee,}
\author[a, c]{and Jaewon Song}

\affiliation[a]{Department of Physics, Korea Advanced Institute of Science and Technology\\
291 Daehak-ro, Yuseong-gu, Daejeon 34141, Republic of Korea.}
\affiliation[b]{Department of Physics and Haifa Center for Physics and Astrophysics, University of Haifa\\ 
Haifa 3498838, Israel}
\affiliation[c]{Walter Burke Institute for Theoretical Physics, California Institute of Technology\\
Pasadena, CA 91125, USA}

\emailAdd{cms1308@kaist.ac.kr}
\emailAdd{khlee11812@gmail.com}
\emailAdd{jaewon.song@kaist.ac.kr}

\abstract
{
We classify four-dimensional $\mathcal{N}=1$ supersymmetric gauge theories with a simple gauge group admitting a large $N$ limit that flow to non-trivial superconformal fixed points in the infrared. 
We focus on the cases where the large $N$ limit can be taken while keeping the flavor symmetry fixed so that the putative holographic dual has a fixed gauge group. 
We find that they can be classified into three types -- Type I, Type II, and Type III -- exhibiting universal behavior. Type I theories have $a \neq c$ in the large $N$ limit and scale linearly in $N$; the gap of scaling dimensions among BPS operators behaves as $1/N$. Type II theories have $a=c$ in the large $N$ limit, and satisfy $a \simeq c \simeq \frac{27}{128} \dim G$, and Type III theories have $a \simeq c \simeq \frac{1}{4} \dim G$. For Type II and Type III theories, the gap of scaling dimensions stays $O(1)$ in the large $N$ limit. 
We enumerate relevant and marginal operators of these theories and find that non-trivial conformal manifolds emerge upon relevant deformations. Moreover, we find that a modified version of the AdS Weak Gravity Conjecture, based on the supersymmetric Cardy formula, holds for all of these theories, even for finite $N$. 
}


\maketitle

\section{Introduction and Summary}

Conformal field theory (CFT) describes the critical point of the phase transition, the gapless phase of matter, and also quantum gravity. Moreover, it serves as a simpler model for more general quantum field theory (QFT) and also arises as an endpoint of the renormalization group flow. Therefore, it is apparent that understanding the precise nature of conformal field theory is crucial to addressing fundamental questions in theoretical physics. 

In this paper, we focus on the four-dimensional superconformal field theory (SCFT), which has played a pivotal role in the modern development of quantum field theory. Thanks to its robustness, it has been an indispensable tool to discover deep non-perturbative phenomena in quantum field theory and string theory, such as S-duality \cite{Sen:1994fa, Seiberg:1994rs}, IR duality \cite{Seiberg:1994pq}, and AdS/CFT correspondence \cite{Maldacena:1997re, Gubser:1998bc, Witten:1998qj}. 
One crucial aspect of the aforementioned phenomena is that they involve intricate connections among a family of theories, rather than a single theory. Therefore, it is important to explore the space of QFTs, involving the classification of SCFTs, a longstanding challenging problem. 

For the case of maximal supersymmetry ($\CN=4$), it is conjectured that they are classified by the simple Lie groups, realized by the $\CN=4$ super Yang-Mills theory. There are partial classification results for the extended supersymmetry $\mathcal{N} \ge 2$ \cite{Bhardwaj:2013qia, Argyres:2015ffa, Argyres:2015gha, Argyres:2016xmc, Argyres:2016xua, Gaiotto:2009we, Chacaltana:2010ks, Chacaltana:2011ze, Chacaltana:2013oka, Chacaltana:2014jba, Chacaltana:2015bna, Chacaltana:2016shw, Chacaltana:2017boe, Chacaltana:2018vhp, Bonetti:2018fqz}, but it is still a challenging problem. There are fewer results for the case of minimal supersymmetry ($\CN=1$), because the constraints imposed by supersymmetry are much weaker, making the problem even more difficult, and there have been fewer attempts along these lines \cite{Razamat:2020pra, Agarwal:2020pol, Gaiotto:2015usa, Razamat:2016dpl, Razamat:2022gpm}. In this paper, we attempt to classify a small subset of the 4d $\CN=1$ SCFT theory space, realized by \emph{large $N$} gauge theories with a \emph{simple gauge group}. The current work is a continuation of the previous paper by two of the authors (KL and JS) \cite{Agarwal:2020pol}. 

We will focus on the large $N$ limit, which is a powerful tool for understanding strongly interacting systems \cite{tHooft:1973alw}. This limit simplifies the theory, providing the small parameter $1/N$, so that we can compute various observables in a perturbative way. Our motivation for the large $N$ limit is the AdS/CFT correspondence \cite{Maldacena:1997re, Gubser:1998bc, Witten:1998qj}. Via AdS/CFT correspondence, quantum gravity in anti-de Sitter space in $d+1$ dimensions is exactly equivalent to a $d$-dimensional conformal field theory. Therefore, we can study AdS quantum supergravity by studying large $N$ superconformal field theories. 
Even though AdS/CFT is best understood when the bulk theory is weakly coupled Einstein gravity, it is important to note that the correspondence holds even if the bulk is exotic.  
The bulk `gravity' might be highly quantum, strongly-coupled with light strings. Therefore, we are naturally motivated to ask the question: for a given CFT, when do we have a good holographic dual, described by a weakly-coupled Einstein-like gravity? This question has been studied in \cite{Heemskerk:2009pn, El-Showk:2011yvt}, for example. By exploring the space of 4d $\CN=1$ SCFTs, we would like to understand how gravity emerges from large $N$ gauge theories. 
As was shown in \cite{Agarwal:2019crm, Agarwal:2020pol}, our classification program yields a set of SCFTs whose spectrum is dense in the sense that the gap in the operator scaling dimensions scales as $1/N$. This set of theories cannot be holographic, since the gap should scale as $O(1)$ for a holographic theory, or the spectrum has to be sparse. We also find many examples whose operator spectrum is sparse, so that they are putatively holographic. 

Another motivation comes from the Swampland program \cite{Vafa:2005ui}. The purpose of the Swampland program is to find universal properties for the low-energy effective field theory in order for it to be consistently coupled to quantum gravity. In the framework of the AdS/CFT correspondence, the swampland conjectures can be translated to the universal properties of conformal field theories. One such swampland conjecture, which has been studied in the context of holography, is the Weak Gravity Conjecture (WGC) \cite{Arkani-Hamed:2006emk, Nakayama:2015hga}. 
The Weak Gravity Conjecture was initially derived in the semi-classical regime, so there is no a priori good reason for it to be true for exotic gravitational theories. However, if we wish to view WGC as a genuine constraint for any consistent theory of quantum gravity, it must be true for a generic AdS/CFT setup, even if the bulk is exotic or the boundary CFT is weakly-coupled. Indeed, in the context of AdS$_3$/CFT$_2$, the AdS Weak Gravity Conjecture can indeed be shown to be a consequence of modular invariance of the partition function \cite{Benjamin:2016fhe, Montero:2016tif, Bae:2018qym}, showing that it comes from the consistency of the theory.  
We can ask if the AdS WGC holds in higher dimensions as well. Indeed, there are numerous non-holographic examples and counter-examples for the AdS WGC \cite{Nakayama:2015hga, Agarwal:2019crm, Agarwal:2020pol}. Using the result of our classification, we have proposed a modified AdS WGC that holds for arbitrary 4d SCFTs \cite{Cho:2023koe}. 

\subsection{Summary of results}
In this paper, we classify all possible 4d ($\CN=1$) SCFTs under the following assumptions: 
\begin{enumerate}
    \item Each of them arises as a renormalization group fixed point of a gauge theory with a simple gauge group and a number of matter chiral multiplets.\footnote{Conformal gauge theories obtained without RG flow have been classified in \cite{Razamat:2020pra}.} 
    \item The rank of the gauge group $N$ can be scaled to be large. Therefore, the gauge group is restricted to the classical ones: $SU(N)$, $SO(N)$ and $Sp(N)$. 
    \item The superpotential is set to zero, $W=0$. 
    \item The flavor symmetry remains fixed as we take the large $N$ limit.
\end{enumerate}
The last condition is motivated by holography, since the flavor symmetry is mapped to the gauge symmetry in the bulk.
This is the continuation of the classification program of large $N$ 4d superconformal gauge theories as outlined in \cite{Agarwal:2020pol}.
Unlike previous work, which focuses on theories with dense spectra, this work comprehensively studies all classes of large-$N$ theories under the aforementioned assumptions. 
Since our paper is inevitably long, we summarize the results in this section and highlight some of our findings. 

\paragraph{Universality of $R$-charges and central charges}
We find that, in the large $N$ limit, the leading-order behavior of the $R$-charges of chiral multiplets, as well as the central charges $a$ and $c$, is universal across certain sets of theories. Based on these universal behaviors, we classify large $N$ theories into three distinct types, which we call Type I, II, and III. They are essentially determined by the effective number of matters in rank-2 tensor representations of the gauge group. For the $SU(N)$ gauge theories, we define the effective number of rank-2 tensors as
\begin{align} \label{eq:Nrank2}
    N_{\text{rank-2}}\equiv N_{\textbf{Adj}} + \frac{N_{\mbox{\tiny\sym}}+N_{\overline{\mbox{\tiny\sym}}}+N_{\mbox{\tiny\antisym}}+ N_{\overline{\mbox{\tiny\antisym}}}}{2}\,,
\end{align}
where $N_\mathbf{R}$ refers to the number of matter (chiral) multiplets in representation $\mathbf{R}$ of the gauge group. 

If $N_{\text{rank-2}}$ is either 2 or 3, the leading behavior of the $R$-charges and central charges becomes universal in the large $N$ limit. 
For theories with $N_{\text{rank-2}}=2$, which we refer to as Type II theories, the asymptotic behavior is
\begin{align}
    R_{\text{rank-2}} \sim \half + O(N^{-1})\,, ~~R_{\mbox{\tiny\fund}} \sim \frac{12-\sqrt{26}}{12} + O(N^{-1})\,, ~~a \sim c \sim \frac{27}{128}\dim(G) + O(N)\,.
\end{align}
For $N_{\text{rank-2}}=3$ theories, which we refer to as Type III theories, we find
\begin{align}
    R_{\text{rank-2}} \sim R_{\mbox{\tiny\fund}} \sim \frac{2}{3} + O(N^{-1})\,,\quad a \sim c \sim \frac{1}{4}\dim(G) + O(N)\,,
\end{align}
The central charges $a$ and $c$ for the Type II and III theories agree in the large $N$ limit, as is required for a holographic theory. 
Notice that the leading behavior of $R$-charges and central charges for the Type III theory is identical to that of $\CN=4$ SYM theory. For the Type II theory, they resemble that of the two-adjoint theory or the $\CN=4$ SYM with one of the chiral adjoints integrated out. We observe $\frac{27}{128} = \frac{27}{32} \times \frac{1}{4}$, where $\frac{27}{32}$ is a universal number that appears when we flow from $\CN=2$ to $\CN=1$ SCFT via integrating out adjoint chiral multiplets in the $\CN=2$ vector multiplet \cite{Tachikawa:2009tt}. 
We also find that in the Veneziano limit -- where $N\goto\infty$ with the number of fundamentals scaling as $O(N)$ -- the leading-order behavior of both $R$-charges and central charges remains universal across theories within each type.

The theories with $N_{\text{rank-2}}=1$, which we refer to as Type I theories, exhibit a dense spectrum, as the rank-2 tensor chiral multiplet has an $R$-charge of order $O(N^{-1})$ \cite{Agarwal:2019crm, Agarwal:2020pol}.
Type I theories generically flow to interacting SCFTs along with $O(N)$ number of decoupled free chiral multiplets. Since the set of decoupled operators varies across theories, we do not find universal formulae for the $R$-charges and central charges. However, if we take additional limit $1\ll N_f \ll N$, we do obtain universal behavior for the $R$-charges:
\begin{align}
    R_{\text{rank-2}}\sim  \frac{(4-\sqrt{3})N_f}{3N} \,,\quad R_{\mbox{\tiny\fund}} \sim \frac{\sqrt{3}-1}{3}\,.
\end{align}
The central charges for these theories scale as $a = O(N), c=O(N)$, and they are, in general, different at large $N$, unlike the case of Type II and III theories. 

For the theories with $N_{\text{rank-2}}=0$ or a half-integer, the number of fundamental chiral multiplets cannot remain finite in the large $N$ limit. Instead, it must scale as $O(N)$, in order to cancel gauge anomalies. Therefore, only the Veneziano limit is allowed. The detailed aspects of these types of theories will be discussed in future work. 
The Veneziano limit is unnatural from the holographic perspective, since it scales the rank of the flavor symmetry group with $N$, which is identical to the bulk gauge symmetry. 
If we demand not to scale the flavor symmetry as we take the large $N$ limit, $N_{\text{rank-2}}$ defined in \eqref{eq:Nrank2} must be an integer.\footnote{The Veneziano-type limit ($N_f, N \to \infty$ with $\alpha \equiv N_f/N$ fixed) is \emph{allowed} for Type I and II, but forbidden for Type III.} We will primarily focus on such cases in this paper. 

\paragraph{Conformal window}
We identify a family of theories parametrized by $N$ (rank of the gauge group) and also $N_f$, which is the number of fundamental/anti-fundamental chiral multiplet pairs. For a given family, there is an allowed range of $N_f$, referred to as a conformal window,  under which the theory flows to a non-trivial superconformal fixed point.
The most basic requirement is the asymptotic freedom, which imposes an upper bound on $N_f$. In most cases, the lower bound of $N_f$ is zero for any $N$. However, there are notable exceptions where a non-zero lower bound arises -- these are studied in Section \ref{sec:adjSQCD}, \ref{sec:a1A1}, \ref{sec:SOadjSQCD}, and \ref{sec:SpadjSQCD}. Additionally, certain exceptions appear only at small values of $N$, as discussed in Section \ref{sec:a2A2} and \ref{sec:Spa2}.

\paragraph{The ratio of central charges $a/c$}
We also examine the range of central charge ratio $a/c$ for theories in the conformal window. For a holographic theory, the ratio has to be $a/c=1$ at least when $N$ is infinite \cite{Henningson:1998gx}. However, for a finite $N$, the ratio $a/c$ is not exactly equal to 1, and it is known that this deviation contributes to the higher curvature corrections in the supergravity effective action \cite{Anselmi:1998zb}. This also modifies the celebrated viscosity-entropy density ratio bound $\eta/s \ge 1/4\pi$ \cite{Kovtun:2004de}. Depending on the sign of $c-a$, it can either strengthen ($a/c>1$) or weaken ($a/c<1$) the bound \cite{Buchel:2008vz, Kats:2007mq}. 
 
Most theories we find have $a/c \leq 1$, but there are exceptions. In particular, some $SU(N)$ gauge theories discussed in Section \ref{sec:s1S1}, \ref{sec:s2S2}, and \ref{sec:adj1s1S1}, and certain $SO(N)$ gauge theories covered in Section \ref{sec:SOs1}, \ref{sec:SOadjSQCD}, \ref{sec:SOs2}, and \ref{sec:SOs1a1} exhibit $a/c>1$ when the number of fundamental chiral multiplets ($N_f$) is small. We do not find such examples in $Sp(N)$ gauge theories.
There also exists a small number of cases where $a/c=1$ exactly at finite $N$. The theories with two and three adjoint chiral multiplets (without any fundamentals) enjoy such a property.\footnote{It is known that there exists a rather large landscape of $a=c$ theories for any value of $N$ once we introduce `non-Lagrangian' components in the theory \cite{Kang:2021lic, Kang:2021ccs, Kang:2022zsl, Kang:2023dsa, Kang:2024inx}.} 

The Hofman-Maldacena unitarity bound \cite{Hofman:2008ar, Hofman:2016awc} restricts the ratio to lie in the window $\frac{1}{2} \le \frac{a}{c} \le \frac{3}{2}$, and the lower and upper bound is saturated by free chiral and vector multiplets, respectively. However, an empirical gap exists in the ratio. Namely, there are no known theories of $1.25 \lesssim a/c \lesssim 1.5$ nor $0.5 \lesssim a/c \lesssim 0.6$. It is therefore interesting to ask if there exists any theory within this window, or to find a reason for it to be forbidden.

The maximum value of the $a/c$ ratio among Type I-III theories is $417/368 \simeq 1.1332$, which arises in the $SO(5)$ gauge theory with a rank-2 symmetric tensor chiral multiplet (Section \ref{sec:SOs1}). 
The current known maximal value for the $a/c$ is found in the $SO(5)$ gauge theory with a rank-2 symmetric and a fundamental and a singlet chiral multiplets, under a specific superpotential, yielding $a/c=507/419\simeq 1.2100$ \cite{Cho:2024civ}.
When the superpotential is turned off, the known maximal value of $a/c$ arises in the ISS model \cite{Intriligator:1994rx}, with $a/c=183/158\simeq 1.158$. Note that this model has a rank-3 tensor matter multiplet, making it impossible to take a large $N$ limit. 

The minimum value of the $a/c$ ratio among Type I-III theories is $\frac{70169619-220676\sqrt{3201}}{83242974}\simeq 0.6930$, found in an $SU(5)$ gauge theory with a pair of rank-2 antisymmetric and conjugate tensors, along with four pairs of fundamental and anti-fundamental chiral multiplets ($N_f=4$). Let us remark that we find the $Sp(2)$ gauge theory with 10 fundamental chiral multiplets ($N_f=5$) exhibits an even smaller value, $a/c=\frac{339}{514}\simeq 0.6595$, and this theory is excluded in our classification. (But see Section \ref{sec:discussion}.) This value is close to $3/5$, which may be the lower bound for an interacting theory \cite{Benini:2015bwz, Bobev:2017uzs}. 

\paragraph{Relevant operators}
We investigate the relevant operators of each theory. For the Type I theories, the rank-2 tensor chiral multiplets $X$ have $R$-charges of order $O(N^{-1})$. Therefore, we have the operators of the form $\Tr O X^n$ ($O$ being some combination of elementary fields or identity) leading to a dense spectrum of operators with gap $1/N$ in the large $N$ limit \cite{Agarwal:2019crm, Agarwal:2020pol}. This gives rise to $O(N)$ amount of relevant operators. 
On the other hand, the Type II and Type III theories contain only a finite number of relevant operators, i.e., of order $O(1)$. For these cases, we write an explicit list of the number and forms of the relevant operators. While additional relevant operators may appear for small values of $N$, we focus on those that exist generically in the large $N$ limit.

\paragraph{Conformal manifolds}
We also search for the conformal manifolds using the methods of \cite{Leigh:1995ep, Green:2010da}. In the case where the one-loop beta function for the gauge coupling vanishes, the theory is not guaranteed to be an interacting conformal field theory unless there exists a non-trivial conformal manifold. Such cases are studied in \cite{Razamat:2020pra}. 
On the other hand, since our gauge theories flow to a non-trivial fixed point, we do not require the existence of a conformal manifold. In fact, among the theories that flow, we find that only the two-adjoint theories \cite{Leigh:1995ep} and certain Type I theories under specific conditions have non-trivial conformal manifolds. 
However, we find that for at least $N_f=0,1$ cases, there always exists a superpotential deformation, such that the deformed theory flows to a superconformal fixed point with a non-trivial conformal manifold for general $N$.

\paragraph{Holographic SCFT}
We find that Type I theories are not holographic, while Type II and III theories are potentially holographic. 
For a large $N$ 4d CFT to be holographic, it has to have identical central charges $a=c$ \cite{Henningson:1998gx} and also have a sparse low-energy spectrum \cite{Heemskerk:2009pn, El-Showk:2011yvt}. For the Type I theories, these conditions are not satisfied, so they cannot be holographic. The Type II and III theories satisfy these conditions, so they are potentially holographic. Indeed, some of these theories have known holographic descriptions, such as $\CN=4$ SYM theory or the two adjoint theory, also referred to as the Leigh-Strassler theory \cite{Leigh:1995ep, Khavaev:1998fb, Karch:1999pv, Freedman:1999gp, Pilch:2000fu}. 

But these conditions do not guarantee a good holographic dual description. However, we find a further clue that supports the notion that the Type II and III theories are indeed holographic in nature. First, we find that upon a suitable relevant deformation, the fixed point theory has exactly marginal operators, or it has a non-trivial conformal manifold, as described in the previous paragraph. This is a promising signal for the theory to be `holographic,' since all the known weakly-coupled bulk descriptions come with a `coupling constant' or moduli (such as the gravitational constant, and string couplings) that can be tuned. These moduli sometimes enable us to move away from singular to smooth internal geometry in the bulk. See, for example \cite{Belin:2020nmp}, for a discussion on this aspect in the context of 2d SCFT. 

Another universal property of a holographic theory is that it should exhibit the Hawking-Page phase transition \cite{Hawking:1982dh}, which is dual to the confinement-deconfinement transition \cite{Witten:1998zw}. Namely, thermodynamic free energy at low energy is dominated by a gas of gravitons (thermal AdS phase), and at high energies, it is dominated by the large AdS black hole. In holography, the free energy at low energies should scale as $F = O(1)$, while at high energies it scales as $F = O(N^2)$. It is rather difficult to carry out such a computation at finite temperature for a strongly-coupled field theory. One interesting development in recent years is the observation of a phase transition even at zero temperature, which is visible in the BPS sector. 

The superconformal index \cite{Kinney:2005ej, Romelsberger:2005eg} counts $(-1)^F$-graded degeneracy of certain BPS states. Because of the Bose-Fermi cancellation, this quantity is robust, invariant under any continuous deformations. One of the original motivations for the index is to account for the Bekenstein-Hawking entropy of the BPS black holes in AdS found in \cite{Gutowski:2004ez, Gutowski:2004yv, Chong:2005hr, Kunduri:2006ek}.  
This goal was achieved later in \cite{Cabo-Bizet:2018ehj, Choi:2018hmj, Benini:2018ywd} through utilizing complexified chemical potentials. 
Especially, it was found that there is a Cardy-like formula for the superconformal index, giving asymptotic density of states at high energies in terms of central charges \cite{Kim:2019yrz, Cabo-Bizet:2019osg}. This formula turns out to account for the black hole entropy in AdS. 
The Cardy-like limit produces the density of states for high energy states $\Delta \gg N^2$ for finite $N$. On the other hand, holography requires the Cardy formula to be valid for $\Delta \sim O(N^2)$ since the black hole states correspond to the ones with scaling dimension $O(N^2)$. Indeed, there exists a large $N$ saddle for the matrix integral, which reproduces the formula without taking the Cardy limit \cite{Choi:2021rxi, Choi:2023tiq}. We note that among our set of theories, this large $N$ saddle works for the Type II and III theories, but not for the Type I theories. This also implies that there is a Hawking-Page-like phase transition as we dial the chemical potential (angular velocity) of the superconformal index. Therefore, Type II and III theories are good candidates to be holographic.

\paragraph{Weak Gravity Conjecture}
Finally, we test the AdS/CFT Weak Gravity Conjecture (WGC). We find that the WGC proposed in \cite{Nakayama:2015hga} does not hold for some theories with finite $N$ and even for a number of large $N$ theories with a sparse spectrum. 
However, we show that the modified WGC proposed in \cite{Cho:2023koe} successfully addresses all the counterexamples identified in our study.

\paragraph{Generalized symmetry}
Let us make a comment that when the number of fundamentals $N_f$ is zero, the theory has a non-trivial one-form symmetry, and the detailed aspects of the symmetry depend on the global structure of the gauge group \cite{Gaiotto:2014kfa}. The full list of such cases is already studied in \cite{Kang:2024elv}, where all possible simple gauge theories with non-trivial one-form symmetry are classified. 

\subsection{Organization of the paper}
The rest of the paper is structured as follows. 
In Section \ref{sec:classification}, we describe our methods and strategy for the classification program. Then we discuss in detail deriving the universal properties of $SU(N)$ in Section \ref{sec:SUclass}, $SO(N)$ in Section \ref{sec:SOclass}, and $Sp(N)$ theories in Section \ref{sec:Spclass}. We also review how we test the AdS version of the Weak Gravity Conjecture in Section \ref{sec:testWGC}. 
We describe the details of our classification result for the $SU(N)$ gauge theories in Section \ref{sec:SU}. We find 20 families of $SU(N)$ theories, and most of them have an additional parameter of $N_f$, which is the number of fundamental/anti-fundamental pairs. 
In Section \ref{sec:SO}, we study $SO(N)$ gauge theories. There are 6 families of theories, and most of them have an additional parameter $N_f$, which is the number of vector representations of $SO(N)$. 
In Section \ref{sec:Sp}, we study $Sp(N)$ gauge theories. There are 9 families of theories, with an additional parameter $N_f$ being twice the number of fundamental representations.\footnote{The number of fundamentals has to be even for $Sp(N)$ theories due to the Witten anomaly \cite{Witten:1982fp}.}
We conclude with future directions in Section \ref{sec:discussion}.

\section{Large N superconformal gauge theories}\label{sec:classification}

\subsection{Method and Strategy}

In this section, we classify four-dimensional $\mathcal{N}=1$ supersymmetric large $N$ gauge theory with a simple gauge group, which flows to an interacting superconformal field theory in the infrared. Since we are interested in theories admitting a large $N$ limit, we restrict the gauge groups to the classical groups $SU(N)$, $SO(N)$, and $Sp(N)$. We will also assume there is no superpotential. Our method is identical to that of \cite{Agarwal:2020pol}. 

Suppose we have a gauge theory with gauge group $G$ with chiral multiplets (labeled by $\chi$) in the representations $\mathbf{R}_\chi$ of the gauge group $G$.  
The gauge theory should satisfy two conditions. The first and the obvious one is that it has to be free of gauge anomaly, which constrains the matter contents of the theory as
\begin{align}\label{eq:Ganom}
    \sum_\chi\mathcal{A}(\mathbf{R}_\chi)=0\, \ , 
\end{align}
where the sum is over the chiral multiplets in the theory, and $\mathcal{A}(\mathbf{R}_\chi)$ represents the anomaly coefficients, defined by:
\begin{align}
    \mathcal{A}(\textbf{R})d^{abc}=\frac{1}{2}\Tr(T_\textbf{R}^a\{T_\textbf{R}^b,T_\textbf{R}^c\})\,.
\end{align}
Here $T^a_\textbf{R}$ denotes the generators of the gauge group in representations \textbf{R}, and the normalization is set such that $\mathcal{A}=1$ for the fundamental representation with totally symmetric rank-3 tensor $d^{abc}$. Let us remark that the anomaly coefficient is non-trivial only for the $SU(N)$ gauge group with $N \geq 3$, since it vanishes for any real or pseudo-real representations. 
For the case of the $Sp(N)$ gauge group, there is a global gauge anomaly, known as the Witten anomaly, that must vanish \cite{Witten:1982fp}. This condition imposes that the number of fundamental representations be even.

Secondly, we want the gauge theory to flow to a non-trivial fixed point in the infrared. This requires the one-loop beta function to be negative:
\begin{align}\label{eq:asympfree}
    b_0=\left(3h^\vee (G) -\sum_\chi T(\textbf{R}_\chi)\right)\geq 0 \,
\end{align}
Here, $h^\vee$ is the dual Coxeter number of the gauge group $G$, and $T(\textbf{R})$ is the Dynkin index of the representation $\textbf{R}$, defined as:
\begin{align}
    \Tr (T_\textbf{R}^a T_\textbf{R}^b) =T(\textbf{R})\d^{ab}\,.
\end{align}
The Dynkin index is normalized as $T=\frac{1}{2}$ for the fundamental representations. It is important to notice that the dual Coxeter number for our gauge groups is linear in $N$. Therefore, for a large $N$ theory to be asymptotically free, the matter fields must be in the representation whose Dynkin indices are either of order $N$ or of order 1. Therefore, only possible representations are the fundamental, adjoint, rank-2 symmetric, rank-2 anti-symmetric, and their complex conjugates, as the Dynkin index of higher-rank tensor representations grows faster than $N$. 
Furthermore, the number of such representations is also restricted. The number of rank-2 tensor representations should be of order one, while the number of fundamental representations can be at most of order $N$.

Asymptotic freedom does not guarantee that the gauge theory will flow to an interacting SCFT at the IR fixed point. One necessary condition is that it should have an (ABJ) anomaly-free $U(1)_R$ symmetry, which implies
\begin{align}\label{eq:ABJ}
    \Tr RGG=h^\vee+\sum_\chi (r_\chi-1)T(\textbf{R}_\chi)=0\,,
\end{align}
where $r_\chi$ denotes the $R$-charge for the chiral multiplet $\chi$. 
If there exists ABJ anomaly free $U(1)$ flavor symmetries $F_i$, we also have
\begin{align}
     \Tr F_iGG=\sum_\chi q_{i,\chi} T(\textbf{R}_\chi)=0,
\end{align}
where $q_{i,\chi}$ is the charge of the chiral matter $\chi$ under $F_i$. 
Upon RG flow, the flavor $F_i$'s can mix with the $R$-symmetry 
\begin{align}
    R=R_0+\sum_i\varepsilon_i F_i\,, 
\end{align}
with $\varepsilon_i$ being the parameters that need to be fixed. 
The superconformal $R$-symmetry in the IR SCFT is determined by the $a$-maximization procedure \cite{Intriligator:2003jj} that maximizes the trial $a$-central charge \cite{Anselmi:1997am}
\begin{align}
    a(R;\varepsilon_i)=\frac{3}{32}\left(3\Tr R^3 -\Tr R\right)\,,\quad \left.\frac{\partial a}{\partial\varepsilon_i}\right|_{\text{IR}}=0\,,\quad\left.\frac{\partial^2a}{\partial\varepsilon_i^2}\right|_{\text{IR}}<0\,.
\end{align}
This is computed by assuming there exists a superconformal fixed point in the IR. If there is no solution to this equation, then the IR physics cannot be described by a SCFT. 
If we do find a sensible solution, then we conjecture that the theory flows to a SCFT in the infrared. This procedure does not guarantee the existence of a SCFT. However, we do not find any argument against it either. 

However, the $a$-maximization procedure can sometimes fail. A typical situation we face is when certain (gauge-invariant) chiral operators have their $R$-charges less than $2/3$ upon $a$-maximization. The chiral operators satisfy the relation between their scaling dimensions $\Delta$ and their $R$-charges 
\begin{align}
    \Delta=\frac{3}{2}R\,. 
\end{align}
Thus, a chiral operator with $R<\frac{2}{3}$ violates the unitarity bound $\Delta\geq 1$, where the equality is saturated when the chiral operator is free. 
When such a phenomenon happens, it may imply that the IR theory is sick. However, another plausible scenario is that along the RG-flow, some operators hit the unitarity bound and decouple from the rest of the system. Therefore, we have an accidental symmetry that only acts on the decoupled free fields. In this case, we have to subtract the contribution from the decoupled fields in the $a$-maximization procedure \cite{Kutasov:2003iy}. 
A well-known prescription to incorporate the decoupled chirals introduces extra ``flip field" $X_{\CO}$ that couple to the decoupled chiral operator $\CO$ by the superpotential \cite{Barnes:2004jj, Benvenuti:2017lle, Maruyoshi:2018nod}
\begin{align}
    W=X_{\CO}\CO\,.
\end{align}
The flip field effectively provides a mass to the would-be decoupled operator $\CO$ and removes it from the spectrum in the IR. Whenever one finds chiral operators below the unitarity bound, one should introduce extra flip fields for those operators and reapply the $a$-maximization procedure. If no unitarity-violating operators remain after repeating the procedure, we stop. We then claim that there exists a unitary interacting SCFT in the IR. 

Quite often, we find that more than one operator violates the unitarity bound upon $a$-maximization. One should be careful not to remove the putative unitarity-violating operators all at once. The correct prescription is to first remove the operator with the lowest scaling dimension, or the one farthest from the unitarity bound. This can cause some of the operators that seem to decouple to remain coupled. This prescription is justified by the $a$-theorem. See \cite{Agarwal:2020pol} for the discussion.

Let us remark that even though there seems to be no unitarity-violating operators in the chiral ring, the putative IR theory can still suffer from unitarity violation and invalidate our analysis. This can happen if there is an extra emergent symmetry not visible at the level of Lagrangian that can be further mixed with $R$-symmetry. Sometimes such cases are visible through the analysis of the superconformal index \cite{Beem:2012yn, Evtikhiev:2017heo, Maruyoshi:2018nod, Kang:2022vab, Cho:2024civ}. In this paper, we do not perform such a refined analysis. However, we will sometimes utilize superconformal index to count the exactly marginal operators.  

\subsection{SU(N) theories}\label{sec:SUclass}
In this section, we classify $SU(N)$ gauge theories admitting large $N$ limits. The theories of our interest satisfy the conditions \eqref{eq:Ganom} and \eqref{eq:asympfree}. As we discussed, this restricts the allowed matter representations to be either rank-1 or rank-2 tensor representations. 
The allowed matter representations and their dimension, Dynkin index, and the anomaly coefficient of each representation are listed in Table \ref{tab:SUdata}.
{\renewcommand\arraystretch{1.5}
\begin{table}[H]
\centering
\begin{tabular}{|c|ccccccc|}\hline
$\mathbf{R}$ & \footnotesize $\fund$ & \footnotesize $\overline{\fund}$ & $\textbf{Adj}$ & \footnotesize $\sym$ & \footnotesize $\overline{\sym}$ & \footnotesize$\antisym$ & \footnotesize $\overline{\antisym}$\\\hline
dim$(\mathbf{R})$ & $N$ & $N$ & $N^2-1$ & $\frac{N(N+1)}{2}$ & $\frac{N(N+1)}{2}$ & $\frac{N(N-1)}{2}$ & 
$\frac{N(N-1)}{2}$\\
$T(\mathbf{R})$ & $\frac{1}{2}$ & $\frac{1}{2}$ & $N$ & $\frac{N+2}{2}$ & $\frac{N+2}{2}$ & $\frac{N-2}{2}$ & $\frac{N-2}{2}$\\
$\CA(\mathbf{R})$ & $1$ & $-1$ & $0$ & $N+4$ & $-(N+4)$ & $N-4$ & $-(N-4)$\\\hline
\end{tabular}
\caption{Matter representations in $SU(N)$ gauge group}
\label{tab:SUdata}
\end{table}}
Let us denote $N_{\mathbf{R}}$ to be the number of chiral multiplets in representation $\mathbf{R}$ under the gauge group. Then the gauge anomaly-free condition \eqref{eq:Ganom} and asymptotic freedom condition \eqref{eq:asympfree} are written as
\begin{equation} \label{eq:SUconditions}
\begin{gathered}
N\left(N_{\mbox{\tiny\sym}}-N_{\overline{\mbox{\tiny\sym}}}+N_{\mbox{\tiny\antisym}}-N_{\overline{\mbox{\tiny\antisym}}}\right)+\left(4N_{\mbox{\tiny\sym}}-4N_{\overline{\mbox{\tiny\sym}}}-4N_{\mbox{\tiny\antisym}}+4N_{\overline{\mbox{\tiny\antisym}}}+N_{\mbox{\tiny\fund}}-N_{\overline{\mbox{\tiny\fund}}}\right)=0\,,\\
N\left(N_{\textbf{Adj}}+\frac{N_{\mbox{\tiny\sym}}}{2}+\frac{N_{\overline{\mbox{\tiny\sym}}}}{2}+\frac{N_{\mbox{\tiny\antisym}}}{2}+\frac{N_{\overline{\mbox{\tiny\antisym}}}}{2}\right)+\left(N_{\mbox{\tiny\sym}}+N_{\overline{\mbox{\tiny\sym}}}-N_{\mbox{\tiny\antisym}}-N_{\overline{\mbox{\tiny\antisym}}}+\frac{N_{\mbox{\tiny\fund}}}{2}+\frac{N_{\overline{\mbox{\tiny\fund}}}}{2}\right)\leq 3N\,.
\end{gathered}
\end{equation}
The second condition restricts the number of rank-2 tensors as
\begin{align}
    N_{\text{rank-}2}\equiv N_{\textbf{Adj}}+\frac{N_{\mbox{\tiny\sym}}+N_{\overline{\mbox{\tiny\sym}}}+N_{\mbox{\tiny\antisym}}+N_{\overline{\mbox{\tiny\antisym}}}}{2}\leq 3\,.
\end{align}
On the other hand, the number of fundamental/anti-fundamental matters $N_{\mbox{\tiny\fund}}$, $N_{\overline{\mbox{\tiny\fund}}}$ can be of order $N$. 
The effective number of rank-2 tensor $N_{\text{rank-}2}$ is, in general, half-integer quantized. However, we see that if $N_{\text{rank-}2}$ is a half-integer, we must take the number of fundamentals/anti-fundamentals to be of order $N$ to cancel the gauge anomaly. 

When we scale the rank of the gauge group ($N$) and the flavor group ($N_f$) with a fixed ratio ($N_f/N=\alpha$), this is referred to as the Veneziano limit. 
In the current paper, we focus on the family of theories which admit the large $N$ limit \emph{without} taking the Veneziano limit, that is, when $N_{\text{rank-}2}$ is an integer. We will focus on the case where we can keep the flavor symmetry fixed as we take the large $N$ limit. This ensures that the putative holographic dual theory has a fixed gauge group in the bulk as we vary $N$. 

  {\renewcommand\arraystretch{1.4}
  \begin{table}[t]\small
    \centering
    \begin{tabular}{|c|c|c|c|}
      \hline
       Type & $SU(N)$ theory & NN-WGC & Conformal window  \\\hline \hline 
    
      \multirow{4}{*}{I} &  1 \textbf{Adj} + $N_f$  ( $\fund$ + $\overline{\fund}$ )&Always  & $1 \leq N_f \leq 2N^*$\\ \cline{2-4}
    
       & 1 $\sym$ + 1 $\overline{\sym}$ + $N_f$  ( $\fund$ + $\overline{\fund}$ ) & $\alpha\lesssim 1.3$  & $0\le N_f < 2N-2$\\\cline{2-4}
    
       & 1 $\antisym$ + 1 $\overline{\antisym}$ + $N_f$ ( $\fund$ + $\overline{\fund}$ ) & $\alpha\lesssim 0.3$  & $4\le N_f< 2N+2$ \\ \cline{2-4}
    
      &  1 $\sym$ + 1 $\overline{\antisym}$ + 8 $\overline{\fund}$ + $N_f$  ( $\fund$ + $\overline{\fund}$ ) &$\alpha\lesssim 1.3$ & $0\le N_f \leq 2N-4^*$  {\rule[-2ex]{0pt}{-3.0ex}} \\ \hline\hline
    
      \multirow{10}{*}{II} &2 $\sym$ + 2 $\overline{\sym}$ + $N_f$  ( $\fund$ + $\overline{\fund}$ ) & $\alpha\lesssim 0.6$ & $0\le N_f< N-4$  \\ \cline{2-4}
    
      & 1 $\sym$ + 2 $\overline{\sym}$ + 1 $\antisym$ + 8 $\fund$ + $N_f$  ( $\fund$ + $\overline{\fund}$ ) & $\alpha \lesssim 0.6$  & $0\le N_f< N-6$ \\ \cline{2-4}
    
      & 1 $\sym$ + 1 $\overline{\sym}$ + 1 $\antisym$ + 1 $\overline{\antisym}$ + $N_f$  ( $\fund$ + $\overline{\fund}$ ) &$\alpha\lesssim 0.6$  & $0\le N_f < N$ \\\cline{2-4}
    
      &  1 $\sym$ + 1 $\antisym$ + 2 $\overline{\antisym}$ + 8 $\overline{\fund}$ + $N_f$  ( $\fund$ + $\overline{\fund}$ ) &$\alpha \lesssim 0.9$  & $0\le N_f<N-2$ \\\cline{2-4}
    
       &2 $\sym$ + 2 $\overline{\antisym}$ + 16 $\overline{\fund}$ + $N_f$  ( $\fund$ + $\overline{\fund}$ ) & Always  & $0\leq N_f<N-8$ \\\cline{2-4}
    
      &  1 \textbf{Adj} + 1 $\sym$ + 1 $\overline{\sym}$ + $N_f$  ( $\fund$ + $\overline{\fund}$ ) &Always  & $0\le N_f \leq N-2$ \\\cline{2-4}
    
      & 2 $\antisym$ + 2 $\overline{\antisym}$ + $N_f$  ( $\fund$ + $\overline{\fund}$ ) & $\alpha\lesssim 0.6$  & $0\le N_f< N+4$\tablefootnote{Except for the case of $(N_f,N)=(0,4)$, which is equivalent to $SO(6)$ theory with four vectors and does not flow to an interacting SCFT.} \\\cline{2-4}
    
       & 1 \textbf{Adj} + 1 $\sym$ + 1 $\overline{\antisym}$ + 8 $\overline{\fund}$ + $N_f$  ( $\fund$ + $\overline{\fund}$ ) & Always  & $0\le N_f \leq N-4^*$ \\\cline{2-4}
    
      &  1 \textbf{Adj} + 1 $\antisym$ + 1 $\overline{\antisym}$ + $N_f$  ( $\fund$ + $\overline{\fund}$ )&$\alpha\lesssim 0.6$  & $0 \leq  N_f \leq N+2^*$ \\\cline{2-4}
    
       &2 \textbf{Adj} + $N_f$  ( $\fund$ + $\overline{\fund}$ )& Always  & $0\le N_f \leq N^*$ \\\hline\hline
    
      \multirow{6}{*}{III}&1 $\sym$ + 1 $\overline{\sym}$ + 2 $\antisym$ + 2 $\overline{\antisym}$ + $N_f$  ( $\fund$ + $\overline{\fund}$ ) & $N_f=0$  & $0 \le N_f < 2$ \\\cline{2-4}
    
       & 3 $\antisym$ + 3 $\overline{\antisym}$ + $N_f$  ( $\fund$ + $\overline{\fund}$ ) &$N_f=0,1,2$  & $0 \le N_f < 6$ \\\cline{2-4}
    
      & 1 \textbf{Adj} + 2 $\antisym$ + 2 $\overline{\antisym}$ + $N_f$  ( $\fund$ + $\overline{\fund}$ ) &$\begin{aligned}[c]&(N_f,N)\neq(1,5),\\&(2,6),(3,6),(3,7) \end{aligned}$   & $0\leq N_f\leq 4$ \\\cline{2-4}
    
     &  1 \textbf{Adj} + 1 $\sym$ + 1 $\overline{\sym}$ + 1 $\antisym$ + 1 $\overline{\antisym}$  &$N\neq 4$  & $*$ \\\cline{2-4}
    
     & 2 \textbf{Adj} + 1 $\antisym$ + 1 $\overline{\antisym}$ + $N_f$  ( $\fund$ + $\overline{\fund}$ ) & Always  & $0\leq N_f \leq 2^*$  \\\cline{2-4}
    
      &3 \textbf{Adj} & Always  & $*$  \\ \hline
    \end{tabular}
    \caption{List of all possible superconformal $SU(N)$ theories with large $N$ limit in Type I, II, III. Each theory satisfies the modified WGC. The third column lists the condition to satisfy the NN-WGC in the Veneziano limit (here $\alpha \equiv N_f/N$). The last column specifies the condition for each theory to flow to a superconformal fixed point. The entries with $*$ (if $N_f$ saturates the upper bound) do not flow but have non-trivial conformal manifolds \cite{Razamat:2020pra, Bhardwaj:2013qia}.\label{tab:SUlist}}
  \end{table}}

Under our assumption, we can simply write all possible solutions to \eqref{eq:Ganom}, \eqref{eq:asympfree}, and perform $a$-maximization to test the existence of the IR SCFT. This will allow us to enumerate the possible matter representations. The result is summarized in Table \ref{tab:SUlist}, and we will discuss each of them in great detail in Section \ref{sec:SU}. 
Under our assumption, we find that whenever $N_{\text{rank-}2}$ is an integer, the number of fundamentals/anti-fundamentals can be taken to be of order 1.\footnote{The conditions of \eqref{eq:SUconditions} may admit some cases with integer $N_{\text{rank-}2}$ and $O(N)$ amounts of fundamentals or anti-fundamentals. For example, $N_{\mbox{\tiny\sym}} = N_{\mbox{\tiny\antisym}} = 1$, $ N_{\overline{\mbox{\tiny\sym}}} = N_{\overline{\mbox{\tiny\antisym}}} = 0$ with $N_{\mbox{\tiny\fund}}=N_f$, $N_{\overline{\mbox{\tiny\fund}}} = 2N + N_f$ with $N_f = 0, 1, \cdots N$ satisfy the conditions.} 
The full list of superconformal fixed points should include the cases with the Veneziano limit. We present the list in Table \ref{table:type0}. The detailed aspects will appear in a future work.

In this subsection, we discuss universal properties shared among these SCFTs. Based on their properties, we find that they can be classified into three distinct types, which we refer to as Type I, Type II, and Type III.
Especially, we find that the $R$-charges of the chiral multiplets at the leading order in the large $N$ limit are solely determined by the total number of rank-2 tensors $N_{\text{rank-}2}$. This follows from the $a$-maximization procedure \cite{Intriligator:2003jj}. The IR $R$-symmetry maximizes the trial $a$-function \cite{Anselmi:1997am}
\begin{align}
a(\{r_\chi\})=\frac{3}{32}\left(2(N^2-1)+\sum_\chi N_\chi\text{dim}(\mathbf{R}_\chi)\left(3(r_\chi-1)^3-(r_\chi-1)\right)\right)\,,
\end{align}
under the anomaly-free condition of the $R$-symmetry
\begin{align}
    C(\{r_\chi\})=N+\sum_\chi N_\chi T(\mathbf{R}_\chi)(r_\chi-1)=0\,.
\end{align}
Then, $a$-maximization requires
\begin{align} \label{eq:amax}
    \frac{\partial a(\{r_\chi\})}{\partial r_i}=\frac{3}{32}N_i\text{dim}(\mathbf{R}_i)(9(r_i-1)^2-1)=\lambda\frac{\partial C(\{r_\chi\})}{\partial r_i}=\lambda N_iT(\mathbf{R}_i)\,,
\end{align}
where $\lambda$ is the Lagrange multiplier. Here, we emphasize that $N_i$ is canceled on each side. The only data that distinguishes the $R$-charge of each chiral multiplet is the ratio $\text{dim}(\mathbf{R})/T(\mathbf{R})$. Notice that rank-2 tensors share the same value of the ratio in the large $N$ limit
\begin{align}
    \left.\frac{\text{dim}(\mathbf{R})}{T(\mathbf{R})}\right|_{\text{rank-}2}=N+O(1)\,,
\end{align}
while the ratio is 
\begin{align}
    \left.\frac{\text{dim}(\mathbf{R})}{T(\mathbf{R})}\right|_{\tiny\fund}=2N+O(1)\,,
\end{align}
for the fundamental/anti-fundamentals. 
Therefore, at the leading order in $N$, the chiral matters in rank-2 tensor representations have the same $R$-charge  as
\begin{align}\label{eq:rrank2}
    r_{\text{rank-}2}=1-\sqrt{\frac{1}{9}+\frac{32\lambda} {27N}} + \cdots
\end{align}
Then, the leading order of the anomaly-free condition for the $R$-symmetry is dominated by the rank-2 tensor chiral multiplets
\begin{align}
    C(\{r_\chi\})=N(1+N_{\text{rank-}2}(r_{\text{rank-}2}-1))+O(1)=0\,,
\end{align}
while the fundamentals/antifundamentals contribute in $O(1)$ since their Dynkin indices are in that order.
In the end, the value of the Lagrange multiplier and the universal leading behavior of $R$-charge are given as 
\begin{equation}\label{eq:univR}
\begin{gathered}
    r_{\text{rank-}2}=1-\frac{1}{N_{\text{rank-}2}}\,+O(N^{-1})\,,\\
    \lambda=\frac{27N}{32}\left(\frac{1}{N_{\text{rank-}2}^2}-\frac{1}{9}\right)+O(1)\,,\\
    r_{\mbox{\tiny\fund}}=1-\sqrt{\frac{1}{18}+\frac{1}{2N_{\text{rank-}2}^2}}+O(N^{-1})\,.
    \end{gathered}
\end{equation}
We see that $R$-charges for the matter fields in the large $N$ limit only depend on the number of rank-2 tensor chiral multiplets $N_{\text{rank-}2}$! 

Let us remark that the universal $R$-charge in equation \eqref{eq:univR} is not valid when $N_{\mbox{\tiny\fund}} =  O(N)$, especially when $N_{\text{rank-}2}$ is a half-integer, since the previous computation assumes the number of fundamentals to be $O(1)$.

\paragraph{Type I ($\mathbf{N_{\text{rank-}2}=1})$}
When $N_{\text{rank-}2}=1$, the universal $R$-charges of the chiral multiplets are given according to the equation \eqref{eq:univR} as
\begin{equation}
\begin{gathered}
    r_{\text{rank-}2}=O(N^{-1})\,,\\
    r_{\mbox{\tiny\fund}}=\frac{3-\sqrt5}{3}+O(N^{-1})\,. 
    \end{gathered}
\end{equation}
We find that our numerical computation of individual SCFTs in this class yields R-charges very close to those above. However, the precise values can differ. This is because the $R$-charges of the chiral multiplets in rank-2 tensor representations are order $O\left(N^{-1}\right)$ for the Type I theories \cite{Agarwal:2019crm, Agarwal:2020pol}. As a result, the spectrum of the theory in this class contains a dense band of low-lying operators.   
The central charges $a$ and $c$ in this class grow linearly in $N$, and their ratio $a/c$ does not asymptote to 1 in general:
\begin{align}
  a = O(N) \ , \quad c= O(N) \ , \quad a \neq c   
\end{align}
This means that this is not a good holographic theory, not having a weakly-coupled Einstein gravity as an AdS dual. 

The $O(1/N)$ behavior of the rank-2 tensor makes some of the gauge-invariant operators violate the unitarity bound and become decoupled along the RG flow. Typically, the number of decoupled operators also scales as $O(N)$, which invalidates our naive computation. Especially, the spectrum of decoupled operators differs for each theory; the $R$-charges in Type I theories are not universal even in the large $N$ limit. 

However, once we take the double-scaling limit $N\gg N_f\gg 1$, the Type I theories exhibit universal $R$-charges. In the double-scaling limit of a Type I theory, symmetric and antisymmetric representations are indistinguishable in the $a$-maximization procedure, because the dimensions and the Dynkin indices of both are $N^2/2$ and $N/2$ in the leading order. Additionally, the fundamental and anti-fundamental chirals have identical $R$-charges, as the number of each is the same, $N_f$, in the leading order of the double-scaling limit. Then, the trial $a$-function can be written as
\begin{align} \label{eq:atrialTypeI}
    a=\frac{3}{32}\left(N^2(3(r_{\text{rank-}2}-1)^3-(r_{\text{rank-}2}-1))+2NN_f(3(r_{\mbox{\tiny\fund}}-1)^3-(r_{\mbox{\tiny\fund}}-1)\right)+a_{\text{flip}}\,.
\end{align}
Here, the last term $a_{\text{flip}}$ denotes the contribution of the flip fields for the decoupled operators, which breaks the universality we discussed above. 
The two variables $r_{\text{rank-}2}$ and $r_{\mbox{\tiny\fund}}$ are not independent, because they are related by the anomaly-free condition of the $R$-charge:
\begin{equation}\label{eq:rfund}    \begin{gathered}N+N(r_{\text{rank-}2}-1)+N_f(r_{\mbox{\tiny\fund}}-1)=0\,\\
    \Rightarrow r_{\mbox{\tiny\fund}}=1-\frac{N}{N_f}r_{\text{rank-}2}\,.
    \end{gathered}
\end{equation}
Since $r_{\mbox{\tiny\fund}}$ is of $O(1)$, the scaling behavior of $r_{\text{rank}-2}$ in the double-scaling limit is
\begin{align}\label{eq:rank2scale}
    r_{\text{rank-}2}\sim\frac{N_f}{N}\,.
\end{align}

Now, to estimate $a_{\text{flip}}$, we need to know the spectrum of decoupled operators. Though the spectrum of decoupled operators for each theory differs in detail, the light operators schematically take the common forms:
\begin{align} \label{eq:decSU}
\begin{split} 
    \Tr \,X^m \,,\quad q X^n q
\end{split}
\end{align}
where $X$ refers to any of the rank-2 tensor fields $\Phi$ (adjoint), $S$ (symmetric), $\widetilde{S}$ (symmetric conjugate), $A$ (anti-symmetric), $\widetilde{A}$ (anti-symmetric conjugate) and $q$ refers to either $Q$ (fundamental) or $\widetilde{Q}$ (anti-fundamental). For example, the light operators take the form $\Tr\,\Phi^{m_1}\,,\Tr(S\widetilde{S})^{m_2}$ or $Q\Phi^{n_1}\widetilde{Q}\,,Q\widetilde{A}(S\widetilde{A})^{n_2}Q$ and so on, depending on group theoretic property of the representation. The detailed form of the operator will be unimportant for our discussion. 

Suppose $\Tr X^{\mu N}$ and $q X^{\nu N} q$ are close but slightly above to the unitarity bound so that barely decoupled, i.e. $R[X^{\mu N}]\sim R[q X^{\nu N} q]\sim 2/3$. Operators of the form $\Tr X^m$ with $m < \mu N$ and $q X^n q$ with $n < \nu N$ are decoupled. Then, the constants $\mu, \nu$ take the form
\begin{align}\label{eq:munucond}
\begin{split}
        &\mu Nr_{\text{rank-}2}=\frac{2}{3}+O(N^{-1})\,,\\
        &\nu Nr_{\text{rank-}2}+2r_{\mbox{\tiny\fund}}=\left(\nu-\frac{2}{N_f}\right)Nr_{\text{rank-}2}+2=\frac{2}{3}+O(N^{-1})\,.
\end{split}
\end{align}
Then, $\mu$ must scale inversely proportional to $N_f$ 
\begin{align}\label{eq:muscale}
    \mu=\frac{2}{3Nr_{\text{rank-}2}}\sim N_f^{-1}\,.
\end{align}
Also, comparing the first and second lines of \eqref{eq:munucond}, we obtain the following relation.
\begin{align}\label{eq:munurel}
    -2\mu=\nu-\frac{2}{N_f}\,\Rightarrow\,\nu=\frac{2}{N_f}-2\mu\,.
\end{align}
The trial $a$-function depends on $\mu$, so that the $R$-charge that maximizes $a$ is also a function of $\mu$. In the end, the condition $\mu Nr_{\text{rank-}2}(\mu)=2/3$ determines $\mu$, which turns out to be universal in the double-scaling limit.

Let us move on to the detailed computation. We first consider $a_{\text{flip}}$.
\begin{align}
\begin{split}
    a_{\text{flip}}=\frac{3}{32}&\left( \sum_{m=d, 2d, \ldots, \mu N} \left(3(1-\,m\,r_{\text{rank-}2})^3-(1-m\,r_{\text{rank-}2})\right)\right.\\
     &\left.+N_f^2\sum_{n=0}^{\nu N}\left(3(1-n\, r_{\text{rank-}2}-2r_{\mbox{\tiny\fund}})^3-(1-n\, r_{\text{rank-}2}-2r_{\mbox{\tiny\fund}})\right)\right)\,.
\end{split}
\end{align}
The first line corresponds to the contribution by $\Tr X^m$ for $m=d, 2d, \cdots \mu N$ with $d$ being some degree to have a non-vanishing gauge-invariant operator. For example, we can have $\Tr (S \tilde{S})^n$ with $d=2$. 
The second line corresponds to the contribution by $q X^n q$. Note that the functional form of the second line is universal in the double-scaling limit of an arbitrary Type I theory. Next, we maximize $a$. Carefully collecting the most dominant terms in $\frac{\partial a}{\partial r_{\text{rank-}2}}$ using \eqref{eq:rfund}, \eqref{eq:rank2scale}, \eqref{eq:muscale}, \eqref{eq:munurel}, we obtain
\begin{align}\label{eq:Iamax}
        \frac{3}{32}N^2\left(10-36\frac{1-N_f^4\mu^4}{N_f^2}N^2r_{\text{rank-}2}^2+48\frac{1-N_f^3\mu^3}{N_f}Nr_{\text{rank-}2}-16(1-N_f^2\mu^2)\right)=O(N^2/N_f^2)\,.
\end{align}
We emphasize that the maximization condition \eqref{eq:Iamax} is universal in all Type I theories, because the model-dependent part containing $d$ in $a_{\text{flip}}$ is suppressed by $1/N_f^2$ after differentiation. It is simply because the operators of the form $q X^n q$ (`generalized mesons') overwhelms $\Tr X^m$ (`Coulomb branch operators') by the number of flavors in the large $N_f$ limit. 
Since the generalized mesons appear in all Type I theories, the whole $a$-maximization procedure becomes universal as well in the double-scaling limit. 
Solving the quadratic equation of $r_{\text{rank-}2}$, we obtain
\begin{align}\label{eq:rank2mu}
    r_{\text{rank-}2}=\frac{N_f}{N}\frac{4N_f^3\mu^3+\sqrt{26N_f^4\mu^4-32N_f^3\mu^3+8N_f^2\mu^2+3}-4}{2N_f^4\mu^4}\,.
\end{align}
Finally, we plug \eqref{eq:rank2mu} back in the relation \eqref{eq:munucond} to obtain the universal $R$-charges in the leading order of the double-scaling limit as
\begin{equation}\label{eq:univRds}
\begin{gathered}
    \mu=\frac{2(4+\sqrt{3})}{13N_f}\,,\\
    r_{\text{rank-}2}=\frac{(4-\sqrt{3})N_f}{3N}\,,\\
    r_{\mbox{\tiny\fund}}=\frac{\sqrt{3}-1}{3}\,.
\end{gathered}
\end{equation}
The asymptotic $R$-charge of fundamental matters is slightly lower than what is expected from \eqref{eq:univR}. In Section \ref{sec:SU}, we confirm the $R$-charges of fundamental matters in the Type I theory indeed approach the value of \eqref{eq:univRds}.

\paragraph{Type II ($\mathbf{N_{\text{rank-}2}=2}$)}
For the Type II theories ($N_{\text{rank-}2}=2$), we find the universal $R$-charges in the leading order to be
\begin{align}
    r_{\text{rank-}2}=\frac{1}{2}+O(N^{-1})\,,\quad r_{\mbox{\tiny\fund}}=\frac{12-\sqrt{26}}{12}+O(N^{-1})\,.
\end{align}
This is true if we scale the number of fundamentals $N_{\mbox{\tiny\fund}}$ to be finite and do not scale with $N$. Consequently, the spectrum of operators is sparse, i.e., the gap of the dimensions between the low-lying operators is of $O(1)$. The central charges of the theories in this class also show the universal behavior
\begin{align}
    a\sim c =\frac{27}{128}N^2+O(N)\,.
\end{align}
Note that when the $G$-gauge theory with two adjoint chiral multiplets has the central charge given as $a=c=\frac{27}{128} \text{dim}\,G$. This is also identical to that of mass-deformed $\CN=4$ SYM theory, integrating out one of the adjoint chirals. 
For a generic Type II theory, there will be $O(N)$ corrections to the central charges, and the difference $a-c$ is of $O(N)$. This term contributes to the higher-curvature correction to the bulk supergravity \cite{Anselmi:1998zb}. 

The Type II theories satisfy a number of conditions that are required to be holographic. Most obviously, we have $a=c$ at large $N$. Also, the low-energy spectrum is sparse in the large $N$ limit. In addition, the superconformal index exhibits a Hawking-Page-like phase transition, and the large $N$ saddle accounts for the black hole entropy in the AdS$_5$ \cite{Choi:2023tiq}. 
In many respects, Type II theories provide a generalization of Klebanov-Witten theory \cite{Klebanov:1998hh} (except that the KW theory has a non-simple gauge group); having superconformal $R$-charges $1/2$ for the chiral multiplets, non-trivial conformal manifolds (upon superpotential deformation), phases of superconformal indices, and so on. It would be interesting to find a holographic dual for other Type II theories. 

\paragraph{Type III ($\mathbf{N_{\text{rank-}2}=3}$)} 
For this class, the universal $R$-charges in the leading order are given as
\begin{align}
    r_{\text{rank-}2}\sim r_{\mbox{\tiny\fund}}=\frac{2}{3}+O(N^{-1})\,.
\end{align}
The leading behavior of the $R$-charges is identical to that of weakly-coupled conformal gauge theory, where the gauge coupling does not run \cite{Razamat:2020pra}. This includes the $\CN=4$ SYM theory. 

The dimensions of chiral operators are almost integer-quantized (since $r \simeq 2/3$ means $\Delta \simeq 1$). Also, the two central charges exhibit universal behavior in the large $N$ limit as
\begin{align}
    a\sim c=\frac{1}{4}N^2+O(N)\,.
\end{align}
This value is identical to that of $\CN=4$ SYM theory, whose central charge is given as $a=c=\frac{1}{4} \dim \, G$. Generally, Type III theories have $O(N)$ corrections to $c-a$, which contribute to the higher-curvature corrections to the bulk supergravity, as in the case of Type II. Type III theories also satisfy the same criterion to be holographic as Type II theories. The crucial difference is that the theory is connected to a relatively weakly-coupled fixed point, in the sense that the scaling dimensions of fields/operators are very close to the engineering dimensions. This means that the flow is `short.' 
We would like to remark that many known holographic theories that are dual to AdS$_5 \times$ SE$_5$ behave as such \cite{Martelli:2004wu, Benvenuti:2004dy} (for example $AdS_5 \times Y^{p, q}$ with $p=q$), except that the gauge groups are non-simple in those cases. 

\paragraph{Veneziano limit}
Before moving on to other gauge groups, let us comment on the Veneziano limit. 
The Veneziano limit is to take the number of the fundamentals $N_f$ to grow with $N$, when taking the large $N$ limit of the gauge group rank, while keeping the ratio fixed: 
\begin{align}
    N_f = \alpha N \,.
\end{align}
The anomaly-free constraint of the $R$-symmetry becomes
\begin{align}
    C(\{r_\chi\})=N(1+N_{\text{rank-}2}(r_{\text{rank-}2}-1))+\alpha N(r_{\mbox{\tiny\fund}}-1)=O(1)\,.
\end{align}
Here, we used the fact that the $R$-charge of fundamental and anti-fundamental chirals is the same. 
Using the $a$-maximization procedure (with the $R$ anomaly constraint) as in \eqref{eq:amax}, the $R$-charge of fundamental matters can be written as
\begin{align}
    r_{\mbox{\tiny\fund}}=1-\sqrt{\frac{1}{9}+\frac{16\lambda}{27N}}\,.
\end{align}
Then, we obtain the equation
\begin{align}\label{eq:vconst}
    1-N_{\text{rank-}2}\sqrt{\frac{1}{9}+\frac{32\lambda}{27N}}-\alpha\sqrt{\frac{1}{9}+\frac{16\lambda}{27N}}=O(N^{-1})\,.
\end{align}
Then the Lagrange multiplier $\lambda$ and the consequent $R$-charges turn out to be
\begin{align}\label{eq:univRvene}
\begin{split}
    \lambda&=3N\frac{3N_{\text{rank-2}}^2(6+\alpha^2)+9\alpha^2-\alpha^4-2N_{\text{rank-}2}^4-6N_{\text{rank-}2}\alpha\sqrt{18+2N_{\text{rank-}2}^2-\alpha^2}}{16(2N_{\text{rank-}2}^2-\alpha^2)^2}+O(1)\,,\\
    r_{\text{rank-}2}&=\frac{6N_{\text{rank-}2}^2-6N_{\text{rank-}2}-3\alpha^2+\alpha\sqrt{18+2N_{\text{rank-}2}^2-\alpha^2}}{3(2N_{\text{rank-}2}^2-\alpha^2)}+O(N^{-1})\,,\\
    r_{\mbox{\tiny\fund}}&=\frac{6N_{\text{rank-}2}^2+3\alpha-3\alpha^2-N_{\text{rank-}2}\sqrt{18+2N_{\text{Rank-}2}^2-\alpha^2}}{3(2N_{\text{rank-}2}^2-\alpha^2)}+O(N^{-1})\,.
    \end{split}
\end{align}
Here, among the two solutions for \eqref{eq:vconst}, we picked the one so that $R$-charges approach $\frac{2}{3}$ near the upper limit of the conformal window: 
\begin{align}
r_{\text{rank-}2}\,,r_{\mbox{\tiny\fund}}\Big|_{\alpha=3-N_{\text{rank-}2}}=\frac{2}{3}\,.
\end{align}
We see that the $R$-charges depend only on $N_{\text{rank-}2}$ and $\alpha = N_f/N$. 

For the Type I theories, \eqref{eq:univRvene} needs to be corrected because of the decoupled operators. To analyze the decoupled operators in the Veneziano limit, it is easier to start from the upper bound of the conformal window $\alpha=2$ and lower the number of flavors. When the number of flavors $N_f$ is large enough, the $R$-charge of rank-2 tensor matters scales as $O(1)$ in the Veneziano limit, so that no operator decouples. 

The first operator that can decouple is $\Tr X^2$ when the $R$-charge for the tensor reaches $r_{\text{rank-}2}=1/3$, which happens at
\begin{align}
    \alpha=\sqrt{\frac{2}{5}}\sim 0.632456\,.
\end{align}
As we lower the value of $\alpha$, the value of $r_{\text{rank-}2}$ gets lowered, and more operators of the form $\Tr X^n$ will decouple. The decoupling will correct the $R$-charges, but the correction will be of $O(1)$ and does not affect the leading order in large $N$, $N_f$.  

The first operator that decouples and can affect the leading order of $R$ is given by the form $q q'$, which has the multiplicity of order $N_f^2=\alpha^2N^2$. When it gets decoupled, we have
\begin{align}
   2r_{\mbox{\tiny\fund}}=\frac{2}{3}\quad\Rightarrow\quad \alpha=\frac{3-\sqrt 7}{2}\sim 0.177124.
\end{align}
After $qq'$ decouples, the $R$-charges are given as 
\begin{align}
\begin{split}
    r_{\text{rank-}2}&=\frac{\alpha(15\alpha-\sqrt{20-48\alpha+87\alpha^2-16\alpha^3})}{3(8\alpha+\alpha^2-2)}\,,\\
    r_{\mbox{\tiny\fund}}&=\frac{9\alpha+3\alpha^2-6+\sqrt{20-48\alpha+87\alpha^2-16\alpha^3}}{3(8\alpha+\alpha^2-2)}\,.
\end{split}
\end{align}
Treating $R$-charge as an order parameter, we can say there is a phase transition as we dial $\alpha$, since its behavior jumps. The transition observed across $\alpha\sim 0.177124$ is a second-order phase transition because the $R$-charges are continuous, but $\frac{d r}{d\alpha}$ jumps. 
The next phase transition occurs when the operator of the form $q X q'$ decouples. When it happens, we have
\begin{align}
    2r_{\mbox{\tiny\fund}}+r_{\text{rank-}2}=\frac{2}{3}\quad\Rightarrow\quad \alpha\sim0.100506\,.
\end{align}
After removing the decoupled operator, we get the $R$-charges as
\begin{align}
\begin{split}    r_{\text{rank-}2}=&\,\frac{\alpha\left(27\alpha-12\alpha^2+3\alpha^3-\sqrt{20-96\alpha+311\alpha^2-228\alpha^3+36\alpha^4+8\alpha^5+\alpha^6}\right)}{3(16\alpha-11\alpha^2+6\alpha^3-\alpha^4-2)}\,,\\
    r_{\mbox{\tiny\fund}}=&\,\frac{21\alpha-21\alpha^2+15\alpha^3-3\alpha^4+\sqrt{20-96\alpha+311\alpha^2-228\alpha^3+36\alpha^4+8\alpha^5+\alpha^6}-6}{3(16\alpha-11\alpha^2+6\alpha^3-\alpha^4-2)}\,.
\end{split}
\end{align}
Again, the transition at $\alpha\sim 0.100506$ is a second-order phase transition. A similar second-order phase transition occurs whenever the operator of the form $q X^n q'$ decouples. Concretely, we have 
\begin{align}
\begin{split}
    q X^2 q' :&\quad\alpha\sim 0.0703675\,,\\
    q X^3 q' :&\quad\alpha\sim 0.0541733\,,\\
    q X^4 q' :&\quad\alpha\sim 0.04405\,,\\
    &\quad\vdots\quad\,.
\end{split}
\end{align}
As we further lower the value of $\alpha$, more and more second-order phase transitions occur. In the end, the limit $\alpha\rightarrow 0$ is not well-defined in the Type I theories because these theories do not have universal $R$-charges in the generic large $N$ limit $N_f\ll N$.

\subsection{SO(N) theories}\label{sec:SOclass}
Similar to the previous subsection, we classify the $SO(N)$ gauge theories that admit the large $N$ limit and flow to interacting SCFTs. One crucial difference between the $SU(N)$ and the $SO(N)$ gauge group is that the $SO(N)$ is a real group. Consequently, any representation in the $SO(N)$ group is real, and the $SO(N)$ gauge theories are free of gauge anomaly, i.e., $A(\mathbf{R})=0$. 
Additionally, the adjoint representation is given by the rank-2 antisymmetric tensor, which reduces the number of possible cases. 
The allowed representations of the chiral multiplets in the large $N$ limit are vector ({\footnotesize{$\fund$}}), rank-2 symmetric ({\footnotesize $\sym$}), and rank-2 antisymmetric representations ({\footnotesize $\antisym$}). We do not include spinor representations in $Spin(N)$ as their degrees of freedom grow much faster than those of rank-2 tensors at large $N$. Some relevant group-theoretic properties of these representations are given in Table \ref{tab:SOdata}.
{\renewcommand\arraystretch{1.5}
\begin{table}[H] 
\centering
\begin{tabular}{|c|ccc|}\hline
$\mathbf{R}$ &  {\footnotesize $\fund$} & {\footnotesize $\sym$} & {\footnotesize $\antisym$} \\\hline
$\text{dim}(\mathbf{R})$ & $N$ & $\frac{N(N+1)}{2}-1$ & $\frac{N(N-1)}{2}$\\
$T(\mathbf{R})$ & $1$ & $N+2$ & $N-2$\\
\hline
\end{tabular}
\caption{Matter representations in $SO(N)$ gauge group}
\label{tab:SOdata}
\end{table}}

Asymptotic freedom constrains the number of matter multiplets to
\begin{align}
    N(N_{\mbox{\tiny\sym}}+N_{\mbox{\tiny\antisym}})+(2N_{\mbox{\tiny\sym}}-2N_{\mbox{\tiny\antisym}}+N_{\mbox{\tiny\fund}})\leq 3N-6\,.
\end{align}
We define the number of rank-2 tensors as
\begin{align}\label{eq:SO_N_rank2}
    N_{\text{rank-}2} \equiv N_{\mbox{\tiny\sym}}+N_{\mbox{\tiny\antisym}}\,,
\end{align}
whose maximum is bounded above by 3 from the asymptotic freedom.

Now we repeat the analysis as done in Section \ref{sec:SUclass}. The full list of classification is given in Table \ref{tab:SOlist}. Here, let us attempt to determine the universal $R$-charges in the large $N$ limit.  
The trial $a$-charge is given as
\begin{align}
    a(\{r_\chi\})=\frac{3}{32}\left(N(N-1)+\sum_\chi N_\chi\text{dim}(\mathbf{R}_\chi)(3(r_\chi-1)^3-(r_\chi-1)\right)\,,
\end{align}
under the $R$-anomaly constraint
\begin{align}
    C(\{r_\chi\})=N-2+\sum_\chi N_\chi T(\mathbf{R}_\chi)(r_\chi-1)=0\,.
\end{align}
Introducing the Lagrange multiplier $\lambda$, the $a$-maximization yields
\begin{align}
    \frac{3}{32}N_i\text{dim}(\mathbf{R}_i)(9(r_i-1)^2-1)=\lambda N_iT(\mathbf{R}_i)\,.
\end{align}
From this, we obtain
\begin{align}
    r_i=1-\sqrt{\frac{1}{9}+\frac{32\,\lambda T(\mathbf{R}_i)}{27\,\text{dim}(\mathbf{R}_i)}}\,.
\end{align}
In the large $N$ limit, all the rank-2 tensor representations have the same ratio of the dimension and the Dynkin index
\begin{align}
    \frac{T(\mathbf{R})}{\text{dim}(\mathbf{R})}\sim\frac{2}{N}\,.
\end{align}
Consequently, the rank-2 tensor matters have same $R$-charge $r_{\text{rank-}2}$ in the leading order in $N$
\begin{align}
    r_{\text{rank-}2}=1-\sqrt{\frac{1}{9}+\frac{64\lambda}{27N}}+O(N^{-1})\,.
\end{align}
We can then easily find the $R$-charge of the chiral multiplets in rank-2 tensor representations by directly solving the anomaly constraint
\begin{align}
    C(\{r_\chi\})= N(1+N_{\text{rank-}2}(r_{\text{rank-}2}-1)+O(1)=0\,,
\end{align}
to yield
\begin{equation}
\begin{gathered}
    r_{\text{rank-}2}=1-\frac{1}{N_{\text{rank-}2}}+O(N^{-1})\,,\\
    \lambda=\frac{27}{64}N\left(\frac{1}{N_{\text{rank-}2}^2}-\frac{1}{9}\right)+O(N^{-1})\,,\\
    r_{\mbox{\tiny\fund}}=1-\sqrt{\frac{1}{18}+\frac{1}{2N_{\text{rank-}2}^2}}+O(N^{-1})\,.
\end{gathered}
\end{equation}
We get the same universal leading order expressions for the $R$-charges of rank-2 tensors and vector (fundamental) matters as in the case of $SU(N)$ gauge theories. 

As in the case of $SU(N)$ gauge theories, the universal expression for the Type I theories is spoiled by the decoupled operators. At the end of the day, we find the same universal $R$-charges \eqref{eq:univRds} in the double-scaling limit $N\gg N_{\text{vec}}\gg1$, with $N_f$ being given by the number of $SO(N)$ vectors. 
One difference is that the number of rank-2 tensor matters \eqref{eq:SO_N_rank2} in $SO(N)$ gauge theories is integer-quantized. 

Since the $R$-charges of chiral multiplets share the same universal behavior as the $SU(N)$ gauge theories in the large $N$ limit, the theories of each type also share the same qualitative behaviors as those of the $SU(N)$ gauge theories of the same type, such as the density of the low-lying operators and asymptotic behavior of the central charges. 
For the Type I theories, some operators decouple from the rest of the theory, which spoils the above universality. The universality is retained in the double-scaling limit, as in the cases of $SU(N)$ theory. In fact, the same schematic form of operators with \eqref{eq:decSU} may decouple in $SO(N)$ gauge theory:
\begin{align}
\begin{split} 
    \Tr \,X^m \,,\quad Q X^n Q
\end{split}
\end{align}
where $X$ refers to any of the rank-2 tensor fields $S$ (symmetric) or $A$ (anti-symmetric). $Q$ refers to the vector (or fundamental) representation of $SO(N)$.  For example, the light operators take the form $\Tr\,S^{m}\,,\Tr A^{2m}$ or $Q S^{n} Q\,, Q A^{n} Q$, and the detailed form will depend on the group theoretic property of the representation. 
By exactly the same argument as for the case of $SU(N)$, the universal $R$-charges of vectors and rank-2 tensors turn out to be exactly the same as \eqref{eq:univRds} in the double-scaling limit.

\subsection{Sp(N) theories} \label{sec:Spclass}
Lastly, let us discuss the $Sp(N)$ gauge theories that flow to interacting SCFTs in the IR. 
Asymptotic freedom condition in the large $N$ limit allows only three representations for the chiral multiplets: fundamental ({\tiny$\fund$}), rank-2 symmetric ({\tiny$\sym$}), and rank-2 antisymmetric ({\tiny$\antisym$}).
There is no local gauge anomaly for the $Sp(N)$ gauge theories i.e., $A(\mathbf{R})=0$. However, $Sp(N)$ gauge theory can have a global gauge anomaly, often referred to as the Witten anomaly \cite{Witten:1982fp}. It restricts the number of fundamental chiral multiplets to be even.
The dimension and the Dynkin index of each representation are listed in Table \ref{tab:Spdata}.

{\renewcommand\arraystretch{1.5}
\begin{table}[H]
\centering
\begin{tabular}{|c|ccc|}\hline
$\mathbf{R}$ & \footnotesize $\fund$ & \footnotesize $\sym$  & \footnotesize$\antisym$ \\\hline
dim$(\mathbf{R})$ & $2N$ & $N(2N+1)$ & $N(2N-1)-1$ \\
$T(\mathbf{R})$ & $\frac{1}{2}$ & $N+1$ & $N-1$ \\\hline
\end{tabular}
\caption{Matter representations in $Sp(N)$ gauge group}
\label{tab:Spdata}
\end{table}
We repeat the same analysis as done in Section \ref{sec:SUclass} and \ref{sec:SOclass}. From the asymptotic freedom, we have
\begin{align}
   N\left(N_{\mbox{\tiny\sym}} + N_{\mbox{\tiny\antisym}}\right) + \left(N_{\mbox{\tiny\sym}} - N_{\mbox{\tiny\antisym}} + \frac{1}{2}N_{\mbox{\tiny\fund}}\right)\leq 3N+3\,.
\end{align}
We define the number of rank-2 tensor matters $N_{\text{rank-}2}$ as
\begin{align}
    N_{\text{rank-}2} \equiv N_{\mbox{\tiny\sym}} + N_{\mbox{\tiny\antisym}}\leq 3\,,
\end{align}
which is integer-valued.
Next, we perform the $a$-maximization to obtain the IR $R$-charges. The trial $a$-charge is written as
\begin{align}
    a(\{r_\chi\})=\frac{3}{32}\left(2N(2N+1)+\sum_\chi N_\chi(3(r_\chi-1)^3-(r_\chi-1))\right) \ , 
\end{align}
under the anomaly-free constraint for the $R$-symmetry
\begin{align}
    C(\{r_\chi\})=N+1+\sum_\chi N_\chi T(\mathbf{R}_\chi)(r_\chi-1)=0\,.
\end{align}
Using the Lagrange multiplier $\lambda$, as we did for the $SU$ and $SO$ theories, we obtain the universal $R$-charges in the large $N$ limit as
\begin{align}
\begin{split}
    r_{\text{rank-}2}=1-\frac{1}{N_{\text{rank-}2}}+O(N^{-1})\,,\\
    \lambda=\frac{27N}{16}\left(\frac{1}{N_{\text{rank-}2}^2}-\frac{1}{9}\right)+O(N^{-1})\,,\\
    r_{\mbox{\tiny\fund}}=1-\sqrt{\frac{1}{18}+\frac{1}{2N_{\text{rank-}2}^2}}+O(N^{-1})\,.
\end{split}
\end{align}
We find that the asymptotic $R$-charges of chiral multiplets are the same as those of $SU(N)$ and $SO(N)$ gauge theories. This justifies our trichotomy of Type I, II, and III. As before, the formula for the $N_{\text{rank-}2}$ has to be corrected due to the decoupled operators. 

The full list of theories we find is given in Table \ref{tab:Splist}.

\subsection{Testing AdS Weak Gravity Conjecture} \label{sec:testWGC}

The Weak Gravity Conjecture (WGC) is one of the well-known Swampland conjectures, which constrains the charged matter spectrum in a quantum theory of gravity \cite{Arkani-Hamed:2006emk}. This conjecture states that, in any consistent theory of gravity, there must be a particle with a mass smaller than its charge. This is motivated by the condition that an extremal black hole can decay without leaving remnants. See \cite{Harlow:2022ich}, for a more comprehensive review on the WGC. 

We test the Weak Gravity Conjecture in AdS space against our list of theories, utilizing the AdS/CFT correspondence. The AdS$_5$/CFT$_4$ version of the WGC was first formulated by Nakayama and Nomura \cite{Nakayama:2015hga}, which we call NN-WGC. 
The NN-WGC states that in a four-dimensional conformal field theory with a single $U(1)$ flavor symmetry, there should exist an operator whose scaling dimension $\D$ is smaller than its $U(1)$ charge $q$:
\begin{align}
    \frac{q^2}{\D^2}\geq \frac{k_F}{12c}\,.\label{eq:cftwgc}
\end{align}
Here, the central charges $k_F$ and $c$ are found in the coefficients of the two-point functions of the conserved current $J_\m$ and the energy-momentum tensor $T_{\m\n}$, respectively:
\begin{align}
    \vev{J_\m(x) J_\n(0)}=\frac{3}{4\pi^4}\frac{k_F}{x^6}I_{\m\n}(x),\quad \vev{T_{\m\n}(x)T_{\r\s}(0)}=\frac{40}{\pi^4}\frac{c}{x^8}I_{\m\n,\r\s}(x)\,,
\end{align}
where $I_{\m\n}(x)=\d_{\m\n}-x_{\m\n}/x^2$ and $I_{\m\n,\r\s}(x)=(I_{\m\r}(x)I_{\n\s}(x)+I_{\m\s}(x)I_{\n\r}(x))/2-\d_{\m\n}\d_{\r\s}/4$.
The NN-WGC (\ref{eq:cftwgc}) is based on a specific black hole solution, the extremal AdS$_5$ Reissner-Nordstr\"om black hole, and it assumes that the scaling dimension of an operator $\D\simeq Lm$ where $L$ is the AdS radius and $m$ is the mass of the particle in the bulk such that $Lm\gg1$. 

However, it is intriguing that the theories with a dense spectrum (Type I) and with a fixed small $N_f$, which are far from being holographic, still satisfy the NN-WGC \cite{Agarwal:2020pol, Agarwal:2019crm}. In fact, it was already observed in \cite{Nakayama:2015hga} that many of the non-holographic theories indeed satisfy the conjecture. However, it was also found in these papers that some non-holographic theories do not satisfy the NN-WGC. These facts motivate us to test the NN-WGC further and see if the conjecture can be `improved' to become a universal property of any 4d SCFT, as was the case of WGC in 2d CFT \cite{Montero:2016tif, Benjamin:2016fhe, Bae:2018qym}. 

To this end, we further extend the survey across the Type I, II, and III theories. Moreover, to test against `non-holographic' theories, we also investigate the Veneziano limit of these theories, which is to take the large $N$ limit with a fixed $N_f/N$. 
We find that some theories fail to meet the NN-WGC condition \eqref{eq:cftwgc}. 

It is very interesting to find that WGC holds for theories that are very far from being holographic. This fact suggests that there might be a version of WGC that holds universally for all consistent CFTs. For the case of AdS$_3$/CFT$_2$, the universality eventually originated from the modular invariance of the partition function. This modular invariance also gives rise to Cardy's formula \cite{Cardy:1986ie} for the asymptotic density of states at high-energy in terms of the central charge $c$. Unfortunately, the modular invariance in higher-dimensional CFT is much less understood and does not seem to give useful constraints on the CFT data. 
Instead, it was found that the Cardy-like limit of the superconformal index exhibits a universal property similar to Cardy's formula \cite{Kim:2019yrz, Cabo-Bizet:2019osg}. Motivated by this, we propose a modified version of WGC
\begin{align}
    \frac{q^2}{\D^2}\geq \frac{k_F}{12(3c-2a)}\,,\label{eq:cftwgc2}
\end{align}
based on the supersymmetric Cardy formula. No counterexamples have been found for this modified WGC, and some examples even saturate this bound (\ref{eq:cftwgc2}). This modification, along with a few representative examples, has been reported in \cite{Cho:2023koe}. In this paper, we provide a comprehensive result.

\begin{figure}[t]
\centering
\begin{subfigure}[b]{0.35\linewidth}
\includegraphics[width=\linewidth]{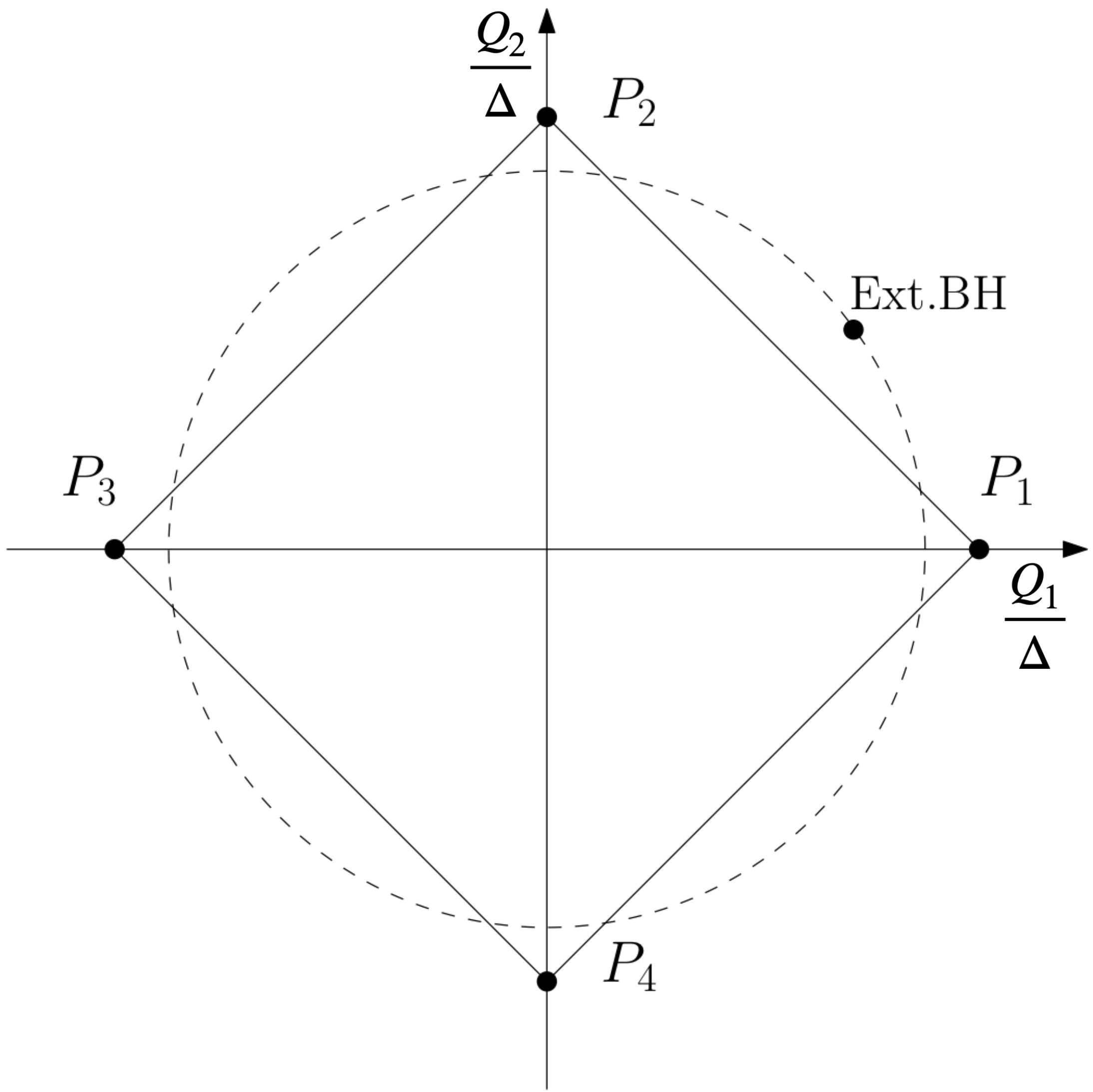}
\subcaption{Inconsistent with WGC}
\end{subfigure}
\hspace{20mm}
\begin{subfigure}[b]{0.35\linewidth}
\includegraphics[width=\linewidth]{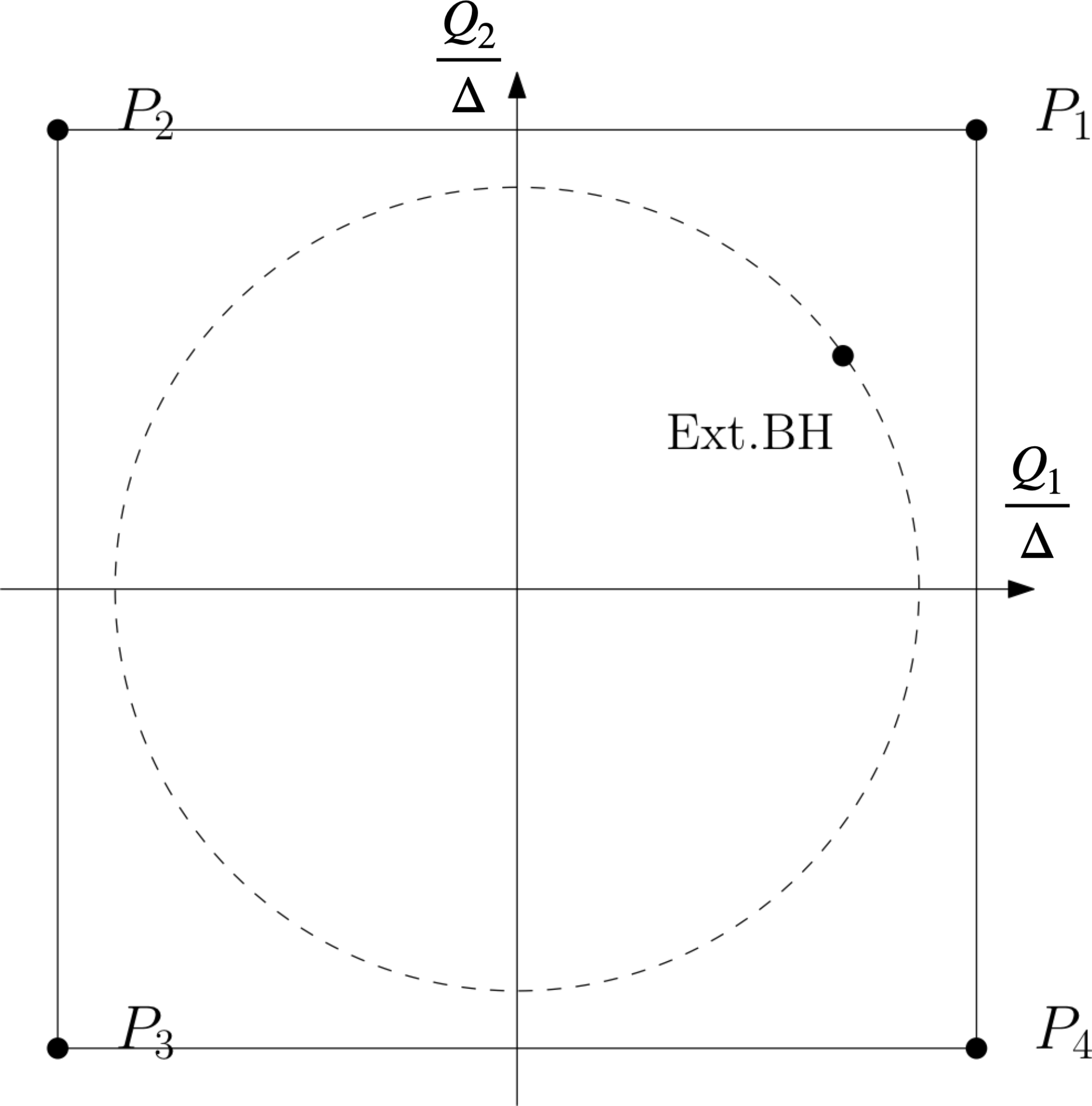}
\subcaption{Consistent with WGC}
\end{subfigure}
\caption{The charge-to-dimension ratio space with two flavor charges $U(1)_1$ and $U(1)_2$. The normalization of the ratio is set by the ratio of an extremal black hole lying on the dotted unit circle. Solid lines represent the convex hull made by gauge-invariant operators $O_1, \dots, O_4$. (a) An extremal black hole outside the convex hull cannot decay by emitting charged particles. (b) When the convex hull encloses the unit ball, any extremal black holes can decay by emitting particles.}
\label{fig:cvh_gen}
\end{figure}

Our theories in general have multiple $U(1)$ flavor symmetries. In such cases, we need to consider the convex hull condition, which is a refinement of WGC when multiple $U(1)$ charges are present \cite{Cheung:2014vva}. For simplicity, let us focus on the scenario where we have two $U(1)$ flavor symmetries $U(1)_1$ and $U(1)_2$, that are orthogonal in the charge-to-dimension ratio space. The term `orthogonal' is defined with respect to
\begin{align}
    \Tr RF_1F_2=\sum_\chi (r_\chi-1)q_1(\chi)q_2(\chi)|\textbf{R}_\chi|=0\,,
\end{align}
where $q_i(\chi)$ denotes the flavor charge of $\chi$. The charge-to-dimension ratios of gauge-invariant operators can be visualized in Figure \ref{fig:cvh_gen}.
Consider the chiral ring operators $O_1$ and $O_2$; the charge-to-dimension ratio of multiparticle states composed of these operators forms a convex hull. This is due to the relation $\D_{O_1O_2}=\D_{O_1}+\D_{O_2}$, leading to the charge-to-dimension ratios aligning on the line segment connecting those of $O_1$ and $O_2$. If the ratio of an extremal black hole is outside the convex hull, this implies there is no state (particle) that an extremal black hole can emit to decay. Therefore, to be consistent with the WGC, the convex hull must look like Figure \ref{fig:cvh_gen} (b), enclosing the unit circle completely. This is known as the convex hull condition, which is more stringent than imposing WGC for each $U(1)$'s. 
In general, there may be $n$ $U(1)$ flavor symmetries. Then, we construct the convex hull in the $n$-dimensional charge-to-dimension ratio space. By determining the minimum distance from the origin to the convex hull, we can check whether the theory satisfies the WGC or not.

\section{Superconformal SU(N) gauge theories with large N limit}
\label{sec:SU}

  In this section, we analyze all possible supersymmetric $SU(N)$ gauge theories that flow to SCFTs and can have a large $N$ limit. As mentioned in Section \ref{sec:classification}, only a small number of representations are allowed. See Table \ref{tab:SUindex}.
  The Young diagrams $\fund\,$, $\sym\,$, and $\antisym\,$ denote the fundamental, rank-2 symmetric, and rank-2 anti-symmetric representations, respectively. In addition, there are complex conjugates of those representations, $\widetilde{Q}(\,\overline{\fund}\,)$, $\widetilde{S}(\,\overline{\sym}\,)$, and $\widetilde{A}(\,\overline{\antisym}\,)$.

{\renewcommand\arraystretch{1.4}
  \begin{table}[t]
    \centering
    \begin{tabular}{|c||c|c|c|c|}
    \hline
    Fields & $SU(N)$ & $|\textbf{R}|$  & $T(\textbf{R})$ & $\mathcal{A}(\textbf{R})$\\\hline

    $Q$ &$\fund$ & $N$ & $\frac{1}{2}$ & 1\\

    $\Phi$ & \textbf{Adj} & $N^2-1$ & $N$ & 0\\
     
    $S$ &$\sym$  & $\frac{1}{2}N(N+1)$ & $\frac{1}{2}(N+2)$ & $N+4$\\
    
    $A$ & $\antisym$  & $\frac{1}{2}N(N-1)$ & $\frac{1}{2}(N-2)$ & $N-4$\\\hline
    \end{tabular}
    \caption{The dimensions, Dynkin indices, and anomaly coefficients of allowed irreducible representations of $SU(N)$. We omit the complex conjugates. \label{tab:SUindex}}
  \end{table}}

  The gauge anomaly-free condition is expressed as:
  \begin{align}
    \sum_\chi\mathcal{A}(\textbf{R}_\chi)=(N_Q - N_{\widetilde{Q}})+(N+4)(N_S-N_{\widetilde{S}})+(N-4)(N_A-N_{\widetilde{A}})=0\,.
  \end{align}
  The asymptotic freedom conditions are given by:
  \begin{align}
    b_0=3N-\frac{1}{2}(N_Q+N_{\widetilde{Q}})-N\times N_\textbf{Adj}-\frac{N+2}{2}(N_S+N_{\widetilde{S}})-\frac{N-2}{2}(N_A+N_{\widetilde{A}})\geq0\,.
  \end{align}
  These two conditions constrain the set of admissible theories to a finite number of families. The result is listed in Table \ref{tab:SUlist}. 
  We also summarize the range of the $a/c$ ratio and the condition for the existence of a non-trivial conformal manifold in Table \ref{tab:SUratio}.

  {\renewcommand\arraystretch{1.4}
  \begin{table}[t]\footnotesize
    \centering
    \begin{tabular}{|c|c|c|c|}
      \hline
     Type & $SU(N)$ theory &  Range of $a/c$ & \makecell{Conformal\\manifold}\\\hline \hline 
      
     \multirow{4}{*}{I} & 1 \textbf{Adj} + $N_f$  ( $\fund$ + $\overline{\fund}$ )& $\begin{aligned}
          0.8214 &\simeq {\scriptstyle\frac{23}{28}}\\&\leq a/c \lesssim 0.9948
      \end{aligned}$  & \makecell{$N_f>1\,,$ \\$2N/N_f \in\IZ$} \\ \cline{2-4}
      
      &1 $\sym$ + 1 $\overline{\sym}$ + $N_f$  ( $\fund$ + $\overline{\fund}$ ) & $\begin{aligned}
          0.875={\scriptstyle\frac{7}{8}}&<a/c\\&\leq {\scriptstyle\frac{3553}{3353}}\simeq1.0597
      \end{aligned}$  & \makecell{$N_f=0\,,$\\$N \in 2\IZ$}\\\cline{2-4}
      
& 1 $\antisym$ + 1 $\overline{\antisym}$ + $N_f$   ( $\fund$ + $\overline{\fund}$ ) & {$\begin{aligned}[c]
          0.6930&\simeq {\scriptstyle\frac{70169619-220676\sqrt{3201}}{83242974}}\\
          &\leq {a/c} \lesssim 0.9159
      \end{aligned}$}  & \makecell{$N_f=4\,,$\\$N \in 2\IZ$} \\ \cline{2-4}
      
      &1 $\sym$ + 1 $\overline{\antisym}$ + 8 $\overline{\fund}$ + $N_f$  ( $\fund$ + $\overline{\fund}$ ) & $0.86=\frac{43}{50}\leq a/c\lesssim0.9196$ & \makecell{$N_f=0\,,$\\$N-2\in 4\IZ$ or\\$N_f=2N-4$}  {\rule[-2ex]{0pt}{-3.0ex}} \\ 
      \hline\hline
      
     \multirow{10}{*}{II} & 2 $\sym$ + 2 $\overline{\sym}$ + $N_f$  ( $\fund$ + $\overline{\fund}$ ) & $\begin{aligned}
          0.9286&\simeq {\scriptstyle\frac{13}{14}}<a/c\\&\leq {\scriptstyle\frac{14001}{13609}}\simeq1.0288
      \end{aligned}$ & $*$  \\ \cline{2-4}
      
&      1 $\sym$ + 2 $\overline{\sym}$ + 1 $\antisym$ + 8 $\fund$ + $N_f$  ( $\fund$ + $\overline{\fund}$ ) & $0.9286\simeq \frac{13}{14}<a/c<1$  & $*$ \\ \cline{2-4}
      
&      1 $\sym$ + 1 $\overline{\sym}$ + 1 $\antisym$ + 1 $\overline{\antisym}$ + $N_f$  ( $\fund$ + $\overline{\fund}$ )& $0.9251\lesssim a/c<1$  & $*$ \\\cline{2-4}
      
&      1 $\sym$ + 1 $\antisym$ + 2 $\overline{\antisym}$ + 8 $\overline{\fund}$ + $N_f$  ( $\fund$ + $\overline{\fund}$ ) & $\begin{aligned}
          0.8520&\simeq {\scriptstyle\frac{660981-6308\sqrt{321}}{641602}}\\&<a/c<1
      \end{aligned}$  & $*$ \\\cline{2-4}
      
&      2 $\sym$ + 2 $\overline{\antisym}$ + 16 $\overline{\fund}$ + $N_f$  ( \fund + $\overline{\fund}$ ) & $0.9286=\frac{13}{14}<a/c<1$  & $*$ \\\cline{2-4}
      
&      1 \textbf{Adj} + 1 $\sym$ + 1 $\overline{\sym}$ + $N_f$  ( $\fund$ + $\overline{\fund}$ ) & $\begin{aligned}
          0.9286&\simeq{\scriptstyle\frac{13}{14}}<a/c\\
          &\leq {\scriptstyle\frac{161399+81\sqrt{2641}}{162580}}\simeq1.0183
      \end{aligned}$  & $N_f=N-2$ \\\cline{2-4}
      
&      2 $\antisym$ + 2 $\overline{\antisym}$ + $N_f$  ( \fund + $\overline{\fund}$ ) & $\begin{aligned}
          0.7901&\simeq{\scriptstyle\frac{29426941-79632\sqrt{1765}}{33008942}}\\
          &\leq a/c<1
      \end{aligned}$  & $*$\\\cline{2-4}
      
&      1 \textbf{Adj} + 1 $\sym$ + 1 $\overline{\antisym}$ + 8 $\overline{\fund}$ + $N_f$  ( $\fund$ + $\overline{\fund}$ ) &  $0.9167\simeq \frac{11}{12}\leq a/c<1$  & $N_f=N-4$ \\\cline{2-4}
      
&      1 \textbf{Adj} + 1 $\antisym$ + 1 $\overline{\antisym}$ + $N_f$  ( $\fund$ + $\overline{\fund}$ ) & $0.875= \frac{7}{8} \leq  a/c <1$ & $N_f=N+2$ \\\cline{2-4}
      
&      2 \textbf{Adj} + $N_f$  ( $\fund$ + $\overline{\fund}$ ) & $0.8913\simeq \frac{41}{46}\leq a/c\leq 1$ & $N_f=0,N$ \\\hline\hline
      
 \multirow{6}{*}{III} &    1 $\sym$ + 1 $\overline{\sym}$ + 2 $\antisym$ + 2 $\overline{\antisym}$ + $N_f$ ( $\fund$ + $\overline{\fund}$ ) & $0.9430\lesssim a/c <1$ & $*$ \\\cline{2-4}
      
&      3 $\antisym$ + 3 $\overline{\antisym}$ + $N_f$  ( $\fund$ + $\overline{\fund}$ ) & $\begin{aligned}
          0.8203&\simeq {\scriptstyle\frac{2835841+8216\sqrt{1869}}{3890246}}\\&\leq a/c <1
      \end{aligned}$ & $*$ \\\cline{2-4}
      
&      1 \textbf{Adj} + 2 $\antisym$ + 2 $\overline{\antisym}$ + $N_f$  ( $\fund$ + $\overline{\fund}$ ) & $0.8879\simeq \frac{103}{116} \leq a/c < 1$ & $N_f=4$ \\\cline{2-4}
      
&      1 \textbf{Adj} + 1 $\sym$ + 1 $\overline{\sym}$ + 1 $\antisym$ + 1 $\overline{\antisym}$ & $0.9891\simeq \frac{91}{92}\leq a/c <1$ & Always \\\cline{2-4}
      
&      2 \textbf{Adj} + 1 $\antisym$ + 1 $\overline{\antisym}$ + $N_f$  ( $\fund$ + $\overline{\fund}$ ) & $0.9369\simeq\frac{193}{206}\leq a/c<1$ & $N_f=2$  \\\cline{2-4}
      
&      3 \textbf{Adj} & $a/c=1$  & Always  \\ \hline
    \end{tabular}
    \caption{The third column lists the range of the ratio $a/c$ for general $N$ and $N_f$ within the conformal window. The last column denotes the condition for the theory to have a non-trivial conformal manifold. The entries with $*$ do not have a non-trivial conformal manifold, in the absence of a superpotential. We observe that a conformal manifold emerges once a suitable superpotential is turned on.}
    \label{tab:SUratio}
  \end{table}}

  In general, the $R$-charges of each chiral multiplet are irrational, determined by the $a$-maximization. Consequently, the $R$-charge of a generic gauge-invariant chiral operator has little chance of being marginal. Nevertheless, there might be certain gauge-invariant operators whose $R$-charge is exactly two, guaranteed by the anomaly-free condition of the $R$-symmetry \eqref{eq:ABJ}. 
  \begin{align}
    h^\vee_G+\sum_{\chi}(r_\chi-1)T(\mathbf{R}_\chi)N_\chi=0\quad\Longleftrightarrow\quad \sum_\chi\frac{2N_\chi T(\mathbf{R}_\chi)}{\sum_{\tilde{\chi}} N_{\tilde{\chi}} T(\mathbf{R}_{\tilde{\chi}})-h^\vee_G}r_\chi=2\,.
  \end{align}
  The above relation implies that an operator of the form
  \begin{align}\label{eq:marginal from anomaly-free}
    \mathcal{O}=\left.\prod_\chi \chi^{p_\chi}\right|_{\text{singlet}}\,,
  \end{align}
  is marginal if
  \begin{align}
    p_\chi\equiv \frac{2N_\chi T(\mathbf{R}_\chi)}{\sum_{\tilde{\chi}} N_{\tilde{\chi}} T(\mathbf{R}_{\tilde{\chi}})-h^\vee_G}\in \IZ \quad\text{for all }\chi\,\text{ appear in } \mathcal{O} \ . 
  \end{align}
  From this, we can extract a condition for a theory to contain a marginal operator of this form:
  \begin{align}
  \label{eq:margcond}
    2N_\chi T(\mathbf{R}_\chi)\geq \sum_{\tilde{\chi}} N_\chi T(\mathbf{R}_{\tilde{\chi}})-h^\vee_G\quad\text{for any }\chi\,.
  \end{align}
  For example, this means that any Type II or III theory containing exactly one chiral multiplet in $\antisym$ or $\overline{\antisym}$ representation, whose spectrum is sparse ($\sum_\chi N_\chi T(\mathbf{R}_\chi)\sim 2h^\vee_G, 3h^\vee_G$), cannot have a marginal operator of this kind. 
  Some of our theories do have marginal operators of this kind, whose condition is summarized in Table \ref{tab:SUratio}. 
  The conformal manifold does not appear frequently in our theories. However, we discover that, once the theory is properly deformed by a sequence of relevant operators, we do find non-trivial conformal manifolds to emerge. 

In this section, we describe each family of theories (parametrized by the number of fundamental/anti-fundamental pairs $N_f$). We identify its $R$-symmetry and central charges, as well as the relevant/marginal operator spectrum. We find the conditions on $N_f$ for the theory to be an interacting SCFT. We also test two versions of the AdS Weak Gravity Conjecture and find that NN-WGC is sometimes violated, whereas our modified WGC is always satisfied, even for a finite $N$.

\subsection{\texorpdfstring{1 \textbf{Adj} + $N_f$  ( $\fund$ + $\overline{\fund}$ )}{1 Adj + Nf ( Q + Qt )}\label{sec:adjSQCD}}

\paragraph{Matter content and symmetry charges}
  Consider an $SU(N)$ gauge theory with an adjoint and $N_f$ pairs of fundamental and anti-fundamental chiral multiplets. The matter fields and their $U(1)$ global charges are listed in Table \ref{tab:adj1}. This theory for a small, fixed $N_f$ is already studied in \cite{Agarwal:2020pol}. 

  {\renewcommand\arraystretch{1.4}
  \begin{table}[h]
    \centering
    \begin{tabular}{|c|c||c|c|c|c|}
    \hline
    \# & Fields & $SU(N)$ & $U(1)_1$ & $U(1)_2$ & $U(1)_R$ \\\hline
     $N_f$ & $Q$ & $\fund$ &  $N$ & 1 & $R_Q$ \\
     $N_f$ & $\widetilde{Q}$ & $\overline{\fund}$  &  $N$ & -1 & $R_{\widetilde{Q}}$ \\
     1 & $\Phi$ & $\textbf{Adj}$ &  $-N_f$ & 0 & $R_\Phi$  \\\hline
    \end{tabular}
    \caption{The matter contents and their corresponding charges in $SU(N)$ gauge theory with 1 adjoint and $N_f$ fundamental/anti-fundamental pairs.\label{tab:adj1}}
  \end{table}}

\paragraph{Gauge-invariant operators}
  Let $I$ and $J$ denote the flavor indices for $Q$. We present a set of single-trace gauge-invariant operators in schematic form:
  \begin{enumerate}
      \item $\Tr \Phi^n\,,\quad n=2,3,\dots, N\,.$
      \item $Q_I\Phi^n \widetilde{Q}_{\tilde{I}}\,,\quad 0,1,\dots,N-1\,.$
      \item $\e\,\CQ_{I_1}^{n_1}\cdots \CQ_{I_N}^{n_N}\,,\quad \e\,\widetilde{\CQ}_{\tilde{I}_1}^{n_1}\cdots\widetilde{\CQ}_{\tilde{I}_N}^{n_N}\,.$
  \end{enumerate}
  We omit various gauge and flavor indices that are contracted. 
  We define the `dressed quark' as $\CQ_I^n=\Phi^n Q_I$ and $\widetilde{\CQ}_{\tilde{I}}^n = \Phi^n\widetilde{Q}_{\tilde{I}}$ to simplify the expression.

\paragraph{$R$-charges and central charges}
  We perform the $a$-maximization to compute the $R$-charges of the matter fields and central charges. In the large $N$ limit with a fixed $N_f$, they are given by
  \begin{align}
      \begin{split}\label{eq:adj1 charges}
          R_Q=R_{\widetilde{Q}}&\sim \frac{3-\sqrt{5}}{3} + \frac{N_f}{2N} + O(N^{-2})\,,\\
          R_\Phi &\sim \frac{\sqrt{5}N_f}{3N} -\frac{N_f^2}{2N^2}+ O(N^{-2})\,,\\
          a & \sim \frac{5\sqrt{5}}{24} N_f N -\frac{15 N_f^2}{32} + O(N^{-1})\,,\\
          c & \sim \frac{11\sqrt{5}}{48}N_f N -\frac{N_f^2}{2} + O(N^{-1})\,,\\
          a/c &\sim \frac{10}{11}-\frac{3\sqrt{5}N_f}{242N} + O(N^{-2})\,.
      \end{split}
  \end{align}
  In the Veneziano limit $N, N_f \to \infty$ with $\a=N_f/N$ fixed, the result of $a$-maximization is as follows:
  \begin{align}
      \begin{split}\label{eq:adj1ven charges}
          R_Q=R_{\widetilde{Q}}&\sim \frac{6-\sqrt{20-\a^2}+3\a-3\a^2}{3(2-\a^2)} + O(N^{-1}) \,,\\
          R_\Phi &\sim \frac{\a\sqrt{20-\a^2}-3\a^2}{3(2-\a^2)} + O(N^{-1})\,,\\
          a & \sim \frac{-18\a^2(5-\a^2)+\a(20-\a^2)^{3/2}}{48(2-\a^2)^2} N^2 + O(N)\,,\\
          c & \sim \frac{-96\a^2+21\a^4+(22\a-2\a^3)\sqrt{20-\a^2}}{48(2-\a^2)^2} N^2 + O(N)\,,\\
          a/c &\sim \frac{440-76\a^2-3\a\sqrt{20-\a^2}}{484-89\a^2}+  O(N^{-1})\,.
      \end{split}
  \end{align}
  
  For a fixed $N_f$, since the $R$-charge of an adjoint chiral multiplet scales as $1/N$, the scaling dimension (proportional to the $R$-charge) can become arbitrarily small for large $N$. As a result, the spectrum of gauge-invariant operators becomes dense and there are $O(N)$ decoupled operators \cite{Agarwal:2019crm, Agarwal:2020pol}. We refer to theories exhibiting this behavior as Type I.
  
  In the Veneziano limit with $N_f/N<\sqrt{2/5}\simeq 0.6325$, the dimension of an adjoint chiral multiplet becomes less than $1/2$. Then, the dimensions of the gauge-invariant operators of the form $\Tr\Phi^n$ and $Q_I\Phi^m\widetilde{Q}_{\tilde{I}}$ can violate the unitarity constraint for some $n\geq2$ and $m\geq0$. These unitary-violating operators become free and decouple along the RG flow.
    \begin{figure}[t]
    \centering
    \begin{subfigure}[b]{0.45\textwidth}
        \includegraphics[width=\linewidth]{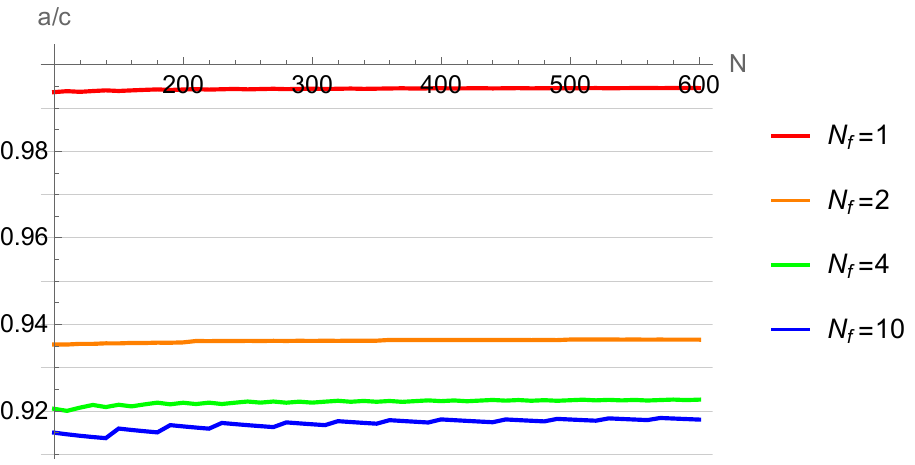}
    \end{subfigure}
    \hspace{4mm}
    \begin{subfigure}[b]{0.45\textwidth}
        \includegraphics[width=\linewidth]{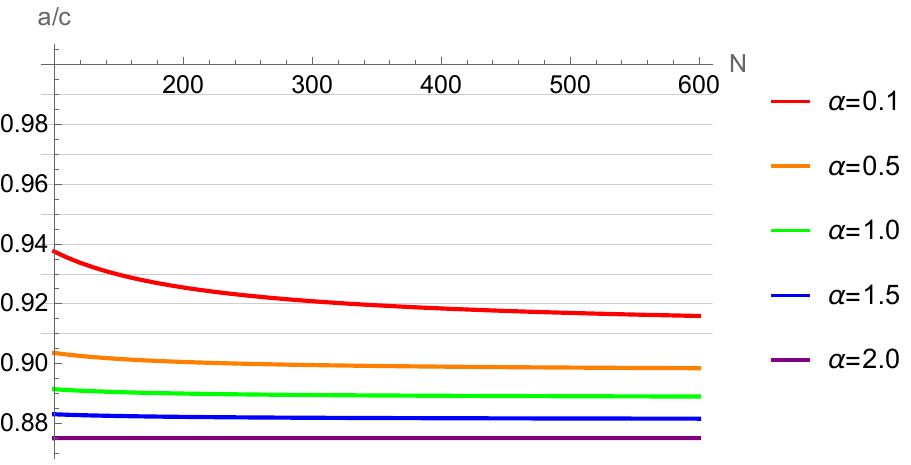}
    \end{subfigure}
    \hfill
     \caption{The central charge ratio for $SU(N)$ theory with 1 \textbf{Adj} + $N_f$  ( $\fund$ + $\overline{\fund}$ ). Left: $a/c$ versus $N$ with a fixed $N_f$. Right: $a/c$ versus $N$ with a fixed $\a=N_f/N$.}
    \label{fig:adj1ratio}   
  \end{figure}
  
  In the presence of decoupled operators, the results of $a$-maximization (\ref{eq:adj1 charges}) are no longer valid, but after flipping decoupled operators and performing $a$-maximization again, one can get the correct answer. For example, in the case of $N_f=1$, after flipping all decoupled operators, the correct result of the $a$-maximization is given by
  \begin{align}
  \begin{split}
    R_\Phi &\sim 0.7156/N + O(N^{-2})\,,\\
    R_Q=R_{\widetilde{Q}} & \sim 0.2844 + O(N^{-1})\,,\\
    a&\sim 0.5008 N + O(1)\,,\\
    c&\sim 0.5034 N + O(1)\,,\\
    a/c &\sim 0.9948 + O(N^{-1})\,.
  \end{split}
  \end{align}
  The plot of $a/c$ is shown in Figure \ref{fig:adj1ratio}. In general, the ratio lies within the range $0.8214\simeq \frac{23}{28}\leq a/c \lesssim 0.9948$. The minimum value of $a/c$ arises when $(N_f,N)=(4,2)$. The maximum values occur in the large $N$ limit with $N_f=1$.

\paragraph{Conformal window}
  The conformal window is already obtained in \cite{Agarwal:2020pol}. 
  The upper bound of the conformal window is simply the asymptotic freedom bound. 
  When $N_f=0$, this theory does not flow to an interacting superconformal fixed point; instead, it is identical to $\CN=2$ SYM theory, whose IR phase is in the Coulomb phase described by Seiberg-Witten theory \cite{Seiberg:1994rs}. 
  When $N_f=2N$, the gauge coupling does not run and is exactly marginal, and gives rise to an interacting $\CN=2$ SCFT (upon turning on the cubic superpotential) \cite{Leigh:1995ep}. 
  For $ 1 \leq N_f \leq 2N$, the $a$-maximization procedure poses no issue apart from the decoupling for a small $N_f$. The conformal window is therefore $1\leq N_f \leq 2N$.

\paragraph{Relevant operators}
  For a fixed $N_f$, since the $R$-charge of $\Phi$ scales as $1/N$, there are $O(N)$ relevant operators of the form $\Tr\Phi^n$ and $Q_I\Phi^m\widetilde{Q}_{\tilde{I}}$. The $O(N)$ amount of them get decoupled along the RG flow.

  The matter content of this theory is identical to that of $\CN=2$ supersymmetric QCD with $N_f$ fundamental hypermultiplet flavors. One can deform the theory by a cubic superpotential $W = Q \Phi \tilde{Q}$ to land on $\CN=2$ supersymmetric theory. This theory is, in general, IR-free and in the Coulomb phase with massive charged particles \cite{Seiberg:1994aj}. Only for $N_f=2N$, it gives a superconformal theory (with vanishing 1-loop beta function).  
  
  One can consider the deformation of the form $W = \Tr \Phi^{k+1}$. In this case, a well-known dual description due to Kutasov-Schwimmer \cite{Kutasov:1995ve, Kutasov:1995np} is given by $SU(kN_f - N)$ theory with a number of flip fields.  
  This theory also has a non-Lagrangian dual description given by gauging multiple Argyres-Douglas theories \cite{Maruyoshi:2023mnv}. 
  
  Let us also note that when $N_f=1$, one can flip all the operators of the form $\Tr \Phi^n$ and $Q \Phi^n \tilde{Q}$. Upon such a deformation, we obtain $(A_1, A_{2N-1})$ Argyres-Douglas theory \cite{Argyres:1995jj, Argyres:1995xn, Eguchi:1996vu, Cecotti:2010fi, Xie:2012hs} in the IR with supersymmetry enhancement \cite{Maruyoshi:2016tqk, Maruyoshi:2016aim}. When $N_f=2$, there also exists a deformation along with flipping which triggers a supersymmetry-enhancing RG flow to $(A_1, D_{2N})$ Argyres-Douglas theory \cite{Agarwal:2016pjo}.

  We plot the $a/c$ against the number of single-trace relevant operators. The behavior of $a/c$ in this case is opposite to the result in \cite{Cho:2024civ}, which examines the ratio across a large set of $\CN=1$ superconformal fixed points of simple, small $N$ gauge theory. In \cite{Cho:2024civ}, $a/c$ tends to be close to 1 when the number of relevant operators is small. However, in these Type I large $N$ theories, the ratio slightly increases as the number of relevant operators increases. 
  \begin{figure}[h]
      \centering
      \includegraphics[width=0.5\linewidth]{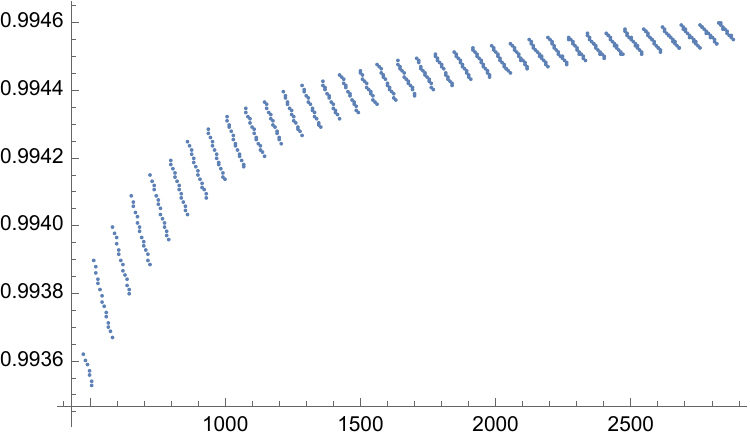}
      \caption{$a/c$ versus the number of single-trace relevant operators in $SU(N)$ with 1 \textbf{Adj} + $N_f$  ( $\fund$ + $\overline{\fund}$ ) with $N_f=1$ and $100<N<600$.}
      \label{fig:adj1reltoratio}
  \end{figure}

\paragraph{Conformal manifold}
  As discussed earlier, the anomaly-free condition for $U(1)_R$ symmetry allows us to identify the combinations of elementary fields in \eqref{eq:marginal from anomaly-free} with $R$-charge 2, corresponding to marginal operators. For this theory, the anomaly-free condition is given by
  \begin{align}
      N + N(R_\Phi -1) + \half N_f(R_Q-1)+ \half N_f (R_{\widetilde{Q}}-1)=0 \implies \frac{2N}{N_f}R_\Phi + R_Q + R_{\widetilde{Q}}=2\,.
  \end{align}
  This implies that the operators of the form
  \begin{align}\label{eq:adj1 marginal ops}
      (\Tr \Phi^n)(Q_{I}\Phi^m\widetilde{Q}_{\tilde{I}})\,,\quad n+m = \frac{2N}{N_f}\,,
  \end{align}
  are marginal for $2N/N_f\in \IZ$.
  The maximum value of $n$ and $m$ is $N-1$, meaning that $n+m$ can never reach $2N$. Thus, in the $N_f=1$ case, one cannot make marginal operators of the form in \eqref{eq:adj1 marginal ops}. 

  In the case of $N_f=2$, for any $N$, there exist (multi-trace) marginal operators of the form
  \begin{align}
      (\Tr\Phi^n)(Q_I\Phi^m \widetilde{Q}_{\tilde{J}})\,,\quad n+m=N\,.
  \end{align}
  Note that $\Tr\Phi^p$ with $p\lesssim 0.4465 N$ and $Q_I\Phi^q \widetilde{Q}_{\tilde{J}}$ with $q\lesssim 0.1069N$ are decoupled (this is an approximation at large $N$). Nevertheless, since there are $O(N)$ possible ways to partition $n$ and $m$ such that $n+m=N$ while satisfying $n>0.4465N$ and $m>0.1069N$, the theory contains $O(N)$ marginal operators, leading to the existence of a large-dimensional conformal manifold.

  For $2<N_f<2N$, there are no marginal operators for general $N$, except for values of $N$ satisfying $2N/N_f\in\IZ$. For such an $N$ with a fixed $N_f$, the theory again has $O(N)$ marginal operators of the form \eqref{eq:adj1 marginal ops}, leading the theory to have an $O(N)$-dimensional conformal manifold.

  When $N_f=2N$, the conformal manifold has dimension one for $N>3$ and dimension seven for $N=3$. The conformal manifold contains a one-dimensional subspace that preserves $\CN=2$ supersymmetry \cite{Bhardwaj:2013qia,Razamat:2020pra}.
  
  When searching for all possible superpotential deformations in $SU(3)$ $N_f=1$ adjoint SQCD, we find certain deformations where the IR fixed points have a non-trivial conformal manifold. However, we were unable to find a superpotential deformation that works for general $N$, as it includes terms that are not relevant for larger $N$.

\paragraph{Weak Gravity Conjecture}
  We examine the AdS WGC using the gauge-invariant operators and $U(1)$ flavor charges identified at the beginning of this section. We find that this theory always satisfies both versions of the WGC. The result is shown in Figure \ref{fig:wgc_adj1ven}. 
  Let us remark that we consider only a subset of low-lying gauge-invariant operators, rather than the full chiral ring. These selected operators are sufficient, as they can form the vertices of the convex hull generated by the full set of chiral ring operators. Since the set of operators that appear as the convex hull vertices can depend on $N$, this leads to the observed discontinuities in the plot.
  \begin{figure}[t]
    \centering
    \begin{subfigure}[b]{0.45\textwidth}
      \centering
      \includegraphics[width=\linewidth]{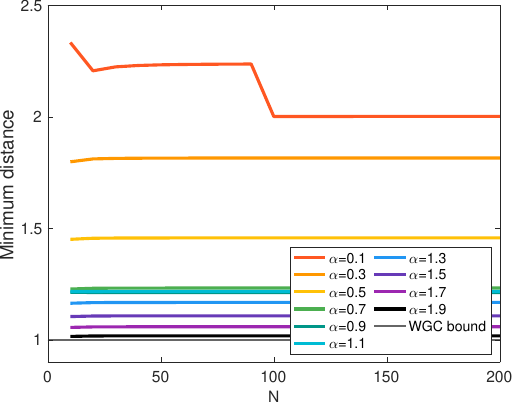}
      \caption{NN-WGC}
    \end{subfigure}
    \hspace{4mm}
    \begin{subfigure}[b]{0.45\textwidth}
      \centering
      \includegraphics[width=\linewidth]{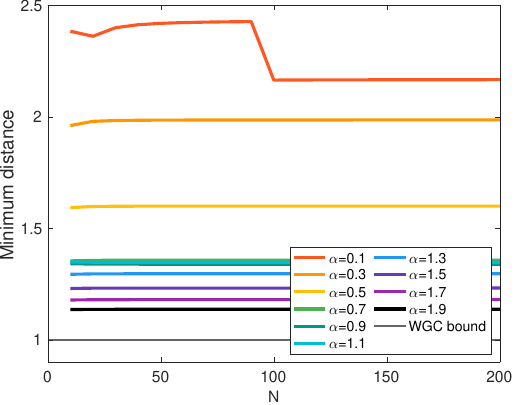}
      \caption{modified WGC}
    \end{subfigure}
    \hfill
    \caption{Testing AdS WGC for $SU(N)$ theory with 1 \textbf{Adj} + $N_f$  ( $\fund$ + $\overline{\fund}$ ).
    The minimum distance from the origin to the convex hull with a fixed $\a=N_f/N$.
    Theories below the solid line of minimum distance 1 do not satisfy the WGC.}
    \label{fig:wgc_adj1ven}
  \end{figure}

\subsection{\texorpdfstring{1 $\sym$ + 1 $\overline{\sym}$ + $N_f$  ( $\fund$ + $\overline{\fund}$ )}{1 S + 1 St + Nf ( Q + Qt )}\label{sec:s1S1}}

\paragraph{Matter content and symmetry charges}
  Consider an $SU(N)$ gauge theory with a pair of rank-2 symmetric and its conjugate, together with $N_f$ pairs of fundamental and anti-fundamental chiral multiplets. The matter fields and their $U(1)$ global charges are listed in Table \ref{tab:s1S1}.
  {\renewcommand\arraystretch{1.4}
  \begin{table}[h]
    \centering
    \begin{tabular}{|c|c||c|c|c|c|c|}
    \hline
    \# & Fields & $SU(N)$ & $U(1)_1$ & $U(1)_2$ & $U(1)_3$ & $U(1)_R$ \\\hline
     $N_f$ & $Q$ & $\fund$ &  $N+2$ & 1 & 0 & $R_Q$ \\
     $N_f$ & $\widetilde{Q}$ & $\overline{\fund}$  &  $N+2$ & -1 & 0 & $R_{\widetilde{Q}}$ \\
     1 & $S$ & $\sym$ &  $-N_f$ & 0 & 1 & $R_S$  \\
     1 & $\widetilde{S}$ & $\overline{\sym}$ &  $-N_f$ & 0 & -1 & $R_{\widetilde{S}}$  \\\hline
    \end{tabular}
    \caption{The matter contents and their corresponding charges in $SU(N)$ gauge theory with a pair of symmetric and its conjugate, together with $N_f$ pairs of fundamentals and anti-fundamental chiral multiplets.\label{tab:s1S1}}
  \end{table}}

\paragraph{Gauge-invariant operators}
  Let $I$ and $J$ denote the flavor indices for $Q$. We present a set of single-trace gauge-invariant operators in schematic form.
  \begin{enumerate}
      \item $\Tr (S\widetilde{S})^n\,,\quad n=1,2,\dots,N-1\,.$
      \item $Q_I(\tilde{S}S)^n \widetilde{Q}_{\tilde{I}}\,,\quad 0,1,\dots,N-1\,.$
      \item $Q_I\widetilde{S}(S\widetilde{S})^n Q_{J}\,,\quad \widetilde{Q}_{\tilde{I}}S(\widetilde{S}S)^n\widetilde{Q}_{\tilde{J}}\,,\quad 0,1,\dots,N-2\,.$
      \item $\e\,\CQ_{I_1}^{n_1}\cdots \CQ_{I_N}^{n_N}\,,\quad \e\,\widetilde{\CQ}_{\tilde{I}_1}^{n_1}\cdots\widetilde{\CQ}_{\tilde{I}_N}^{n_N}\,.$
      \item $\e\,\e\, S^{N-k} (Q_{I_1}Q_{J_1})\cdots(Q_{I_k}Q_{J_k})\,,\quad \e\,\e\, \widetilde{S}^{N-k} (\widetilde{Q}_{\tilde{I}_1}\widetilde{Q}_{\tilde{J}_1})\cdots(\widetilde{Q}_{\tilde{I}_k}\widetilde{Q}_{\tilde{J}_k})\,,\quad 0,1,\dots, N_f\,.$
  \end{enumerate}
  Here, we define the dressed quarks as $\CQ_I^n=(S\widetilde{S})^n Q_I$ and $\widetilde{\CQ}_{\tilde{I}}^n = (S\widetilde{S})^n\widetilde{Q}_{\tilde{I}}$. 

\paragraph{$R$-charges and central charges}
  We perform the $a$-maximization to compute the $R$-charges of the matter fields and central charges. In the large $N$ limit with a fixed $N_f$, they are given by
  \begin{align}
      \begin{split}\label{eq:s1 charges}
          R_Q=R_{\widetilde{Q}}&\sim \frac{3-\sqrt{5}}{3} + \frac{22\sqrt{5}+ 15 N_f}{30N} + O(N^{-2})\,,\\
          R_S=R_{\widetilde{S}} &\sim \frac{6 + \sqrt{5}N_f}{3N} - \frac{5N_f^2 + 14\sqrt{5} N_f + 40}{10N^2}+ O(N^{-3})\,,\\
          a & \sim \frac{63+10\sqrt{5}N_f}{48} N -\frac{15 N_f^2 + 44 \sqrt{5} N_f + 162}{32} + O(N^{-1})\,,\\
          c & \sim \frac{60+11\sqrt{5}N_f}{48} N -\frac{20N_f^2 + 56\sqrt{5}N_f + 195}{40} + O(N^{-1})\,,\\
          a/c &\sim \frac{63+10\sqrt{5}N_f}{60+11\sqrt{5}N_f}-\frac{3(25\sqrt{5}N_f^3 + 360 N_f^2 +198\sqrt{5}N_f -540)}{10(60+11\sqrt{5}N_f)^2N} + O(N^{-2})\,.
      \end{split}
  \end{align}
  For a fixed $N_f$, the $R$-charge of a rank-2 symmetric chiral multiplet scales as $1/N$. As a result, the spectrum of gauge-invariant operators becomes dense, with $O(N)$ operators decoupling, and hence this theory belongs to the Type I class.
  
  In the Veneziano limit $N,N_f\goto\infty$ with $\a=N_f/N$ fixed, the leading-order result of $a$-maximization coincides with (\ref{eq:adj1ven charges}) in the previous section, where $R_S=R_{\widetilde{S}}\sim R_\Phi$. This leading behavior of the $R$-charges of rank-2 tensor chiral multiplets is universal across Type I theories. However, since the set of decoupled operators differs from theory to theory, the final corrected $R$-charges do not exhibit universality. As discussed in Section \ref{sec:SUclass}, we can obtain a universal result if we take the additional limit $1\ll N_f\ll N$.
  
  In this theory, operators of the form $\Tr (S\widetilde{S})^n$, $Q_I(\widetilde{S}S)^n \widetilde{Q}_{\tilde{I}}$, $\widetilde{Q}_{\tilde{I}}(S\widetilde{S})^nS\widetilde{Q}_{\tilde{J}}$ are decoupled along the RG flow. For example, in the case of $N_f=0$, after flipping all decoupled operators, the correct value of $a/c$ is given by
  \begin{align}
      a/c\sim \frac{19}{18} + \frac{7}{108N} + O(N^{-2})\,.
  \end{align}
  The plot of $a/c$ is shown in Figure \ref{fig:s1S1ratio}. 
  We find that the ratio lies within the range $0.875=7/8< a/c \leq 3553/3353\simeq 1.0597$. 
  For $N_f=0,1$ and sufficiently large $N$, the ratio $a/c$ is greater than 1. The minimum value of $a/c$ arises in the Veneziano limit with $\a\goto2$. The maximum value of $a/c$ occurs when $(N_f,N)=(0,10)$.
  \begin{figure}[t]
    \centering
    \begin{subfigure}[b]{0.45\textwidth}
        \includegraphics[width=\linewidth]{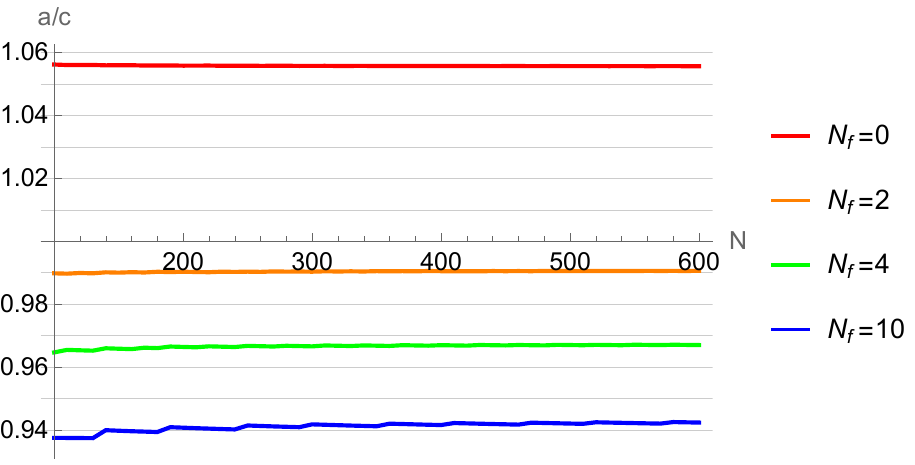}
    \end{subfigure}
    \hspace{4mm}
    \begin{subfigure}[b]{0.45\textwidth}
        \includegraphics[width=\linewidth]{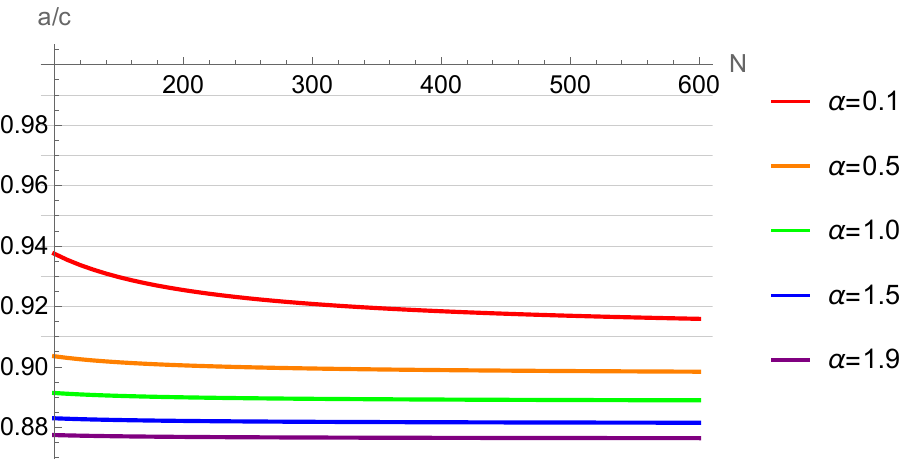}
    \end{subfigure}
    \hfill
     \caption{The central charge ratio for $SU(N)$ theory with 1 $\sym$ + 1 $\overline{\sym}$ + $N_f$  ( $\fund$ + $\overline{\fund}$ ). 
     Left: $a/c$ versus $N$ with a fixed $N_f$. Right: $a/c$ versus $N$ with a fixed $\a=N_f/N$.}
    \label{fig:s1S1ratio}   
  \end{figure}

\paragraph{Conformal window}
  The conformal window is obtained in \cite{Agarwal:2020pol}. The upper bound is the asymptotic freedom bound. 
  When $N_f=2N-2$, even though it has a vanishing beta function for the gauge coupling, it is marginally irrelevant without a non-trivial conformal manifold \cite{Green:2010da}. Therefore, it does not give an interacting SCFT. The same argument works for all the theories whose upper bound of the asymptotic freedom is excluded (strict inequality) from the conformal window.
  
  For a small $N_f$, the $a$-maximization procedure poses no issues apart from the decoupling for a small $N_f$. Notably, we find that the $N_f=0$ theory flows to a good interacting SCFT upon removing the decoupled operators. 
   The conformal window of this theory is therefore $0\leq N_f < 2N-2$. 

\paragraph{Relevant operators}
  For a fixed $N_f$, since the $R$-charge of a rank-2 symmetric chiral multiplet scales as $1/N$, there are $O(N)$ relevant operators of the form $\Tr (S\widetilde{S})^n$, $Q_I(\widetilde{S}S)^n \widetilde{Q}_{\tilde{I}}$, $\widetilde{Q}_{\tilde{I}}(S\widetilde{S})^nS\widetilde{Q}_{\tilde{J}}$. The $O(N)$ amount of them get decoupled along the RG flow.

  One can consider the deformation of the form $W=\Tr (S\widetilde{S})^{k+1}$. In this case, the theory has a dual description given by $SU((2k + 1)N_f + 4k - N)$ gauge theory \cite{Intriligator:1995ax} with a number of flip fields.

\paragraph{Conformal manifold}
  In the case of $N_f=0$, there is only one $U(1)$ flavor symmetry, under which $S$ carries charge $+1$ and $\widetilde{S}$ carries charge $-1$. This symmetry does not mix with the $U(1)_R$ symmetry because it is baryonic or traceless. 
  As a result, the $U(1)_R$ symmetry is uniquely determined by the $R$-anomaly cancellation condition as
  \begin{align}
      R_S=R_{\widetilde{S}}=\frac{2}{N+2}\,.
  \end{align}
  For even $N$, except for $N=2$, there exists marginal operators of the form
  \begin{align}
      \Tr(S\widetilde{S})^{(N+2)/2}\,,\quad \Tr(S\widetilde{S})^n \Tr(S\widetilde{S})^m \,,\quad n+m=(N+2)/2 \,.
  \end{align}
  Note that $n$ and $m$ has to be greater than $\lfloor (N+2)/6\rfloor$ since the operators of the form $\Tr (S\widetilde{S})^n$ with $n\leq \lfloor (N+2)/6\rfloor$ are decoupled. Consequently, there exists an $O(N)$ number of marginal operators, leading the theory to have an $O(N)$-dimensional conformal manifold. 
  For odd $N$, a non-trivial conformal manifold appears only when $N=3$, where the marginal operators have the forms of $(\Tr S\widetilde{S}) (\e\,\e\, S^3)$ and $(\Tr S\widetilde{S} )(\e\,\e\, \widetilde{S}^3)$. 
  For odd $N>3$, since the operator of the form $\Tr S\widetilde{S}$ is decoupled, the operators of the form $(\Tr S\widetilde{S})(\e\,\e\,S^N)$, which have the marginal dimension, do not exist in the chiral ring.
  
  Let us now consider the case of $N_f\neq 0$. From the anomaly-free condition for the $U(1)_R$ symmetry, we obtain
  \begin{align}
  \begin{split}
      &N + \frac{N+2}{2}(R_S - 1)+ \frac{N+2}{2}(R_{\widetilde{S}} - 1) + \half N_f(R_Q-1)+ \half N_f (R_{\widetilde{Q}}-1)=0\\
      &\quad \quad \implies \quad \frac{N+2}{N_f+2}R_S + \frac{N+2}{N_f+2}R_{\widetilde{S}} + \frac{N_f}{N_f+2}R_Q + \frac{N_f}{N_f+2}R_{\widetilde{Q}}=2\,.
  \end{split}
  \end{align}
  Taking into account $R_S=R_{\widetilde{S}}$ and $R_Q=R_{\widetilde{Q}}$, each of the coefficients does not need to be an integer. Rather, it implies the possibility that an operator of the form
 \begin{align}
      S^p\widetilde{S}^qQ^r\widetilde{Q}^s\,,\quad p+q=\frac{2(N+2)}{N_f+2}\,,\quad r+s=\frac{2N_f}{N_f+2}\,,
  \end{align}
  might be marginal if $p+q$ and $r+s$ are integers. However, $r+s$ can be an integer only when $N_f=2$ with $r+s=1$. In that case, the operator cannot be gauge-invariant with a single fundamental representation.  
Empirically, we have not identified any marginal operators for generic $N$ and $N_f>0$.
  
\paragraph{Weak Gravity Conjecture}
 
\begin{figure}[h]
    \centering
     \begin{subfigure}[b]{0.45\textwidth}
     \centering
    \includegraphics[width=\linewidth]{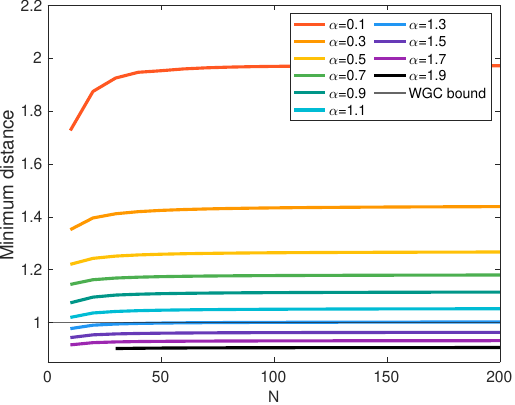}
    \caption{NN-WGC}
     \end{subfigure}
     \hspace{4mm}
     \begin{subfigure}[b]{0.45\textwidth}
     \centering
        \includegraphics[width=\linewidth]{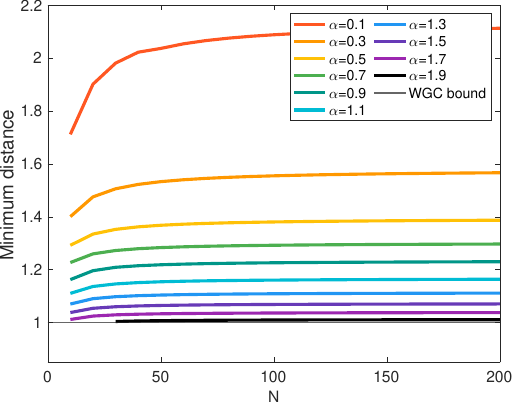}
        \caption{modified WGC}
     \end{subfigure}
     \hfill
        \caption{Testing AdS WGC for $SU(N)$ theory with 1 $\sym$ + 1 $\overline{\sym}$ + $N_f$  ( $\fund$ + $\overline{\fund}$ ). 
        The minimum distance from the origin to the convex hull with a fixed $\a=N_f/N$. Theories below the solid line of minimum distance 1 do not satisfy the WGC.\label{fig:wgc_s1S1ven}}
 \end{figure}

 We test the AdS WGC using the gauge-invariant operators and $U(1)$ flavor charges identified at the beginning of this section. We find that this theory does not satisfy the NN-WGC for large values of $\a$ in the Veneziano limit, whereas the modified WGC always holds. The result is shown in Figure \ref{fig:wgc_s1S1ven}.

\subsection{\texorpdfstring{1 $\antisym$ + 1 $\overline{\antisym}$ + $N_f$   ( $\fund$ + $\overline{\fund}$ )}{1 A + 1 At + Nf ( Q + Qt )}}\label{sec:a1A1}
\paragraph{Matter content and symmetry charges}
  The third entry of Type I theories is an $SU(N)$ gauge theory with a pair of rank-2 anti-symmetric and $N_f$ pairs of fundamental and anti-fundamental chiral multiplets. The matter fields and their $U(1)$ global charges are listed in Table \ref{tab:a1A1}.
  {\renewcommand\arraystretch{1.4}
  \begin{table}[h]
    \centering
    \begin{tabular}{|c|c||c|c|c|c|c|}
    \hline
    \# & Fields & $SU(N)$ & $U(1)_1$ & $U(1)_2$ & $U(1)_3$ & $U(1)_R$ \\\hline
     $N_f$ & $Q$ & $\fund$ &  $N-2$ & 1 & 0 & $R_Q$ \\
     $N_f$ & $\widetilde{Q}$ & $\overline{\fund}$  &  $N-2$ & -1 & 0 & $R_{\widetilde{Q}}$ \\
     1 & $A$ & $\antisym$ &  $-N_f$ & 0 & 1 & $R_A$  \\
     1 & $\widetilde{A}$ & $\overline{\antisym}$ &  $-N_f$ & 0 & -1 & $R_{\widetilde{A}}$  \\\hline
    \end{tabular}
    \caption{The matter contents and their corresponding charges in $SU(N)$ gauge theory with a pair of anti-symmetric tensor and its conjugate and $N_f$ fundamental/anti-fundamental pairs. \label{tab:a1A1}}
  \end{table}}
  
\paragraph{Gauge-invariant operators}
  Let $I$ and $J$ denote the flavor indices for $Q$. We present the set of single-trace gauge-invariant operators in schematic form as follows:
  \begin{enumerate}
      \item $\Tr (A\widetilde{A})^n\,,\quad n=1,2,\dots,\lfloor\frac{N-1}{2}\rfloor\,.$
      \item $Q_I(\widetilde{A}A)^n \widetilde{Q}_{\tilde{I}}\,,\quad 0,1,\dots,\lfloor\frac{N}{2}\rfloor-1\,.$
      \item $Q_I\widetilde{A} (A\widetilde{A})^n Q_J\,,\widetilde{Q}_{\tilde{I}}A (\widetilde{A}A)^n \widetilde{Q}_{\tilde{J}}\,,\quad 0,1,\dots,\lfloor\frac{N-1}{2}\rfloor-1\,.$
      \item $\e\, A^n Q_{I_1}\cdots Q_{I_{N-2n}}\,,\quad \e\, \widetilde{A}^n \widetilde{Q}_{\tilde{I}_1}\cdots \widetilde{Q}_{\tilde{I}_{N-2n}}\,,\quad n=\lceil\frac{N-N_f}{2}\rceil, \cdots, \lfloor \frac{N}{2}\rfloor \,.$
  \end{enumerate}

\paragraph{$R$-charges and central charges}
  For $N=3$, the rank-2 anti-symmetric tensor is the same as the anti-fundamental representation. Thus, we only consider cases with $N>3$ here.
  
  We perform the $a$-maximization to compute the $R$-charges of the matter fields and central charges. In the large $N$ limit with a fixed $N_f$, they are given by
  \begin{align}
      \begin{split}\label{eq:a1 charges}
          R_Q=R_{\widetilde{Q}}&\sim \frac{3-\sqrt{5}}{3} + \frac{-22\sqrt{5}+ 15 N_f}{30N} + O(N^{-2})\,,\\
          R_A=R_{\widetilde{A}} &\sim \frac{-6+\sqrt{5}N_f}{3N} - \frac{5N_f^2-14\sqrt{5}+40}{10N^2} + O(N^{-3})\,,\\
          a & \sim \frac{-63+10\sqrt{5}N_f}{48} N -\frac{15 N_f^2 - 44 \sqrt{5} N_f + 162}{32} + O(N^{-1})\,,\\
          c & \sim \frac{-60+11\sqrt{5}N_f}{48} N -\frac{20N_f^2 - 56\sqrt{5}N_f + 195}{40} + O(N^{-1})\,,\\
          a/c &\sim \frac{-63+10\sqrt{5}N_f}{-60+11\sqrt{5}N_f}-\frac{3(25\sqrt{5}N_f^3 - 360 N_f^2 +198\sqrt{5}N_f + 540)}{10(-60+11\sqrt{5}N_f)^2N} + O(N^{-2})\,.
      \end{split}
  \end{align}
  For a fixed $N_f$, the $R$-charge of the rank-2 anti-symmetric chiral multiplet scales as $1/N$. Therefore, this theory belongs to Type I. 
  
  In the Veneziano limit with a fixed $\a=N_f/N$, the leading-order result of $a$-maximization is the same as \eqref{eq:adj1ven charges} in the previous section, with $R_A=R_{\widetilde{A}}\sim R_\Phi$. This result is universal across Type I theories when we neglect the effect of decoupled operators. 
  
  To perform $a$-maximization correctly, we must take into account the decoupled operators of the form $\Tr (A\widetilde{A})^n$, $Q_I(\widetilde{A}A)^n \widetilde{Q}_{\tilde{I}}$, $Q_I\widetilde{A}(A\widetilde{A})^nQ_J$, and $\widetilde{Q}_{\tilde{I}}(A\widetilde{A})^nA\widetilde{Q}_{\tilde{J}}$. 
  Additionally, for even $N$, operators of the form $\e\,A^{N/2}$ and $\e\, \widetilde{A}^{N/2}$ exist, and if $N_f=4$, these operators are also decoupled. The plot of the correct value of $a/c$ is shown in Figure \ref{fig:a1A1ratio}. For example, in the case of $N_f=4$, after flipping all the decoupled operators, the ratio $a/c$ is approximately $a/c\sim 0.7538 -0.1221/N$ at large $N$, which is obtained via fitting the result for $N$ from 100 to 600. 

  We find that the ratio of central charges for this theory lies within the range $ 0.6930\simeq \frac{70169619-220676\sqrt{3201}}{83242974}\leq  a/c \lesssim 0.9159 $. The minimum value of $a/c$ arises when $(N_f,N)=(4,5)$. We numerically find that the maximum value of $a/c$ occurs in the Veneziano limit with $\a\sim 0.1$.
  \begin{figure}[t]
    \centering
    \begin{subfigure}[b]{0.45\textwidth}
        \includegraphics[width=\linewidth]{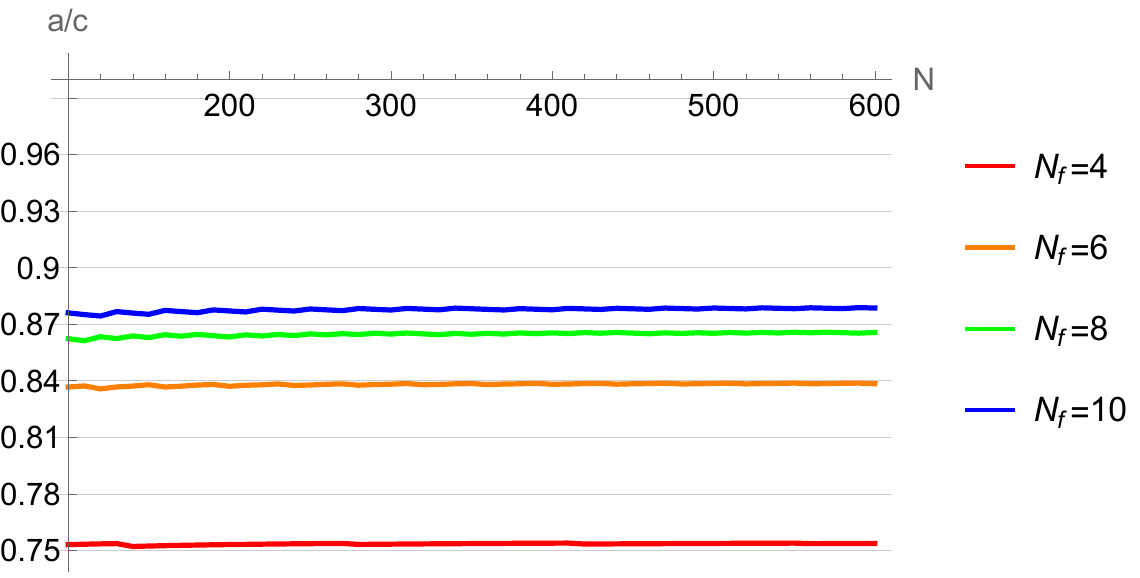}
    \end{subfigure}
    \hspace{4mm}
    \begin{subfigure}[b]{0.45\textwidth}
        \includegraphics[width=\linewidth]{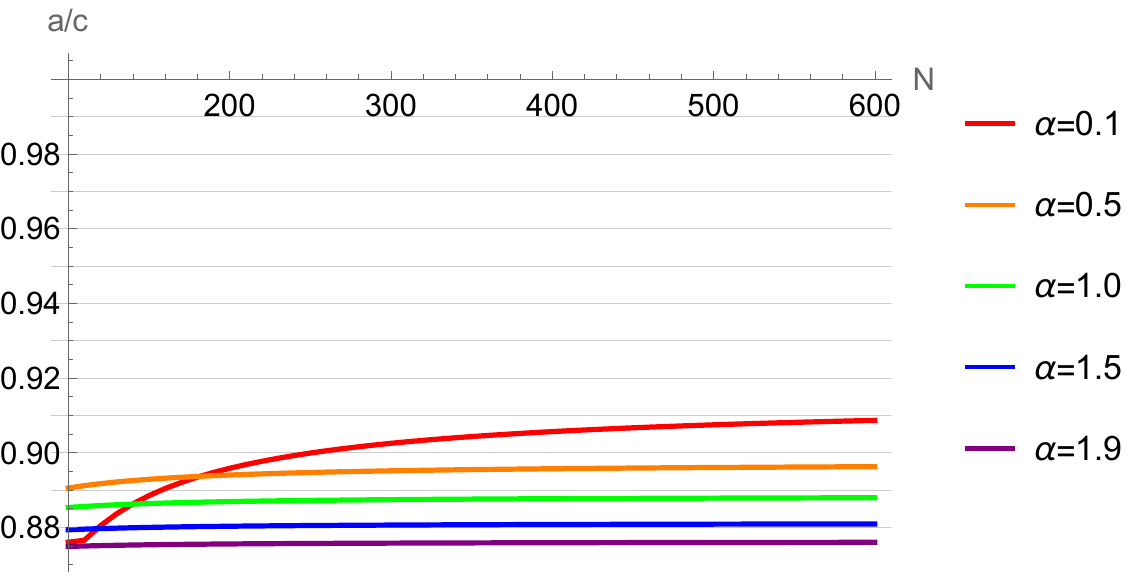}
    \end{subfigure}
    \hfill
     \caption{The central charge ratio for $SU(N)$ theory with 1 $\antisym$ + 1 $\overline{\antisym}$ + $N_f$ ($\fund$ + $\overline{\fund}$). 
     Left: $a/c$ versus $N$ with a fixed $N_f$. Right: $a/c$ versus $N$ with a fixed $\a=N_f/N$.}
    \label{fig:a1A1ratio}   
  \end{figure}

\paragraph{Conformal window}
  The conformal manifold for this theory is already found in \cite{Agarwal:2020pol}. The upper bound comes from the asymptotic freedom.
  When $N_f\leq 3$, this theory does not flow to an interacting superconformal fixed point, as there is no solution that locally maximizes $a$ or satisfies the Hofman-Maldacena bound for the ratio $a/c$. Otherwise, the $a$-maximization procedure poses no issues. The conformal window of this theory is therefore $4\leq N_f < 2N+2$.

\paragraph{Relevant operators}
  For a fixed $N_f$, since the $R$-charge of a rank-2 anti-symmetric chiral multiplet scales as $1/N$, there is $O(N)$ number of relevant operators of the form $\Tr (A\widetilde{A})^n$, $Q_I(\widetilde{A}A)^n \widetilde{Q}_{\tilde{J}}$, $Q_I\widetilde{A}(A\widetilde{A})^n Q_J$, $\widetilde{Q}_{\tilde{I}}A(\widetilde{A}A)^n\widetilde{Q}_{\tilde{J}}$, $\e\, A^{n}Q_{I_1}\cdots Q_{I_{N-2n}}$, $\e\,\widetilde{A}^{n}\widetilde{Q}_{\tilde{I}_1}\cdots \widetilde{Q}_{\tilde{I}_{N-2n}}$ and $O(N)$ decoupled operators among them. This is a generic feature of Type I theory.

  Upon deforming the theory by the superpotential $W = \Tr (A \tilde A)^{k+1}$, the theory admits a dual description given by $SU((2k+1)N_f -4k- N)$ gauge theory with the same set of charged matters with certain superpotential and flip fields \cite{Intriligator:1995ax}. 

\paragraph{Conformal manifold}
  From the anomaly-free condition for $U(1)_R$ symmetry, we obtain
  \begin{align}
  \begin{split}\label{eq:a1A1 marginal}
      &N + \frac{N-2}{2}(R_A - 1)+ \frac{N-2}{2}(R_{\widetilde{A}} - 1) + \half N_f(R_Q-1)+ \half N_f (R_{\widetilde{Q}}-1)=0\\
      &\quad \quad \implies \quad \frac{N-2}{N_f-2}R_A + \frac{N-2}{N_f-2}R_{\widetilde{A}} + \frac{N_f}{N_f-2}R_Q + \frac{N_f}{N_f-2}R_{\widetilde{Q}}=2\,.
  \end{split}
  \end{align}
  When $N_f=4$, equation (\ref{eq:a1A1 marginal}) becomes
  \begin{align}
      \frac{N-2}{2}R_A + \frac{N-2}{2}R_{\widetilde{A}} + 2R_Q + 2R_{\widetilde{Q}}=2\,.
  \end{align}
  For even $N$, there exist marginal operators of the form
  \begin{align}
      \Tr(A\widetilde{A})^n (Q_{I_1}(A\widetilde{A})^m \widetilde{Q}_{\tilde{J}_1})(Q_{I_2}(A\widetilde{A})^l \widetilde{Q}_{\tilde{J}_2})\,,\quad m,l\geq 1\,,\quad n+m+l=\frac{N-2}{2}\,.
  \end{align}
  Note that operators of the form $\Tr (A\widetilde{A})^p$ with $p\lesssim 0.3070 N$, as well as $Q (A\widetilde{A})^q \widetilde{Q}$, $Q\widetilde{A}(A\widetilde{A})^q\widetilde{Q}$, and $\widetilde{Q}A(\widetilde{A}A)^q\widetilde{Q}$ with $q\lesssim 0.0965 N$, are decoupled -- this is an approximation at large $N$. 
  Nevertheless, since there are $O(N)$ possible ways to partition $n$, $m$, and $l$ such that $n+m+l=(N-2)/2$ while satisfying $n>0.3070 N$ and $m,l > 0.0965 N$, the theory contains $O(N)$ marginal operators. This leads to the theory having an $O(N)$-dimensional conformal manifold.

  For $N_f>4$, there are no integer combinations of $R$-charges that sum up to 2 in equation \eqref{eq:a1A1 marginal}. As a result, marginal operators cannot be constructed using this equation. Empirically, we have not identified any marginal operators for generic $N$ and $N_f>4$.

\paragraph{Weak Gravity Conjecture}
  We examine the AdS WGC using the gauge-invariant operators and $U(1)$ flavor charges identified at the beginning of this section. We find that this theory does not satisfy the NN-WGC for large values of $\a$ in the Veneziano limit, whereas the modified WGC always holds. The result is shown in Figure \ref{fig:wgc_a1A1ven}.
\begin{figure}[t]
    \centering
     \begin{subfigure}[b]{0.45\textwidth}
     \centering
    \includegraphics[width=\linewidth]{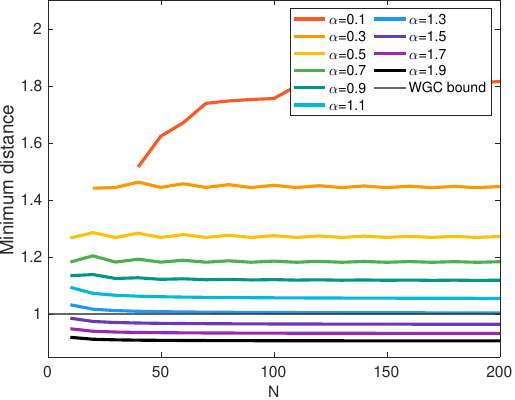}
    \caption{NN-WGC}
     \end{subfigure}
     \hspace{4mm}
     \begin{subfigure}[b]{0.45\textwidth}
     \centering
        \includegraphics[width=\linewidth]{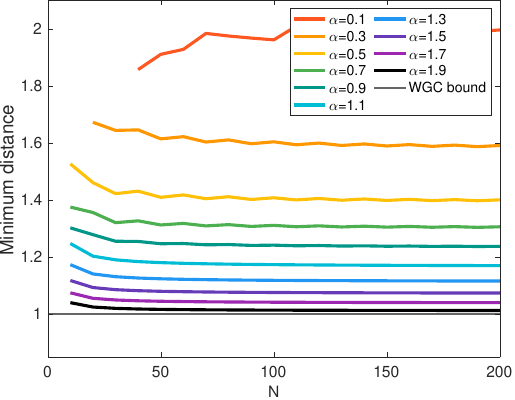}
        \caption{modified WGC}
     \end{subfigure}
     \hfill
        \caption{Testing AdS WGC for $SU(N)$ theory with 1 $\antisym$ + 1 $\overline{\antisym}$ + $N_f$ ($\fund$ + $\overline{\fund}$). The minimum distance from the origin to the convex hull with a fixed $\a=N_f/N$. Theories below the solid line of minimum distance 1 do not satisfy the WGC.\label{fig:wgc_a1A1ven}}
 \end{figure}

\subsection{\texorpdfstring{1 $\sym$ + 1 $\overline{\antisym}$ + 8 $\overline{\fund}$ + $N_f$  ( $\fund$ + $\overline{\fund}$ )}{1 S + 1 At + 8 Qt + Nf ( Q + Qt )}}\label{sec:s1A1F8}
\paragraph{Matter content and symmetry charges}
  The fourth entry in Type I is an $SU(N)$ gauge theory with a rank-2 symmetric, a conjugate of rank-2 anti-symmetric, $N_f$ fundamental, and $N_f+8$ anti-fundamental chiral multiplets. The matter fields and their $U(1)$ global charges are listed in Table \ref{tab:s1A1F8}. This is a chiral theory, with different numbers of fermions with each chirality. 
  {\renewcommand\arraystretch{1.4}
  \begin{table}[h]
    \centering
    \begin{tabular}{|c|c||c|c|c|c|c|}
    \hline
    \# & Fields & $SU(N)$ & $U(1)_1$ & $U(1)_2$ & $U(1)_3$ & $U(1)_R$ \\\hline
     $N_f$ & $Q$ & $\fund$ &  $N+2$ & $N-2$ & $N_f+8$ & $R_Q$ \\
     $N_f+8$ & $\widetilde{Q}$ & $\overline{\fund}$  &  $N+2$ & $N-2$ & $-N_f$ & $R_{\widetilde{Q}}$ \\
     1 & $S$ & $\sym$ & $-2N_f-8$ & 0 & 0 & $R_S$  \\
     1 & $\widetilde{A}$ & $\overline{\antisym}$ &  0 & $-2N_f-8$ & 0 & $R_{\widetilde{A}}$  \\\hline
    \end{tabular}
    \caption{The matter contents and their corresponding charges in $SU(N)$ gauge theory with 1 $\sym$ + 1 $\overline{\antisym}$ + 8 $\overline{\fund}$ + $N_f$  ( $\fund$ + $\overline{\fund}$).\label{tab:s1A1F8}}
  \end{table}}

\paragraph{Gauge-invariant operators}
  Let $I$ and $J$ denote the flavor indices for $Q$. The set of single-trace gauge-invariant operators in schematic form is given as follows:
  \begin{enumerate}
      \item $\Tr (S\widetilde{A})^{2n}\,,\quad n=1,2,\dots,\lfloor\frac{N-1}{2}\rfloor\,.$
      \item $Q_{I} (\widetilde{A}S)^n \widetilde{Q}_{\tilde{J}}\,,\quad n=0,1,\dots, N-2 \,.$
      \item $Q_I\widetilde{A}(S\widetilde{A})^n Q_J\,,\quad n=0,1,\dots, N-2\,.$
      \item $\widetilde{Q}_{\tilde{I}}S(\widetilde{A}S)^n \widetilde{Q}_{\tilde{J}}\,,\quad n=0,1,\dots,N-2\,.$
      \item $\e\, \widetilde{A}^n \widetilde{Q}_{\tilde{I}_1}\cdots \widetilde{Q}_{\tilde{I}_{N-2n}}\,,\quad n=\lceil\frac{N-N_f-8}{2}\rceil, \cdots, \lfloor \frac{N}{2}\rfloor \,.$
      \item $\e\,\e\, S^{N-n}(Q_{I_1}Q_{J_1})\cdots (Q_{I_n}Q_{J_n})\,,\quad n=0,1,\dots, N_f\,.$
      \item $\e\, (S\widetilde{A}S)^{(N-n)/2} \CQ_{I_1}^{k_1}\cdots \CQ_{I_n}^{k_n}\,.$
  \end{enumerate}
  Here, we define the dressed quarks as
  \begin{align}
      \CQ_I^n=\begin{cases}
          (S\widetilde{A})^{n/2}Q_I & n=0,2,4,\dots\,,\\
          (S\widetilde{A})^{(n-1)/2}S\widetilde{Q} & n=1,3,5,\dots\,.
      \end{cases}
  \end{align}
  
\paragraph{$R$-charges and central charges}
  We perform the $a$-maximization to compute the $R$-charges of the matter fields and central charges. In the large $N$ limit with a fixed $N_f$, they are given by
  \begin{align}
      \begin{split}\label{eq:s1A1 charges}
          R_Q=R_{\widetilde{Q}}&\sim \frac{3-\sqrt{5}}{3} + \frac{ N_f + 4 }{2N} + O(N^{-2})\,,\\
          R_S &\sim \frac{-4+3(N_f+4)\sqrt{5}}{9N} -\frac{27 N_f^2 + (216-10(N_f+4))\sqrt{5} + 384}{54N^2} + O(N^{-3})\,,\\
          R_{\widetilde{A}} & \sim  \frac{4+3(N_f+4)\sqrt{5}}{9N} -\frac{27 N_f^2 + (216 + 10(N_f+4))\sqrt{5} + 384}{54N^2} + O(N^{-3})\, \\
          a & \sim \frac{5(N_f+4)\sqrt{5}}{24} N - \frac{45 N_f^2 + 360 N_f + 722}{96} + O(N^{-1})\,,\\
          c & \sim \frac{11(N_f+4)\sqrt{5}}{48} N -\frac{36N_f^2 +288 N_f + 575}{72} + O(N^{-1})\,,\\
          a/c &\sim \frac{10}{11} - \frac{45N_f^2 + 360 N_f + 826}{726(N_f+4)\sqrt{5}N} + O(N^{-2})\,.
      \end{split}
  \end{align}
  For fixed $N_f$, the $R$-charges of rank-2 symmetric and rank-2 anti-symmetric chiral multiplets scale as $1/N$, which is a characteristic feature of Type I theories. 

  In the Veneziano limit with a fixed $\a=N_f/N$, the leading-order result of $a$-maximization is the same as (\ref{eq:adj1ven charges}) in the previous section, with $R_S\sim R_{\widetilde{A}}\sim R_\Phi$. This result is universal across the Type I theories.
  
  Precise $a$-maximization requires removing the decoupled operators, take the form $\Tr(S\widetilde{A})^{2n}$, $Q_{I} (\widetilde{A}S)^n \widetilde{Q}_{\tilde{J}}$, $Q_I\widetilde{A}(S\widetilde{A})^n Q_J$, and $\widetilde{Q}_{\tilde{I}}S(\widetilde{A}S)^n \widetilde{Q}_{\tilde{J}}$.
  The plot of $a/c$ is shown in Figure \ref{fig:s1A1F8ratio}. 
  For example, in the case of $N_f=0$, after flipping decoupled operators, the ratio $a/c$ is approximately $a/c\sim 0.9196-0.3246/N$, where we fit the result for $N$ from 100 to 600. 

  We find the central charge ratio lies within the range $ 0.86 \leq  a/c \lesssim 0.9196 $. The minimum value of $a/c$ arises when $(N_f,N)=(4,4)$. The maximum value of $a/c$ occurs in the large $N$ limit with $N_f=0$.
  \begin{figure}[t]
    \centering
    \begin{subfigure}[b]{0.45\textwidth}
        \includegraphics[width=\linewidth]{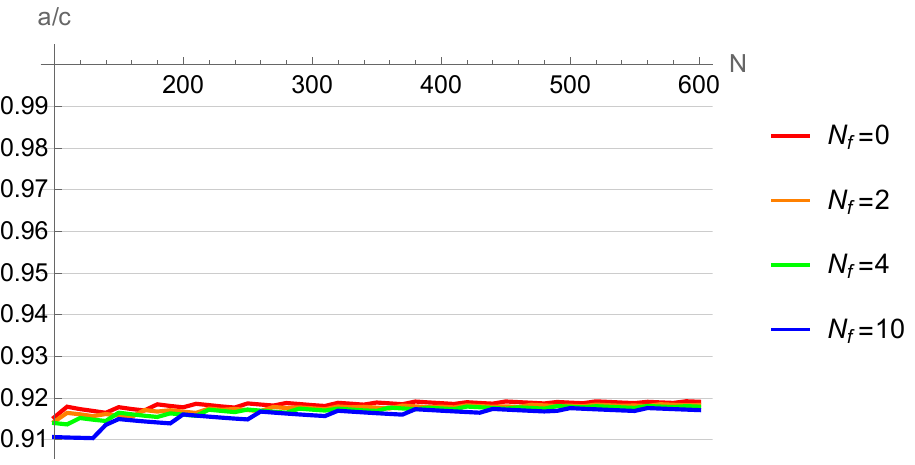}
    \end{subfigure}
    \hspace{4mm}
    \begin{subfigure}[b]{0.45\textwidth}
        \includegraphics[width=\linewidth]{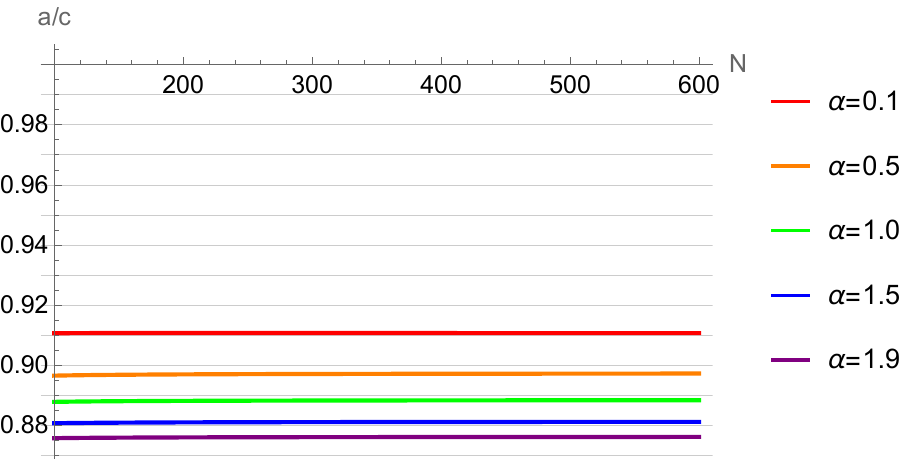}
    \end{subfigure}
    \hfill
     \caption{The central charge ratio for $SU(N)$ theory with 1 $\sym$ + 1 $\overline{\antisym}$ + 8 $\overline{\fund}$ + $N_f$  ( $\fund$ + $\overline{\fund}$ ). 
     Left: $a/c$ versus $N$ with a fixed $N_f$. Right: $a/c$ versus $N$ with a fixed $\a=N_f/N$.}
    \label{fig:s1A1F8ratio}   
  \end{figure}

\paragraph{Conformal window}
   The conformal window is already identified in \cite{Agarwal:2020pol}. The upper bound comes from the asymptotic freedom. For this theory, when it is at the top of the conformal window $N_f = 2N-4$, there is a non-trivial conformal manifold \cite{Razamat:2020pra} so that it is an interacting SCFT. 
   For small $N_f$, the $a$-maximization procedure poses no issue. For instance, $a$-maximization always yields a unique real solution, and the resulting central charges lie within the Hofman-Maldacena bound. The conformal window of this theory is therefore $0\leq N_f \leq 2N-4$.

\paragraph{Relevant operators}
  For a fixed $N_f$, since the $R$-charges of rank-2 tensor chiral multiplets scale as $1/N$, there are $O(N)$ relevant operators of the form $\Tr (A\widetilde{A})^n$, $Q_I(\widetilde{A}A)^n \widetilde{Q}_{\tilde{J}}$, $Q_I\widetilde{A}(A\widetilde{A})^n Q_J$, $\widetilde{Q}_{\tilde{I}}A(\widetilde{A}A)^n\widetilde{Q}_{\tilde{J}}$, $\e\, A^{n}Q_{I_1}\cdots Q_{I_{N-2n}}$, $\e\,\widetilde{A}^{n}\widetilde{Q}_{\tilde{I}_1}\cdots \widetilde{Q}_{\tilde{I}_{N-2n}}$ and $O(N)$ decoupled operators among them.
  
  One may consider the deformation $W=\Tr (S\widetilde{A})^{2(k+1)}$. In this case, the theory admits a dual description as an $SU\left((4k+3)(N_f+4)+4k-N\right)$ gauge theory \cite{Intriligator:1995ax} with a set of flip fields.

\paragraph{Conformal manifold}
  Consider the anomaly-free condition for $U(1)_R$ symmetry to obtain
  \begin{align}
  \begin{split}\label{eq:s1A1F8 marginal}
      &N + \frac{N+2}{2}(R_S - 1)+ \frac{N-2}{2}(R_{\widetilde{A}} - 1) + \half N_f(R_Q-1)+ \half (N_f+8) (R_{\widetilde{Q}}-1)=0\\
      &\quad \quad \implies \quad \frac{N+2}{N_f+4}R_S + \frac{N-2}{N_f+4}R_{\widetilde{A}} + \frac{N_f}{N_f+4}R_Q + \frac{N_f+8}{N_f+4}R_{\widetilde{Q}}=2\,.
  \end{split}
  \end{align} 
  First, let us consider the case of $N_f=0$. The equation \eqref{eq:s1A1F8 marginal} implies that the operators of the form
  \begin{align}
      \Tr(S\widetilde{A})^{2n} (\widetilde{Q} S (\widetilde{A}S)^m\widetilde{Q})\,,\quad \text{with } 4(2n+m)+2 = N
  \end{align}
  are marginal. Here, the operators of the form $\Tr(S\widetilde{A})^{2p}$ with $p\lesssim 0.05519N$ and $\widetilde{Q}S(\widetilde{A}S)^q\widetilde{Q}$ with $q\lesssim 0.02944 N$ are decoupled and not present in the interacting SCFT.
  Nevertheless, for a given $N=4k+2$, there are $O(N)$ possible ways to partition $n$ and $m$ such that $2n+m=k$ while satisfying $n>0.05519 N$ and $m > 0.02944 N$. Thus, the theory contains $O(N)$ marginal operators, leading to an $O(N)$-dimensional conformal manifold. 
  
  For $N_f>0$, there are no integer combinations of $R$-charges that sum up to 2 in equation \eqref{eq:s1A1F8 marginal}. As a result, there is no marginal operator of this form. We have not identified any marginal operators for generic $N$ and $N_f>0$.

  When $N_f=2N-4$, the one-loop beta function for the gauge-coupling vanishes. At this value, the theory possesses a non-trivial conformal manifold \cite{Razamat:2020pra}.
  
\paragraph{Weak Gravity Conjecture}
  We test the AdS WGC using the gauge-invariant operators and $U(1)$ flavor charges identified at the beginning of this section. We find that this theory does not satisfy the NN-WGC for large values of $\a$ in the Veneziano limit, whereas the modified WGC is always satisfied. The result is shown in Figure \ref{fig:wgc_s1A1F8ven}.
\begin{figure}[h]
    \centering
     \begin{subfigure}[b]{0.45\textwidth}
     \centering
    \includegraphics[width=\linewidth]{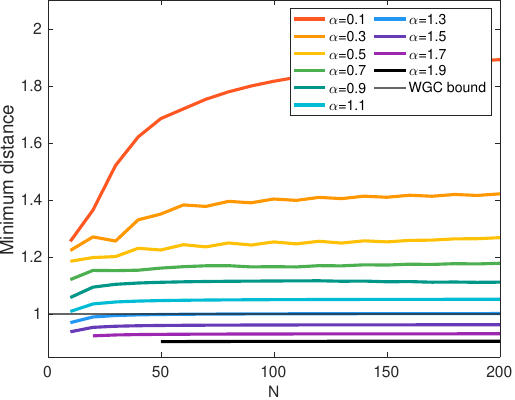}
    \caption{NN-WGC}
     \end{subfigure}
     \hspace{4mm}
     \begin{subfigure}[b]{0.45\textwidth}
     \centering
        \includegraphics[width=\linewidth]{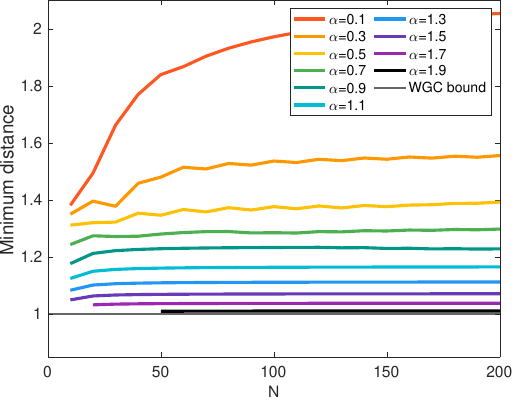}
        \caption{modified WGC}
     \end{subfigure}
     \hfill
    \caption{Testing AdS WGC for $SU(N)$ theory with 1 $\sym$ + 1 $\overline{\antisym}$ + 8 $\overline{\fund}$ + $N_f$  ( $\fund$ + $\overline{\fund}$ ).
    The minimum distance from the origin to the convex hull with a fixed $\a=N_f/N$.
    Theories below the solid line of minimum distance 1 do not satisfy the WGC.}
        \label{fig:wgc_s1A1F8ven}
 \end{figure}

\subsection{\texorpdfstring{2 $\sym$ + 2 $\overline{\sym}$ + $N_f$ ( $\fund$ + $\overline{\fund}$ )}{2 S + 2 St + Nf ( Q + Qt )}\label{sec:s2S2}}

\paragraph{Matter content and symmetry charges}
  The first example of the Type II theory is the $SU(N)$ gauge theory with two pairs of rank-2 anti-symmetric chiral multiplets and their conjugates, as well as $N_f$ pairs of fundamental and anti-fundamental chiral multiplets. The matter fields and their $U(1)$ global charges are listed in Table \ref{tab:s2S2}.
  {\renewcommand\arraystretch{1.4}
    \begin{table}[h]
        \centering
        \begin{tabular}{|c|c||c|c|c|c|c|}
        \hline
        \# & Fields & $SU(N)$  & $U(1)_1$ & $ U(1)_2$ & $U(1)_3$& $U(1)_R$ \\\hline
         $ N_f$ & $Q$ & $\fund$  & $2N+4$& 1& 0 & $R_Q$  \\ 
         $ N_f$ & $\widetilde{Q}$ & $\overline{\fund}$&  $2N+4$ &-1 & 0  & $R_{\widetilde{Q}}$ \\
         2 &$S$ & $\sym$ &  $-N_f$ &0 & 1& $R_S$ \\
         2 & $\widetilde{S}$ & $\overline{\sym}$  &  $-N_f$ &0 & -1& $R_{\widetilde{S}}$  \\\hline
        \end{tabular}
        \caption{The matter contents and their corresponding charges in $SU(N)$ gauge theory with 2 $\sym$ + 2 $\overline{\sym}$ + $N_f$ ( $\fund$ + $\overline{\fund}$) .\label{tab:s2S2}}
    \end{table}
  }

\paragraph{Gauge-invariant operators}
  Let $I$ and $J$ denote the flavor indices for $Q$, and $K$ and $L$ denote the flavor indices for $S$. We present a sample of single-trace gauge-invariant operators in schematic form, rather than providing an exhaustive list of such operators. The ellipsis indicates that only the low-lying operators have been listed. This subset is sufficient to identify relevant operators or to test the Weak Gravity Conjecture.
  \begin{enumerate}
    \item $\Tr (S_{K_1}\widetilde{S}_{\widetilde{K}_1}\cdots S_{K_n}\widetilde{S}_{\widetilde{K}_n})\,,\quad n=1,2,\dots\,.$ 
    \item $Q_{I}(\widetilde{S}_{\widetilde{K}_1}S_{K_1})\cdots(\widetilde{S}_{\widetilde{K}_n}S_{K_n})\widetilde{Q}_{\tilde{J}}\,,\quad n=0,1,\dots\,.$
    \item $Q_{I}\widetilde{S}_{\widetilde{L}}(S_{K_1}\widetilde{S}_{\widetilde{K}_1})(S_{K_n}\widetilde{S}_{\widetilde{K}_n})Q_{J},\quad n=0,1,\dots\,.$
    \item $\e (S_{(K_1} \widetilde{S}S_{K_2)})^i\CQ^{n_1}_{I_1}\cdots \CQ^{n_{N-2i}}_{I_{N-2i}},\quad$ 
    \item $\e\e S_{K_1}\cdots S_{K_{N-i}} (\CQ_{I_1}^{n_1}\CQ_{J_1}^{m_1})\cdots(\CQ_{I_i}^{n_i}\CQ_{J_i}^{m_i})$.
    \item The conjugates of the above-listed operators.
    
    $\vdots$
  \end{enumerate}
  Here, the dressed quarks are defined as
    \begin{align}
    \CQ^n_I=\begin{cases}
        (S_{K_1}\widetilde{S}_{\widetilde{K}_1} \cdots S_{K_{n/2}}\widetilde{S}_{\widetilde{K}_{n/2}})Q_I & n=0,2,4,\dots\,,\\
        (S_{K_1}\widetilde{S}_{\widetilde{K}_1} \cdots S_{K_{(n-1)/2}}\widetilde{S}_{\widetilde{K}_{(n-1)/2}})S_L\widetilde{Q}_{\tilde{I}} & n=1,3,5,\dots\,.
        \end{cases}
    \end{align}

\paragraph{$R$-charges and central charges}
  There are no decoupled operators along the RG flow, which is a generic feature of Type II (and also Type III) theory. 
  Upon $a$-maximization, we obtain the $R$-charges of the matter fields and central charges, in the large $N$ limit with a fixed $N_f$ as follows: 
  \begin{align}
    \begin{split}\label{eq:s2S2 rcharges}
        R_Q = R_{\widetilde{Q}} &\sim \frac{12-\sqrt{26}}{12}+\frac{39 N_f + 41\sqrt{26}}{312}\frac{1}{N}+O(N^{-2})\,,\\
        R_S = R_{\widetilde{S}} &\sim \frac{1}{2}+\frac{24+N_f\sqrt{26}}{24}\frac{1}{N}+O(N^{-2})\,,        
    \end{split}
  \end{align}
  The central charges $a$ and $c$ of the theory are given by 
  \begin{align} \label{eq:s2S2 central charges}
  \begin{split}
    a & \sim \frac{27}{128}N^2+\frac{198+13N_f\sqrt{26} }{768}N+O(N^0)\,, \\
    c & \sim \frac{27}{128}N^2+\frac{150+17N_f\sqrt{26} }{768}N+O(N^0)\,, \\
    \frac{a}{c} & \sim 1+\frac{2(12-N_f\sqrt{26})}{81}\frac{1}{N}+O(N^{-2})\,.
  \end{split}
  \end{align}
  When $N_f\leq  2$, the ratio $a/c$ is greater than one for sufficiently large $N$. Since a pair of rank-2 symmetric tensors contributes the same as an adjoint chiral multiplet to the central charges in the leading order in the large $N$ limit, the coefficient $27/128$ for the $N^2$ is identical to that of the two adjoint theory (which will be discussed in Section \ref{sec:sutwoadj}). We classify theories whose central charges exhibit this behavior $a\sim c \sim \frac{27}{128} \dim G$ as Type II theories.

  With the ratio $\a=N_f/N$ fixed, the $R$-charges of the matter fields, in the Veneziano limit, are given by
  \begin{align}\label{eq:s2S2ven rcharges}
  \begin{split}
    R_Q =R_{\widetilde{Q}} & \sim \frac{24-2\sqrt{26-\a^2}+3\a-3\a^2}{3(8-\a^2)}+O(N^{-1})\,,\\
    R_S =R_{\widetilde{S}} & \sim \frac{12-3\a^2+\a\sqrt{26-\a^2}}{3(8-\a^2)}+O(N^{-1})\,,
  \end{split}
  \end{align}
  and the central charges $a$ and $c$ of the theory are given by
  \begin{align}
  \begin{split}\label{eq:s2S2ven central charges}
    a & \sim \frac{648-279\a^2 +18\a^4+2\a(26-\a^2)^{3/2}}{48(8-\a^2)^2}N^2+O(N)\,,\\
    c & \sim \frac{648-303\a^2 +21\a^4+4\a(17-\a^2)\sqrt{26-\a^2}}{48(8-\a^2)^2}N^2+O(N)\,,\\
    a/c & \sim 1-\frac{13\a(8-\a^2)}{162\sqrt{26-\a^2}+\a (685-71\a^2-30\a \sqrt{26-\a^2})}+O(N^{-1})\,.
  \end{split}
  \end{align}
  These leading-order behavior of central charges in the Veneziano limit is universal across Type II theories.
  
  The ratio $a/c$ lies within the range $13/14\simeq 0.9286<a/c\leq 14001/13609\simeq1.0288$. The minimum value of $a/c$ arises in the Veneziano limit with $\a \goto 1$. The maximum value of $a/c$ occurs when $(N_f=N)=(0,5)$. Figure \ref{fig:s2S2ratio} illustrates the behavior of the ratio $a/c$.

  \begin{figure}[t]
    \centering
    \begin{subfigure}[b]{0.45\textwidth}
        \includegraphics[width=\linewidth]{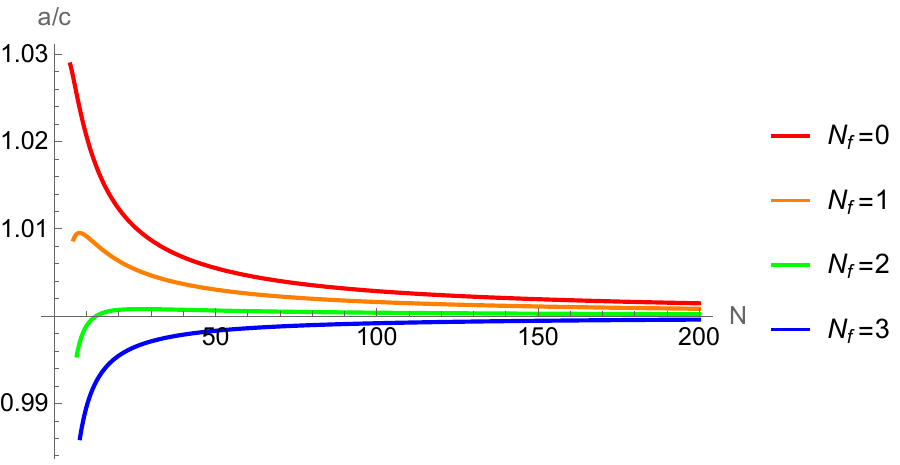}
    \end{subfigure}
    \hspace{4mm}
    \begin{subfigure}[b]{0.45\textwidth}
        \includegraphics[width=\linewidth]{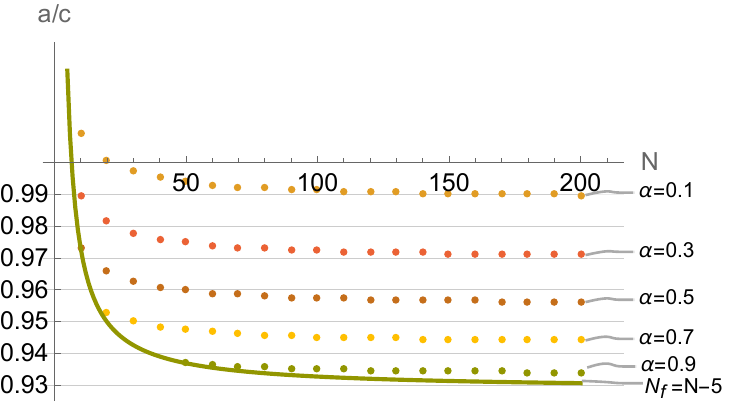}
    \end{subfigure}
    \hfill
     \caption{The central charge ratio for $SU(N)$ theory with 2 $\sym$ + 2 $\overline{\sym}$ + $N_f$ ( $\fund$ + $\overline{\fund}$ ). 
     Left: $a/c$ versus $N$ with a fixed $N_f$. Right: $a/c$ versus $N$ with a fixed $\a=N_f/N$.}
    \label{fig:s2S2ratio}   
  \end{figure}

\paragraph{Conformal window}
  As before, the upper bound of the conformal window is determined by the asymptotic freedom. 
  For small $N_f$, we find that the $a$-maximization procedure poses no issue. For instance, $a$-maximization always yields a unique solution, and the resulting central charges lie within the Hofman-Maldacena bound. 
  Therefore, we conjecture that this theory flows to a good interacting SCFT for $ 0\leq N_f<N-4$.

\paragraph{Relevant operators}
  For generic $N$ and $N_f$, there exist the following relevant operators:
  \begin{itemize}
    \item four operators of the form $\Tr S_{K}\widetilde{S}_{\widetilde{K}}$ with a dimension $\frac{3}{2} < \D < 2\,,$
    \item $N_f^2$ operators of the form $Q_I\widetilde{Q}_{\tilde{J}}$ with a dimension $1.7253\simeq \frac{12-\sqrt{26}}{4}  < \D < 2\,,$
    \item $N_f(N_f+1)$ operators of the form $ Q_I\widetilde{S}_{\widetilde{K}}Q_J$ and their conjugates with a dimension $2.4753\simeq \frac{15-\sqrt{26}}{4} < \D < 3\,.$
  \end{itemize}
  As we see, the number of relevant operators does not scale with $N$ for a fixed $N_f$. The low-lying operator spectrum is sparse (or the gap in the operator dimensions is of $O(1)$) in the large $N$ limit. 

  Upon a deformation of the form 
  \begin{align}
      W=M_1\Tr (S_1\widetilde{S}_1)^{k_1+1}+M_2\Tr(S_2\widetilde{S}_2)^{k_2+1}+M_3(\Tr S_1\widetilde{S}_2 + \Tr S_2\widetilde{S}_1)\,.
  \end{align}
  where $M_i$ are the gauge-singlet chiral superfields, a dual description is proposed \cite{Abel:2009ty}. It is given by an $SU\left((2k^*+1)N_f+4k^*-N\right)$ gauge theory, with a set of flip fields, where $k^*=\half\left[(2k_1+1)(2k_2+1)-1\right]$.
  However, we note that the operator of the form $M (S \tilde{S})^{k+1}$ is not relevant either at the $W=0$ fixed point or at the $W=M_3(\Tr S_1\widetilde{S}_2+\Tr S_2\widetilde{S}_1)$ point for $k \ge 1$, which casts a doubt on the proposed duality. 

\paragraph{Conformal manifold}
  There are no marginal operators for generic $N_f$ and $N$ in the absence of a superpotential. However, we find that with a suitable superpotential deformation, the theory can flow to a superconformal fixed point with a non-trivial conformal manifold.

  For the $N_f=0$ case, the following superpotential deformation allows the theory to possess a non-trivial conformal manifold:
  \begin{align}
    W = M_1 S_1\widetilde{S}_1 + M_2 S_1\widetilde{S}_2  + M_1 S_2\widetilde{S}_2  + M_1 M_2\,.
  \end{align}
  Here, $M_1$ and $M_2$ are flip fields, which are gauge-singlet chiral superfields. 
  We find that this superpotential deformation is relevant for all $N$. Upon deformation, we find that the IR SCFT has a sufficient number of marginal operators so that we have an exactly marginal operator. One way of checking this is to use the superconformal index \cite{Beem:2012yn}. 
  For instance, the reduced superconformal index\footnote{Reduced index is defined as $\CI_{\text{red}} = (1-t^3 y)(1-t^3/y)(\CI - 1)$ so that it removes the descendants. Here we use the standard fugacities $(t, y)$ so that the trace formula for the index is given as $\CI = \Tr (-1)^F t^{3(R+j_2)} y^{2j_1}$.} for the $SU(5)$ gauge theory with this superpotential is given by
  \begin{align}
    \CI_{\text{red}}=t^{2.14}+2t^{3.86}+2t^{5.36}+t^{5.57}+t^6 + \cdots \,.
  \end{align}
  The coefficient of $t^6$ term gives the number of marginal operators minus the number of conserved currents. If this coefficient $n$ is positive, the theory has to have at least $n$ exactly marginal operators \cite{Green:2010da, Beem:2012yn}. 
  The marginal operators take the form $\Tr S_1\widetilde{S}_2 S_1\widetilde{S}_1$ and $\Tr S_1\widetilde{S}_2 S_2 \widetilde{S}_2$. These operators remain marginal for general $N$ with a fixed $N_f=0$.

  For the $N_f=1$ case, upon the following superpotential deformation, the IR theory possesses a non-trivial conformal manifold:
  \begin{align}
    W = Q_1\widetilde{S}_1 Q_1 + \widetilde{Q}_1 S_1 \widetilde{Q}_1 + M_1 S_2\widetilde{S}_1 + M_1 S_2\widetilde{S}_2 + M_1^2\,.
  \end{align}
  The reduced index for the $SU(6)$ gauge theory with this superpotential is given by
  \begin{align}
    \begin{split}
        \CI_{\text{red}}&=2t^3+t^{3.6}+2t^{4.8}+2t^6+\cdots\,,
    \end{split}
  \end{align}
  From the coefficient of $t^6$, we see that there exist at least two exactly marginal operators. 
  The marginal operators take the form $\Tr S_{1}\widetilde{S}_{\widetilde{K}_1}S_{1}\widetilde{S}_{\widetilde{K}_2}$. This deformation and the presence of marginal operators of these forms works for general $N$ with a fixed $N_f=1$.

  We note that the superpotential that provides a non-trivial conformal manifold is not unique; other deformations can also endow the theory with a conformal manifold.

\paragraph{Weak Gravity Conjecture}
  We test the AdS WGC using the gauge-invariant operators and $U(1)$ flavor charges identified at the beginning of this section. We find that this theory does not satisfy the NN-WGC for large values of $\a$ in the Veneziano limit, whereas the modified WGC always holds. The result is shown in Figure \ref{fig:wgc_s2S2ven}.
  \begin{figure}[t]
    \centering
    \begin{subfigure}[b]{0.45\textwidth}
      \centering
      \includegraphics[width=\linewidth]{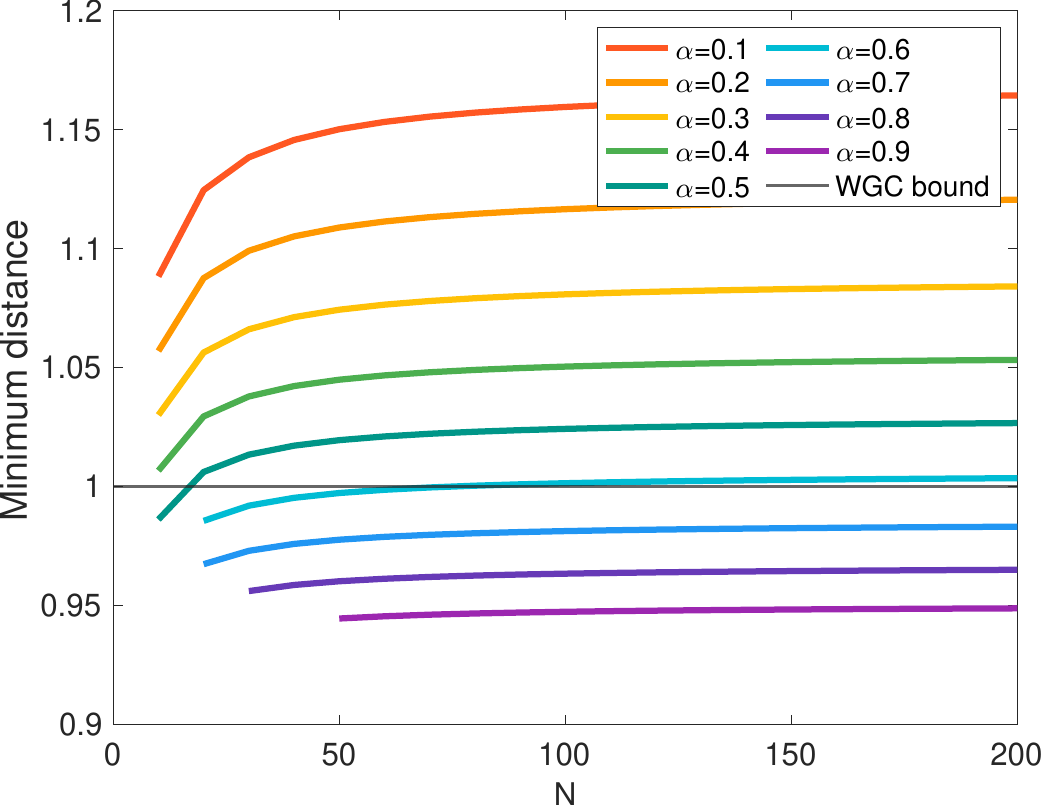}
      \caption{NN-WGC}
    \end{subfigure}
    \hspace{4mm}
    \begin{subfigure}[b]{0.45\textwidth}
      \centering
      \includegraphics[width=\linewidth]{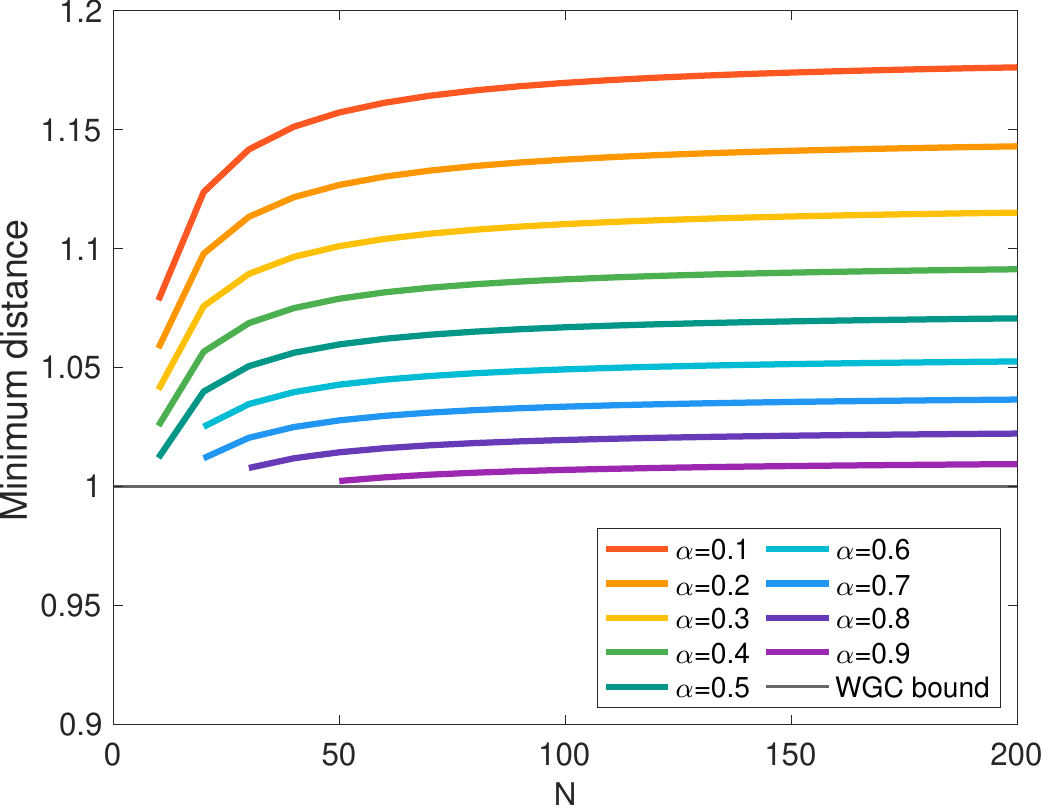}
      \caption{modified WGC}
    \end{subfigure}
    \hfill
    \caption{Testing AdS WGC for $SU(N)$ theory with 2 $\sym$ + 2 $\overline{\sym}$ + $N_f$ ( $\fund$ + $\overline{\fund}$ ). The minimum distance from the origin to the convex hull with a fixed $\a=N_f/N$. Theories below the solid line of minimum distance 1 do not satisfy the WGC.\label{fig:wgc_s2S2ven}}
   \end{figure}

\subsection{\texorpdfstring{1 $\sym$ + 2 $\overline{\sym}$ + 1 $\antisym$ + 8 $\fund$ + $N_f$ ( $\fund$ + $\overline{\fund}$ )}{1 S + 2 St + 1 A + 8 Q + Nf ( Q + Qt )}}

\paragraph{Matter content and symmetry charges}
  The next example of the Type II theory is the $SU(N)$ gauge theory with a rank-2 symmetric tensor, two conjugate of rank-2 symmetric tensors, a rank-2 anti-symmetric tensor, $N_f+8$ fundamental, and $N_f$ anti-fundamental chiral multiplets. The matter fields and their $U(1)$ global charges are listed in Table \ref{tab:s1S2a1f8}. 
  {\renewcommand\arraystretch{1.6}
  \begin{table}[h]
    \centering
    \begin{tabular}{|c|c||c|c|c|c|c|c|}
    \hline
    \# & Fields & $SU(N)$& $U(1)_1$ & $U(1)_2$ & $U(1)_3$ & $ U(1)_4$ & $U(1)_R$ \\\hline
     $ N_f+8$ & $Q$ & $\fund$ & $N+2$ & $N+2$ & $N-2$ & $N_f$ & $R_Q$\\
     $ N_f$ & $\widetilde{Q}$ & $\overline{\fund}$ & $N+2$ & $N+2$ & $N-2$ & $-N_f-8$ & $R_{\widetilde{Q}}$ \\
     1 &$S$ & $\sym$ & $-2N_f-8$ & 0 & 0 & 0 & $R_S$ \\
     2 & $\widetilde{S}$ & $\overline{\sym}$& 0 & $-N_f-4$ & 0 & 0 & $R_{\widetilde{S}}$\\
     1 & $A$& $\antisym$& 0 & 0 & $-2N_f-8$ & 0 & $R_A$\\\hline
    \end{tabular}
    \caption{The matter contents and their corresponding charges in $SU(N)$ gauge theory with 1 $\sym$ + 2 $\overline{\sym}$ + 1 $\antisym$ + 8 $\fund$ + $N_f$ ( $\fund$ + $\overline{\fund}$ ).\label{tab:s1S2a1f8}}
  \end{table}}

\paragraph{Gauge-invariant operators}
  Let $I$ and $J$ denote the flavor indices for $Q$, and $\widetilde{K}$ denote the flavor indices for $\widetilde{S}$. We present a sample of single-trace gauge-invariant operators in schematic form, rather than providing an exhaustive list of all such operators as follows: 
  \begin{enumerate}
    \item $\Tr (S\widetilde{S}_{\widetilde{K}_1}\cdots S\widetilde{S}_{\widetilde{K}_{n}})\,,\quad \Tr(A\widetilde{S}_{\widetilde{K}_1}\cdots A\widetilde{S}_{\widetilde{K}_{2n}}) \,,\quad n=1,2,\dots\,.$
    \item $\Tr (S\widetilde{S}_{\widetilde{K}_1}\cdots S\widetilde{S}_{\widetilde{K}_{n}})(A\widetilde{S}_{\widetilde{K}_1}\cdots A\widetilde{S}_{\widetilde{K}_{m}})\,,\quad n, m=1,2,\dots\,.$
    \item ${Q}_I(\widetilde{S}_{\widetilde{K}_1}S\cdots \widetilde{S}_{\widetilde{K}_{n}}S)(\widetilde{S}_{\widetilde{K}_1}A\cdots \widetilde{S}_{\widetilde{K}_{m}}A)\widetilde{Q}_{\tilde{J}}\,,\quad n, m=0,1,\dots\,.$
    \item ${Q}_I(\widetilde{S}_{\widetilde{K}_1}S\cdots \widetilde{S}_{\widetilde{K}_{n}}S)(\widetilde{S}_{\widetilde{K}_1}A\cdots \widetilde{S}_{\widetilde{K}_{m}}A)\widetilde{S}{Q}_{{J}}\,,\quad n, m=0,1,\dots\,.$
    \item $\widetilde{Q}_{\tilde{I}}S(\widetilde{S}_{\widetilde{K}_1}S\cdots \widetilde{S}_{\widetilde{K}_{n}}S)(\widetilde{S}_{\widetilde{K}_1}A\cdots \widetilde{S}_{\widetilde{K}_{m}}A)\widetilde{Q}_{\tilde{J}}\,,\quad n, m=0,1,\dots\,.$
    \item $\widetilde{Q}_{\tilde{I}}A(\widetilde{S}_{\widetilde{K}_1}S\cdots \widetilde{S}_{\widetilde{K}_{n}}S)(\widetilde{S}_{\widetilde{K}_1}A\cdots \widetilde{S}_{\widetilde{K}_{m}}A)\widetilde{Q}_{\tilde{J}}\,,\quad n, m=0,1,\dots\,.$
    \item $\e\, A^i \CQ_{{I}_1}^{n_1}\cdots \CQ_{{I}_{N-i}}^{n_{N-i}},\quad \e \,(\widetilde{S}_{\widetilde{K}_1}A\widetilde{S}_{\widetilde{K}_2})^i\widetilde{\CQ}_{I_1}^{n_1}\cdots\widetilde{\CQ}_{I_{N-2i}}^{n_{N-2i}}\,.$
    \item $\e\,\e\, S^i A^{2j}  (\CQ_{I_1}^{n_1}\CQ_{J_1}^{m_1})\cdots (\CQ_{I_k}^{n_k}\CQ_{J_k}^{m_k}),\quad i+2j+k=N$.
    \item $\e\,\e\, \widetilde{S}_{\widetilde{K}_1}\cdots\widetilde{S}_{\widetilde{K}_i} (\widetilde{\CQ}_{I_1}^{n_1}\widetilde{\CQ}_{J_1}^{m_1})\cdots (\widetilde{\CQ}_{I_{N-i}}^{n_{N-i}}\widetilde{\CQ}_{J_{N-i}}^{m_{N-i}})$.

    $\vdots$
  \end{enumerate}
  The ellipsis indicates that only the low-lying operators have been listed. This subset is sufficient to identify relevant operators or to test the Weak Gravity Conjecture.
  Here, the dressed quarks are defined as
  \begin{align}
    \CQ^n_I=\begin{cases}
        (S\widetilde{S}_{\widetilde{K}_1} \cdots S\widetilde{S}_{\widetilde{K}_i})(A\widetilde{S}_{\widetilde{K}_{i+1}} \cdots A\widetilde{S}_{\widetilde{K}_{n/2}})Q_I & n=0,2,4,\dots\\
        (S\widetilde{S}_{\widetilde{K}_1} \cdots S\widetilde{S}_{\widetilde{K}_i})(A\widetilde{S}_{\widetilde{K}_{i+1}} \cdots A\widetilde{S}_{\widetilde{K}_{(n-1)/2}})S\widetilde{Q}_{\tilde{I}} & n=1,3,5,\dots
    \end{cases}\,,
  \end{align}
  and
  \begin{align}
    \widetilde{\CQ}^n_I=\begin{cases}
        (\widetilde{S}_{\widetilde{K}_1}S)\cdots(\widetilde{S}_{\widetilde{K}_i}S)(\widetilde{S}_{\widetilde{K}_{i+1}}A)\cdots(\widetilde{S}_{\widetilde{K}_{n/2}}A)\widetilde{Q}_{\tilde{I}} & n=0,2,4,\dots\\
        (\widetilde{S}_{\widetilde{K}_1}S)\cdots(\widetilde{S}_{\widetilde{K}_i}S)(\widetilde{S}_{\widetilde{K}_{i+1}}A)\cdots(\widetilde{S}_{\widetilde{K}_{(n-1)/2}}A)\widetilde{S}{Q}_{{I}} & n=1,3,5,\dots
    \end{cases}\,,
  \end{align}

\paragraph{$R$-charges and central charges}
  We perform the $a$-maximization to compute the $R$-charges. No operators are decoupled along the RG flow, as is the case for all the Type II theories. 
  The $R$-charges of the matter fields and central charges, in the large $N$ limit with a fixed $N_f$, are given by
  \begin{align}
    \begin{split}
        R_Q=R_{\widetilde{Q}}&\sim\frac{12-\sqrt{26}}{12}+\frac{78(N_f+4)+41\sqrt{26}}{624N}+O(N^{-2})\,,\\
        R_S=R_{\widetilde{S}}&\sim\half+\frac{31+3(N_f+4)\sqrt{26}}{72N}+O(N^{-2})\,,\\
        R_A&\sim \half+\frac{17+(N_f+4)\sqrt{26}}{24N}+O(N^{-2})\,,\\
        a&\sim \frac{27}{128}N^2+ \frac{99+13(N_f+4)\sqrt{26}}{768}N+O(N^0)\,,\\
        c&\sim\frac{27}{128} N^2 + \frac{75 + 17(N_f+4)\sqrt{26}}{768} N+O(N^0)\,,\\
        a/c&\sim 1 - \frac{-12 + (N_f+4)\sqrt{26}}{81 N} + O(N^{-2})\,.
    \end{split}
  \end{align}
  The ratio $a/c$ is always less than one and it asymptotes to 1 in large $N$. The leading-order terms of the $R$-charges and central charges are universal across the Type II theories, as we have seen in the case of two pairs of rank-2 symmetric tensors, (\ref{eq:s2S2 rcharges}) and (\ref{eq:s2S2 central charges}).

  With a fixed $\a=N_f/N$, the $R$-charges of the matter fields, in the Veneziano limit, are given by
  \begin{align}
    \begin{split}
        R_Q =R_{\widetilde{Q}} & \sim\frac{24-2\sqrt{26-\a^2}+3\a-3\a^2}{3(8-\a^2)}+O(N^{-1})\,,\\
        R_S =R_{\widetilde{S}} & \sim \frac{12-3\a^2 + \a\sqrt{26-\a^2}}{3(8-\a^2)}+O(N^{-1})\,,\\
        R_A & \sim \frac{12-3\a^2 + \a\sqrt{26-\a^2}}{3(8-\a^2)}+O(N^{-1})\,,\\
    \end{split}
  \end{align}
  which exhibit the same leading behavior as in (\ref{eq:s2S2ven rcharges}) for the theory with two pairs of rank-2 symmetric tensors. Similarly, the leading-order term of the central charges matches (\ref{eq:s2S2ven central charges}):
  \begin{align}
    \begin{split}
        a & \sim \frac{648-279\a^2 +18\a^4+2\a(26-\a^2)^{3/2}}{48(8-\a^2)^2}N^2 + O(N)\,,\\
        c & \sim \frac{648-303\a^2 + 21\a^4 + 4\a(17-\a^2)\sqrt{26-\a^2}}{48(8-\a^2)^2} N^2 + O(N)\,,\\
        a/c & \sim 1-\frac{13\a(8-\a^2)}{162\sqrt{26-\a^2}+\a(685-71\a^2-30\a\sqrt{26-\a^2})}+ O(N^{-1})\,.
    \end{split}
  \end{align}
  These leading-order terms in the Veneziano limit are universal across entire Type II theories.
  
  The ratio of central charges $a/c$ lies within the range $13/14\simeq 0.9286<a/c<1$. The minimum value of $a/c$ arises in the Veneziano limit with $\a \goto 1$. The maximum value of $a/c$ occurs in the large $N$ limit with $N_f=0$. Figure \ref{fig:s1S2a1f8ratio} illustrates the behavior of the ratio $a/c$.

\begin{figure}[t]
    \centering
    \begin{subfigure}[b]{0.45\textwidth}
        \includegraphics[width=\linewidth]{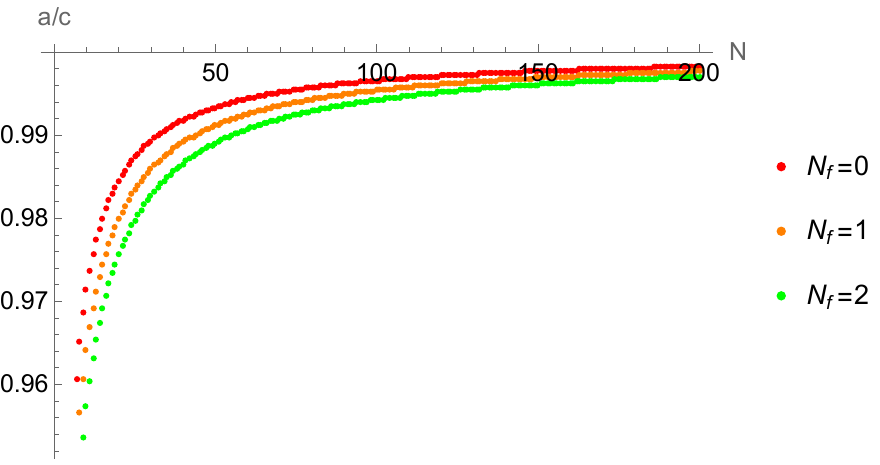}
    \end{subfigure}
    \hspace{4mm}
    \begin{subfigure}[b]{0.45\textwidth}
        \includegraphics[width=\linewidth]{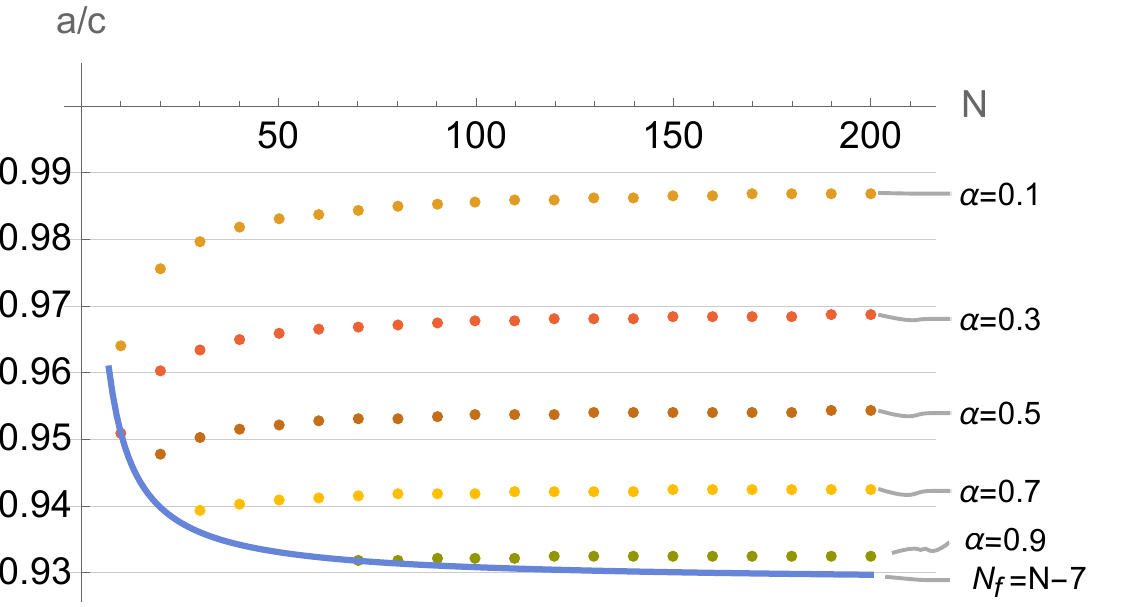}
    \end{subfigure}
    \hfill
     \caption{The central charge ratio for $SU(N)$ theory with 1 $\sym$ + 2 $\overline{\sym}$ + 1 $\antisym$ + 8 $\fund$ + $N_f$ ( $\fund$ + $\overline{\fund}$ ). 
     Left: $a/c$ versus $N$ with a fixed $N_f$. Right: $a/c$ versus $N$ with a fixed $\a=N_f/N$.}
    \label{fig:s1S2a1f8ratio}   
\end{figure}

\paragraph{Conformal window}
  The asymptotic freedom bounds the upper limit of the conformal windows to be $N_f < N-6$. 
  For small $N_f$, we find that the $a$-maximization always yields a unique solution, and the resulting central charges lie within the Hofman-Maldacena bound. Hence we conjecture that this theory flows to an interacting SCFT for $0 \leq N_f < N-6$.

\paragraph{Relevant operators}
  For generic $N$ and $N_f$, there exist following relevant operators:
  \begin{itemize}
    \item two operators of the form $\Tr S\widetilde{S}_{\widetilde{K}}$ with a dimension $\frac{3}{2}<\D<2\,,$
    \item $N_f(N_f+8)$ operators of the form $Q_I\widetilde{Q}_{\tilde{J}}$ with a dimension $1.7253\simeq\frac{12-\sqrt{26}}{4}<\D<2\,,$
    \item $(N_f+8)(N_f+9)$ operators of the form $Q_I\widetilde{S}_{\widetilde{K}}Q_J$ and $\frac{N_f(N_f+1)}{2}$ operators of the form $\widetilde{Q}_{\tilde{I}}S\widetilde{Q}_{\tilde{J}}$ with a dimension $ 2.4753\simeq \frac{15-\sqrt{26}}{4}<\D<3\,,$
    \item $\frac{N_f(N_f-1)}{2}$ operators of the form $\widetilde{Q}_{\tilde{I}}A\widetilde{Q}_{\tilde{J}}$ with a dimension $ 2.4753\simeq\frac{15-\sqrt{26}}{4}<\D<3\,.$
  \end{itemize}
  As we see, the number of relevant operators does not scale with $N$ for a fixed $N_f$. The low-lying operator spectrum is sparse (or the gap in the operator dimensions is of $O(1)$) in the large $N$ limit. 

\paragraph{Conformal manifold}
  There are no marginal operators for generic $N_f$ and $N$ in the absence of a superpotential. However, we find that upon a suitable superpotential deformation, the theory flows to a superconformal fixed point with a non-trivial conformal manifold.

  For the $N_f=0$ case, upon the following superpotential deformation, a non-trivial conformal manifold emerges at the IR fixed point: 
  \begin{align}
    W=Q_1\widetilde{S}_1 Q_2 + Q_3 \widetilde{S}_1 Q_4 + Q_5 \widetilde{S}_2 Q_6 + Q_7 \widetilde{S}_2 Q_8\,.
  \end{align}
  We can test this by computing the reduced superconformal index. For example, the superconformal index for the $SU(7)$ gauge theory with this superpotential is given by
  \begin{align}
    \begin{split}
        \CI_{\text{red}}&=2 t^{3.85}+3 t^{6} + \cdots \,.
    \end{split}
  \end{align}
  The positivity of the coefficient at the $t^6$ term indicates the existence of a non-trivial conformal manifold. The marginal operators include, for example, $Q_5\widetilde{S}_1 Q_5$, $Q_5\widetilde{S}_1 Q_7$, $Q_1\widetilde{S}_2 Q_1$, $Q_1\widetilde{S}_2 Q_3$, and others. This deformation and the presence of marginal operators of these forms are universal for general $N$ with a fixed $N_f=0$. 

  For the $N_f=1$ case, consider the following superpotential deformation: 
  \begin{align}
    W= M_1 Q_1\widetilde{Q}_1 + Q_2\widetilde{S}_1 Q_3 + Q_4 \widetilde{S}_1 Q_5 + Q_1\widetilde{S}_2 Q_6 + M_1 M_2 + Q_7\widetilde{S}_2 Q_8\,.
  \end{align}
  Here, $M_1$ and $M_2$ are flip fields, which are gauge-singlet chiral superfields. 
  The IR fixed point theory possesses a non-trivial conformal manifold. 
  We can verify this using the superconformal index. For instance, the reduced index for the $SU(8)$ gauge theory with this superpotential is given by
  \begin{align}
  \begin{split}
    \CI_{\text{red}}&=2 t^{3.87}+t^{3.93}+8 t^{3.99}+t^{5.85}+2 t^{5.88}+8 t^{5.94}+t^{6}+\cdots \,.
  \end{split}
  \end{align}
  See the positive coefficient for $t^6$. 
  The marginal operators include, for example, $Q_1\widetilde{S}_1 Q_1$, $Q_1\widetilde{S}_1 Q_7$, $Q_2\widetilde{S}_2 Q_2$, $Q_2\widetilde{S}_2 Q_4$, and others. Such a deformation and marginal operators exist for general $N$ with a fixed $N_f=1$.

\paragraph{Weak Gravity Conjecture}
  
  \begin{figure}[t]
    \centering
    \begin{subfigure}[b]{0.45\textwidth}
      \centering
      \includegraphics[width=\linewidth]{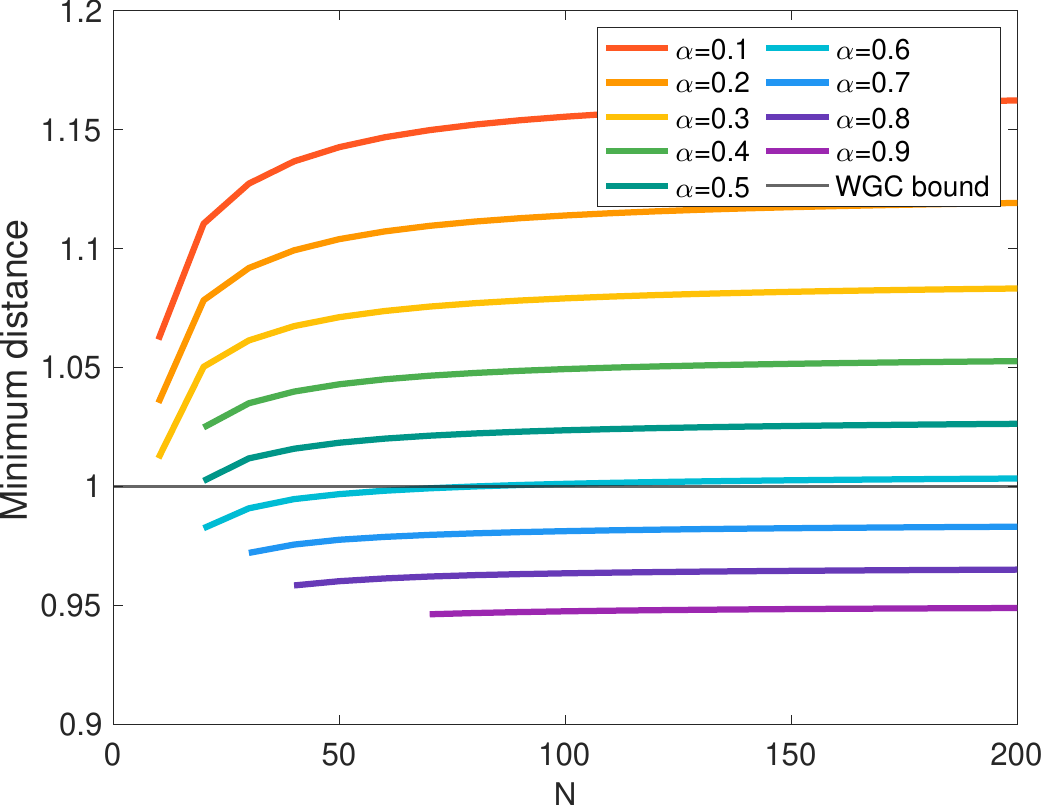}
      \caption{NN-WGC}
    \end{subfigure}
    \hspace{4mm}
    \begin{subfigure}[b]{0.45\textwidth}
      \centering
      \includegraphics[width=\linewidth]{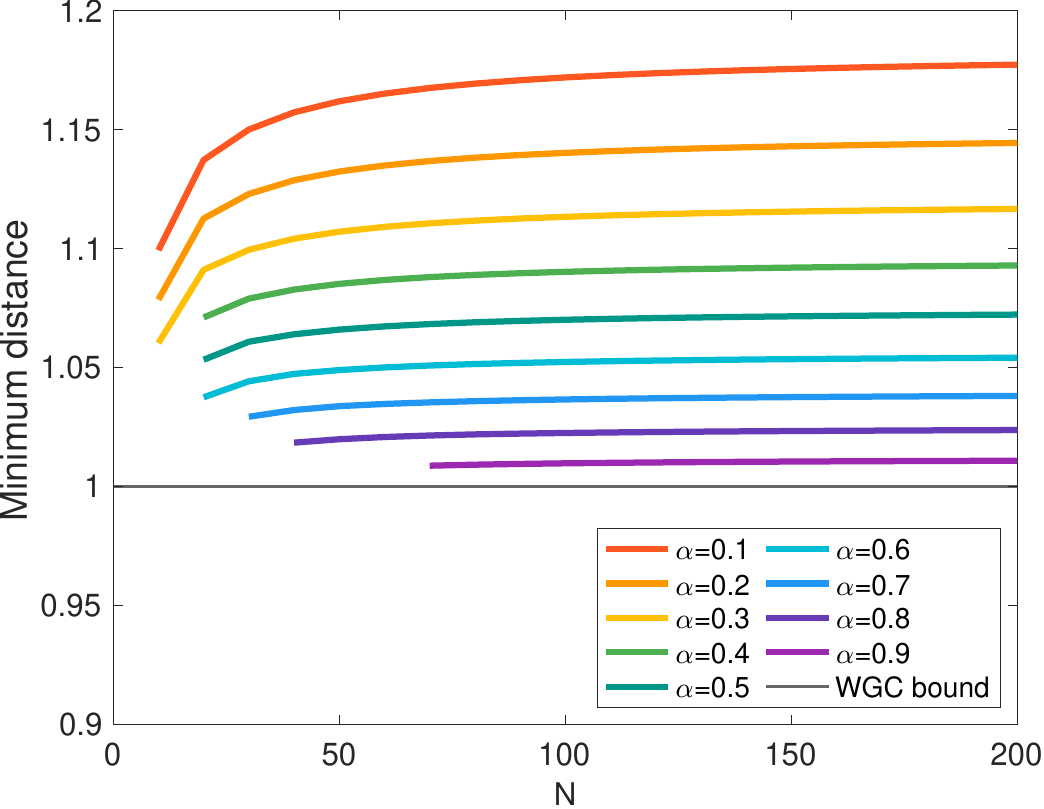}
      \caption{modified WGC}
    \end{subfigure}
    \hfill
    \caption{ 
    Testing AdS WGC for $SU(N)$ theory with 1 $\sym$ + 2 $\overline{\sym}$ + 1 $\antisym$ + 8 $\fund$ + $N_f$ ( $\fund$ + $\overline{\fund}$ ). 
    The minimum distance from the origin to the convex hull with a fixed $\a=N_f/N$. Theories below the solid line at the minimum distance 1 do not satisfy the WGC.} \label{fig:wgc_s1S2a1f8ven}
  \end{figure}
  
We test the AdS WGC using the gauge-invariant operators and $U(1)$ flavor charges identified at the beginning of this section. We find that this theory does not satisfy the NN-WGC for large values of $\a$ in the Veneziano limit, whereas the modified WGC always holds. The result is shown in Figure \ref{fig:wgc_s1S2a1f8ven}.

\subsection{\texorpdfstring{1 $\sym$ + 1 $\overline{\sym}$ + 1 $\antisym$ + 1 $\overline{\antisym}$ + $N_f$ ( $\fund$ + $\overline{\fund}$ )}{1 S + 1 St + 1 A + 1 At + Nf ( Q + Qt )}}

\paragraph{Matter content and symmetry charges}
  The third entry of the Type II theory is an $SU(N)$ gauge theory with a rank-2 symmetric, a rank-2 anti-symmetric, and their conjugate chiral multiplets, as well as $N_f$ pairs of fundamental and anti-fundamental chiral multiplets. The matter fields and their $U(1)$ global charges are listed in Table \ref{tab:s1S1a1A1}.
  {\renewcommand\arraystretch{1.6}
  \begin{table}[h]
    \centering
    \begin{tabular}{|c|c||c|c|c|c|c|c|c|}
    \hline
    \# & Fields & $SU(N)$ & $U(1)_{1}$ & $U(1)_{2}$ & $U(1)_{3}$ & $U(1)_{4}$ & $ U(1)_5$ & $U(1)_R$ \\\hline
     $N_f$ & $Q$& $\fund$ & $N+2$ & $N-2$ & $1$ & $0$ & $0$ &$R_Q$\\
     
     $N_f$ & $\widetilde{Q}$ & $\overline{\fund}$ & $N+2$ & $N-2$ & $-1$ & 0 & 0 & $R_{\widetilde{Q}}$ \\
     
     1 &$S$ &$\sym$& $-N_f$ & 0 & 0 & 1 &  0 & $R_S$ \\
     1 & $\widetilde{S}$ &$\overline{\sym}$& $-N_f$ &  0 & 0 &-1 & 0 &  $R_{\widetilde{S}}$  \\
     1 & $A$&$\antisym$ & 0 & $-N_f$ & 0 & 0 & 1 & $R_A$\\
     1 & $\widetilde{A}$ &$\overline{\antisym}$ & 0 & $-N_f$ &  0 & 0 & -1 &  $R_{\widetilde{A}}$\\\hline
    \end{tabular}
    \caption{The matter contents and their corresponding charges in $SU(N)$ gauge theory with 1 $\sym$ + 1 $\overline{\sym}$ + 1 $\antisym$ + 1 $\overline{\antisym}$ + $N_f$ ( $\fund$ + $\overline{\fund}$ ).\label{tab:s1S1a1A1}}
  \end{table}}

\paragraph{Gauge-invariant operators}
  Let $I$ and $J$ denote the flavor indices for $Q$. We present a sample of single-trace gauge-invariant operators in schematic form:
  \begin{enumerate}
    \item $\Tr (S\widetilde{S})^n\,,\quad \Tr (A\widetilde{A})^m\,,\quad \Tr (S\widetilde{A})^{2m}\,,\quad n=1,2,\dots,N-1\,,\quad m=1,2,\dots, \lfloor \frac{N-1}{2}\rfloor\,.$
    \item $\Tr (S\widetilde{S})^k(A\widetilde{A})^l(S\widetilde{A})^{m}(A\widetilde{S})^{n}\,,\quad k,l,m,n=1,2,\dots$.
    \item $Q_I(S\widetilde{S})^k(A\widetilde{A})^l(S\widetilde{A})^{m}(A\widetilde{S})^{n}\widetilde{Q}_{\tilde{J}}\,,\quad k,l,m,n=0,1,\dots$.
    \item $Q_I\widetilde{S}(S\widetilde{S})^k(A\widetilde{A})^l(S\widetilde{A})^{m}(A\widetilde{S})^{n}Q_J,\quad k,l,m,n=0,1,\dots$.
    \item $Q_I\widetilde{A}(S\widetilde{S})^k(A\widetilde{A})^l(S\widetilde{A})^{m}(A\widetilde{S})^{n}Q_J,\quad k,l,m,n=0,1,\dots$.
    \item $\e A^i \CQ_{{I}_1}^{n_1}\cdots \CQ_{{I}_{N-2i}}^{n_{N-2i}},\quad \e \widetilde{A}^i \widetilde{\CQ}_{I_1}^{n_1}\cdots \widetilde{\CQ}_{I_{N-2i}}^{n_{N-2i}}$.
    \item $\e\,\e\, S^i A^{2j}  (\CQ_{I_1}\CQ_{J_1})\cdots (\CQ_{I_k}\CQ_{J_k}),\quad i+2j+k=N$.
    \item The conjugates of the above-listed operators.

    $\vdots$
  \end{enumerate}
  The ellipsis indicates that only the low-lying operators have been listed. Here, the dressed quarks are defined as:
  \begin{align}
    \CQ^n_I=\begin{cases}
        (S\widetilde{S})^i(S\widetilde{A})^j(A\widetilde{S})^{n/2-i-j}Q_I & n=0,2,4,\dots\,,\\
        (S\widetilde{S})^i(S\widetilde{A})^j(A\widetilde{S})^{(n-1)/2-i-j} S\widetilde{Q}_{\tilde{I}} & n=1,3,5,\dots\,.
    \end{cases}
  \end{align}
 This subset is sufficient to identify relevant operators or to test the Weak Gravity Conjecture.
  
\paragraph{$R$-charges and central charges}
  We perform the $a$-maximization to compute the $R$-charges. There are no decoupled operators along the RG flows. The $R$-charges of the matter fields and central charges, in the large $N$ limit with a fixed $N_f$, are given by
  \begin{align}
  \begin{split}
    R_Q = R_{\widetilde{Q}} & \sim \frac{12-\sqrt{26}}{12}+\frac{N_f}{8 N} +O(N^{-2})\\
    R_S = R_{\widetilde{S}} & \sim \half + \frac{-10+3N_f\sqrt{26}}{72N} +O(N^{-2})\\
    R_A = R_{\widetilde{A}} & \sim \half + \frac{10+3N_f\sqrt{26}}{72N} +O(N^{-2})\\
    a & \sim \frac{27}{128}N^2 + \frac{13\sqrt{26}}{768} N_f N - \frac{526+117N_f^2}{3072} O(N^{-1})\\
    c & \sim \frac{27}{128}N^2 + \frac{17\sqrt{26}}{768} N_f N -\frac{1162+423N_f^2}{9216} + O(N^{-1})\\
    a/c & \sim 1 -\frac{2N_f\sqrt{26}}{81N} +\frac{-1404+685 N_f^2}{6561N^2} + O(N^{-3})
  \end{split}
  \end{align}
  The ratio $a/c$ is always less than one, and it asymptotes to 1 in large $N$. The leading-order terms of the $R$-charges and central charges are universal across Type II theories.

  This theory contains an equal number of symmetric and anti-symmetric tensors. In such cases, the sub-leading $O(N)$ contribution to the central charges is proportional to $N_f$, and thus vanishes when $N_f=0$. Consequently, for $N_f=0$, we have $a-c\sim O(1)$, which is an uncommon feature.

  In the Veneziano limit with a fixed $\a=N_f/N$, the leading-order result of $a$-maximization is the same as \eqref{eq:s2S2ven rcharges} and \eqref{eq:s2S2ven central charges} in the previous section, with $R_A=R_{\widetilde{A}}\sim R_S$. This result is universal across the Type II theories.

  We find that the ratio of central charges for this theory lies within the range $0.9251\lesssim a/c<1$. The minimum value of $a/c$ arises when $(N_f,N)=(3,4)$. The maximum value of $a/c$ occurs in the large $N$ limit with $N_f=0$. Figure \ref{fig:s1S1a1A1ratio} illustrates the behavior of the ratio $a/c$. 
  \begin{figure}[t]
    \centering
    \begin{subfigure}[b]{0.4\textwidth}
      \includegraphics[width=\linewidth]{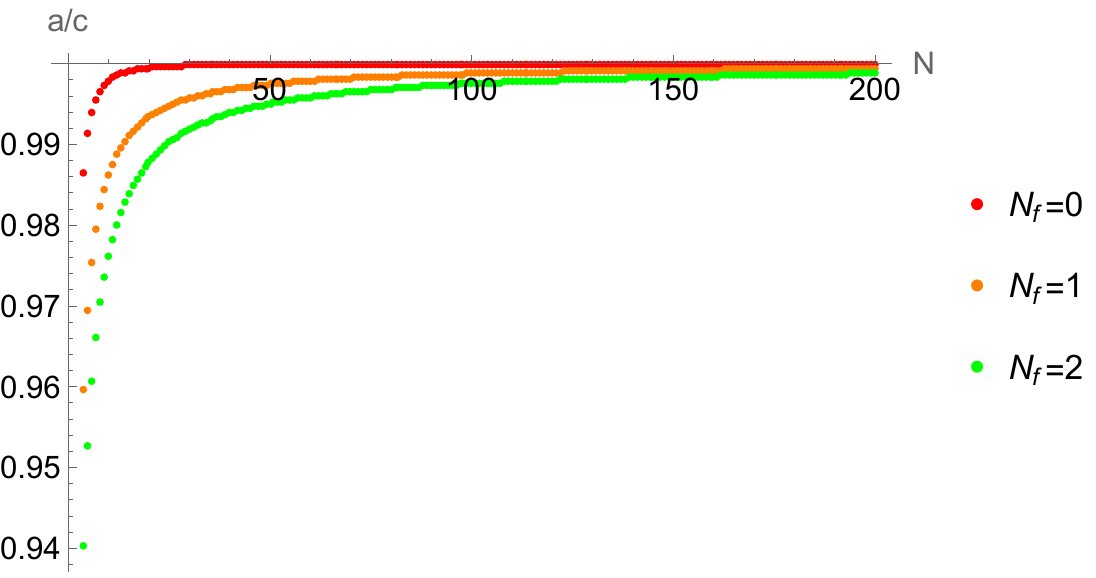}
    \end{subfigure}
    \hspace{4mm}
    \begin{subfigure}[b]{0.5\textwidth}
      \includegraphics[width=\linewidth]{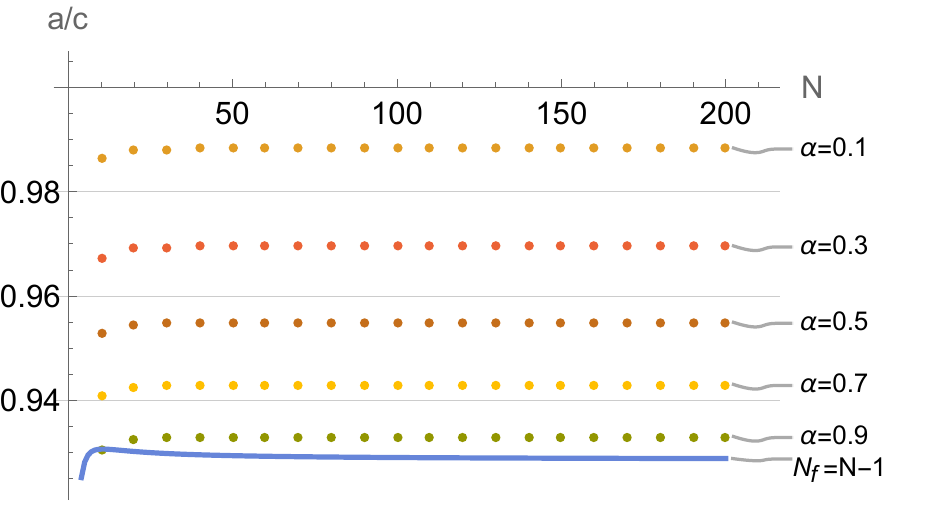}
    \end{subfigure}
    \hfill
    \caption{The central charge ratio for $SU(N)$ theory with 1 $\mbox{\small\sym}$ + 1 $\overline{\mbox{\small\sym}}$ + 1 $\mbox{\small\antisym}$ + 1 $\overline{\mbox{\small\antisym}}$ + $N_f$ ( $\mbox{\small\fund}$ + $\overline{\mbox{\small\fund}}$ ). 
     Left: $a/c$ versus $N$ with a fixed $N_f$. Right: $a/c$ versus $N$ with a fixed $\a=N_f/N$.}
    \label{fig:s1S1a1A1ratio} 
    
  \end{figure}

\paragraph{Conformal window}
  The asymptotic freedom bounds the upper limit of the conformal window to be $N_f<N$. For small $N_f$, we find that the $a$-maximization always yields a unique solution, and the resulting central charges lie within the Hofman-Maldacena bound. Hence, we conjecture that this theory flows to an interacting SCFT for $0\leq N_f< N$.

\paragraph{Relevant operators}
  For generic $N$ and $N_f$, there exist the following relevant operators:
  \begin{itemize}
    \item an operator of the form $\Tr S\widetilde{S}$ with a dimension $1.4456  \lesssim \D \leq 2\,,$ 
    \item (if $N_f=0$) an operator of the form $\Tr (S\widetilde{S})^2$ and $(\Tr S\widetilde{S})^2$ with a dimension $2.8911\lesssim \D < 3\,,$
  \end{itemize}
  \noindent where the left equality holds when $(N_f,N)=(0,4)$.
  \begin{itemize}
    \item an operator of the form $\Tr A\widetilde{A}$ with a dimension $\frac{3}{2}<\D < 2\,,$
    \item $N_f^2$ operators of the form $Q_I\widetilde{Q}_{\tilde{J}}$ with a dimension $1.7253\simeq \frac{12-\sqrt{26}}{4}<\D < 2\,,$
    \item $N_f(N_f+1)/2$ operators of the form $Q_{I}\widetilde{S}Q_J$ and their conjugate with a dimension $2.4753\simeq\frac{15-\sqrt{26}}{4} <\D\ < 3\,,$
    \item $N_f(N_f-1)/2$ operators of the form $Q_I\widetilde{A}Q_J$ and their conjugate with a dimension $2.4753\simeq \frac{15-\sqrt{26}}{4} <\D <  3\,,$
  \end{itemize}
  For certain small values of $N$, additional relevant operators not listed above exist. However, since their presence depends on the specific values of $N$ and $N_f$, we will only list the generic relevant operators that appear for general $N$.

\paragraph{Conformal manifold}
  There are no marginal operators for generic $N_f$ and $N$ in the absence of a superpotential. However, upon a suitable superpotential deformation, we find that this theory flows to a superconformal fixed point with a non-trivial conformal manifold.

  When $N_f=0$, the following superpotential deformation allows the theory to possess a non-trivial conformal manifold:
  \begin{align}
    W=\Tr(S\widetilde{S})^2\,.
  \end{align}
  We can test this by computing the reduced superconformal index. For example, the superconformal index for the $SU(5)$ gauge theory with this superpotential is given by
  \begin{align}
    \CI_{\text{red}}=2 t^{3}+t^{6}\left(5-2y-\frac{2}{y}\right)+\cdots\,.
  \end{align}
  The positivity of the coefficient at the $t^6$ term indicates that there is a non-trivial conformal manifold. 
  The marginal operators take the form $\Tr (A\widetilde{A})^2$, $(\Tr A\widetilde{A})^2$, $\Tr (S\widetilde{A})^2$, $\Tr (A\widetilde{S})^2$, $\Tr A\widetilde{A}S\widetilde{S}$, and $\Tr A\widetilde{S}S\widetilde{A}$. 
  This deformation and the marginal operators of these form exist for general $N$ with a fixed $N_f=0$.

  For the $N_f=1$ case, consider the following superpotential deformation:
  \begin{align}
    W=Q\widetilde{S}Q + \widetilde{Q}S\widetilde{Q} + M_1\Tr A\widetilde{A} + M_2 \Tr S\widetilde{S} + M_1 M_2\,.
  \end{align}
  Here, $M_1$ and $M_2$ are flip fields, which are gauge-singlet chiral superfields. It turns out that the IR fixed point theory possesses a non-trivial conformal manifold. We can verify this using the superconformal index.
  For instance, the reduced index for the $SU(5)$ gauge theory with this superpotential has a positive coefficient at the $t^6$ term:
  \begin{align}
    \CI_{\text{red}}=t^{2.57}+t^{3.43}+t^{4.29}+2 t^{4.71}+2 t^{5.14}+t^{6}\left(1-2y-\frac{2}{y}\right)+ \cdots \,.
  \end{align}
  The marginal operators take the form $\Tr (S\widetilde{A})^2$, $\Tr(A\widetilde{S})^2$, and $\Tr (S\widetilde{S}A\widetilde{A})$. Such a deformation and marginal operators exist for general $N$ with a fixed $N_f=1$.

\paragraph{Weak Gravity Conjecture}
\begin{figure}[t]
    \centering
     \begin{subfigure}[b]{0.45\textwidth}
     \centering
    \includegraphics[width=\linewidth]{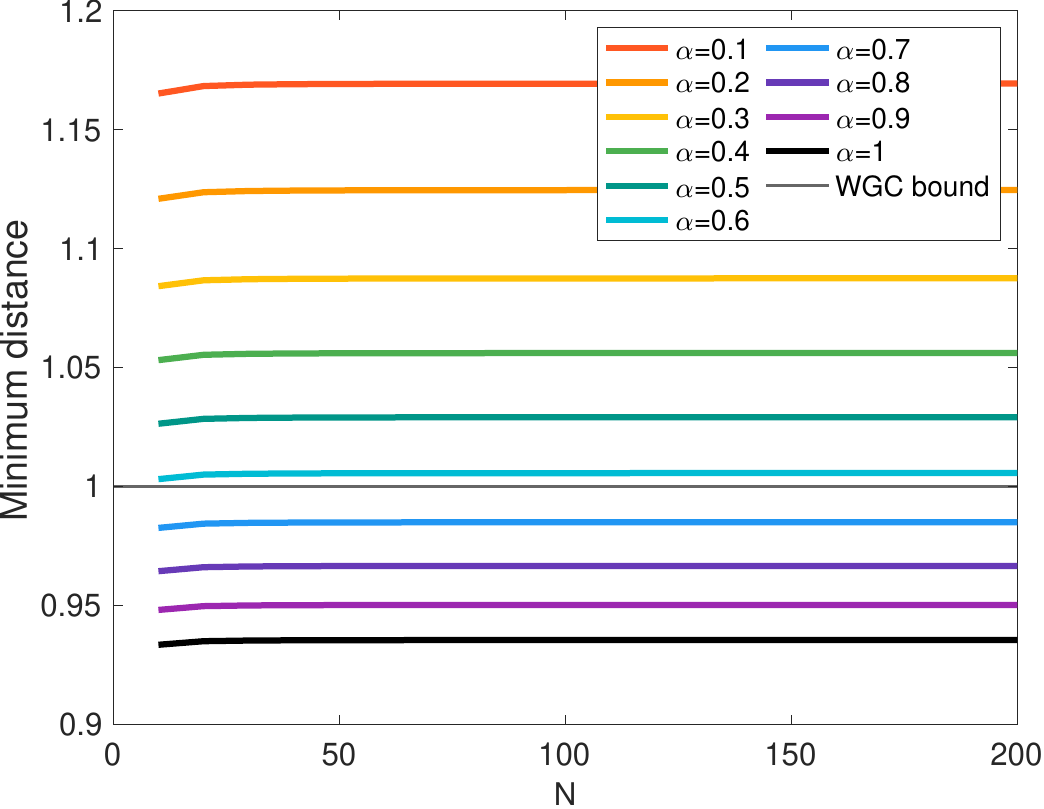}
    \caption{original WGC}
     \end{subfigure}
     \hspace{4mm}
     \begin{subfigure}[b]{0.45\textwidth}
     \centering
        \includegraphics[width=\linewidth]{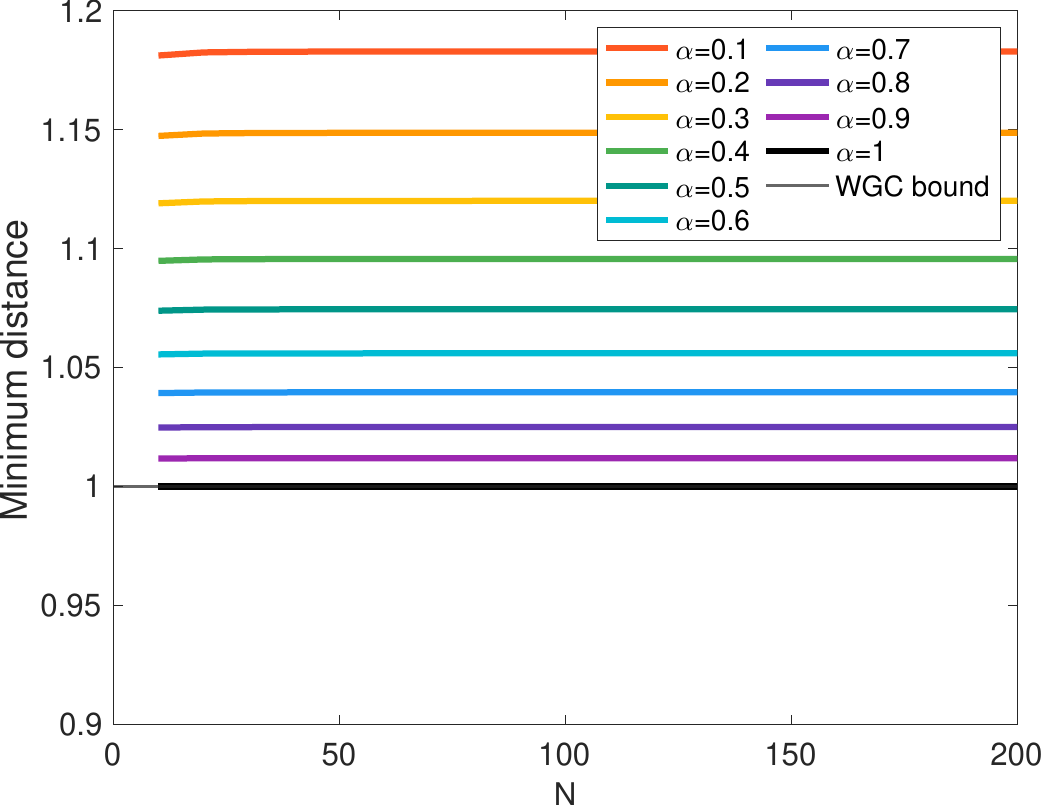}
        \caption{modified WGC}
     \end{subfigure}
     \hfill
     \caption{ 
    Testing AdS WGC for $SU(N)$ theory with 1 $\mbox{\small\sym}$ + 1 $\overline{\mbox{\small\sym}}$ + 1 $\mbox{\small\antisym}$ + 1 $\overline{\mbox{\small\antisym}}$ + $N_f$ ($\mbox{\small\fund}$ + $\overline{\mbox{\small\fund}}$). 
    The minimum distance from the origin to the convex hull with a fixed $\a=N_f/N$. Theories below the solid line at the minimum distance 1 do not satisfy the WGC.}
    \label{fig:wgc_s1S1a1A1ven}
 \end{figure}

  We test the AdS WGC using the gauge-invariant operators and $U(1)$ flavor charges identified at the beginning of this section. We find that this theory does not satisfy the NN-WGC for large values of $\a = N_f/N$ in the Veneziano limit, whereas the modified WGC always holds. The result is shown in Figure \ref{fig:wgc_s1S1a1A1ven}.

\subsection{\texorpdfstring{1 $\sym$ + 1 $\antisym$ + 2 $\overline{\antisym}$ + 8 $\overline{\fund}$ + $N_f$ ( $\fund$ + $\overline{\fund}$ )}{1 S + 1 A + 2 At + 8 Qt + Nf ( Q + Qt )}\label{subsec:s1a1A2F8}}

\paragraph{Matter content and symmetry charges}
  The fourth entry of the Type II theory is an $SU(N)$ gauge theory with a rank-2 anti-symmetric, a rank-2 symmetric, two conjugate rank-2 symmetric, $N_f$ fundamental, and $N_f+8$ anti-fundamental chiral multiplets. The matter fields and their $U(1)$ global charges are listed in Table \ref{tab:s1a1A2F8}.
  {\renewcommand\arraystretch{1.6}
  \begin{table}[h]
    \centering
    \begin{tabular}{|c|c||c|c|c|c|c|c|}
    \hline
    \#  & Fields &$SU(N)$ & $U(1)_1$ & $U(1)_2$ & $U(1)_3$ & $ U(1)_4$ & $U(1)_R$ \\\hline
    
     $N_f$ & $Q$ &$\fund$ & $N+2$ & $N-2$ & $N_f+8$ &  0 &$R_Q$\\
     
     $N_f+8$ & $\widetilde{Q}$ &$\overline{\fund}$& $N+2$  &$N-2$ &  $-N_f$ & 0 & $R_{\widetilde{Q}}$ \\
     
     1 &$S$ &$\sym$& $-2N_f-8$ & 0 & 0 & 0 & $R_S$ \\
     
     1 & $A$&$\antisym$& 0 &  $-2N_f-8$  & 0 & $2$ & $R_A$\\
     
     2 & $\widetilde{A}$ &$\overline{\antisym}$& 0 & 0 & 0 & $-1$ & $R_{\widetilde{A}}$\\\hline
    \end{tabular}
    \caption{The matter contents and their corresponding charges in $SU(N)$ gauge theory with 1 $\sym$ + 1 $\antisym$ + 2 $\overline{\antisym}$ + 8 $\overline{\fund}$ + $N_f$ ( $\fund$ + $\overline{\fund}$ ).\label{tab:s1a1A2F8}}
  \end{table}}

\paragraph{Gauge-invariant operators}
  Let $I$ and $J$ denote the flavor indices for $Q$, and $\widetilde{M}$ denote the flavor indices for $\widetilde{A}$. We present a sample of single-trace gauge-invariant operators in schematic form, rather than providing an exhaustive list of all such operators as follows: 
  \begin{enumerate}
    \item $\Tr(S\widetilde{A}_{\widetilde{M}_1}\cdots S\widetilde{A}_{\widetilde{M}_{2n}})\,,\quad \Tr (A\widetilde{A}_{\widetilde{M}_1}\cdots A\widetilde{A}_{\widetilde{M}_m})\,,\quad n=1,2,\dots\,, \quad m=1,2,\dots\,.$
  
    \item $\Tr (S\widetilde{A}_{\widetilde{M}_1}\cdots S\widetilde{A}_{\widetilde{M}_{n}})(A\widetilde{A}_{\widetilde{M}_1}\cdots A\widetilde{A}_{\widetilde{M}_{m}})\,,\quad n,m=1,2,\dots\,.$

    \item ${Q}_I(\widetilde{A}_{\widetilde{M}_1}A\cdots \widetilde{A}_{\widetilde{M}_{n}}A) (\widetilde{A}_{\widetilde{M}_1}S\cdots \widetilde{A}_{\widetilde{M}_{m}}S) \widetilde{Q}_{\tilde{J}}\,, \quad n,m = 0,1,\dots\,.$

    \item ${Q}_I(\widetilde{A}_{\widetilde{M}_1}A\cdots \widetilde{A}_{\widetilde{M}_{n}}A)(\widetilde{A}_{\widetilde{M}_1}S\cdots \widetilde{A}_{\widetilde{M}_{m}}S)\widetilde{A}{Q}_{{J}}\,,\quad n, m=0,1,\dots\,.$
    
    \item $\widetilde{Q}_{\tilde{I}}S(\widetilde{A}_{\widetilde{M}_1}A\cdots \widetilde{A}_{\widetilde{M}_{n}}A)(\widetilde{A}_{\widetilde{M}_1}S\cdots \widetilde{A}_{\widetilde{M}_{m}}S)\widetilde{Q}_{\tilde{J}}\,,\quad n, m=0,1,\dots\,.$
    
    \item $\widetilde{Q}_{\tilde{I}}A(\widetilde{A}_{\widetilde{M}_1}A\cdots \widetilde{A}_{\widetilde{M}_{n}}A)(\widetilde{A}_{\widetilde{M}_1}S\cdots \widetilde{A}_{\widetilde{M}_{m}}S)\widetilde{Q}_{\tilde{J}}\,,\quad n, m=0,1,\dots\,.$
    
    \item $\e\, A^i \CQ_{{I}_1}^{n_1}\cdots \CQ_{{I}_{N-i}}^{n_{N-i}},\quad \e \,(\widetilde{A}_{\widetilde{M}_1}\cdots \widetilde{A}_{\widetilde{M}_i})\widetilde{\CQ}_{I_1}^{n_1}\cdots\widetilde{\CQ}_{I_{N-2i}}^{n_{N-2i}}\,.$
    
    \item $\e\,\e\, S^i A^{2j}  (\CQ_{I_1}^{n_1}\CQ_{J_1}^{m_1})\cdots (\CQ_{I_k}^{n_k}\CQ_{J_k}^{m_k}),\quad i+2j+k=N$.
    
    \item $\e\,\e\,\widetilde{A}_{\widetilde{M}_1}^{ \lfloor N/2\rfloor}(\widetilde{A}_{(\widetilde{M}_1}A)\widetilde{A}_{\widetilde{M}_2)}^{N - \lfloor N/2\rfloor},\quad\epsilon\,\epsilon\, \widetilde{A}_{\widetilde{M}_1}^{\lfloor N/2\rfloor}(\widetilde{A}_{(\widetilde{M}_1}S)\widetilde{A}_{\widetilde{M}_2)}^{N - \lfloor N/2\rfloor}\quad$ if $N$ is odd.

    $\vdots$
  \end{enumerate}
  The ellipsis indicates that only the low-lying operators have been listed. This subset is sufficient to identify relevant operators or to test the Weak Gravity Conjecture.
  Here, the dressed quarks are defined by:
  \begin{align}
    \CQ^n_I=\begin{cases}
        (S\widetilde{A}_{\widetilde{M}_{1}} \cdots S\widetilde{A}_{\widetilde{M}_{n/2}})Q_I & n=0,2,4,\dots\\
        (S\widetilde{A}_{\widetilde{M}_{1}} \cdots S\widetilde{A}_{\widetilde{M}_{(n-1)/2}})S\widetilde{Q}_{\tilde{I}} & n=1,3,5,\dots
    \end{cases}\,,
  \end{align}
  and
  \begin{align}
    \widetilde{\CQ}^n_I=\begin{cases}
        (\widetilde{A}_{\widetilde{M}_{1}}S\cdots\widetilde{A}_{\widetilde{M}_{i}}S)(\widetilde{A}_{\widetilde{M}_{i+1}}A\cdots\widetilde{A}_{\widetilde{M}_{n/2}}A)\widetilde{Q}_{\tilde{I}} & n=0,2,4,\dots\\
        (\widetilde{A}_{\widetilde{M}_{1}}S\cdots\widetilde{A}_{\widetilde{M}_{i}}S)(\widetilde{A}_{\widetilde{M}_{i+1}}A\cdots\widetilde{A}_{\widetilde{M}_{n/2}}A)\widetilde{A}{Q}_{{I}} & n=1,3,5,\dots
    \end{cases}\,,
  \end{align}

\paragraph{$R$-charges and central charges}
  We perform the $a$-maximization to compute the $R$-charges. There are no decoupled operators along the RG flows. The $R$-charges of the matter fields and central charges, in the large $N$ limit with a fixed $N_f$, are given by
  \begin{align}
  \begin{split}
    R_Q = R_{\widetilde{Q}} & \sim \frac{12-\sqrt{26}}{12} + \frac{-41\sqrt{26} + 78(N_f+4)}{624N}+ O(N^{-2})\,, \\
    R_S & \sim \half + \frac{-17 + (N_f+4)\sqrt{26}}{24N}+ O(N^{-2})\,, \\
    R_A = R_{\widetilde{A}} & \sim \half + \frac{-31 + 3(N_f+4)\sqrt{26}}{72N}+ O(N^{-2})\,, \\
    a & \sim \frac{27}{128}N^2 + \frac{-99+13(N_f+4)\sqrt{26}}{768}N+O(N^0)\,,\\
    c & \sim \frac{27}{128}N^2 + \frac{-75+17(N_f+4)\sqrt{26}}{768}N+O(N^0)\,,\\
    a/c & \sim 1- \frac{12 + 2(N_f+4)\sqrt{26}}{81N}+O(N^{-2})
  \end{split}
  \end{align}
  The ratio $a/c$ is always less than one, and it asymptotes to 1 in large $N$. The leading-order behavior of the $R$-charges and central charges is universal across Type II theories.
  
  In the Veneziano limit with a fixed $\a=N_f/N$, the leading-order result of $a$-maximization is the same as \eqref{eq:s2S2ven rcharges} and \eqref{eq:s2S2ven central charges} in the previous section, with $R_A=R_{\widetilde{A}}\sim R_S$. The result is universal across the Type II theories.

  We find that the ratio of central charges for this theory lies within the range $(660981-6380\sqrt{321})/641602\simeq 0.8520 < a / c < 1$. The minimum value of $a/c$ arises when $(N_f,N)=(0,3)$. The maximum value of $a/c$ occurs in the large $N$ limit with $N_f=0$. Figure \ref{fig:s1a1A2F8ratio} illustrates the behavior of the ratio $a/c$.
  \begin{figure}[t]
    \centering
    \begin{subfigure}[b]{0.45\textwidth}
        \includegraphics[width=\linewidth]{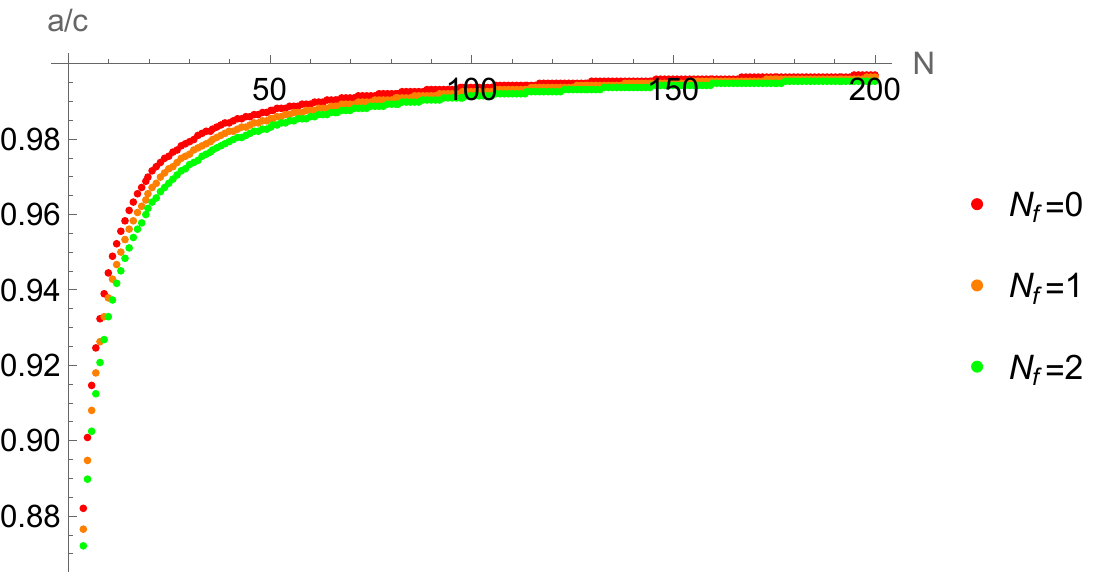}
    \end{subfigure}
    \hspace{4mm}
    \begin{subfigure}[b]{0.45\textwidth}
        \includegraphics[width=\linewidth]{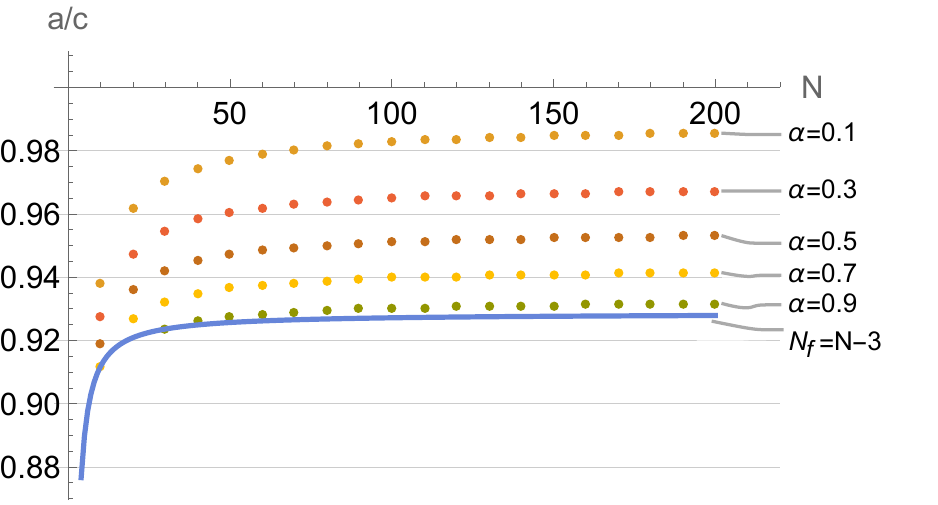}
    \end{subfigure}
    \hfill
     \caption{
     The central charge ratio for $SU(N)$ theory with 1 $\sym$ + 1 $\antisym$ + 2 $\overline{\antisym}$ + 8 $\overline{\fund}$ + $N_f$ ( $\fund$ + $\overline{\fund}$ ). 
     Left: $a/c$ versus $N$ with a fixed $N_f$. Right: $a/c$ versus $N$ with a fixed $\a=N_f/N$.}
    \label{fig:s1a1A2F8ratio}   
  \end{figure}

\paragraph{Conformal window}
  The asymptotic freedom bounds the upper limit of the conformal windows to be $N_f<N-2$. For small $N_f$, we find that the $a$-maximization always yields a unique solution, and the resulting central charges lie within the Hofman-Maldacena bound. Hence, we conjecture that this theory flows to an interacting SCFT for $0\leq N_f<N-2$.

\paragraph{Relevant operators}
  For generic $N$ and $N_f$, there exist the following relevant operators:
  \begin{itemize}
    \item an operator of the form $\Tr A\widetilde{A}$ with a dimension $\frac{3}{2}<\D <2\,,$
    \item $N_f(N_f+8)$ operators of the form $Q_I\widetilde{Q}_{\tilde{J}}$ with a dimension $1.7253\simeq \frac{12-\sqrt{26}}{4}<\D<2\,,$
    \item $\frac{(N_f+8)(N_f+9)}{2}$ operators of the form $\widetilde{Q}_{\tilde{I}}S\widetilde{Q}_{\tilde{J}}$ with a dimension $2.4753\simeq \frac{15-\sqrt{26}}{4}<\D<3\,,$
    \item $\frac{(N_f+7)(N_f+8)}{2}$ operators of the form $\widetilde{Q}_{\tilde{I}} A \widetilde{Q}_{\tilde{J}}$ with a dimension $2.4753\simeq \frac{15-\sqrt{26}}{4}<\D<3\,,$
    \item $N_f(N_f-1)$ operators of the form $Q_{I}\widetilde{A}_{\widetilde{M}} Q_{J}$ with a dimension $2.4753\simeq \frac{15-\sqrt{26}}{4}<\D<3\,.$    
  \end{itemize}
  The number of relevant operators does not scale with $N$ for a fixed $N_f$. The low-lying operator spectrum is sparse in the large $N$ limit.

\paragraph{Conformal manifold}
  There are no marginal operators for generic $N_f$ and $N$ in the absence of a superpotential. However, we find that upon a suitable superpotential deformation, the theory flows to a superconformal fixed point with a non-trivial conformal manifold.

  For the $N_f=0$ case, upon the following superpotential deformation, a non-trivial conformal manifold emerges at the IR fixed point:
  \begin{align}
    W= M_1\Tr A\widetilde{A}_1 + M_2 \Tr A\widetilde{A}_2 + M_1 M_2 + \widetilde{Q}_1 S\widetilde{Q}_2 + \widetilde{Q}_1 A\widetilde{Q}_2 \,.
  \end{align}
  Here, $M_1$ and $M_2$ are flip fields, which are gauge-singlet chiral superfields. We can test this by computing the superconformal index. For example, the reduced superconformal index for the $SU(5)$ gauge theory with this superpotential is given by
  \begin{align}
    \CI_{\text{red}}=2 t^{3}+24 t^{5}+t^{6}\left( 5-2y-\frac{2}{y}\right) + \cdots \,.
  \end{align}
  The positivity of the coefficient at the $t^6$ term indicates the existence of a non-trivial conformal manifold. 
  The marginal operators include, for example, $\Tr(A\widetilde{A}_1)^2$, $\Tr A \widetilde{A}_1 S\widetilde{A}_2$, $M_1^2$, $\widetilde{Q}_1 A\widetilde{Q}_3$, $\widetilde{Q}_3 S\widetilde{Q}_3$, and others. 
  This deformation and the marginal operators of such form exist for general $N$ with $N_f=0$.

  For the $N_f=1$ case, consider the following superpotential deformation:
  \begin{align}
    W=M_1\Tr A\widetilde{A}_1 + \widetilde{Q}_1 A \widetilde{Q}_2 + \widetilde{Q}_1 S \widetilde{Q}_2 + M_2\Tr A\widetilde{A}_2 + Q\widetilde{Q}_1 + M_1M_2\,.
  \end{align}
  We find that a non-trivial conformal manifold emerges at the IR fixed point. We can verify this using the superconformal index. For instance, the reduced index for the $SU(5)$ gauge theory with this superpotential has a positive coefficient at the $t^6$ term:
  \begin{align}
    \CI_{\text{red}}=2t^3 + 24 t^5 + t^6 \left(5-2y-2\frac{1}{y}\right) + \cdots\,.
  \end{align}
  The marginal operators include, for example, $\Tr (A\widetilde{A}_1)^2$, $\Tr A\widetilde{A}_1 S\widetilde{A}_2$, $M_1^2$, $Q\widetilde{Q}_3$, $\widetilde{Q}_3 A\widetilde{Q}_4$, $\widetilde{Q}_3 S \widetilde{Q}_3$, and others. Such a deformation and the marginal operators exist for general $N$ with a fixed $N_f=1$.

  \begin{figure}[t]
    \centering
    \begin{subfigure}[b]{0.45\textwidth}
      \centering
      \includegraphics[width=\linewidth]{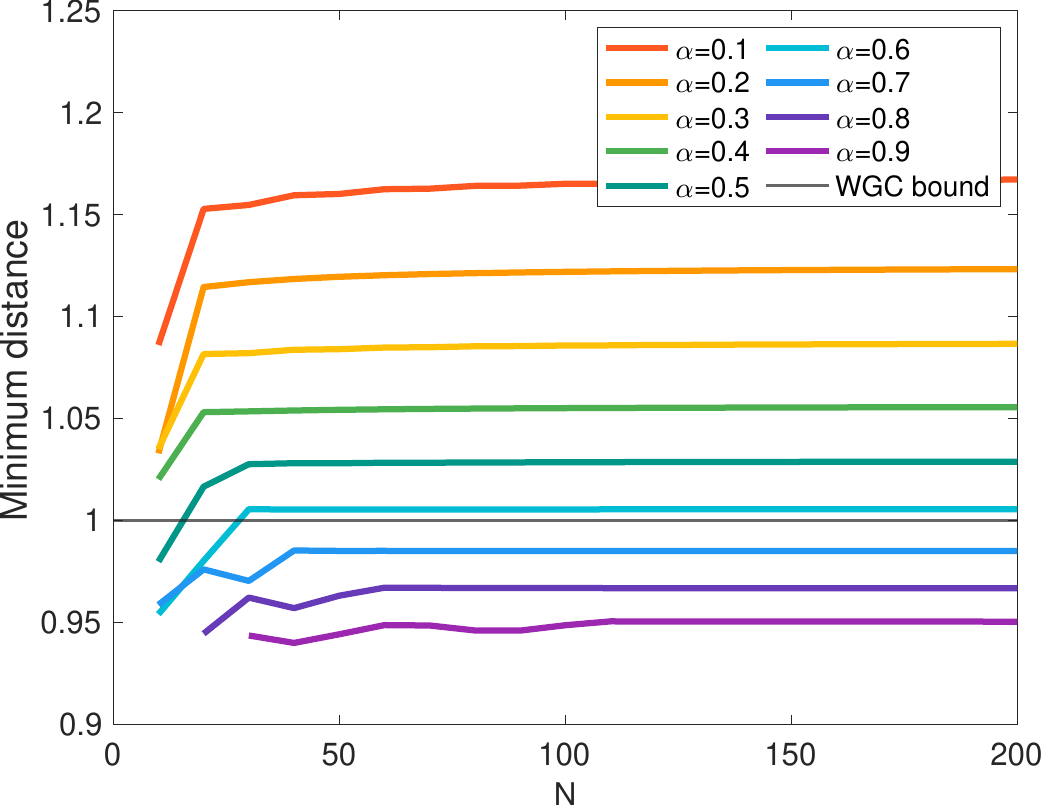}
      \caption{original WGC}
    \end{subfigure}
    \hspace{4mm}
    \begin{subfigure}[b]{0.45\textwidth}
      \centering
      \includegraphics[width=\linewidth]{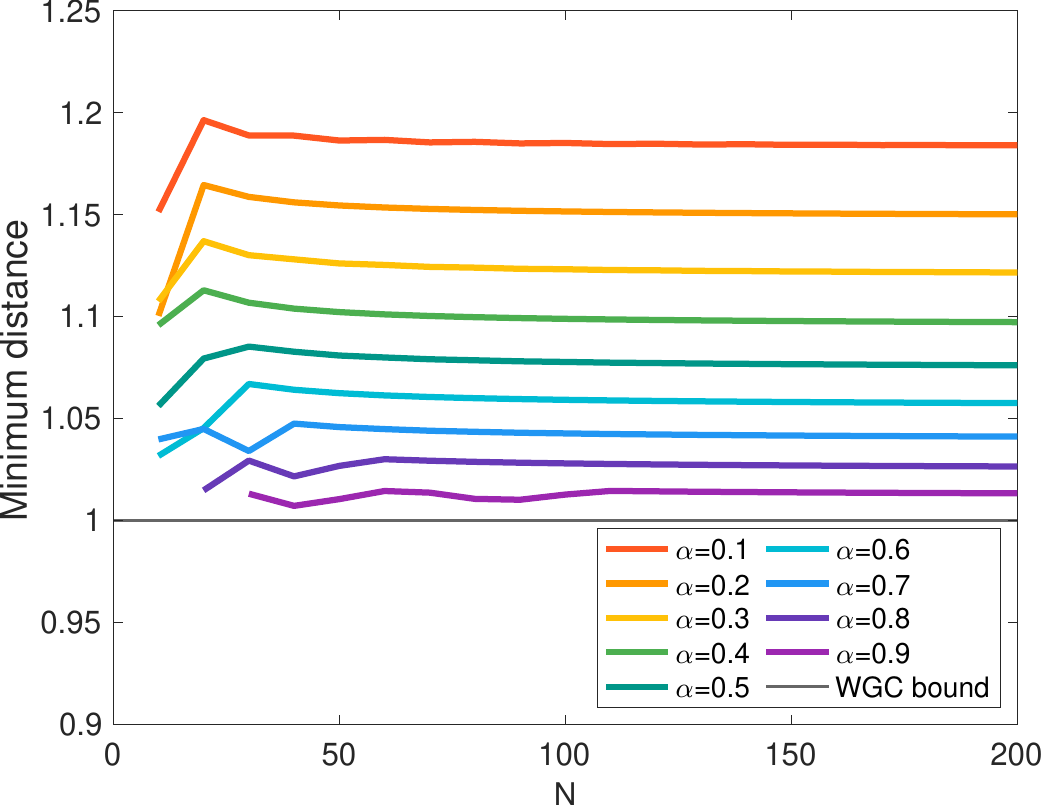}
      \caption{modified WGC}
    \end{subfigure}
    \hfill
    \caption{Testing AdS WGC for $SU(N)$ theory with 1 $\sym$ + 1 $\antisym$ + 2 $\overline{\antisym}$ + 8 $\overline{\fund}$ + $N_f$ ( $\fund$ + $\overline{\fund}$ ). The minimum distance from the origin to the convex hull with a fixed $\a=N_f/N$. Theories below the solid line at the minimum distance 1 do not satisfy the WGC.\label{fig:wgc_s1a1A2F8ven}}
 \end{figure}

\paragraph{Weak Gravity Conjecture}

  We examine the AdS WGC using the gauge-invariant operators and $U(1)$ flavor charges identified at the beginning of this section. We find that this theory does not satisfy the NN-WGC for large values of $\a = N_f/N$ in the Veneziano limit, whereas the modified WGC always holds. The result is shown in Figure \ref{fig:wgc_s1a1A2F8ven}.

\subsection{\texorpdfstring{2 $\sym$ + 2 $\overline{\antisym}$ + 16 $\overline{\fund}$ + $N_f$ ( $\fund$ + $\overline{\fund}$ )}{2 S + 2 At + 16 Qt + Nf ( Q + Qt )}}

\paragraph{Matter content and symmetry charges}
  Next case of the Type II theory is given by $SU(N)$ gauge theory with two rank-2 symmetric, two conjugate rank-2 anti-symmetric, $N_f$ fundamental, and $N_f+16$ anti-fundamental chiral multiplets. The matter fields and their $U(1)$ global charges are listed in Table \ref{tab:s2A2F16}.
  {\renewcommand\arraystretch{1.6}
  \begin{table}[h]
    \centering
    \begin{tabular}{|c|c||c|c|c|c|c|}
    \hline
    \# & Fields &$SU(N)$&$U(1)_1$ & $U(1)_2$ & $ U(1)_3$ & $U(1)_R$ \\\hline
    
     $N_f$ & $Q$ &$\fund$& $N+2$ & $N-2$ & $N_f+16$ & $R_Q$\\
     
     $N_f+16$ & $\widetilde{Q}$ &$\overline{\fund}$& $N+2$ & $N-2$ & $-N_f$ & $R_{\widetilde{Q}}$ \\
     
     2 &$S$ &$\sym$& $-N_f-8$ & 0  & 0 & $R_S$ \\
     
     2 & $\widetilde{A}$ &$\overline{\antisym}$& 0  & $-N_f-8$ & 0 & $R_{\widetilde{A}}$\\\hline
    \end{tabular}
    \caption{The matter contents and their corresponding charges in $SU(N)$ gauge theory with 2 $\sym$ + 2 $\overline{\antisym}$ + 16 $\overline{\fund}$ + $N_f$ ( $\fund$ + $\overline{\fund}$ ).} 
    \label{tab:s2A2F16}
  \end{table}}

\paragraph{Gauge-invariant operators}
  Let $I$ and $J$ denote the flavor indices for $Q$, $K$ and $L$ denote the flavor indices for $S$, and $\widetilde{M}$ denote the flavor indices for $\widetilde{A}$. We present a sample of single-trace gauge-invariant operators in schematic form as follows: 
  \begin{enumerate}
    \item $\Tr(S_{K_1}\widetilde{A}_{\widetilde{M}_1}\cdots S_{K_n}\widetilde{A}_{\widetilde{M}_n})\,,\quad n=2,3,\dots\,.$
    \item $Q_I(\widetilde{A}_{\widetilde{M}_1}S_{K_1}\cdots\widetilde{A}_{\widetilde{M}_n} S_{K_n})\widetilde{Q}_{\tilde{J}},\quad n=0,1,\dots\,.$
    \item $Q_{I}\widetilde{A}_{\widetilde{M}_0}(S_{K_1}\widetilde{A}_{\widetilde{M}_1}\cdots S_{K_n}\widetilde{A}_{\widetilde{M}_n} )Q_{J}\,,\quad n=0,1,\dots\,.$
    \item $\widetilde{Q}_{\tilde{I}}S_L(\widetilde{A}_{\widetilde{M}_1}S_{K_1}\cdots \widetilde{A}_{\widetilde{M}_n}S_{K_n})\widetilde{Q}_{\tilde{J}},\quad n=0,1,\dots\,.$
    \item $\e\,(S_{K_1}\widetilde{A}_{\widetilde{M_1}}S_{K_2})\cdots(S_{L_1}\widetilde{A}_{\widetilde{M_i}}S_{L_2})\CQ^{n_1}_{I_1}\cdots\CQ^{n_{N-2i}}_{I_{N-2i}},\quad \e\, \widetilde{A}_{\widetilde{M}_1}\cdots\widetilde{A}_{\widetilde{M}_i} \widetilde{\CQ}^{n_1}_{\tilde{I}_1}\cdots\widetilde{\CQ}^{n_{N-2i}}_{\tilde{I}_{N-2i}}\,.$
    \item $\e\,\e\,  S_{K_1}\cdots S_{K_{i}} (\CQ_{I_1}^{n_1}\CQ_{J_1}^{m_1})\cdots (\CQ^{n_{N-i}}_{I_{N-i}}\CQ^{n_{N-i}}_{J_{N-i}})\,.$
    \item $\e\,\e\,\widetilde{A}_{\widetilde{M}_1}^{ \lfloor N/2\rfloor}(\widetilde{A}_{(\widetilde{M}_1}A)\widetilde{A}_{\widetilde{M}_2)}^{N - \lfloor N/2\rfloor},\quad\epsilon\,\epsilon\, \widetilde{A}_{\widetilde{M}_1}^{\lfloor N/2\rfloor}(\widetilde{A}_{(\widetilde{M}_1}S)\widetilde{A}_{\widetilde{M}_2)}^{N - \lfloor N/2\rfloor}\quad$ if $N$ is odd.

    $\vdots$
  \end{enumerate}
  The ellipsis indicates that only the low-lying operators have been listed. This subset is sufficient to identify relevant operators or to test the Weak Gravity Conjecture.
  Here, we define the dressed quarks as
  \begin{align}
    \CQ_I^n=\begin{cases}
        (S_{K_1}\widetilde{A}_{\widetilde{M}_1}\cdots S_{K_{n/2}}\widetilde{A}_{\widetilde{M}_{n/2}}) Q_I & n=0,2,4,\dots\,,\\
        (S_{K_1}\widetilde{A}_{\widetilde{M}_1}\cdots S_{K_{(n-1)/2}}\widetilde{A}_{\widetilde{M}_{(n-1)/2}})S\widetilde{Q}_{\tilde{I}} & n=1,3,5,\dots\,.
    \end{cases}
  \end{align}
  \begin{align}
    \widetilde{\CQ}^n_{\tilde{I}}=\begin{cases}
        (\widetilde{A}_{\widetilde{M}_1}S_{K_1}\cdots \widetilde{A}_{\widetilde{M}_{n/2}}S_{K_{n/2}})\widetilde{Q}_{\tilde{I}} & n=0,2,4,\dots\,,\\
        (\widetilde{A}_{\widetilde{M}_1}S_{K_1}\cdots \widetilde{A}_{\widetilde{M}_{(n-1)/2}}S_{K_{(n-1)/2}})\widetilde{A}Q_I & n=1,3,5,\dots\,.
    \end{cases}
  \end{align}

\paragraph{$R$-charges and central charges}
  We perform the $a$-maximization to compute the $R$-charges of the matter fields and central charges. In the large $N$ limit with a fixed $N_f$, they are given by
  \begin{align}
  \begin{split}
    R_Q = R_{\widetilde{Q}} & \sim \frac{12-\sqrt{26}}{12} + \frac{N_f+8}{8N}+ O(N^{-2})\,, \\
    R_S & \sim \half + \frac{-10 + 3(N_f+8)\sqrt{26}}{72N}+ O(N^{-2})\,, \\
    R_{\widetilde{A}} & \sim \half + \frac{10 + 3(N_f+8)\sqrt{26}}{72N}+ O(N^{-2})\,, \\
    a & \sim \frac{27}{128}N^2 + \frac{13(N_f+8)\sqrt{26}}{768}N+O(N^0)\,,\\
    c & \sim \frac{27}{128}N^2 + \frac{17(N_f+8)\sqrt{26}}{768}N+O(N^0)\,,\\
    a/c & \sim 1- \frac{2(N_f+8)\sqrt{26}}{81N} + O(N^{-2})\,.
  \end{split}
  \end{align}
  We again find that none of the gauge-invariant operators decouple along the RG flow. 
  The leading-order terms of the $R$-charges and central charges are universal across the Type II theories. Similarly, in the Veneziano limit with a fixed $\a=N_f/N$, the leading terms of $R$-charges and central charges are also universal across Type II theories.

  We find that the ratio of central charges $a/c$ is always less than one, taking a value within the range $ 13/14 \simeq 0.9286 < a/c < 1$. The minimum value of $a/c$ can be obtained in the Veneziano limit with $\a \goto 1$. The maximum value of $a/c$ occurs in the large $N$ limit with $N_f=0$. Figure \ref{fig:s2A2F16ratio} illustrates the behavior of the ratio $a/c$.
  \begin{figure}[t]
    \centering
    \begin{subfigure}[b]{0.45\textwidth}
        \includegraphics[width=\linewidth]{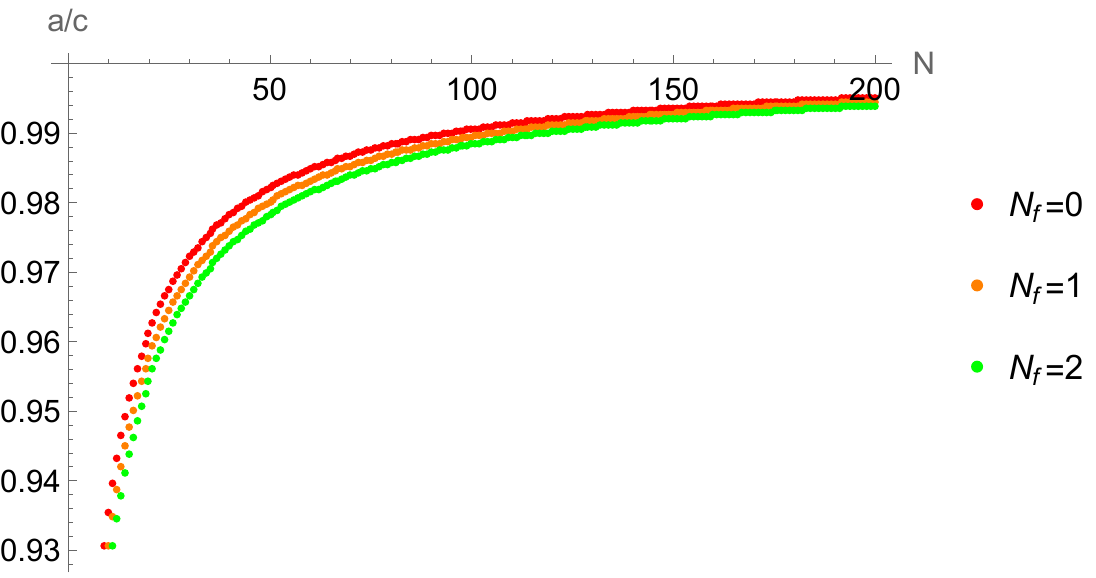}
    \end{subfigure}
    \hspace{4mm}
    \begin{subfigure}[b]{0.45\textwidth}
        \includegraphics[width=\linewidth]{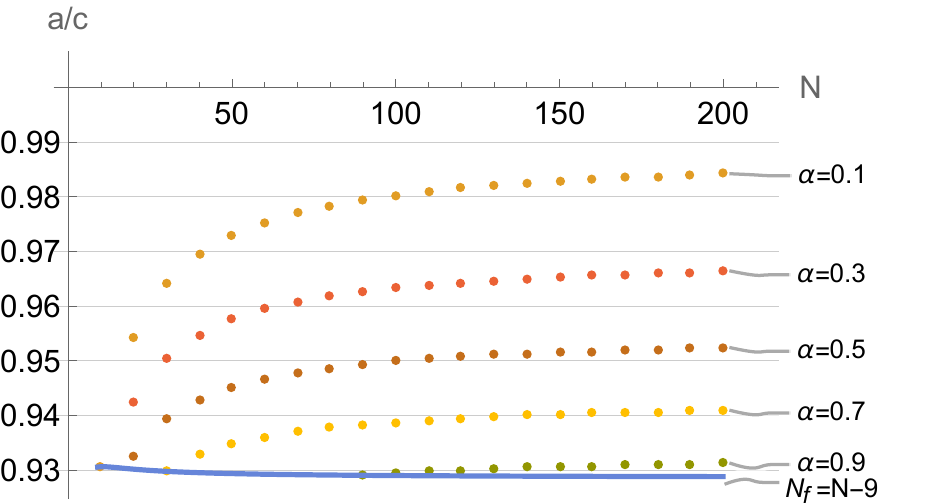}
    \end{subfigure}
    \hfill
     \caption{The central charge ratio for $SU(N)$ theory with 2 $\sym$ + 2 $\overline{\antisym}$ + 16 $\overline{\fund}$ + $N_f$ ( $\fund$ + $\overline{\fund}$ ). Left: $a/c$ versus $N$ with a fixed $N_f$. Right: $a/c$ versus $N$ with a fixed $\a=N_f/N$.}
    \label{fig:s2A2F16ratio}   
  \end{figure}

\paragraph{Conformal window}
  The upper bound of the conformal window is determined by the asymptotic freedom. For small $N_f$, we find that the $a$-maximization always yields a unique solution, and the resulting central charges lie within the Hofman-Maldacena bound. Therefore, we conjecture that this theory flows to an interacting SCFT for $0\leq N_f<N-8$.

\paragraph{Relevant operators}
  For generic $N$ and $N_f$, there exists the following set of relevant operators:
  \begin{itemize}
    \item $N_f(N_f+16)$ operators of the form $Q_I\widetilde{Q}_{\tilde{J}}$ with a dimension $1.7253\simeq \frac{12-\sqrt{26}}{4}<\D<2\,,$
    \item $(N_f+16)(N_f+17)$ operators of the form $\widetilde{Q}_{\tilde{I}}S_K\widetilde{Q}_{\tilde{J}}$ with a dimension $2.4753\simeq\frac{15-\sqrt{26}}{4} < \D < 3\,,$
    \item $N_f(N_f-1)$ operators of the form $Q_{I}\widetilde{A}_{\widetilde{M}} Q_{J}$ with a dimension $2.4753\simeq \frac{15-\sqrt{26}}{4}<\D<3\,.$
  \end{itemize}
  The number of relevant operators does not depend on $N$ for a fixed $N_f$. The low-lying operator spectrum is sparse in the large $N$ limit.

\paragraph{Conformal manifold}
  There are no marginal operators for generic $N_f$ and $N$ when the superpotential is turned off. However, upon a suitable superpotential deformation, the theory flows to a superconformal fixed point with a non-trivial conformal manifold.

  For the $N_f=0$ case, upon the following superpotential deformation, a non-trivial conformal manifold emerges at the IR fixed point:
  \begin{align}
    \begin{split}
        W & = \widetilde{Q}_1 S_1 \widetilde{Q}_2 + \widetilde{Q}_3 S_1 \widetilde{Q}_4  + \widetilde{Q}_5 S_2 \widetilde{Q}_6  + \widetilde{Q}_7 S_2 \widetilde{Q}_8  + \widetilde{Q}_9 S_2 \widetilde{Q}_{10} + \widetilde{Q}_{11} S_1 \widetilde{Q}_{11} +  \widetilde{Q}_{12} S_1\widetilde{Q}_{13}\\
        &\quad + \widetilde{Q}_{12} S_2\widetilde{Q}_{12} \,.
    \end{split}
  \end{align}
  We can test this by computing the superconformal index. For example, the reduced superconformal index for the $SU(9)$ gauge theory with this superpotential is given by
  \begin{align}
    \CI_{\text{red}}=6 t^{5.85} + 28 t^{5.93} + 3 t^6 + \cdots\,.
  \end{align}
  The positivity of the coefficient at the $t^6$ term indicates the existence of a non-trivial conformal manifold. 
  The marginal operators include, for example, $\widetilde{Q}_{10}S_1\widetilde{Q}_{10}$, $\widetilde{Q}_{10}S_1\widetilde{Q}_{12}$, $\widetilde{Q}_1S_2\widetilde{Q}_1$, $\widetilde{Q}_1S_2\widetilde{Q}_{11}$, and others. This deformation and the marginal operators of these forms persist for general $N$ with a fixed $N_f=0$.

  For the $N_f=1$ case, consider the following superpotential deformation:
  \begin{align}
  \begin{split}
      W&=\widetilde{Q}_1 S_1 \widetilde{Q}_2 + \widetilde{Q}_3 S_1 \widetilde{Q}_4 + \widetilde{Q}_5 S_1 \widetilde{Q}_6 + \widetilde{Q}_7 S_1 \widetilde{Q}_8\\
    &\quad +\widetilde{Q}_9 S_2 \widetilde{Q}_{10} + \widetilde{Q}_{11} S_2 \widetilde{Q}_{12}  + \widetilde{Q}_{13} S_2 \widetilde{Q}_{14}  + \widetilde{Q}_{15} S_2 \widetilde{Q}_{16}  \,.
  \end{split}
  \end{align}
  At the IR fixed point, this theory possesses a non-trivial conformal manifold. We can again verify this from the superconformal index. For instance, the reduced index for the $SU(10)$ gauge theory with this superpotential has a positive coefficient at the $t^6$ term:
  \begin{align}
    \CI_{\text{red}}=t^{3.92}+16 t^{3.98}+2 t^{5.87}+16 t^{5.93}+7 t^{6}+\cdots\,.
  \end{align}
  The marginal operators includes, for example, $\widetilde{Q}_1 S_2\widetilde{Q}_1$, $\widetilde{Q}_1 S_2\widetilde{Q}_3$, $\widetilde{Q}_{9} S_1\widetilde{Q}_{9}$, $\widetilde{Q}_{9} S_1\widetilde{Q}_{11}$, and others. Such a deformation and marginal operators exist for general $N$ with a fixed $N_f=1$.

\paragraph{Weak Gravity Conjecture}
  We examine the AdS WGC using the gauge-invariant operators and $U(1)$ flavor charges identified at the beginning of this section. We find that this theory does not satisfy the NN-WGC for some finite $N$ with large values of $\a$, whereas the modified WGC always holds. The result is shown in Figure \ref{fig:wgc_s2A2F16ven}. 
  \begin{figure}[t]
    \centering
    \begin{subfigure}[b]{0.45\textwidth}
      \centering
      \includegraphics[width=\linewidth]{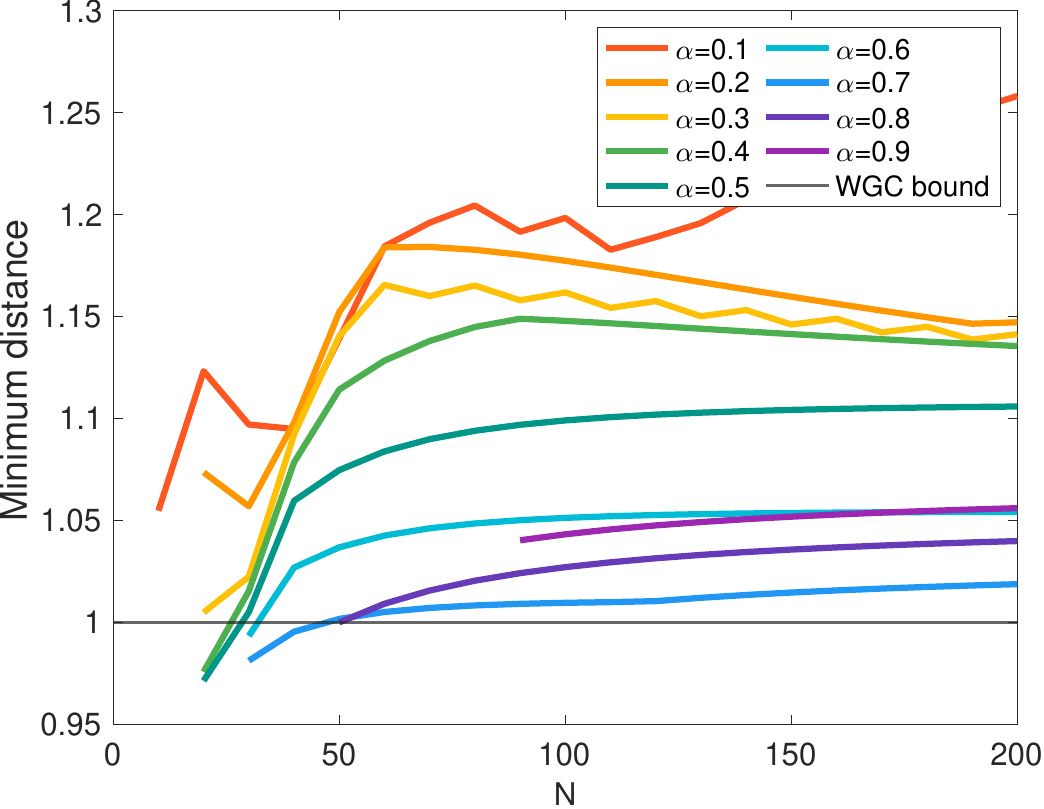}
      \caption{NN-WGC}
    \end{subfigure}
    \hspace{4mm}
    \begin{subfigure}[b]{0.45\textwidth}
      \centering
      \includegraphics[width=\linewidth]{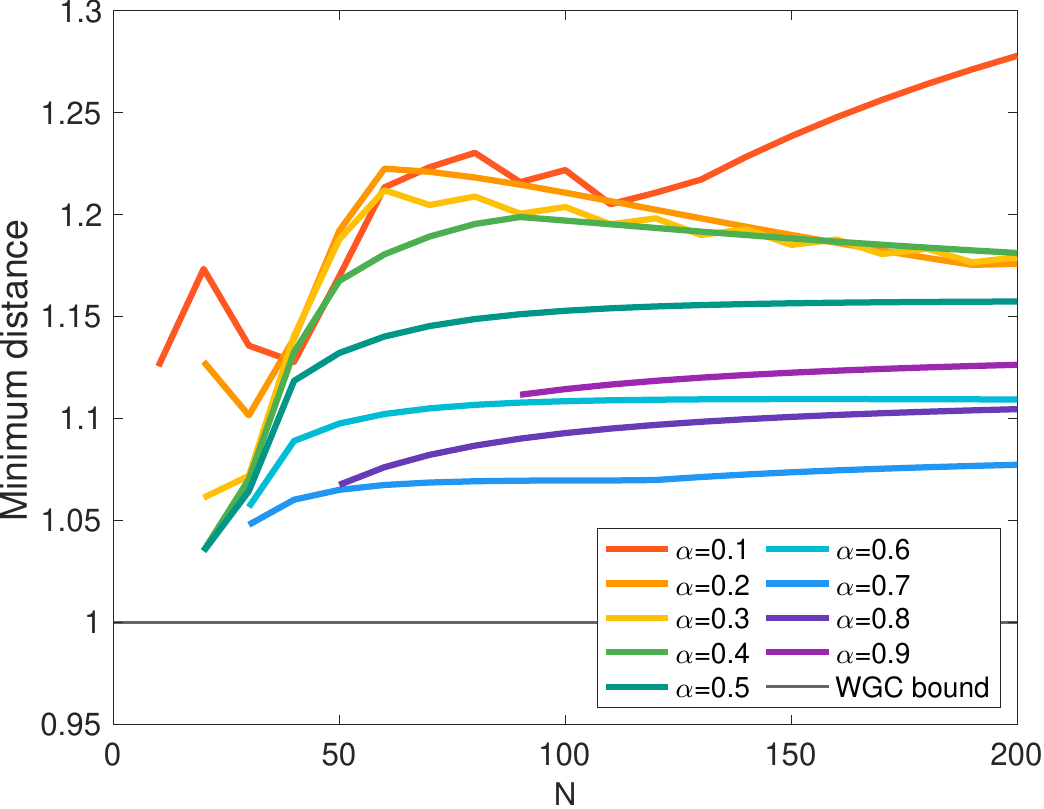}
      \caption{modified WGC}
    \end{subfigure}
    \hfill
    \caption{
    Testing AdS WGC for $SU(N)$ theory with 2 $\sym$ + 2 $\overline{\antisym}$ + 16 $\overline{\fund}$ + $N_f$($\fund$ + $\overline{\fund}$).
    The minimum distance from the origin to the convex hull with a fixed $\a=N_f/N$. Theories below the solid line at the minimum distance 1 do not satisfy the WGC.\label{fig:wgc_s2A2F16ven}}
  \end{figure}

\subsection{\texorpdfstring{1 \textbf{Adj} + 1 $\sym$ + 1 $\overline{\sym}$ + $N_f$ ( $\fund$ + $\overline{\fund}$ )}{1 Adj + 1 S + 1 St + Nf ( Q + Qt )}\label{sec:adj1s1S1}}

\paragraph{Matter content and symmetry charges}
  The next example of the Type II theory is an $SU(N)$ gauge theory with an adjoint and a pair of rank-2 symmetric and its conjugate, as well as $N_f$ pairs of fundamental and anti-fundamental chiral multiplets. The matter fields and their $U(1)$ global charges are listed in Table \ref{tab:adj1s1S1nf1}.
  {\renewcommand\arraystretch{1.6}
  \begin{table}[h]
    \centering
    \begin{tabular}{|c|c||c|c|c|c|c|c|}
    \hline
    \# & Fields &$SU(N)$ & $U(1)_1$ & $U(1)_{2}$ &  $U(1)_{3}$ &  $ U(1)_4$  & $U(1)_R$ \\\hline
    
     $N_f$ & $Q$  &$\fund$& $N$ & 0 & 1 &  0  & $R_Q$\\
     
     $N_f$ & $\widetilde{Q}$ &$\overline{\fund}$ & $N$ & 0 & -1 &  0 & $R_{\widetilde{Q}}$ \\
     
     $1$ & $\Phi$  &\textbf{Adj}&  $-N_f$ & $N+2$ &  0 & 0 & $R_\Phi$\\
     
     1 & $S$ & $\sym$&  0 &  $-N$  & 0 & 1 &$R_S$ \\
     
     1 & $\widetilde{S}$ & $\overline{\sym}$ & 0 &  $-N$ & 0  & -1 & $R_{\widetilde{S}}$\\\hline
    \end{tabular}
    \caption{The matter contents and their corresponding charges in $SU(N)$ gauge theory with 1 \textbf{Adj} + 1 $\sym$ + 1 $\overline{\sym}$ + $N_f$ ( $\fund$ + $\overline{\fund}$ ).\label{tab:adj1s1S1nf1}}
  \end{table}}

\paragraph{Gauge-invariant operators}
  Let $I$ and $J$ denote the flavor indices for $Q$. We present a sample of single-trace gauge-invariant operators in schematic form as follows:
  \begin{enumerate}
    \item $\Tr\Phi^n\,,\quad n=2,3,\dots, N\,.$
    \item $\Tr\Phi^n(S\widetilde{S})^m\,,\quad n=0,1,\dots, N,\quad m=1,2,\dots,N-1\,.$
    \item $Q(\widetilde{S}S)^n\Phi^m\widetilde{Q},\quad n,m=0,1,\dots, N-1\,.$
    \item $Q\widetilde{S}(S\widetilde{S})^n\Phi^m Q,\quad n=0,1,\dots,N-2\,,\quad m=0,1,\dots, N-1\,.$
    \item $\e\, \CQ^{n_1}_{I_1}\cdots \CQ^{n_N}_{I_N}$.
    \item $\e\,\e\, S^{i} (\CQ_{I_1}^{n_1}\CQ_{J_1}^{m_1})\cdots(\CQ_{I_{N-i}}^{n_{N-i}}\CQ_{J_{N-i}}^{m_{N-i}})$.
    \item The conjugates of the above-listed operators.

    $\vdots$
  \end{enumerate}
  The ellipsis indicates that only the low-lying operators have been listed. This subset is sufficient to identify relevant operators or to test the Weak Gravity Conjecture. Here, we define the dressed quarks as:
  \begin{align}
    \CQ^n_I=\begin{cases}
        \Phi^i(S\widetilde{S})^{n/2-i}Q_I & n=0,2,4,\dots\,,\\
        \Phi^i(S\widetilde{S})^{(n-1)/2-i}S\widetilde{Q}_I & n=1,3,5,\dots\,.
    \end{cases}
  \end{align}

\paragraph{$R$-charges and central charges}
  We perform the $a$-maximization to compute the $R$-charges of the matter fields and central charges. In the large $N$ limit with a fixed $N_f$, they are given by
  \begin{align}
  \begin{split}
    R_Q = R_{\widetilde{Q}} & \sim \frac{12-\sqrt{26}}{12} + \frac{41\sqrt{26}+ 78N_f}{624N}+ O(N^{-2})\,, \\
    R_{\Phi} & \sim \half + \frac{41 + 3N_f\sqrt{26}}{72N}+ O(N^{-2})\,, \\
    R_S = R_{\widetilde{S}} & \sim \half + \frac{31 + 3N_f\sqrt{26}}{72N}+ O(N^{-2})\,, \\
    a & \sim \frac{27}{128}N^2 + \frac{99+13N_f\sqrt{26}}{768}N+O(N^0)\,,\\
    c & \sim \frac{27}{128}N^2 + \frac{75+ 17 N_f\sqrt{26}}{768}N+O(N^0)\,,\\
    a/c & \sim 1- \frac{-12+2N_f\sqrt{26}}{81N} + O(N^{-2})\,.
  \end{split}
  \end{align}
  We again find that no gauge-invariant operators decouple along the RG flow. The leading-order behavior of the $R$-charges and central charges is universal across Type II theories. Similarly, in the Veneziano limit with a fixed $\a=N_f/N$, the leading terms of $R$-charges and central charges are also universal across Type II theories.

  For $N_f=0,1$ and sufficiently large $N$, we find that the ratio of central charges $a/c$ is greater than 1. In general, the ratio $a/c$ lies within the range $13/14\simeq 0.9286 < a / c \leq  (161399+81\sqrt{2641})/162580 \simeq 1.0183$. The minimum value of $a/c$ can be obtained in the Veneziano limit with $\a \goto 1$. The maximum value of $a/c$ occurs when $(N_f,N)=(0,5)$. Figure \ref{fig:adj1s1S1ratio} illustrates the behavior of the ratio $a/c$.
  \begin{figure}[t]
    \centering
    \begin{subfigure}[b]{0.45\textwidth}
        \includegraphics[width=\linewidth]{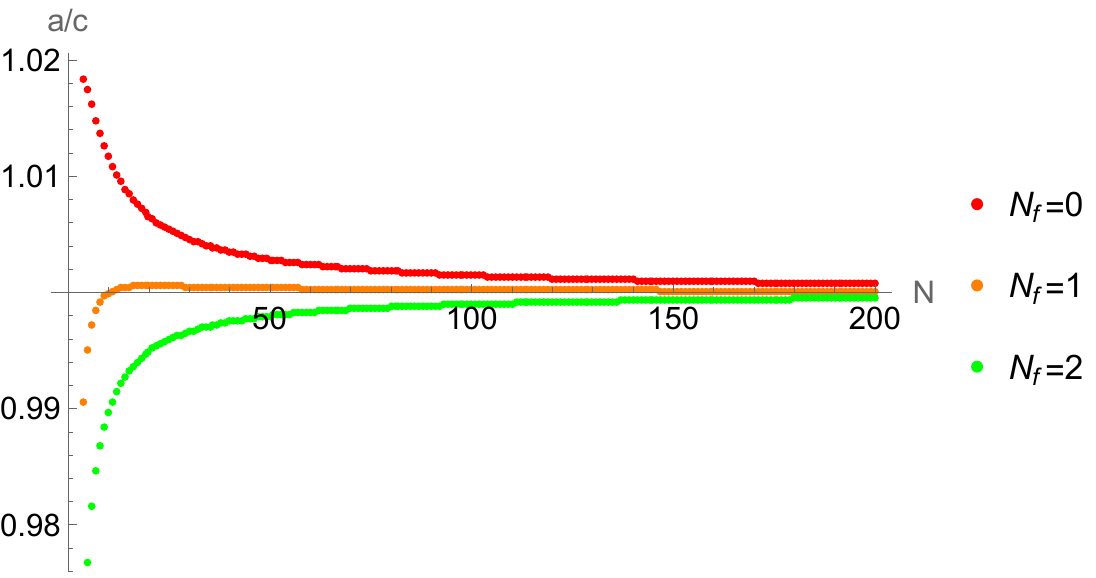}
    \end{subfigure}
    \hspace{4mm}
    \begin{subfigure}[b]{0.45\textwidth}
        \includegraphics[width=\linewidth]{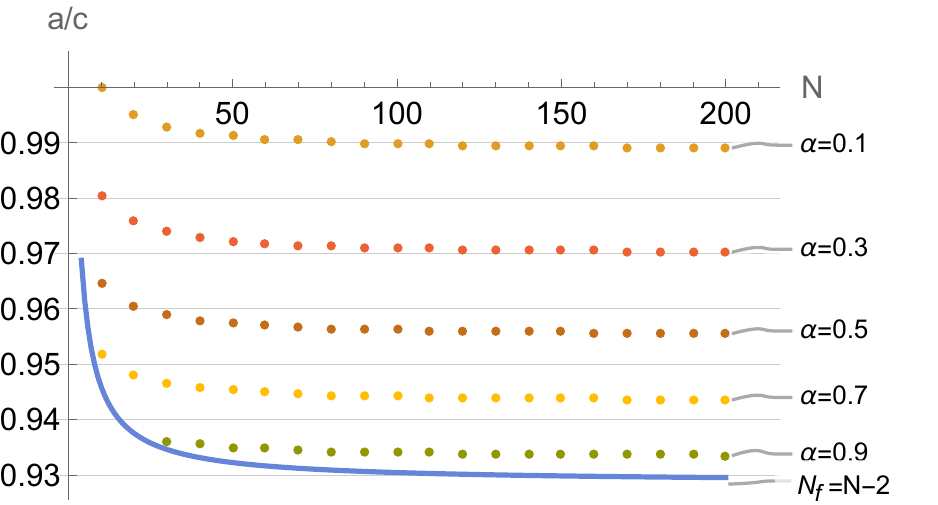}
    \end{subfigure}
    \hfill
    \caption{
    The central charge ratio for $SU(N)$ theory with 1 \textbf{Adj} + 1 $\sym$ + 1 $\overline{\sym}$ + $N_f$ ( $\fund$ + $\overline{\fund}$ ).
    Left: $a/c$ versus $N$ with a fixed $N_f$. Right: $a/c$ versus $N$ with a fixed $\a=N_f/N$.\label{fig:adj1s1S1ratio}}
  \end{figure}

\paragraph{Conformal window}
  The upper bound of the conformal window comes from the asymptotic freedom. For this theory, when it is at the top of the conformal window $N_f=N-2$, there is a non-trivial conformal manifold \cite{Razamat:2020pra} so that it is an interacting SCFT. For small $N_f$, we find that $a$-maximization always yields a unique solution, and the resulting central charges lie within the Hofman-Maldacena bound. Therefore, we conjecture that this theory flows to an interacting SCFT for $0\leq N_f\leq N-2$.

\paragraph{Relevant operators}
  For generic $N$ and $N_f$, there exist the following relevant operators:
  \begin{itemize}
    \item an operator of the form $\Tr \Phi^2$ with a dimension $\frac{3}{2} < \D < 2\,,$
    \item an operator of the form $\Tr S\widetilde{S}$ with a dimension $\frac{3}{2} < \D < 2\,,$
    \item $N_f^2$ operators of the form $Q_I\widetilde{Q}_{\tilde{J}}$ with a dimension $1.7253\simeq \frac{12-\sqrt{26}}{4}<\D<2\,,$
    \item an operator of the form $\Tr \Phi S \widetilde{S}$ with a dimension $\frac{9}{4}<\D < 3\,,$
    \item $N_f^2$ operators of the form $Q_I\Phi\widetilde{Q}_{\tilde{J}}$ with a dimension $2.4753\simeq\frac{15-\sqrt{26}}{4}<\D<3\,,$
    \item $\frac{N_f(N_f+1)}{2}$ operators of the form ${Q}_{I}\widetilde{S}Q_{J}$ and their conjugate with a dimension $2.4753\simeq \frac{15-\sqrt{26}}{4} < \D < 3\,.$
  \end{itemize}
  The number of relevant operators does not depend on $N$ for a fixed $N_f$. The low-lying operator spectrum is sparse in the large $N$ limit.

  It was proposed in \cite{Brodie:1996xm} that deforming the theory by the superpotential $W=\Tr\Phi^{k+1}+\Tr\Phi S\widetilde{S}$ leads to a dual description given by $SU(3kN_f+4-N)$ gauge theory with certain superpotential and a set of flip fields. However, the superpotential deformation is not relevant at the $W=0$ fixed point for $k>1$. 
  Notably, near the fixed point described by $W = \Tr \Phi S \widetilde{S}$, the first term can be relevant for arbitrary $k$ if $N$ is sufficiently large (and $N_f$ is fixed).

\paragraph{Conformal manifold}
  When $N_f=N-2$, which lies at the upper bound of the conformal window, the theory possesses a non-trivial conformal manifold that contains a one-dimensional subspace that preserves $\CN=2$ supersymmetry \cite{Razamat:2020pra}. At this value, the theory is a conformal gauge theory with no running coupling.
  
  When $N_f<N-2$, there are no marginal operators for generic $N$ in the absence of a superpotential. However, upon a suitable superpotential deformation, this theory flows to a superconformal fixed point with a non-trivial conformal manifold.

  For the $N_f=0$ case, consider the following superpotential deformation:
  \begin{align}
      W= M_1\Tr\Phi^2 + M_1^2\,.
  \end{align}
  At the IR fixed point, there exists a marginal operator of the form $\Tr\Phi^4$. Upon this deformation, the theory possesses a single $U(1)$ flavor symmetry, under which $S$ carries charge $+1$ and $\widetilde{S}$ carries charge $-1$. Since the operator $\Tr\Phi^4$ is neutral under this flavor symmetry, it is exactly marginal. This deformation works for general $N$ with a fixed $N_f=0$.
  
  For the $N_f=1$ case, consider the following superpotential deformation:
  \begin{align}
    W=M_1 Q\widetilde{Q} + M_1 \Tr\Phi^2 + M_1^2\,.
  \end{align}
  Here, $M_1$ is a flip field, which is a gauge-singlet chiral superfield. 
  We can test this by computing the superconformal index. For example, the reduced superconformal index for the $SU(5)$ gauge theory with this superpotential is given by
  \begin{align}
    \CI_{\text{red}}=2 t^{3}+t^{4.29} + t^{4.5}\left(2-y-\frac{1}{y}\right)+2
   t^{5.14}+t^{5.79}+t^{6}\left(1-y-\frac{1}{y}\right)+ \cdots\,.
  \end{align}
  The positivity of the coefficient at the $t^6$ term indicates the existence of a non-trivial conformal manifold. 
  The marginal operators take the form $\Tr\Phi^4$ and $Q\Phi^2\widetilde{Q}$. Such a deformation and marginal operators exist for general $N$ with a fixed $N_f=1$.

\paragraph{Weak Gravity Conjecture}
  We examine the AdS WGC using the gauge-invariant operators and $U(1)$ flavor charges identified at the beginning of this section. We find that this theory always satisfies both versions of the WGC. The result is shown in Figure \ref{fig:wgc_adj1s1S1ven}. 
  \begin{figure}[t]
    \centering
    \begin{subfigure}[b]{0.45\textwidth}
      \centering
      \includegraphics[width=\linewidth]{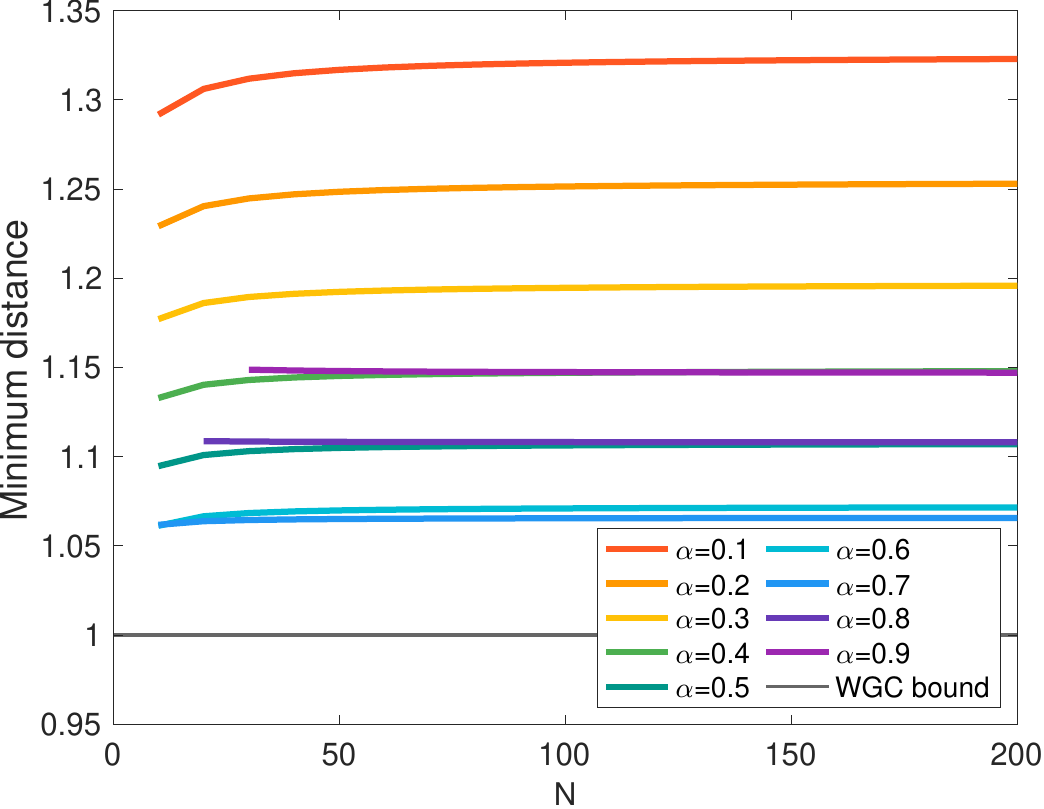}
      \caption{NN-WGC}
    \end{subfigure}
    \hspace{4mm}
    \begin{subfigure}[b]{0.45\textwidth}
      \centering
      \includegraphics[width=\linewidth]{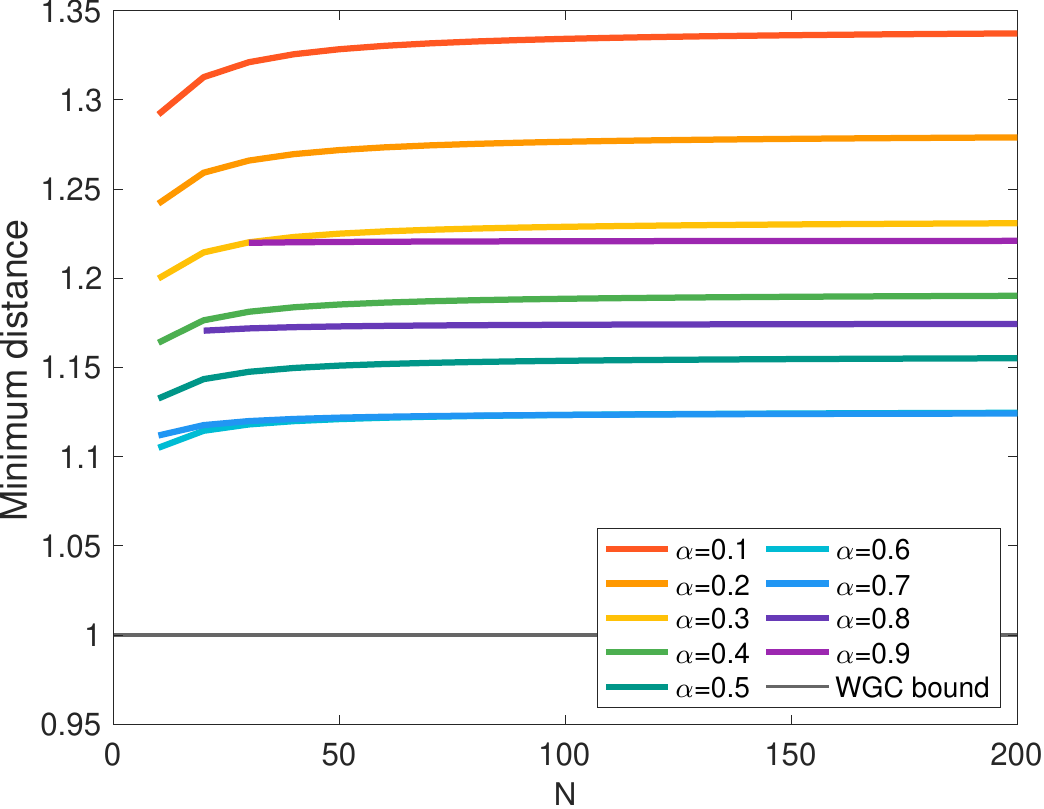}
      \caption{modified WGC}
    \end{subfigure}
    \hfill
    \caption{
    Testing AdS WGC for $SU(N)$ theory with 1 \textbf{Adj} + 1 $\sym$ + 1 $\overline{\sym}$ + $N_f$ ($\fund$ + $\overline{\fund}$).
    The minimum distance from the origin to the convex hull with a fixed $\a=N_f/N$. \label{fig:wgc_adj1s1S1ven}}
 \end{figure}

\subsection{\texorpdfstring{2 $\antisym$ + 2 $\overline{\antisym}$ + $N_f$ ( $\fund$ + $\overline{\fund}$ )}{2 A + 2 At + Nf ( Q + Qt )}}\label{sec:a2A2}

\paragraph{Matter content and symmetry charges}
  Next entry of the Type II theory is an $SU(N)$ gauge theory with two pairs of rank-2 anti-symmetric tensors and conjugates, along with $N_f$ pairs of fundamental and anti-fundamental chiral multiplets. The matter fields and their $U(1)$ global charges are listed in Table \ref{tab:a2A2}.
  {\renewcommand\arraystretch{1.6}
  \begin{table}[h]
    \centering
    \begin{tabular}{|c|c||c|c|c|c|c|c|}
    \hline
    \# & Fields &$SU(N)$&  $U(1)_1$ & $ U(1)_2$  & $U(1)_3$ & $U(1)_R$ \\\hline
    
     $N_f$ & $Q$ &$\fund$&  $2N-4$ & $1$  &0 & $R_Q$\\
     
     $N_f$ & $\widetilde{Q}$ &$\overline{\fund}$& $2N-4$ & $-1$ &  0 &$R_{\widetilde{Q}}$ \\
     
     2 &$A$&$\antisym$  &  $-N_f$ & 0 & 1  & $R_A$ \\
     
     2 & $\widetilde{A}$ &$\overline{\antisym}$&  $-N_f$ & 0 & -1 & $R_{\widetilde{A}}$\\\hline
    \end{tabular}
    \caption{
    The matter contents and their corresponding charges in $SU(N)$ gauge theory with 2 $\antisym$ + 2 $\overline{\antisym}$ + $N_f$ ( $\fund$ + $\overline{\fund}$ ).\label{tab:a2A2}}
  \end{table}}

\paragraph{Gauge-invariant operators}
  Let $I$ and $J$ denote the flavor indices for $Q$, and $M$ denote the flavor indices for $A$. We present a sample of single-trace gauge-invariant operators in schematic form as follows:
  \begin{enumerate}
    \item $\Tr(A_{M_1}\widetilde{A}_{\widetilde{M}_1}\cdots A_{M_n}\widetilde{A}_{\widetilde{M}_n}),\quad n=1,2,\dots\,.$
    \item $Q_{I}(\widetilde{A}_{\widetilde{M}_1}A_{M_1}\cdots\widetilde{A}_{\widetilde{M}_n}A_{M_n})\widetilde{Q}_{\tilde{J}}\,,\quad n=0,1,\dots\,.$
    \item $Q_{I}\widetilde{A}_{\widetilde{M}_0}(\widetilde{A}_{\widetilde{M}_1}A_{M_1}\cdots\widetilde{A}_{\widetilde{M}_n}A_{M_n}) Q_{J},\quad n=0,1,\dots\,.$
    \item $\e\, A_{M_1}\cdots A_{M_i} \CQ_{I_1}^{n_1}\cdots\CQ_{I_{N-2i}}^{n_{N-2i}}\,.$
    \item $\e\,\e\, A_{M_1}^{\lfloor N/2\rfloor}(A_{(M_1}\widetilde{A})A_{M_2)}^{N-\lfloor N/2\rfloor}\quad$ if $N$ is odd.
    \item The conjugates of the above-listed operators.

    $\vdots$
  \end{enumerate}
  The ellipsis indicates that only the low-lying operators have been listed. This subset is sufficient to identify relevant operators or to test the Weak Gravity Conjecture.
  Here, we define the dressed quarks as
  \begin{align}
    \CQ_I^n=\begin{cases}
        (A_{M_1}\widetilde{A}_{\widetilde{M}_1}\cdots A_{M_{n/2}}\widetilde{A}_{\widetilde{M}_{n/2}})Q_I & n=0,2,4,\dots\,,\\
        (A_{M_1}\widetilde{A}_{\widetilde{M}_1}\cdots A_{M_{n/2}}\widetilde{A}_{\widetilde{M}_{n/2}})A\widetilde{Q}_{\tilde{I}} & n=1,3,5,\dots\,.
    \end{cases}
  \end{align}

\paragraph{$R$-charges and central charges}
  For $N=3$, the rank-2 anti-symmetric tensor is the same as the anti-fundamental representation. Therefore, we only consider cases with $N > 3$ here.
  
  We perform the $a$-maximization to compute the $R$-charges of the matter fields and central charges. We find that there are special cases for small $N$ that require separate discussion. 
  For the $SU(4)$ theory, the rank-2 antisymmetric tensor is nothing but the vector representation of $SO(6)$. When $N_f=0$, this is simply identical to $SO(6)$ theory with four vectors, which does not flow to an interacting SCFT. Instead, this theory is in the Coulomb phase with a massless photon supermultiplet \cite{Intriligator:1995id}. When $N_f = 1$, we find a set of gauge-invariant operators that decouple along the flow. Upon removing them, we get an interacting SCFT. 
  
  When $(N_f,N)=(0,5)$, we find that the gauge-invariant operators of the form
  \begin{align}
    \Tr A_{M_1}\widetilde{A}_{\widetilde{M}_2}\,,\quad \Tr A_{M_1}\widetilde{A}_{\widetilde{M}_2} A_{M_3}\widetilde{A}_{\widetilde{M}_4}
  \end{align}
  hit the unitarity bound along the RG flow and decouple from the rest of the system.
  Therefore, we must flip these decoupled operators to perform the $a$-maximization correctly. This can be done by introducing the following superpotential:
  \begin{align}
  \begin{split}
    W &= X_1\Tr A_1\widetilde{A}_1+ X_2\Tr A_1\widetilde{A}_2+ X_3\Tr A_2\widetilde{A}_1 + X_4\Tr A_2\widetilde{A}_2+X_5\Tr(A_1\widetilde{A}_1)^2 \\
    &\quad + X_6\Tr(A_1\widetilde{A}_1 A_1\widetilde{A}_2)+ X_7\Tr(A_1\widetilde{A}_1 A_2\widetilde{A}_1)+X_8\Tr(A_1\widetilde{A}_1 A_2\widetilde{A}_2)\\
    &\quad + X_9\Tr(A_1\widetilde{A}_2)^2+ X_{10}\Tr(A_1\widetilde{A}_2 A_2\widetilde{A}_1) + X_{11}\Tr(A_1\widetilde{A}_2 A_2\widetilde{A}_2) \\
    &\quad  + X_{12}\Tr(A_2\widetilde{A}_1)^2 + X_{13}\Tr(A_2\widetilde{A}_2)^2\,,
  \end{split}
  \end{align}
  where $X_i$'s are gauge-singlet chiral superfields. Upon flipping decoupled operators, we obtain the correct result of the $a$-maximization given by
  \begin{align}
      R_A = R_{\widetilde{A}}=\frac{1}{6}\,,\quad a=\frac{97}{96}\simeq 1.0104\,,\quad c=\frac{119}{96}\simeq 1.2396\,,\quad a/c=\frac{97}{119}\simeq 0.8151\,.
  \end{align}
  
  For $(N_f,N)=(1,4),(0,5),(0,6),(0,7),(0,8),(1,5),(1,6),(2,4)$, we find the only decoupled operators are of the form $\Tr A_{M_1}\widetilde{A}_{\widetilde{M}_2}$. They must likewise be flipped in order to perform the $a$-maximization correctly.

  For all the other cases (when $N$ is large enough), none of the gauge-invariant operators decouple along the RG flow. The $R$-charges of the matter fields and central charges, in the large $N$ limit with a fixed $N_f$, are given by
  \begin{align}
  \begin{split}
    R_Q = R_{\widetilde{Q}} & \sim \frac{12-\sqrt{26}}{12} + \frac{-41\sqrt{26}+ 39N_f}{312N}+ O(N^{-2})\,, \\
    R_A = R_{\widetilde{A}} & \sim \half + \frac{-24 + N_f\sqrt{26}}{24N}+ O(N^{-2})\,, \\
    a & \sim \frac{27}{128}N^2 + \frac{-198+13N_f\sqrt{26}}{768}N+O(N^0)\,,\\
    c & \sim \frac{27}{128}N^2 + \frac{-150 + 17 N_f\sqrt{26}}{768}N+O(N^0)\,,\\
    a/c & \sim 1- \frac{-24+2N_f\sqrt{26}}{81N} + O(N^{-2})\,.
  \end{split}
  \end{align}
  The leading-order behavior of the $R$-charges and central charges is universal across the Type II theories. Similarly, in the Veneziano limit with a fixed $\a=N_f/N$, the leading terms of $R$-charges and central charges are also universal across Type II theories.

  We find that the ratio of central charges $a/c$ is always less than one, taking a value within the range $0.7901 \simeq (29426941-79632\sqrt{1765})/33008942\leq a / c < 1$. The minimum value of $a/c$ arises when $(N_f,N)=(3,4)$. The maximum value of $a/c$ can be obtained in the large $N$ limit with $N_f=0$. Figure \ref{fig:a2A2ratio} illustrates the behavior of the ratio $a/c$.
  \begin{figure}[t]
    \centering
    \begin{subfigure}[b]{0.45\textwidth}
        \includegraphics[width=\linewidth]{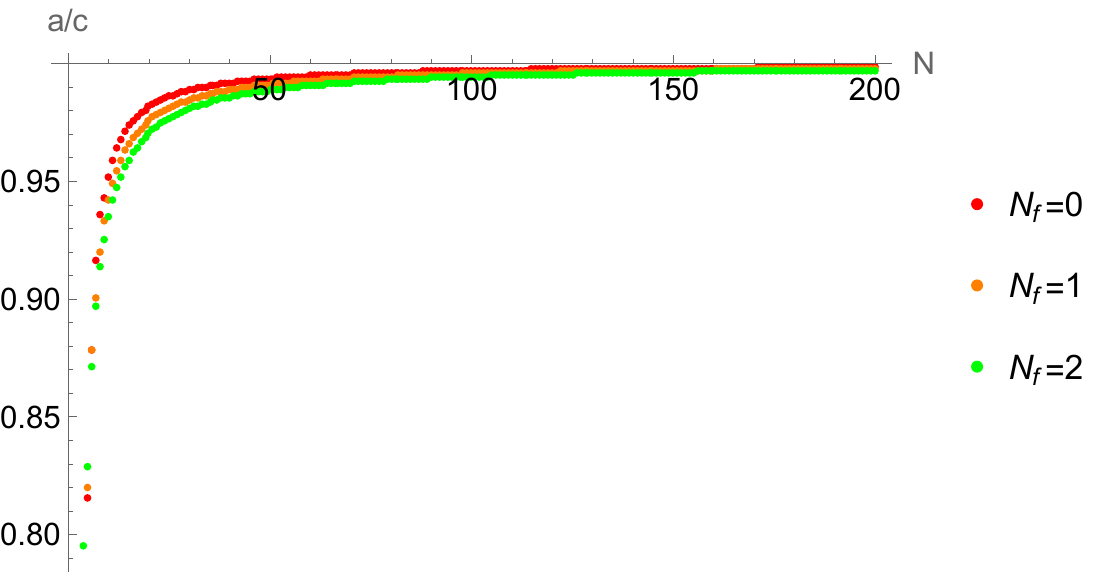}
    \end{subfigure}
    \hspace{4mm}
    \begin{subfigure}[b]{0.45\textwidth}
        \includegraphics[width=\linewidth]{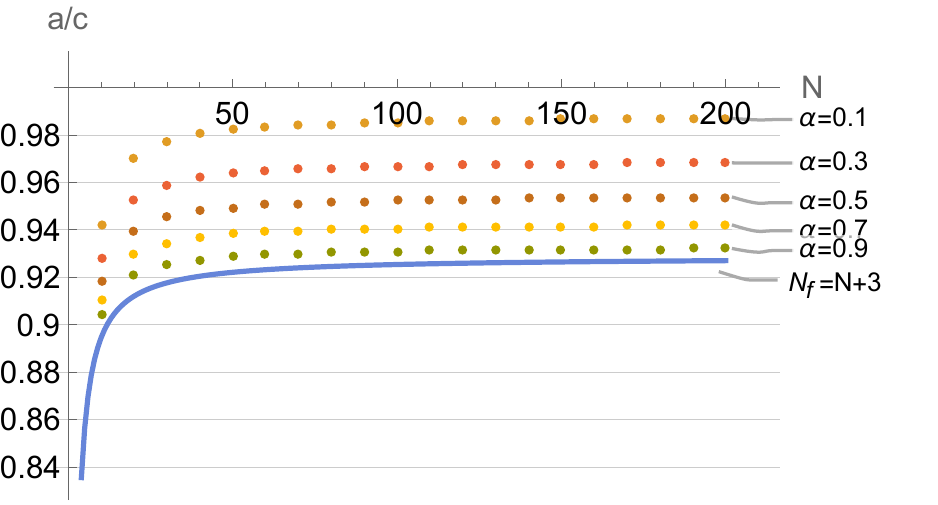}
    \end{subfigure}
    \hfill
    \caption{
    The central charge ratio for $SU(N)$ theory with 2 $\antisym$ + 2 $\overline{\antisym}$ + $N_f$ ( $\fund$ + $\overline{\fund}$ ).
    Left: $a/c$ versus $N$ with a fixed $N_f$. Right: $a/c$ versus $N$ with a fixed $\a=N_f/N$.\label{fig:a2A2ratio}}
  \end{figure}

\paragraph{Conformal window}
  The upper bound of the conformal window comes from the asymptotic freedom. When $N_f < N+4$, the $a$-maximization procedure presents no issue (except for $N=4, N_f=0$, which is a confining theory): it always yields a unique solution, and the resulting central charges fall within the Hofman-Maldacena bound. We therefore conjecture that this theory flows to an interacting SCFT for $0 \leq N_f<N+4$, with the exception of $(N_f, N)=(0,4)$.

\paragraph{Relevant operators}
  For generic values of $N$ and $N_f$ (except for the cases where there are decoupled operators), there exist the following relevant operators:
  \begin{itemize}
    \item four operators of the form $\Tr A_{M_1}\widetilde{A}_{\widetilde{M}_2}$ with a dimension $1.0616\lesssim \D <2\,,$
    \item $N_f^2$ operators of the form $Q_I\widetilde{Q}_{\tilde{J}}$ with a dimension $1.3153\lesssim\D<2\,,$
    \item $N_f(N_f-1)$ operators of the form $Q_I\widetilde{A}_MQ_J$ and their conjugates with a dimension $1.8461\lesssim\D<3\,,$
    \item (if $N_f\leq 4$) nine operators of the form $\Tr A_{M_1}\widetilde{A}_{\widetilde{M}_2}A_{M_3}\widetilde{A}_{\widetilde{M}_4}$ and ten operators of the form $(\Tr A_{M_1}\widetilde{A}_{\widetilde{M}_2})(\Tr A_{M_3}\widetilde{A}_{\widetilde{M}_4})$ with a dimension $2.1232\lesssim \D < 3\,,$
  \end{itemize}
  The number of relevant operators does not scale with $N$ for a fixed $N_f$. The low-lying operator spectrum is sparse in the large $N$ limit.

  Upon a deformation of the form 
  \begin{align}
      W=M_1\Tr (A_1\widetilde{A}_1)^{k_1+1}+M_2\Tr(A_2\widetilde{A}_2)^{k_2+1}+M_3(\Tr A_1\widetilde{A}_2 + \Tr A_2\widetilde{A}_1)\,.
  \end{align}
  where $M_i$ are the gauge-singlet chiral superfields, a dual description is proposed \cite{Abel:2009ty}. It is given by an $SU\left((2k^*+1)N_f-4k^*-N\right)$ gauge theory, with a number of flip fields, where $k^*=\half\left[(2k_1+1)(2k_2+1)-1\right]$.
  However, we note that the operator of the form $M (A \tilde{A})^{k+1}$ is not relevant either at the $W=0$ fixed point or at the $W=M_3(\Tr A_1\widetilde{A}_2+\Tr A_2\widetilde{A}_1)$ point for $k \ge 1$, which casts a doubt on the proposed duality.

\paragraph{Conformal manifold}
  There are no marginal operators for generic $N_f$ and $N$ in the absence of a superpotential. However, upon a suitable superpotential deformation, the theory flows to a superconformal fixed point with a non-trivial conformal manifold.

  For the $N_f=0$ case, upon the following superpotential deformation, a non-trivial conformal manifold emerges at the IR fixed point:
  \begin{align}
    W=\Tr A_1\widetilde{A}_1 A_1 \widetilde{A}_2 + X_1 \Tr A_2\widetilde{A}_1 + X_2 \Tr A_2\widetilde{A}_2\,.
  \end{align}
  Here, $X_i$'s are flip fields, which are gauge-singlet chiral superfields. 
  We can test this by computing the superconformal index. For example, the reduced superconformal index for the $SU(9)$ gauge theory with this superpotential is given by
  \begin{align}
    \CI_{\text{red}}=3 t^{2.57}+2 t^{3}+4 t^{3.86}+2 t^{4.29}+2 t^{4.71}+12 t^{5.14}+10 t^{5.57}+t^{6}+\cdots\,.
  \end{align}
  The positivity of the coefficient at the $t^6$ term indicates the existence of a non-trivial conformal manifold. The marginal operators take the form $\Tr (A_1\widetilde{A}_1)^2$ and $\Tr (A_1\widetilde{A}_2)^2$. This deformation works for general $N\geq 9$ with a fixed $N_f=0$.

  For the $N_f=1$ case, consider the following superpotential deformation:
  \begin{align}
    W=\Tr A_1\widetilde{A}_1 A_1\widetilde{A}_2 + M_1 \Tr A_2\widetilde{A}_1A_2\widetilde{A}_2\,.
  \end{align}
  Here, $M_1$ is a flip field, which is a gauge-singlet chiral superfield. At the IR fixed point, this theory possesses a non-trivial conformal manifold. We can verify this using the superconformal index. For instance, the reduced index for the $SU(7)$ gauge theory with this superpotential has a positive coefficient at the $t^6$ term:
  \begin{align}
    \CI_{\text{red}}=2 t^{2.13} + 2 t^{2.89} + \cdots + 8 t^{5.89} + t^6 + \cdots\,.
  \end{align}
  The marginal operators take the form $\Tr (A_1\widetilde{A}_1)^2$, $\Tr (A_1\widetilde{A}_2)^2$, $M_1 \Tr (A_2\widetilde{A}_1)^2$, $M_1 \Tr (A_2\widetilde{A}_2)^2$. Such a deformation--along with possible flipping of decoupled operators--and marginal operators exist for general $N\geq 7$ with a fixed $N_f = 1$.

\paragraph{Weak Gravity Conjecture}
  We test the AdS WGC using the gauge-invariant operators and $U(1)$ flavor charges identified at the beginning of this section. We find that this theory does not satisfy the NN-WGC for large values of $\a = N_f/N$ in the Veneziano limit, whereas the modified WGC always holds. The result is shown in Figure \ref{fig:wgc_a2A2ven}. 
  \begin{figure}[t]
    \centering
    \begin{subfigure}[b]{0.45\textwidth}
      \centering
      \includegraphics[width=\linewidth]{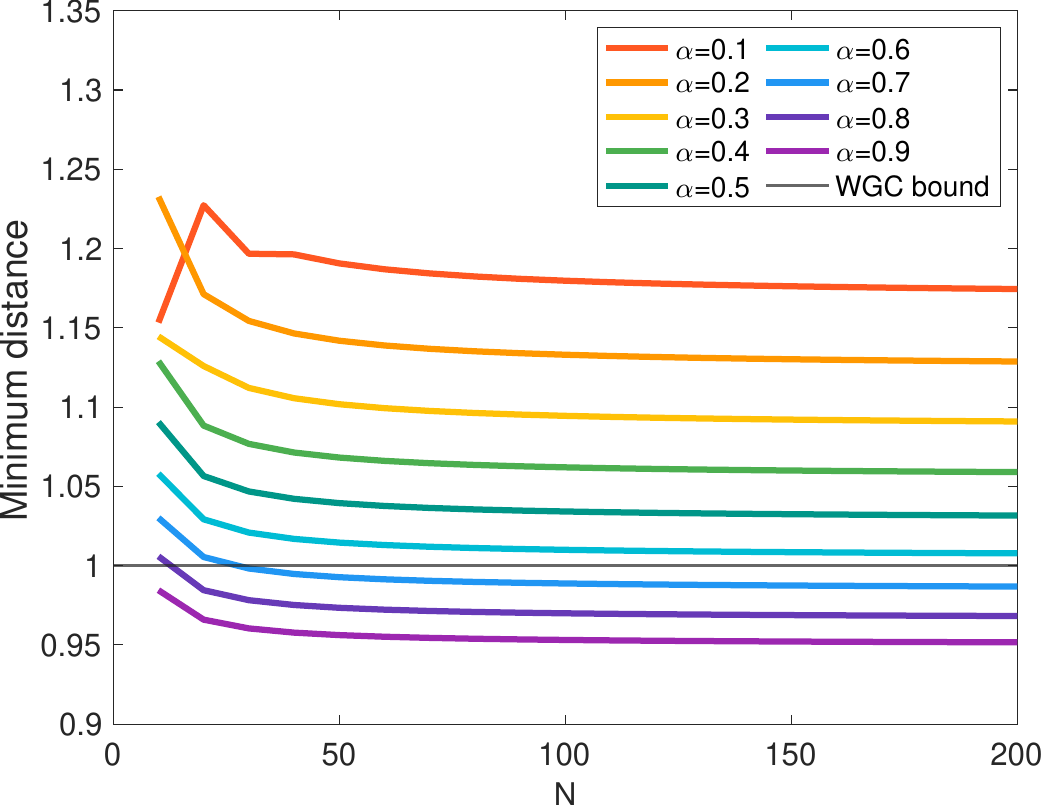}
      \caption{NN-WGC}
    \end{subfigure}
    \hspace{4mm}
    \begin{subfigure}[b]{0.45\textwidth}
      \centering
      \includegraphics[width=\linewidth]{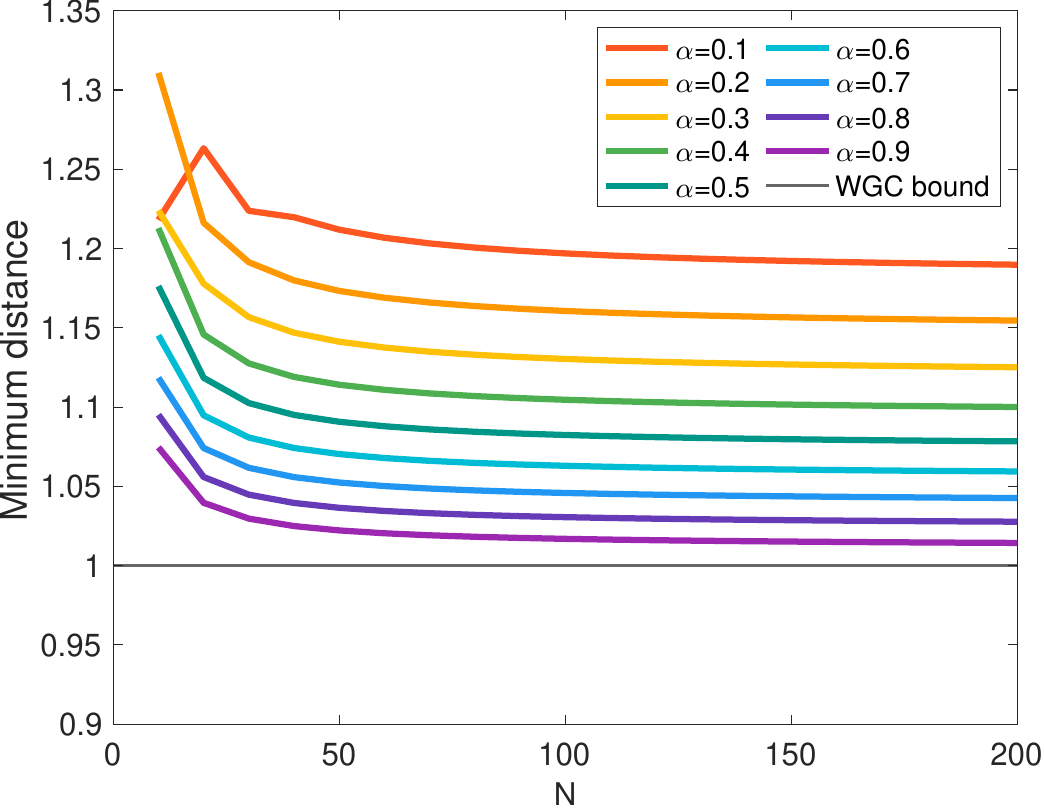}
      \caption{modified WGC}
    \end{subfigure}
    \hfill
    \caption{
    Testing AdS WGC for $SU(N)$ theory with 2 $\antisym$ + 2 $\overline{\antisym}$ + $N_f$ ( $\fund$ + $\overline{\fund}$ ).
    The minimum distance from the origin to the convex hull with a fixed $\a=N_f/N$. Theories below the solid line at the minimum distance 1 do not satisfy the WGC.\label{fig:wgc_a2A2ven}}
 \end{figure}

\subsection{\texorpdfstring{1 \textbf{Adj} + 1 $\sym$ + 1 $\overline{\antisym}$ + 8 $\overline{\fund}$ + $N_f$ ( $\fund$ + $\overline{\fund}$ )}{1 Adj + 1 S + 1 At + 8 Qt + Nf ( Q + Qt )}\label{sec:adj1s1A1F8}}

\paragraph{Matter content and symmetry charges}
  Next case of the Type II theory is an $SU(N)$ gauge theory with an adjoint, a rank-2 symmetric, a conjugate of rank-2 anti-symmetric, $N_f$ fundamental, and $N_f+8$ anti-fundamental chiral multiplets. The matter fields and their $U(1)$ global charges are listed in Table \ref{tab:adj1s1A1F8}.
  {\renewcommand\arraystretch{1.6}
  \begin{table}[h]
    \centering
    \begin{tabular}{|c|c||c|c|c|c|c|c|c|}
    \hline
    \# & Fields &$SU(N)$& $U(1)_1$ & $U(1)_2$ & $U(1)_3$  & $ U(1)_4$  & $U(1)_R$ \\\hline
    
     $N_f$ & $Q$&$\fund$ & $N$ & $N+2$ & $N-2$  & $N_f+8$  & $R_Q$\\
     
     $N_f+8$ & $\widetilde{Q}$  & $\overline{\fund}$ & $N$ & $N+2$ & $N-2$ & $-N_f$& $R_{\widetilde{Q}}$ \\
     
     1 & $\Phi$ &\textbf{Adj} & $-N_f-4$ & 0 & 0 & 0 & $R_\Phi$ \\
     
     1 & $S$&$\sym$ & 0  & $-2N_f-8$  & 0 & 0 & $R_S$ \\
     
     1 & $\widetilde{A}$  &$\overline{\antisym}$ & 0 & 0 & $-2N_f-8$  & 0 & $R_{\widetilde{A}}$\\\hline
    \end{tabular}
    \caption{The matter contents and their corresponding charges in $SU(N)$ gauge theory with 1 \textbf{Adj} + 1 $\sym$ + 1 $\overline{\antisym}$ + 8 $\overline{\fund}$ + $N_f$ ( $\fund$ + $\overline{\fund}$ ).\label{tab:adj1s1A1F8}}
  \end{table}}

\paragraph{Gauge-invariant operators}
  Let $I$ and $J$ denote the flavor indices for $Q$. We present a sample of single-trace gauge-invariant operators in schematic form as follows:
  \begin{enumerate}
    \item $\Tr\Phi^n,\quad \Tr (S\widetilde{A})^{2m}\,,\quad n=2,3,\dots, N-1\,,\quad m=1,2,\dots,\lfloor\frac{N-1}{2}\rfloor\,.$
    \item $\Tr\Phi^n(S\widetilde{A})^{m},\quad n=1,2,\dots, N-1,\quad m=1,2,\dots,\lfloor\frac{N-1}{2}\rfloor\,.$
    \item $Q_I\Phi^n(\widetilde{A}S)^m\widetilde{Q}_{\tilde{J}},\quad n=0,1,\dots,N-1\,,\quad  m=0,1,\dots,N-2\,.$
    \item $Q_I\widetilde{A}\Phi^n(S\widetilde{A})^mQ_J\,,\quad \widetilde{Q}_{\tilde{I}}S\Phi^n(\widetilde{A}S)^m\widetilde{Q}_{\tilde{J}},\quad n=0,1,\dots,N-1\,,\quad  m=0,1,\dots,\lfloor \frac{N+1}{2}\rfloor \,.$
    \item $\e\,(S\widetilde{A}S)^i\mathcal{\CQ}^{n_1}_{I_1}\dots\mathcal{\CQ}^{n_{N-2i}}_{I_{N-2i}},\quad \e\, \widetilde{A}^i \widetilde{\CQ}_{{I}_1}^{n_1}\cdots \widetilde{\CQ}_{I_{N-2i}}^{n_{N-2i}}$.
    \item $\e\,\e\, S^{i}(\CQ^{n_1}_{I_1}\CQ^{m_1}_{J_1})\cdots(\CQ^{n_{N-i}}_{I_{N-i}}\CQ^{m_{N-i}}_{J_{N-i}})$.
    \item $\e\,\e \widetilde{A}^N\Phi^3\quad$ if $N$ is odd.

    $\vdots$
  \end{enumerate}
  The ellipsis indicates that only the low-lying operators have been listed. This subset is sufficient to identify relevant operators or to test the Weak Gravity Conjecture.
  Here, we define the dressed quarks as:
  \begin{align}
    \CQ_I^n=\begin{cases}
        \Phi^i(S\widetilde{A})^{n/2-i} Q_I & n=0,2,4\dots\,,\\
        \Phi^i(S\widetilde{A})^{(n-1)/2-i}S\widetilde{Q}_{\tilde{I}} & n=1,3,5\dots\,,
    \end{cases}
  \end{align}
  and
  \begin{align}
    \widetilde{\CQ}_{\tilde{I}}^n=\begin{cases}
        \Phi^i(\widetilde{A}S)^{n/2-i}\widetilde{Q}_{\tilde{I}} & n=0,2,4\dots\,,\\
        \Phi^i(\widetilde{A}S)^{(n-1)/2-i}\widetilde{A}Q_I & n=1,3,5\dots\,,
    \end{cases}
  \end{align}

\paragraph{$R$-charges and central charges}
  We perform the $a$-maximization to compute the $R$-charges of the matter fields and central charges. In the large $N$ limit with a fixed $N_f$, they are given by
  \begin{align}
  \begin{split}
    R_Q = R_{\widetilde{Q}} & \sim \frac{12-\sqrt{26}}{12} + \frac{N_f+4}{8N}+ O(N^{-2})\,, \\
    R_\Phi & \sim \half +\frac{(N_f+4)\sqrt{26}}{24N} + O(N^{-2}) \,,\\
    R_S & \sim  \half + \frac{-10 + 3 (N_f+4)\sqrt{26}}{72N} + O(N^{-2})\,,\\
    R_{\widetilde{A}} & \sim \half + \frac{10 + 3(N_f+4)\sqrt{26}}{72N}+ O(N^{-2})\,, \\
    a & \sim \frac{27}{128}N^2 + \frac{13(N_f + 4)\sqrt{26}}{768}N+O(N^0)\,,\\
    c & \sim \frac{27}{128}N^2 + \frac{17 (N_f+4)\sqrt{26}}{768}N+O(N^0)\,,\\
    a/c & \sim 1- \frac{2(N_f+4)\sqrt{26}}{81N} + O(N^{-2})\,.
  \end{split}
  \end{align}
  We find that no gauge-invariant operators decouple along the RG flow. The leading-order behavior of the $R$-charges and central charges is universal across Type II theories. Similarly, in the Veneziano limit with a fixed $\a=N_f/N$, the leading terms of $R$-charges and central charges are also universal across Type II theories.

  We observe that the ratio of central charges $a/c$ is always less than one, taking a value within the range $11/12\simeq 0.9167 \leq  a / c < 1$. The minimum value of $a/c$ arises when $(N_f,N)=(0,4)$. The maximum value of $a/c$ can be obtained in the large $N$ limit with $N_f=0$. Figure \ref{fig:adj1s1A1F8ratio} illustrates the behavior of the ratio $a/c$.
  \begin{figure}[h]
    \centering
    \begin{subfigure}[b]{0.45\textwidth}
        \includegraphics[width=\linewidth]{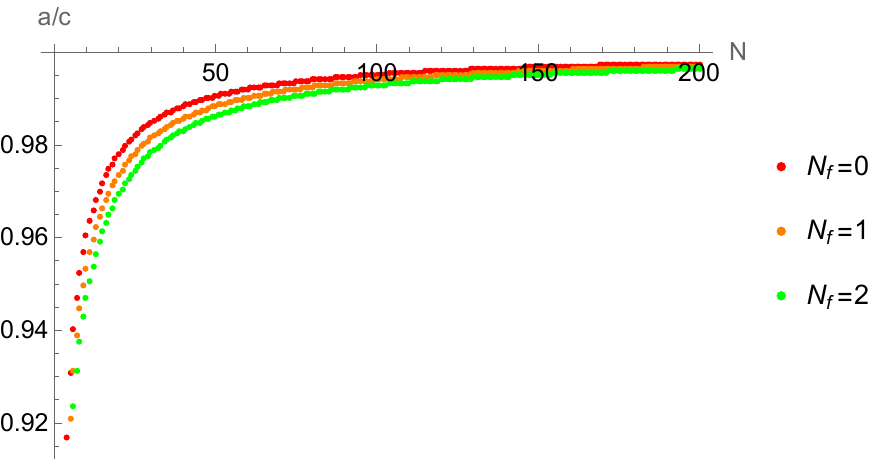}
    \end{subfigure}
    \hspace{4mm}
    \begin{subfigure}[b]{0.45\textwidth}
        \includegraphics[width=\linewidth]{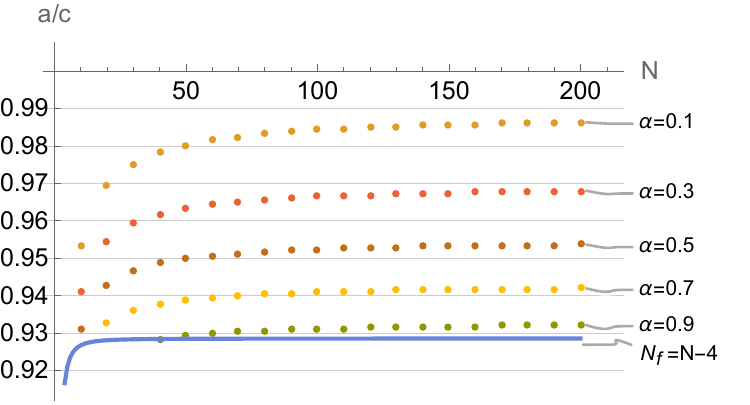}
    \end{subfigure}
    \hfill
    \caption{
    The central charge ratio for $SU(N)$ theory with 1 \textbf{Adj} + 1 $\sym$ + 1 $\overline{\antisym}$ + 8 $\overline{\fund}$ + $N_f$ ( $\fund$ + $\overline{\fund}$ ).
    Left: $a/c$ versus $N$ with a fixed $N_f$. Right: $a/c$ versus $N$ with a fixed $\a=N_f/N$.\label{fig:adj1s1A1F8ratio}}
  \end{figure}

\paragraph{Conformal window}
  The upper bound of the conformal window is determined by the asymptotic freedom. For this theory, when it is at the top of the conformal window $N_f=N-4$, there is a non-trivial conformal manifold \cite{Razamat:2020pra} so that it is an interacting SCFT. For small $N_f$, the $a$-maximization always yields a unique solution, and the resulting central charges lie within the Hofman-Maldacena bound. We therefore conjecture that this theory flows to an interacting SCFT for $0\leq N_f\leq N-4$.

\paragraph{Relevant operators}
  For generic $N$ and $N_f$, there exist the following relevant operators:
  \begin{itemize}
    \item an operator of the form $\Tr\Phi^2$ with a dimension $\frac{3}{2} < \D \leq 2\,,$
    \item an operator of the form $\Tr\Phi^3$ with a dimension $\frac{9}{4} < \D \leq 3\,,$
    \item an operator of the form $\Tr \Phi S\widetilde{A}$ with a dimension $\frac{9}{4} < \D \le 3\,,$
    \item $N_f(N_f+8)$ operators of the form $Q_I\widetilde{Q}_{\tilde{J}}$ with a dimension $1.7253\simeq \frac{12-\sqrt{26}}{4}< \D \leq 2\,,$
    \item $N_f^2$ operators of the form $Q_I\Phi \widetilde{Q}_{\tilde{J}}$ with a dimension $2.4753\simeq \frac{15-\sqrt{26}}{4}< \D \leq 3\,,$
    \item $\frac{(N_f+8)(N_f+9)}{2}$ operators of the form $\widetilde{Q}_{\tilde{I}} S \widetilde{Q}_{\tilde{J}}$ with a dimension $2.4753\simeq \frac{15-\sqrt{26}}{4} < \D \leq 3\,,$
    \item $\frac{N_f(N_f-1)}{2}$ operators of the form $Q_I \widetilde{A} Q_{J}$ with a dimension $2.4753\simeq \frac{15-\sqrt{26}}{4} < \D \leq 3\,,$
  \end{itemize}
  The number of relevant operators does not depend on $N$ for a fixed $N_f$. The low-lying operator spectrum is sparse in the large $N$ limit.
  
  It was proposed in \cite{Brodie:1996xm} that upon deforming the theory by the superpotential $W = \Tr \Phi^{k+1}+\Tr\Phi S\widetilde{A}$, the theory admits a dual description given by $SU(3k(N_f+4)-N)$ gauge theory with certain superpotential and flip fields. However, we note that the superpotential deformation is relevant at $W=0$ fixed point only for $k=1, 2$. Therefore, the theory does not flow to the new interacting fixed point for $k>2$ at least from the $W=0$ fixed point. However, near the fixed point described by $W = \Tr \Phi S \widetilde{A}$, the first term can be relevant for arbitrary $k$ if $N$ is large enough (and $N_f$ is fixed).
  
\paragraph{Conformal manifold}

   When $N_f=N-4$, the one-loop beta function for the gauge coupling vanishes. At this value, the theory possesses a non-trivial conformal manifold \cite{Razamat:2020pra}.
  
  When $N_f<N-4$, there are no marginal operators for generic $N$ in the absence of a superpotential. However, upon a suitable superpotential deformation, this theory flows to a superconformal fixed point with a non-trivial conformal manifold.

  For the $N_f=0$ case, consider the following superpotential deformation:
  \begin{align}
      W= M_1\Tr\Phi^2 + M_1^2\,.
  \end{align}
  At the IR fixed point, there exists a marginal operator of the form $\Tr\Phi^4$. This deformation breaks the $U(1)_1$ flavor symmetry listed in Table \ref{tab:adj1s1A1F8}. The operator $\Tr\Phi^4$ is exactly marginal since it is neutral under the remaining flavor symmetries. This deformation works for general $N$ with a fixed $N_f=0$.

  For the $N_f=1$ case, upon the following superpotential deformation, a non-trivial conformal manifold emerges at the IR fixed point:
  \begin{align}
    W=\widetilde{Q}_1 S \widetilde{Q}_1 + \widetilde{Q}_2 S \widetilde{Q}_3 + \widetilde{Q}_4 S \widetilde{Q}_5 + \Tr\Phi S\widetilde{A} + M_1 Q\widetilde{Q}_1 + M_1^2 + Q_1\widetilde{Q}_6Q_1\widetilde{Q}_7\,.
  \end{align}
  We can test this by computing the superconformal index. For example, the reduced superconformal index for the $SU(6)$ gauge theory with this superpotential is given by
  \begin{align}
    \CI_{\text{red}}=t^{2.97} + t^3 + \cdots + 2 t^6 + \cdots \,.
  \end{align}
  The positivity of the coefficient at the $t^6$ term indicates the existence of a non-trivial conformal manifold. 
  The marginal operators include, for example, $M_1 Q\widetilde{Q}_2$, $(Q_1\widetilde{Q}_2)^2$, $Q_1\widetilde{Q}_2 Q_1\widetilde{Q}_3$, $\widetilde{Q}_6 S \widetilde{Q}_6$, $\widetilde{Q}_6 S \widetilde{Q}_7$, and others. Such a deformation exists for general $N$ with a fixed $N_f=1$.

\paragraph{Weak Gravity Conjecture}
  We test the AdS WGC using the gauge-invariant operators and $U(1)$ flavor charges identified at the beginning of this section. We find that this theory does not satisfy the NN-WGC for large values of $\a=N_f/N$ in the Veneziano limit, whereas the modified WGC always holds. The result is shown in Figure \ref{fig:wgc_adj1s1A1F8ven}.
  \begin{figure}[t]
    \centering
    \begin{subfigure}[b]{0.45\textwidth}
      \centering
      \includegraphics[width=\linewidth]{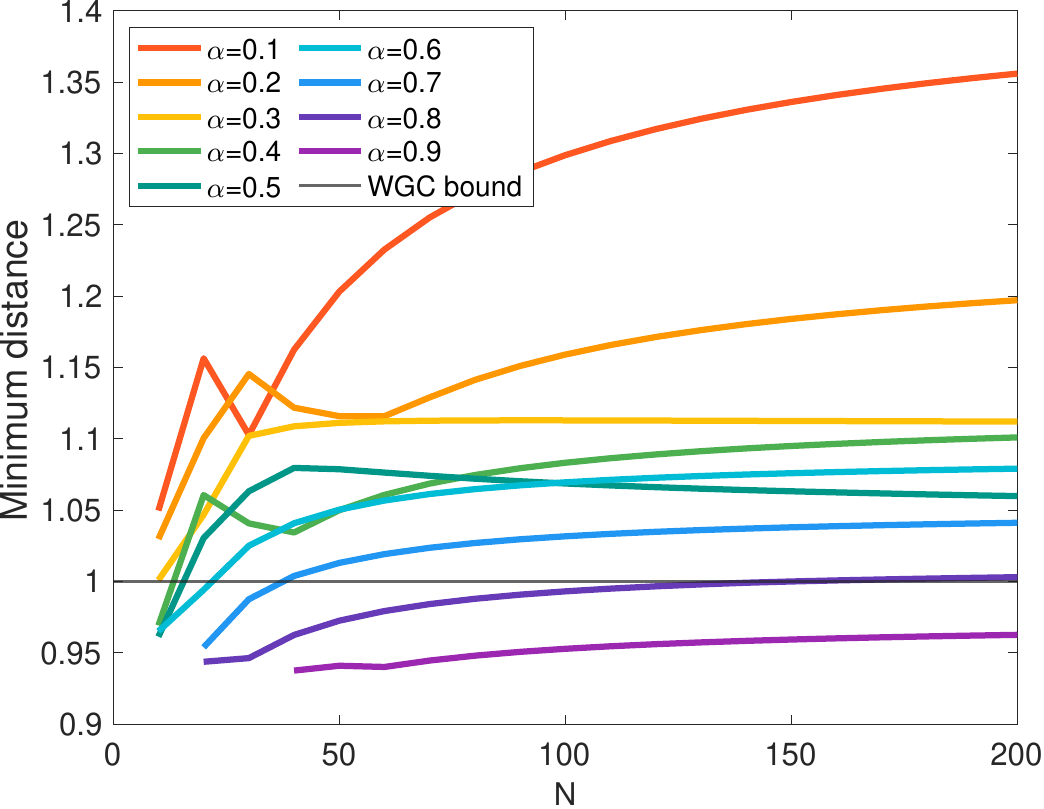}
      \caption{NN-WGC}
    \end{subfigure}
    \hspace{4mm}
    \begin{subfigure}[b]{0.45\textwidth}
      \centering
      \includegraphics[width=\linewidth]{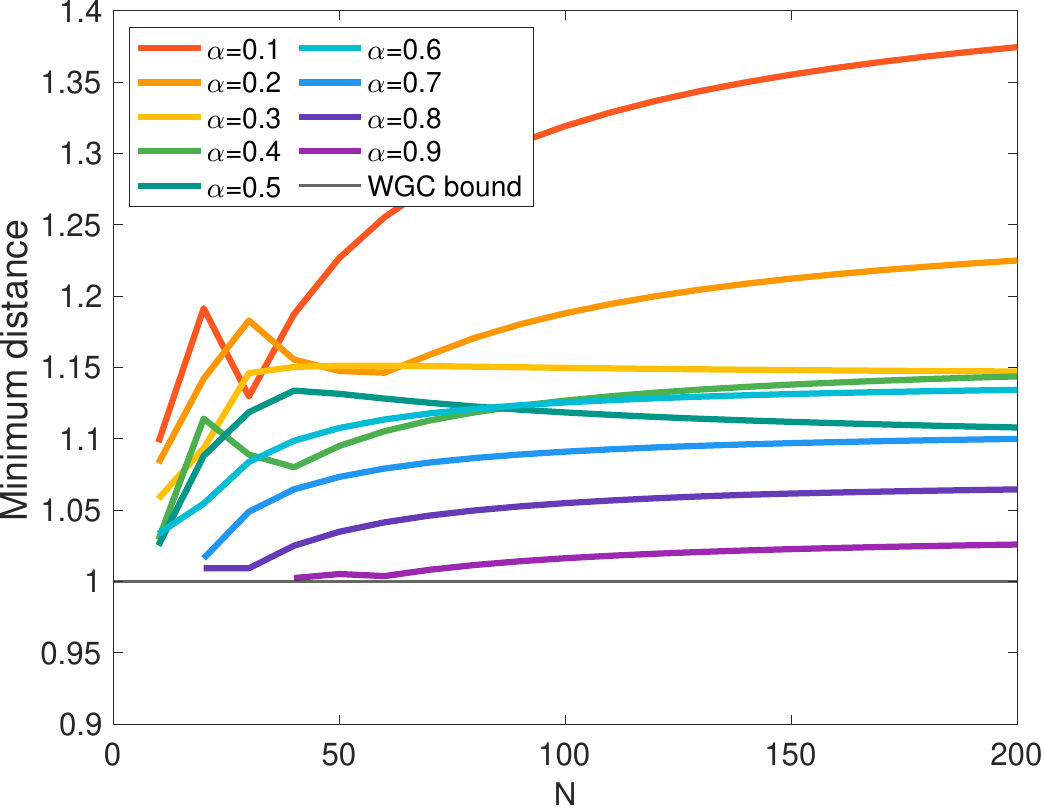}
      \caption{modified WGC}
    \end{subfigure}
    \hfill
    \caption{Testing AdS WGC for $SU(N)$ theory with 1 \textbf{Adj} + 1 $\sym$ + 1 $\overline{\antisym}$ + 8 $\overline{\fund}$ + $N_f$ ( $\fund$ + $\overline{\fund}$ ).
    The minimum distance from the origin to the convex hull with a fixed $\a=N_f/N$. Theories below the solid line at the minimum distance 1 do not satisfy the WGC.\label{fig:wgc_adj1s1A1F8ven}}    
  \end{figure}

\subsection{\texorpdfstring{1 \textbf{Adj} + 1 $\antisym$ + 1 $\overline{\antisym}$ + $N_f$ ( $\fund$ + $\overline{\fund}$ )}{1 Adj + 1 A + 1 At + Nf ( Q + Qt )}}\label{sec:adj1a1A1}

\paragraph{Matter content and symmetry charges}
  Next case of the Type II theory is given by $SU(N)$ gauge theory with an adjoint and a pair of rank-2 anti-symmetric and its conjugate, as well as $N_f$ pairs of fundamental and anti-fundamental chiral multiplets. The matter fields and their $U(1)$ global charges are listed in Table \ref{tab:adj1a1A1}.
  {\renewcommand\arraystretch{1.6}
  \begin{table}[h]
    \centering
    \begin{tabular}{|c|c||c|c|c|c|c|c|c|}
    \hline
    \# & Fields & $SU(N)$ & $U(1)_1$ & $U(1)_2$ & $U(1)_3$ & $ U(1)_4$  & $U(1)_R$ \\\hline
    
     $N_f$ & $Q$ & $\fund$  & $N$ & $N-2$ & 1  & 0 & $R_Q$\\
     
     $N_f$ & $\widetilde{Q}$ & $\overline{\fund}$ & $N$ & $N-2$ & -1 & 0 & $R_{\widetilde{Q}}$ \\
     
     1 & $\Phi$ & \textbf{Adj} & $-N_f$ & 0 & 0 & 0  & $R_\Phi$ \\
     
     1 & $A$  &$\antisym$& 0  & $-N_f$& 0 & 1  & $R_A$ \\
     
     1 & $\widetilde{A}$  &$\overline{\antisym}$& 0 & $-N_f$ & 0  & -1 & $R_{\widetilde{A}}$\\\hline
    \end{tabular}
    \caption{The matter contents and their corresponding charges in $SU(N)$ gauge theory with 1 \textbf{Adj} + 1 $\antisym$ + 1 $\overline{\antisym}$ + $N_f$ ( $\fund$ + $\overline{\fund}$ ).\label{tab:adj1a1A1}}
  \end{table}}
For $SU(3)$, the rank-2 anti-symmetric tensor is the same as the anti-fundamental representations. Thus, we only consider the case with $N>3$ here.

\paragraph{Gauge-invariant operators}
  Let $I$ and $J$ denote the flavor indices for $Q$. We present a sample of single-trace gauge-invariant operators in schematic form as follows: 
  \begin{enumerate}
    \item $\Tr\Phi^n\,,\quad n=2,3,\dots,N,\quad\,.$
    \item $\Tr\Phi^n(A\widetilde{A})^m,\quad n=0,1,\dots,N,\quad m=1,2,\dots,\lfloor \frac{N-1}{2}\rfloor\,.$
    \item $Q_I\Phi^n(\widetilde{A}A)^m\widetilde{Q}_{\tilde{I}},\quad n=0,1,\dots,N-1,\quad m=0,1,\dots,\lfloor\frac{N}{2}\rfloor-1\,.$
    \item $Q_I\widetilde{A}(A\widetilde{A})^n Q_I\,, \quad n=0,1,\dots,\lfloor\frac{N-1}{2}\rfloor-1\,.$
    \item $\e\,A^{i}\CQ_{I_1}^{n_1}\cdots\CQ_{I_{N-2i}}^{n_{N-2i}}\,.$
    \item $\e\,\e\, A^N\Phi^3\,,\quad$ if $N$ is odd.
    \item The conjugates of the above-listed operators.

    $\vdots$
  \end{enumerate}
  The ellipsis indicates that only the low-lying operators have been listed. This subset is sufficient to identify relevant operators or to test the Weak Gravity Conjecture. Here, we define the dressed quarks as:
  \begin{align}
    \CQ_I^n=\begin{cases}
        \Phi^{n/2} Q_I & n=0,2,4\dots\,,\\
        \Phi^{(n+1)/2}A\widetilde{Q}_{\tilde{I}} & n=1,3,5\dots\,.
    \end{cases}
  \end{align}

\paragraph{$R$-charges and central charges}
  
  We perform the $a$-maximization to compute the $R$-charges of the matter fields and central charges.
  When $(N_f,N)=(0,4)$, we find that the gauge-invariant operator $\Tr \Phi^2$ hits the unitarity bound along the RG flow and decouples. In this case, we must flip the decoupled operators in order to perform the $a$-maximization correctly. This can be done by introducing the following superpotential:
  \begin{align}
      W=X\Tr\Phi^2\,, 
  \end{align}
  where $X$ is a flip field. Now, upon correcting the $a$-maximization, we obtain
  \begin{align}
  \begin{split}
  R_\Phi&=\frac{-33+\sqrt{12481}}{267}\simeq 0.2948,\qquad R_A=\frac{333-2\sqrt{12481}}{267}\simeq0.4103\,,\\
  a&=\frac{-528552+12481\sqrt{12481}}{380208}\simeq 2.2772,\quad  c=\frac{-292311+6730\sqrt{12481}}{190104}\simeq 2.4174\,,\\
  \end{split}
  \end{align}
  so that the ratio of central charges is $a/c \simeq 0.9420$. 
  
  For all the other cases, none of the gauge-invariant operators decouple along the RG flow. We obtain the $R$-charges of the matter fields and central charges, in the large $N$ limit with a fixed $N_f$ as
  \begin{align}
    \begin{split}
        R_Q=R_{\widetilde{Q}}&\sim\frac{12-\sqrt{26}}{12}+\frac{78N_f-41\sqrt{26}}{624N}+O(N^{-2})\,,\\
        R_A&\sim \half+\frac{-31+3N_f\sqrt{26}}{72N}+O(N^{-2})\,,\\
        R_\Phi&\sim \half + \frac{-41+3N_f \sqrt{26}}{72N}+ O(N^{-2})\,,\\
        a&\sim \frac{27}{128}N^2+ \frac{-99+13N_f\sqrt{26}}{768}N+O(N^0)\,,\\
        c&\sim\frac{27}{128} N^2 + \frac{-75 + 17N_f\sqrt{26}}{768} N+O(N^0)\,,\\
        a/c&\sim 1 - \frac{12 + 2N_f\sqrt{26}}{81 N} + O(N^{-2})\,.
    \end{split}
  \end{align}
  The leading-order behavior of the $R$-charges and central charges is universal across the Type II theories. Similarly, in the Veneziano limit with a fixed $\a=N_f/N$, the leading terms of $R$-charges and central charges are also universal across Type II theories.

  We observe that the ratio of central charges $a/c$ is always less than one, taking a value within the range $0.875= 7/8 \leq  a / c < 1$. The minimum value of $a/c$ arises when $(N_f,N)=(6,4)$. The maximum value of $a/c$ can be obtained in the large $N$ limit with $N_f=0$. Figure \ref{fig:adj1a1A1ratio} illustrates the behavior of the ratio $a/c$.
  \begin{figure}[h]
    \centering
    \begin{subfigure}[b]{0.45\textwidth}
        \includegraphics[width=\linewidth]{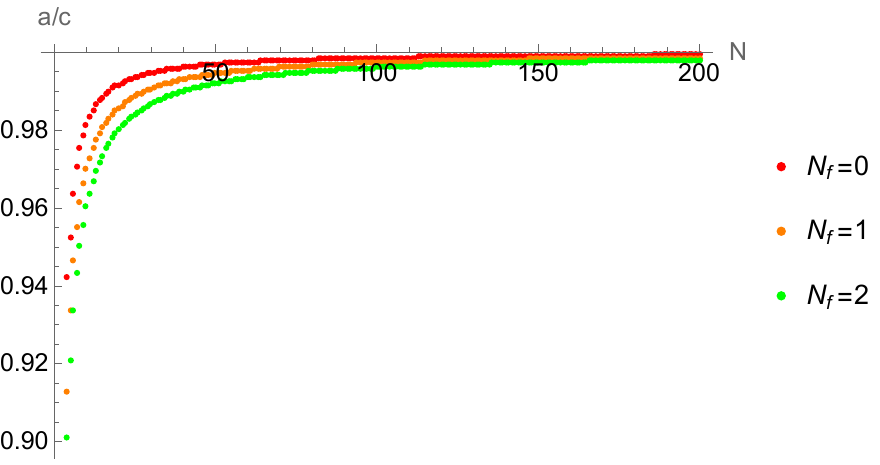}
    \end{subfigure}
    \hspace{4mm}
    \begin{subfigure}[b]{0.45\textwidth}
        \includegraphics[width=\linewidth]{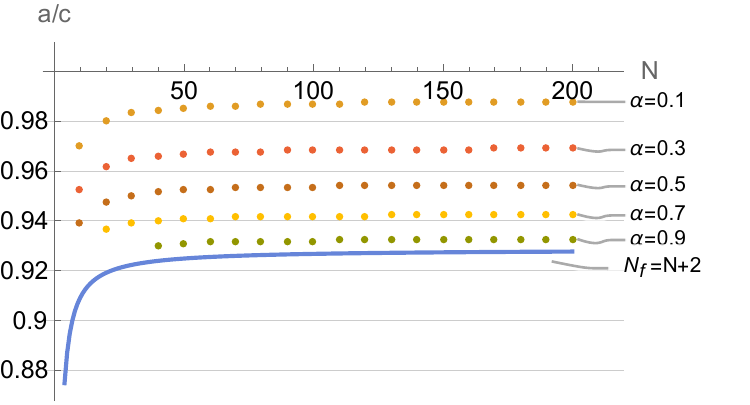}
    \end{subfigure}
    \hfill
    \caption{
    The central charge ratio for $SU(N)$ theory with 1 \textbf{Adj} + 1 $\antisym$ + 1 $\overline{\antisym}$ + $N_f$ ( $\fund$ + $\overline{\fund}$ ).
    Left: $a/c$ versus $N$ with a fixed $N_f$. Right: $a/c$ versus $N$ with a fixed $\a=N_f/N$.\label{fig:adj1a1A1ratio}}
  \end{figure}

\paragraph{Conformal window}
  The upper bound of the conformal window is determined by the asymptotic freedom. For this theory, when it is at the top of the conformal window $N_f=N+2$, there is a non-trivial conformal manifold \cite{Razamat:2020pra} so that it is an interacting SCFT. For small $N_f$, the $a$-maximization always yields a unique solution, and the resulting central charges lie within the Hofman-Maldacena bound. We therefore conjecture that this theory flows to an interacting SCFT for $0\leq N_f \leq N+2$.

\paragraph{Relevant operators}
  For generic $N$ and $N_f$, except for the cases where some operators get decoupled, we have the following set of relevant operators:
  \begin{itemize}
    \item an operator of the form $\Tr\Phi^2$ with a dimension $ 1.0448 \lesssim \D < 2\,,$
    \item an operator of the form $\Tr A\widetilde{A}$ with a dimension $1.2588\lesssim \D <2\,,$
    \item an operator of the form $\Tr \Phi A\widetilde{A}$ with a dimension $1.7810 \lesssim \D < 3\,,$
    \item an operator of the form $\Tr\Phi^3$ with a dimension $1.5672 \lesssim \D < 3\,,$
    \item $N_f^2$ operators of the form $Q_I\widetilde{Q}_{\tilde{J}}$ with a dimension $1.5418 \lesssim \D < 2\,,$
    \item $N_f^2$ operators of the form $Q_I\Phi \widetilde{Q}_{\tilde{J}}$ with a dimension $ 2.1195 \lesssim \D < 3 $,
    \item $\frac{N_f(N_f-1)}{2}$ operators of the form $Q_I \widetilde{A} Q_{J}$ with a dimension $2.4495 < \D < 3$, 
    \item (if $N_f \leq 2$) an operator of the form $\Tr\Phi^4$ and $(\Tr\Phi^2)^2$ with a dimension $2.0896 \lesssim \D < 3\,,$
    \item (if $N_f \leq 2$) an operator of the form $\Tr\Phi^2 A\widetilde{A}$ and $(\Tr\Phi^2 \cdot \Tr A\widetilde{A})$ with a dimension $2.3035 \lesssim \D < 3\,,$
    \item (if $N_f \leq 2$) an operator of the form $\Tr (A\widetilde{A})^2$ and $(\Tr A\widetilde{A})^2$ with a dimension $2.5173 \lesssim \D < 3 $. 
  \end{itemize}
  The number of relevant operators does not depend on $N$ for a fixed $N_f$. The low-lying operator spectrum is sparse in the large $N$ limit.

  Upon deforming the theory by the superpotential $W = \Tr \Phi^{k+1}+\Tr\Phi A\widetilde{A}$, a dual description is proposed in \cite{Brodie:1996xm}, which is given by $SU(3kN_f-4-N)$ gauge theory with a certain superpotential and flip fields. We find the first term in the superpotential for $k>2$ is irrelevant near $W=0$, but at the fixed point of $W=\Phi A \tilde{A}$, it is relevant for large enough $N$ (and small $N_f$). 

\paragraph{Conformal manifold}
   When $N_f=N+2$, the one-loop beta function for the gauge coupling vanishes. At this value, the theory possesses a non-trivial conformal manifold that contains a one-dimensional subspace that preserves $\CN=2$ supersymmetry \cite{Bhardwaj:2013qia, Razamat:2020pra}.
   
   When $N_f<N+2$, there are no marginal operators for generic $N$ in the absence of a superpotential. However, upon a suitable superpotential deformation, this theory flows to a superconformal fixed point with a non-trivial conformal manifold.

   For the $N_f=0$ case, upon the following superpotential deformation, a non-trivial conformal manifold emerges at the IR fixed point:
   \begin{align}
       W= M_1\Tr\Phi^2 + \Tr\Phi^4\,.
   \end{align}
   Here, $M_1$ is a flip field, which is a gauge-singlet chiral superfield. In this deformed theory, the only remaining flavor symmetry is the $U(1)$ symmetry under which the $A$ and $\widetilde{A}$ fields carry charges $+1$ and $-1$, respectively, while all other fields are neutral. There exists a marginal operator of the form $M_1^2$, which is uncharged under the flavor symmetry. Thus, this deformation leads to a one-dimensional conformal manifold for general $N$.

  For the $N_f=1$ case, consider the following superpotential deformation:
  \begin{align}
    W=\Tr \Phi^2 A\widetilde{A} + X_1 Q\widetilde{Q}\,,
  \end{align}
  where $X_1$ is a flip field. At the IR fixed point, this theory possesses a non-trivial conformal manifold. 
  We can verify this using the superconformal index. For instance, the reduced index for the $SU(5)$ gauge theory with this superpotential has a positive coefficient at the $t^6$ term:
  \begin{align}
    \CI_{\text{red}}=t^{2.09}+ t^{2.86}+\cdots + t^{5.86} + 2t^6 + \cdots\,.
  \end{align}
  The marginal operators take the form $Q\Phi^4 \widetilde{Q}$, $(\Tr\Phi^2) (Q\Phi^2\widetilde{Q})$, and $(\Tr\Phi^3) (Q\Phi\widetilde{Q})$. Such a deformation exists for general $N$ with a fixed value of $N_f=1$.

\paragraph{Weak Gravity Conjecture}
  \begin{figure}[t]
    \centering
    \begin{subfigure}[b]{0.45\textwidth}
      \centering
      \includegraphics[width=\linewidth]{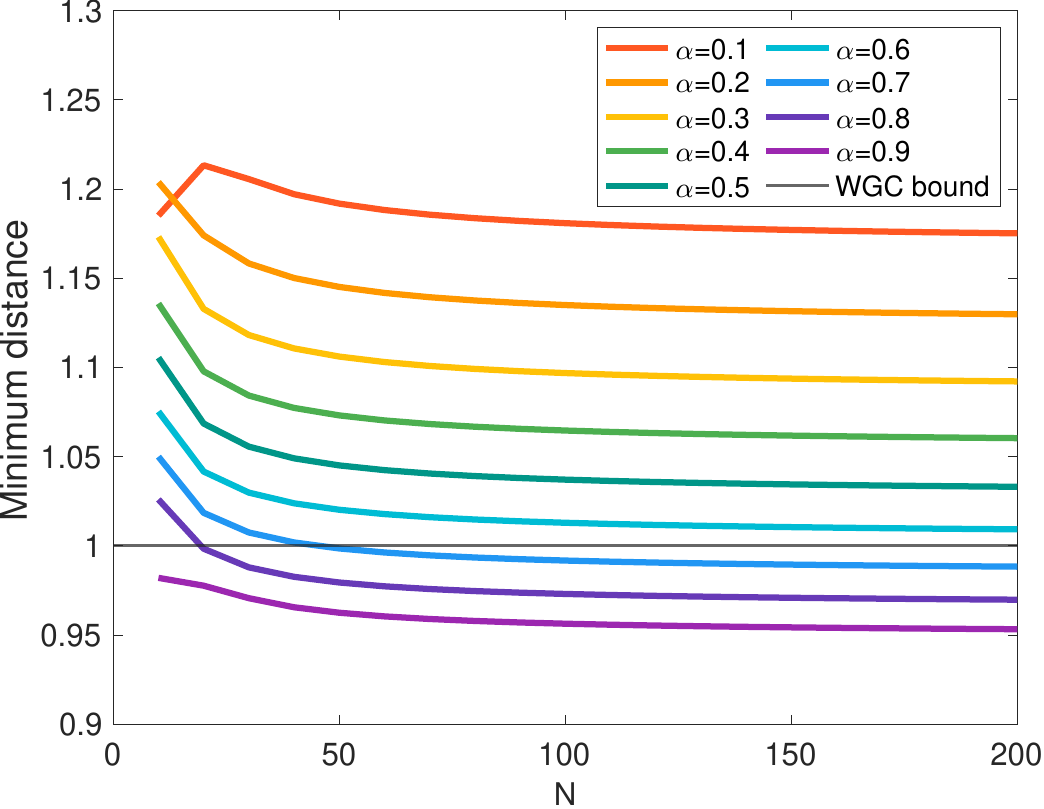}
      \caption{original WGC}
    \end{subfigure}
    \hspace{4mm}
    \begin{subfigure}[b]{0.45\textwidth}
      \centering
      \includegraphics[width=\linewidth]{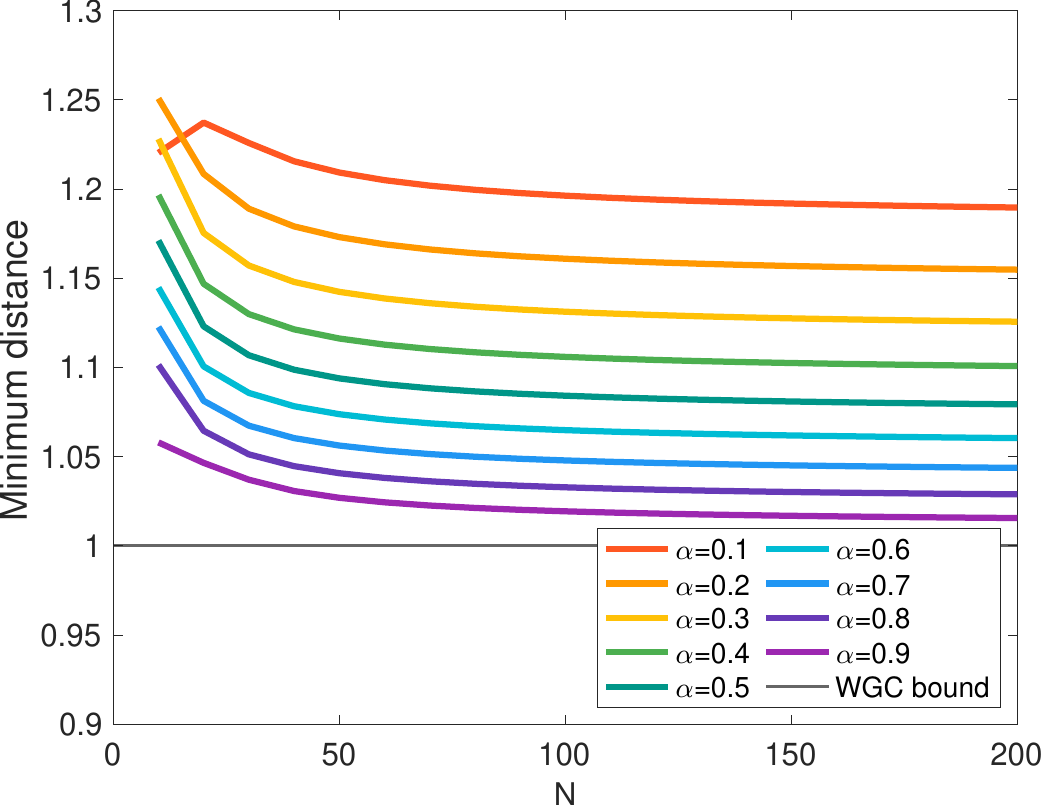}
      \caption{modified WGC}
    \end{subfigure}
    \hfill
    \caption{Testing AdS WGC for $SU(N)$ theory with 1 \textbf{Adj} + 1 $\antisym$ + 1 $\overline{\antisym}$ + $N_f$ ( $\fund$ + $\overline{\fund}$ ).
    The minimum distance from the origin to the convex hull with a fixed $\a=N_f/N$. Theories below the solid line at the minimum distance 1 do not satisfy the WGC.\label{fig:wgc_adj1a1A1ven}}
  \end{figure}
  We test the AdS WGC using the gauge-invariant operators and $U(1)$ flavor charges identified at the beginning of this section. We find that this theory does not satisfy the NN-WGC for large values of $\a=N_f/N$ in the Veneziano limit, whereas the modified WGC always holds. The result is shown in Figure \ref{fig:wgc_adj1a1A1ven}.

\subsection{\texorpdfstring{2 \textbf{Adj} + $N_f$ ( $\fund$ + $\overline{\fund}$ )}{2 Adj + Nf ( Q + Qt )}\label{sec:adj2}} \label{sec:sutwoadj}

\paragraph{Matter content and symmetry charges}
  The last Type II example is an $SU(N)$ gauge theory with two adjoints and $N_f$ pairs of fundamental and anti-fundamental chiral multiplets. The matter fields and their $U(1)$ global charges are listed in Table \ref{tab:adj2}.
  {\renewcommand\arraystretch{1.6}
  \begin{table}[h]
    \centering
    \begin{tabular}{|c|c||c|c|c|c|}
    \hline
    \# & Fields & $SU(N)$ &$U(1)_1$ & $U(1)_2$ & $U(1)_R$ \\\hline
    
     $N_f$ & $Q$ & $\fund$ & $2N$  & $1$  & $R_Q$\\
     
     $N_f$ & $\widetilde{Q}$ & $\overline{\fund}$ & $2N$  & $-1$ & $R_{\widetilde{Q}}$ \\
     
     2 & $\Phi$ &\textbf{Adj}& $-N_f$  & 0 & $R_\Phi$ \\\hline
    \end{tabular}
    \caption{The matter contents and their corresponding charges in $SU(N)$ gauge theory with 2 \textbf{Adj} + $N_f$ ( $\fund$ + $\overline{\fund}$ )}
    \label{tab:adj2}
  \end{table}}

\paragraph{Gauge-invariant operators}
  Let $I$ and $J$ denote the flavor indices for $Q$, and $P$ denote the flavor indices for $\Phi$. We present a sample of single-trace gauge-invariant operators in schematic form as follows:
  \begin{enumerate}
    \item $\Tr(\Phi_{P_1}\cdots\Phi_{P_n}),\quad n=2,3,\dots\,.$
    \item $Q_I\Phi_{P_1}\cdots\Phi_{P_n}\widetilde{Q}_{\tilde{I}},\quad n=0,1,\dots\,.$
    \item $\epsilon\mathcal{Q}_{I_1}^{n_1}\cdots\mathcal{Q}_{I_N}^{n_N},\quad\e\widetilde{\mathcal{Q}}_{I_1}^{n_1}\cdots\widetilde{\mathcal{Q}}_{I_N}^{n_N}$

    $\vdots$
  \end{enumerate}
  The ellipsis indicates that only the low-lying operators have been listed. This subset is sufficient to identify relevant operators or to test the Weak Gravity Conjecture.
  Here, we define the dressed quarks as $\CQ_I^n=\Phi_{P_1}\cdots\Phi_{P_n} Q_I$ and $\widetilde{\CQ}_{I}^n=\Phi_{P_1}\cdots\Phi_{P_n}\widetilde{Q}_{\tilde{I}}$.

\paragraph{$R$-charges and central charges}
  We perform the $a$-maximization to compute the $R$-charges of the matter fields and central charges. In the large $N$ limit with a fixed $N_f$, they are given by
  \begin{align}
  \begin{split}
    R_Q = R_{\widetilde{Q}} & \sim \frac{12-\sqrt{26}}{12} + \frac{N_f}{8N}+ O(N^{-2})\,, \\
    R_\Phi & \sim \half +\frac{N_f\sqrt{26}}{24N} + O(N^{-2}) \,,\\
    a & \sim \frac{27}{128}N^2 + \frac{13N_f\sqrt{26}}{768}N+O(N^0)\,,\\
    c & \sim \frac{27}{128}N^2 + \frac{17 N_f\sqrt{26}}{768}N+O(N^0)\,,\\
    a/c & \sim 1- \frac{2N_f\sqrt{26}}{81N} + O(N^{-2})\,.
  \end{split}
  \end{align}
  We find that no gauge-invariant operators decouple along the RG flow. The leading-order behavior of the $R$-charges and central charges is universal across Type II theories. Similarly, in the Veneziano limit with a fixed $\a=N_f/N$, the leading terms of $R$-charges and central charges are also universal across Type II theories.

  We observe that the ratio of central charges $a/c$ lies within the range $41/46\simeq 0.8913 \leq  a / c \leq 1$. The minimum value of $a/c$ arises when $(N_f,N)=(2,2)$. 
  If $N_f=0$, $a/c$ is exactly 1. This is a prototypical example of $a=c$ theory \cite{Kang:2021ccs}. This theory is known to have a non-Lagrangian dual theory \cite{Kang:2024elv}. 
  Figure \ref{fig:adj2ratio} illustrates the behavior of the ratio $a/c$.
  \begin{figure}[t]
    \centering
    \begin{subfigure}[b]{0.45\textwidth}
        \includegraphics[width=\linewidth]{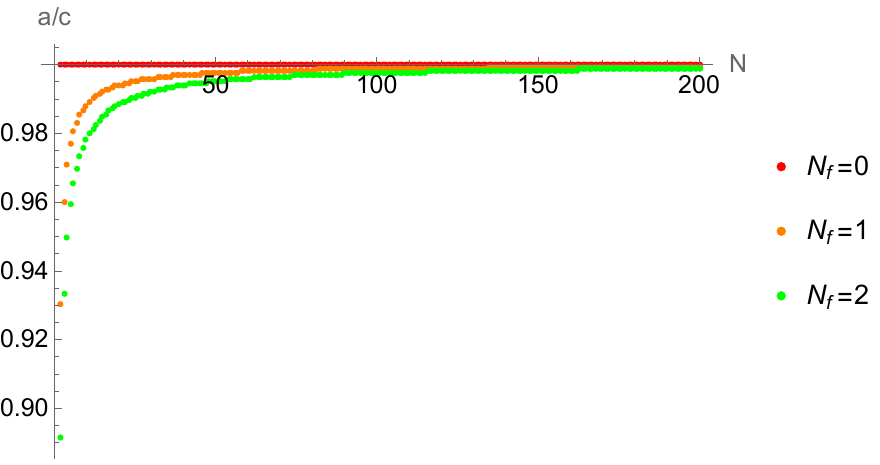}
    \end{subfigure}
    \hspace{4mm}
    \begin{subfigure}[b]{0.45\textwidth}
        \includegraphics[width=\linewidth]{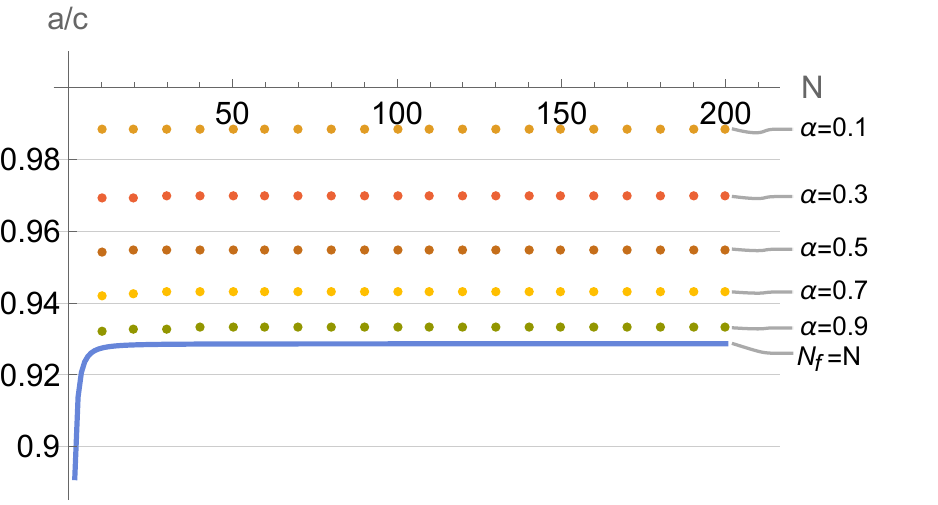}
    \end{subfigure}
    \hfill
    \caption{
    The central charge ratio for $SU(N)$ theory with 2 \textbf{Adj} + $N_f$ ( $\fund$ + $\overline{\fund}$ ).
    Left: $a/c$ versus $N$ with a fixed $N_f$. Right: $a/c$ versus $N$ with a fixed $\a=N_f/N$.\label{fig:adj2ratio}}
  \end{figure}

\paragraph{Conformal window}
  The upper bound of the conformal window is determined by the asymptotic freedom. For this theory, when it is at the top of the conformal window $N_f=N$, there is a non-trivial conformal manifold \cite{Razamat:2020pra} so that it is an interacting SCFT. 
  For small $N_f$, the $a$-maximization always yields a unique solution, and the resulting central charges lie within the Hofman-Maldacena bound. We therefore conjecture that this theory flows to an interacting SCFT for $0\leq N_f \leq N$.
  
\paragraph{Relevant operators}
  For generic $N$ and $N_f$, we have the following set of relevant operators:
  \begin{itemize}
    \item three operators of the form $\Tr\Phi_{P_1}\Phi_{P_2}$ with a dimension $\frac{3}{2}\leq \D \leq 2\,,$
    \item four operators of the form $\Tr\Phi_{P_1}\Phi_{P_2}\Phi_{P_3}$ with a dimension $\frac{9}{4}\leq \D \leq 3\,,$
    \item $N_f^2$ operators of the form $Q_I\widetilde{Q}_{\tilde{J}}$ with a dimension $1.7253\simeq \frac{12-\sqrt{26}}{4}< \D \leq2\,,$
    \item $2N_f^2$ operators of the form $Q_I\Phi_P \widetilde{Q}_{\tilde{J}}$ with a dimension $2.4753\simeq\frac{15-\sqrt{26}}{4}<\D \leq 3\,.$
  \end{itemize}
  We get the lower bound of operator dimensions above when $N_f=0$, while the upper bound appears when $N_f=N$. The number of relevant operators does not depend on $N$ for a fixed $N_f$. The low-lying operator spectrum is sparse in the large $N$ limit, which is a feature of Type II theory. 

  Upon deforming the theory by the superpotential $W = \Tr \Phi_1 ^{k+1}+\Tr\Phi_1\Phi_2^2$, the theory admits a dual description given by $SU(3kN_f-N)$ gauge theory with certain superpotential and flip fields \cite{Brodie:1996vx}. The first term in the superpotential is irrelevant at the $W=0$ fixed point for $k>2$. However, the second term makes the first term relevant for large enough $N$ (and small $N_f$). Therefore, we have a non-trivial fixed-point SCFT described by this superpotential. 
  In fact, it was shown in \cite{Intriligator:2003mi} that there exist superconformal fixed points described by the superpotentials given by polynomials describing ADE singularities. 
  Another duality has been proposed in \cite{Kutasov:2014yqa} for $W=\Tr \Phi_1^3 + \Tr\Phi_2^5$ and tested in \cite{Intriligator:2016sgx}.  
  
\paragraph{Conformal manifold}
  When $N_f=N$, the one-loop beta function for the gauge coupling vanishes. At this value, the theory possesses a non-trivial conformal manifold \cite{Razamat:2020pra}.
  
  When $N_f=0$, there exist twelve quartic marginal operators and the theory has a three-dimensional non-trivial conformal manifold. For details, see \cite{Kang:2022vab, Kang:2024elv}. 
  
  For $0<N_f< N$, there are no marginal operators for generic $N$ in the absence of a superpotential. However, upon a suitable superpotential deformation, this theory flows to a superconformal fixed point with a non-trivial conformal manifold. 
  
  For example, when $N_f=1$, upon the following superpotential deformation, a non-trivial conformal manifold emerges at the IR fixed point:
  \begin{align}
      W=Q\Phi_1\widetilde{Q} + M_1 \Phi_1\Phi_2 + M_2\Phi_2^2\,.
  \end{align}
  Here, $M_i$'s are flip fields, which are gauge-singlet chiral superfields. We can test this by computing the superconformal index. For example, the reduced superconformal index for the $SU(2)$ gauge theory with this superpotential is given by
  \begin{align}
      \CI_{\text{red}}=2t^{2.57} + t^{3.43}+ t^{4.29} - 2 t^{4.71}\left(y+\frac{1}{y}\right) + 3 t^{5.14}+ t^6 + \cdots\,.
  \end{align}
  The positivity of the coefficient at the $t^6$ term indicates the existence of a non-trivial conformal manifold. The exact marginal operator takes the form $M_2\Tr\Phi_1^2$, which is neutral under the remaining flavor symmetry after the deformation. This deformation works for general $N$ with a fixed $N_f=1$.

\paragraph{Weak Gravity Conjecture}
  We examine the AdS WGC using the gauge-invariant operators and $U(1)$ flavor charges identified at the beginning of this section. We find that this theory always satisfies both versions of the WGC. The result is shown in Figure \ref{fig:wgc_adj2ven}. 
  \begin{figure}[t]
    \centering
    \begin{subfigure}[b]{0.45\textwidth}
      \centering
      \includegraphics[width=\linewidth]{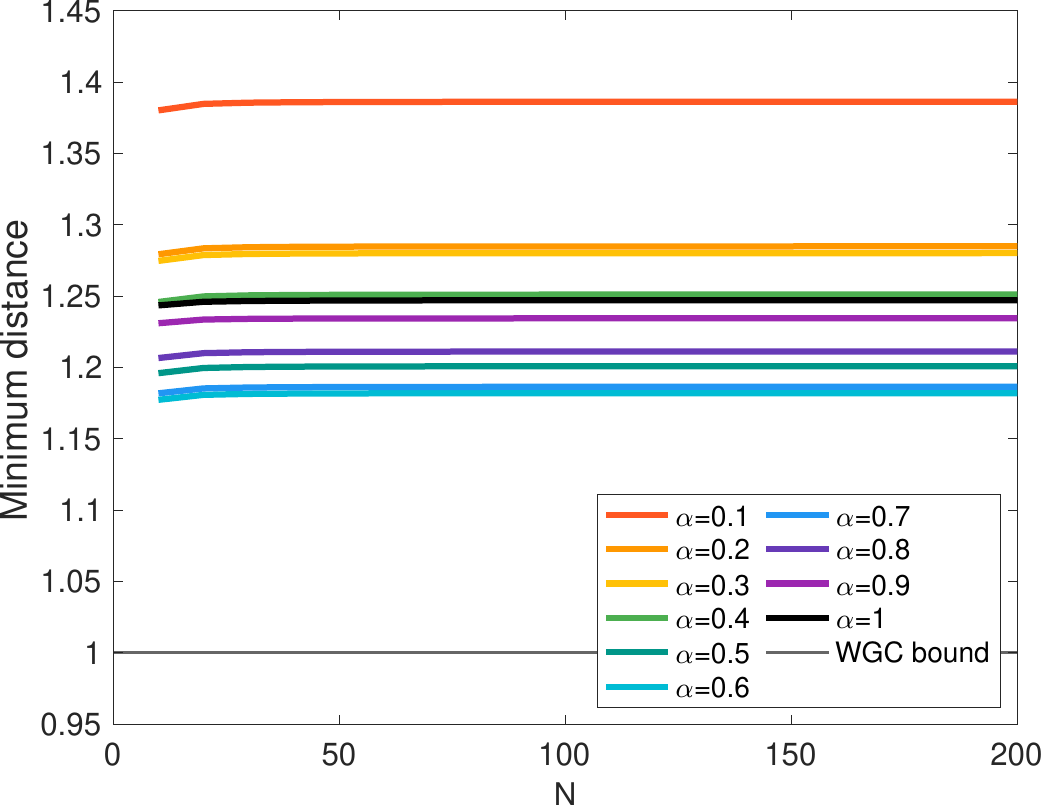}
      \caption{original WGC}
    \end{subfigure}
    \hspace{4mm}
    \begin{subfigure}[b]{0.45\textwidth}
      \centering
      \includegraphics[width=\linewidth]{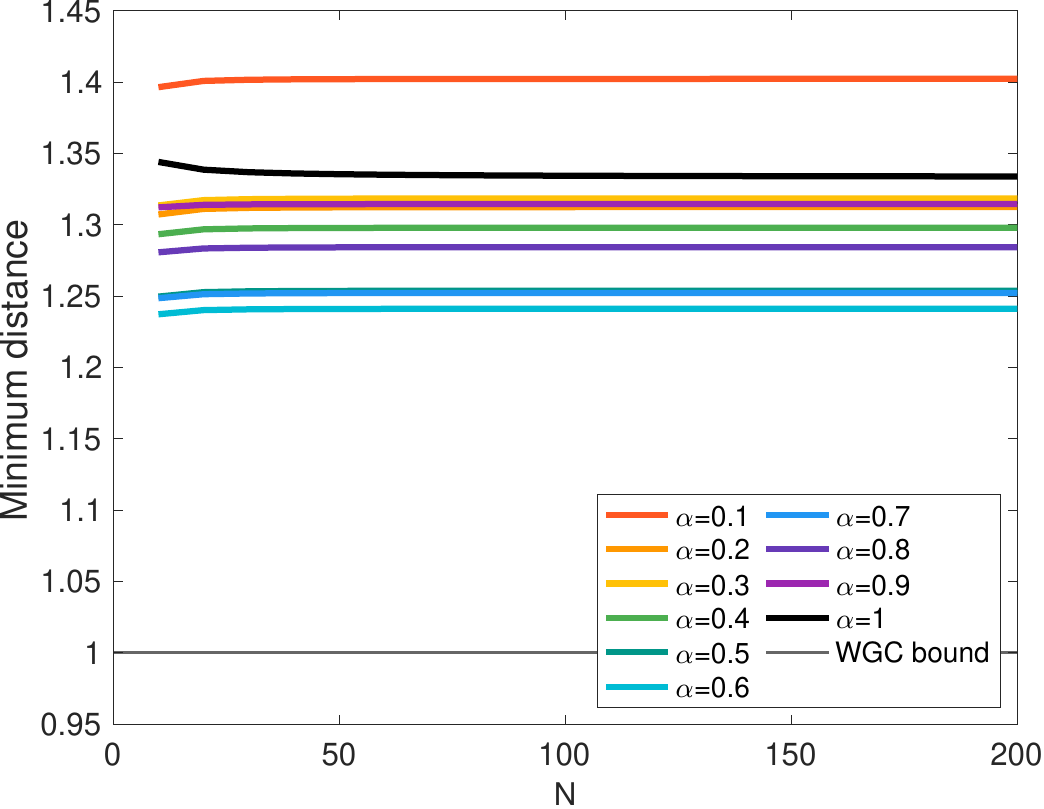}
      \caption{modified WGC}
    \end{subfigure}
    \hfill
    \caption{Testing AdS WGC for $SU(N)$ theory with 2 \textbf{Adj} + $N_f$ ( $\fund$ + $\overline{\fund}$ ).
    The minimum distance from the origin to the convex hull with a fixed $\a=N_f/N$. Theories below the solid line at the minimum distance 1 do not satisfy the WGC.\label{fig:wgc_adj2ven}}
   \end{figure}

\subsection{\texorpdfstring{1 $\sym$ + 1 $\overline{\sym}$ + 2 $\antisym$ + 2 $\overline{\antisym}$ + $N_f$ ( $\fund$ + $\overline{\fund}$ )}{1 S + 1 St + 2 A + 2 At + Nf ( Q + Qt )}}
\paragraph{Matter content and symmetry charges}
  The first example of the Type III theory is an $SU(N)$ gauge theory with a pair of rank-2 symmetric and its conjugate, two pairs of rank-2 anti-symmetric and its conjugate, and $N_f$ pairs of fundamental and anti-fundamental chiral multiplets. The matter fields and their $U(1)$ global charges are listed in Table \ref{tab:s1S1a2A2}.
  {\renewcommand\arraystretch{1.6}
  \begin{table}[h]
    \centering
    \begin{tabular}{|c|c||c|c|c|c|c|c|c|c|}
    \hline
    \# & Fields & $SU(N)$ & $U(1)_1$ & $U(1)_2$ & $U(1)_3$ & $U(1)_4$ & $U(1)_5$   & $U(1)_R$ \\\hline
    
     $N_f$ & $Q$ & $\fund$ & $2N-4$  & 0 & 1 & 0 & 0 & $R_Q$\\
     
     $N_f$ & $\widetilde{Q}$ &$\overline{\fund}$ & $2N-4$  & 0  & $-1$ & 0 & 0  & $R_{\widetilde{Q}}$ \\
     
     1 & $S$&$\sym$ & 0 & $2N-4$ & 0 & 1 & 0   &  $R_S$ \\
     
     1 & $\widetilde{S}$ &$\overline{\sym}$ & 0 & $2N-4$ & 0& $-1$ & 0  & $R_{\widetilde{S}}$\\
     
     2 & $A$ & $\antisym$  & $-N_f$ &$-N-2$  & 0  & 0 & 1 & $R_A$ \\
     
     2 & $\widetilde{A}$ &$\overline{\antisym}$ & $-N_f$  & $-N-2$ & 0 & 0 & $-1$ & $R_{\widetilde{A}}$\\\hline
    \end{tabular}
    \caption{The matter contents and their corresponding charges in $SU(N)$ gauge theory with 1 $\sym$ + 1 $\overline{\sym}$ + 2 $\antisym$ + 2 $\overline{\antisym}$ + $N_f$ ( $\fund$ + $\overline{\fund}$ ).}
    \label{tab:s1S1a2A2}
  \end{table}}
  
For $N=3$, the rank-2 anti-symmetric tensor is the same as the anti-fundamental representations. Therefore, we only consider cases with $N > 3$ here.

\paragraph{Gauge-invariant operators}
  Let $I$ and $J$ denote the flavor indices for $Q$, and $M$ denote the flavor indices for $A$. We present a sample of single-trace gauge-invariant operators in schematic form as follows:
  \begin{enumerate}
    \item $\Tr (S\widetilde{S})^n,\quad \Tr (S\widetilde{A}_{\widetilde{M}_1}\cdots S\widetilde{A}_{\widetilde{M}_{2m}}),\quad n=1,2,\dots, N-1,\quad m=1,2,\dots\,.$
    \item $\Tr (A_{M_1}\widetilde{A}_{\widetilde{M}_1}\cdots A_{M_n}\widetilde{A}_{\widetilde{M}_n}),\quad n=1,2,\dots\,.$
    \item $\Tr(S\widetilde{A}_{\widetilde{M}_1}\cdots S\widetilde{A}_{\widetilde{M}_n})(A\widetilde{A}_{\widetilde{M}_{n+1}} \cdots A\widetilde{A}_{\widetilde{M}_{n+m}})(S\widetilde{S})^{l}\,,\quad n=0,1,\dots, \quad m,l=1,2,\dots\,.$
    \item $Q_I(\widetilde{S}S)^k (A_{M_1}\widetilde{A}_{\widetilde{M}_1}\cdots A_{M_l}\widetilde{A}_{\widetilde{M}_l})(S\widetilde{A}_{\widetilde{M}_{l+1}}\cdots S\widetilde{A}_{\widetilde{M}_{l+n}})(A_{M_{l+1}}\widetilde{S}\cdots A_{M_{l+m}}\widetilde{S})\widetilde{Q}_{\tilde{J}}\,.$
    \item $Q_I\widetilde{S}(S\widetilde{S})^k (A_{M_1}\widetilde{A}_{\widetilde{M}_1}\cdots A_{M_l}\widetilde{A}_{\widetilde{M}_l})(S\widetilde{A}_{\widetilde{M}_{l+1}}\cdots S\widetilde{A}_{\widetilde{M}_{l+n}})(A_{M_{l+1}}\widetilde{S}\cdots A_{M_{l+m}}\widetilde{S})Q_{J}\,.$
    \item $Q_I\widetilde{A}_{\widetilde{M}_0}(S\widetilde{S})^k (A_{M_1}\widetilde{A}_{\widetilde{M}_1}\cdots A_{M_l}\widetilde{A}_{\widetilde{M}_l})(S\widetilde{A}_{\widetilde{M}_{l+1}}\cdots S\widetilde{A}_{\widetilde{M}_{l+n}})(A_{M_{l+1}}\widetilde{S}\cdots A_{M_{l+m}}\widetilde{S})Q_{J}\,.$
    \item $\e A_{M_1}\cdots A_{M_i} \CQ_{{I}_1}^{n_1}\cdots \CQ_{{I}_{N-2i}}^{n_{N-2i}}\,.$
    \item $\e\,\e\, S^i A_{M_1}\cdots A_{M_{2j}}  (\CQ_{I_1}\CQ_{J_1})\cdots (\CQ_{I_k}\CQ_{J_k}),\quad i+2j+k=N$.
    \item $\e\,\e\, A_{M_1}^{\lfloor N/2\rfloor}(A_{(M_1}\widetilde{A})A_{M_2)}^{N-\lfloor N/2\rfloor}\quad$ if $N$ is odd.
    \item The conjugates of the above-listed operators.

    $\vdots$
  \end{enumerate}
  The ellipsis indicates that only the low-lying operators have been listed. This subset is sufficient to identify relevant operators or to test the Weak Gravity Conjecture.
  Here, the dressed quarks are defined as
  \begin{align}
    \CQ_I^n=\begin{cases}
        (S\widetilde{S})^i(S\widetilde{A}_{\widetilde{M}_1}\cdots S\widetilde{A}_{\widetilde{M}_{n/2-i}}) Q & n=0,2,4,\dots\,,\\
        (S\widetilde{S})^i(S\widetilde{A}_{\widetilde{M}_1}\cdots S\widetilde{A}_{\widetilde{M}_{(n-1)/2-i}}) S\widetilde{Q} & n=1,3,5\dots\,.
    \end{cases}
  \end{align}

\paragraph{$R$-charges and central charges}
  We perform the $a$-maximization to compute the $R$-charges of the matter fields and central charges. In the large $N$ limit with a fixed $N_f$, we obtain
  \begin{align}
  \begin{split}
    R_Q=R_{\widetilde{Q}}&\sim \frac{2}{3}+ \frac{N_f-2}{18N}+O(N^{-2}) \,,\\
    R_S=R_{\widetilde{S}}&\sim \frac{2}{3}+ \frac{N_f-2}{9N}+O(N^{-2})\,,\\
    R_A=R_{\widetilde{A}}&\sim \frac{2}{3}+\frac{N_f-2}{9N}+O(N^{-2})\,,\\
    a&\sim \frac{1}{4}N^2+ \frac{2N_f-1}{48}N+O(N^0)\,,\\
    c&\sim \frac{1}{4}N^2 + \frac{N_f}{16}N + O(N^0)\,,\\
    a/c&\sim 1-\frac{N_f+1}{12N}+O(N^{-2})\,.
  \end{split}
  \end{align}
  We find that no gauge-invariant operators decouple along the RG flow. The leading-order behavior of the $R$-charges and central charges is universal across Type III theories.

  We observe that the ratio of central charges $a/c$ is always less than one, taking a value within the range $0.9430 \leq  a / c < 1$. The minimum value of $a/c$ arises when $(N_f,N)=(1,4)$. The maximum value of $a/c$ can be obtained in the large $N$ limit with $N_f=0$. Figure \ref{fig:s1S1a2A2ratio} illustrates the behavior of the ratio $a/c$.
  \begin{figure}[t]
    \centering
    \includegraphics[width=.5\linewidth]{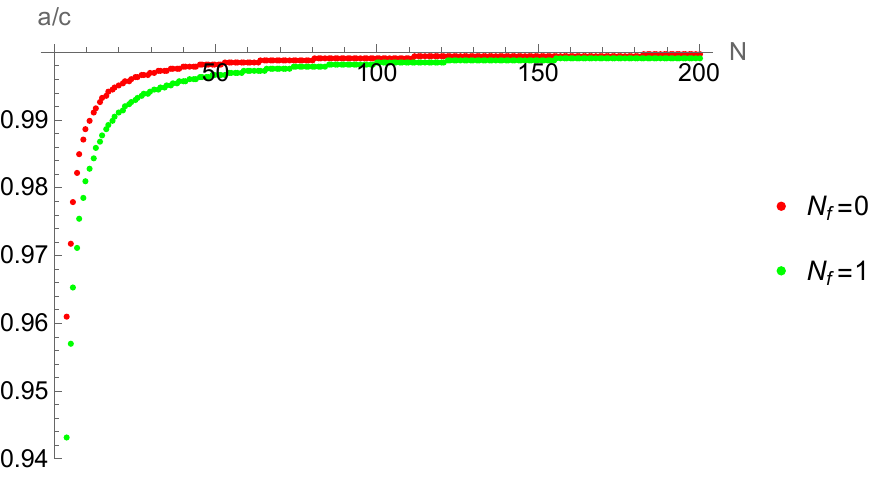}
    \caption{
    The central charge ratio $a/c$ versus $N$ for $SU(N)$ theory with 1 $\sym$ + 1 $\overline{\sym}$ + 2 $\antisym$ + 2 $\overline{\antisym}$ + $N_f$ ( $\fund$ + $\overline{\fund}$ ).
    \label{fig:s1S1a2A2ratio}}
  \end{figure}

\paragraph{Conformal window}
  The asymptotic freedom bounds the upper limit of the conformal windows to be $N_f<2$. The $a$-maximization always yields a unique solution, and the resulting central charges lie within the Hofman-Maldacena bound. We therefore conjecture that this theory flows to an interacting SCFT for $0\leq N_f < 2$. Therefore, there is no Veneziano-type limit. 

\paragraph{Relevant operators}
  For generic $N$ and $N_f$, there exist the following relevant operators:
  \begin{itemize}
    \item an operator of the form $\Tr S\widetilde{S}$ with a dimension $1.7598\lesssim \D < 2\,,$
    \item four operators of the form $\Tr A_{M_1}\widetilde{A}_{\widetilde{M}_2}$ with a dimension $1.8603\lesssim \D < 2\,,$
    \item $N_f^2$ operators of the form $Q_I\widetilde{Q}_{\tilde{J}}$ with a dimension $1.9506 \lesssim \D <2\,,$
    \item $\frac{N_f(N_f+1)}{2}$ operators of the form $Q_I\widetilde{S}Q_J$ and $\widetilde{Q}_{\tilde{I}}S\widetilde{Q}_{\tilde{J}}$ with a dimension $2.8932\lesssim\D<3\,,$
  \end{itemize}
   The number of relevant operators does not depend on $N$ for a fixed $N_f$. The low-lying operator spectrum is sparse in the large $N$ limit, which is a universal feature of Type III theories.

\paragraph{Conformal manifold}
  There are no marginal operators for generic $N$ and $N_f=0,1$ in the absence of a superpotential. However, upon a suitable superpotential deformation, the theory flows to a superconformal fixed point with a non-trivial conformal manifold.

  For any $N_f=0,1$, upon the following superpotential deformation, a non-trivial conformal manifold emerges at the IR fixed point:
  \begin{align}
    W=M\Tr A_1\widetilde{A}_1 + M\Tr A_2\widetilde{A}_2 + M^2\,,
  \end{align}
  where $M$ is a flip field, which is a gauge-singlet chiral superfield.\footnote{Here, one should be careful not to integrate out $M$ too early. The order of deformation is crucial, since $(A \tilde{A})^2$ is irrelevant at $W=0$ fixed point, but its flip deformation is relevant. See \cite{Cho:2024civ} for more discussion of the deformation and RG flow of this type.} At the IR fixed point, $R_A=R_{\widetilde{A}}=1/2$. This theory contains marginal quartic operators of the form $\Tr A_{M_1}\widetilde{A}_{\widetilde{M}_2}A_{M_3}\widetilde{A}_{\widetilde{M}_4}$, and it possesses a non-trivial conformal manifold. Such a deformation exists for general $N$.

\paragraph{Weak Gravity Conjecture}

\begin{figure}[t]
    \centering
     \begin{subfigure}[b]{0.45\textwidth}
     \centering
         \includegraphics[width=\textwidth]{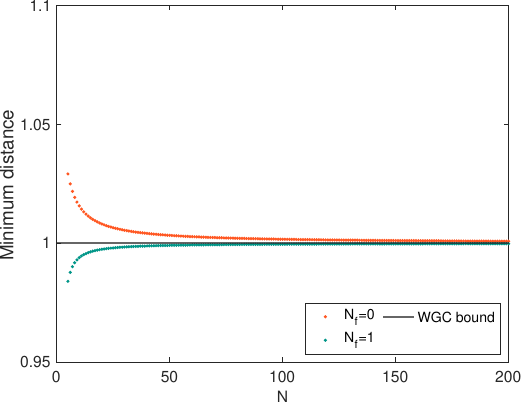}
         \caption{NN-WGC}
     \end{subfigure}
     \hspace{4mm}
     \begin{subfigure}[b]{0.45\textwidth}
     \centering
         \includegraphics[width=\textwidth]{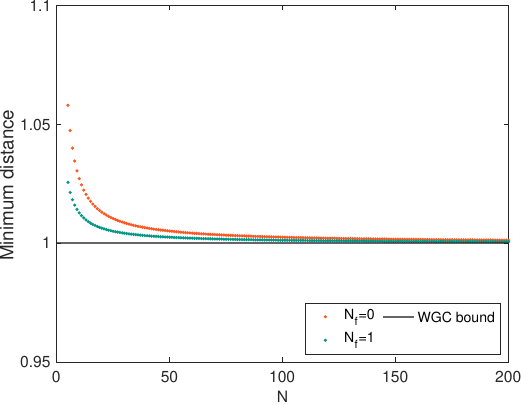}
         \caption{modified WGC}
     \end{subfigure}
     \hfill
     \caption{Testing AdS WGC for $SU(N)$ theory with 1 $\sym$ + 1 $\overline{\sym}$ + 2 $\antisym$ + 2 $\overline{\antisym}$ + $N_f$ ( $\fund$ + $\overline{\fund}$ ).
    The minimum distance from the origin to the convex hull with a fixed $N_f$. Theories below the solid line at the minimum distance 1 do not satisfy the WGC.\label{fig:wgc_s1S1a2A2ven}}
\end{figure}

We examine the AdS WGC using the gauge-invariant operators and $U(1)$ flavor charges identified at the beginning of this section. We find that, for $N_f=1$, this theory does not satisfy the NN-WGC for any $N$, whereas the modified WGC always holds. The result is shown in Figure \ref{fig:wgc_s1S1a2A2ven}.

\subsection{\texorpdfstring{3 $\antisym$ + 3 $\overline{\antisym}$ + $N_f$ ( $\fund$ + $\overline{\fund}$ )}{3 A + 3 At + Nf ( Q + Qt )}}

\paragraph{Matter content and symmetry charges}
  The second example of the Type III theory is an $SU(N)$ gauge theory with three pairs of rank-2 anti-symmetric and its conjugate and $N_f$ pairs of fundamental and anti-fundamental chiral multiplets. The matter fields and their $U(1)$ global charges are listed in Table \ref{tab:a3A3}.
  {\renewcommand\arraystretch{1.6}
  \begin{table}[t]
    \centering
    \begin{tabular}{|c|c||c|c|c|c|c|}
    \hline
    \# & Fields& $SU(N)$ &$U(1)_1$ & $U(1)_2$ & $ U(1)_3$ & $U(1)_R$ \\\hline
    
     $N_f$ & $Q$ & $\fund$ & $3N-6$ & $1$& 0 & $R_Q$\\
     
     $N_f$ & $\widetilde{Q}$ & $\overline{\fund}$ & $3N-6$ & $-1$  & 0& $R_{\widetilde{Q}}$ \\
     
     3 &$A$ & $\antisym$&  $-N_f$  & 0 &  1 & $R_A$ \\
     
     3 & $\widetilde{A}$ & $\overline{\antisym}$ & $-N_f$ & 0 &  -1  &  $R_{\widetilde{A}}$\\\hline
    \end{tabular}
    \caption{The matter contents and their corresponding charges in $SU(N)$ gauge theory with 3 $\antisym$ + 3 $\overline{\antisym}$ + $N_f$ ( $\fund$ + $\overline{\fund}$ ).\label{tab:a3A3}}
  \end{table}}

For $N=3$, the rank-2 anti-symmetric tensor is the same as the anti-fundamental representations. Thus, we only consider cases with $N > 3$ here. 

\paragraph{Gauge-invariant operators}
  Let $I$ and $J$ denote the flavor indices for $Q$, and $M$ denote the flavor indices for $A$. We present a sample of single-trace gauge-invariant operators in schematic form as follows:
  \begin{enumerate}
    \item $\Tr(A_{M_1}\widetilde{A}_{\widetilde{M}_1}\cdots A_{M_n}\widetilde{A}_{\widetilde{M}_n}),\quad n=1,2,\dots\,.$
    \item $Q_{I}(\widetilde{A}_{\widetilde{M}_1}A_{M_1}\cdots\widetilde{A}_{\widetilde{M}_n}A_{M_n})\widetilde{Q}_{\tilde{J}}\,,\quad n=0,1,\dots\,.$
    \item $Q_{I}\widetilde{A}_{\widetilde{M}_0}(A_{M_1}\widetilde{A}_{\widetilde{M}_1}\cdots A_{M_n}\widetilde{A}_{\widetilde{M}_n}) Q_{J},\quad n=0,1,\dots\,.$
    \item $\e\, A_{M_1}\cdots A_{M_i} \CQ_{I_1}^{n_1}\cdots\CQ_{I_{N-2i}}^{n_{N-2i}}\,.$
    \item $\e\,\e\, A_{M_1}^{n_1} A_{M_2}^{n_2} A_{M_3}^{n_3}\,,\quad n_1+n_2+n_3=N\,,\quad n_1,n_2,n_3=1,2,\dots\quad$ if $N$ is odd.
    \item The conjugates of the above-listed operators.

    $\vdots$
  \end{enumerate}
  The ellipsis indicates that only the low-lying operators have been listed. This subset is sufficient to identify relevant operators or to test the Weak Gravity Conjecture.
  Here, the dressed quarks are defined as
  \begin{align}
    \CQ_I^n=\begin{cases}
        (A_{M_1}\widetilde{A}_{\widetilde{M}_1}\cdots A_{M_{n/2}}\widetilde{A}_{\widetilde{M}_{n/2}})Q_I & n=0,2,4,\dots\,,\\
        (A_{M_1}\widetilde{A}_{\widetilde{M}_1}\cdots A_{M_{(n-1)/2}}\widetilde{A}_{\widetilde{M}_{(n-1)/2}})A\widetilde{Q}_{\tilde{I}} & n=1,3,5,\dots\,.
    \end{cases}
  \end{align}

\paragraph{$R$-charges and central charges}

  We perform the $a$-maximization to compute the $R$-charges of the matter fields and central charges. When $(N_f,N)=(0,4)$, we find that nine gauge-invariant operators of the form $\Tr A_{M_1}\widetilde{A}_{\widetilde{M}_1}$ hit the unitarity bound along the RG flow and decouple. After flipping these operators, the $R$-charges of the matter fields and central charges are given by
  \begin{align}
      R_A=R_{\widetilde{A}}=\frac{1}{3}\,,\quad a=\frac{15}{8}\,,\quad  c=\frac{9}{4}\,,\quad  a/c=\frac{5}{6}\,.
  \end{align}

  For all other cases, none of the gauge-invariant operators decouple along the RG flow. The $R$-charges of the matter fields and central charges, in the large $N$ limit with a fixed $N_f$, are given by
  \begin{align}
  \begin{split}
    R_Q=R_{\widetilde{Q}}&\sim \frac{2}{3}+ \frac{N_f-6}{18N} + O(N^{-2}) \,,\\
    R_A=R_{\widetilde{A}}&\sim \frac{2}{3}+\frac{N_f-6}{9N} + O(N^{-2})\,,\\
    a&\sim \frac{1}{4}N^2+ \frac{2N_f-3}{48}N+O(N^0)\,,\\
    c&\sim \frac{1}{4}N^2 + \frac{N_f}{16}N + O(N^0)\,,\\
    a/c&\sim 1-\frac{N_f+3}{12N}+O(N^{-2})\,.
  \end{split}
  \end{align}
  The leading-order behavior of the $R$-charges and central charges is universal across Type III theories.

  We observe that the ratio of central charges $a/c$ is always less than one, taking a value within the range $0.8203 \simeq (2835841+8216\sqrt{1869})/{3890246}\leq  a / c < 1$. The minimum value of $a/c$ arises when $(N_f,N)=(1,4)$. The maximum value of $a/c$ can be obtained in the large $N$ limit with $N_f=0$. Figure \ref{fig:a3A3ratio} illustrates the behavior of the ratio $a/c$.
  \begin{figure}[t]
    \centering
    \includegraphics[width=.5\linewidth]{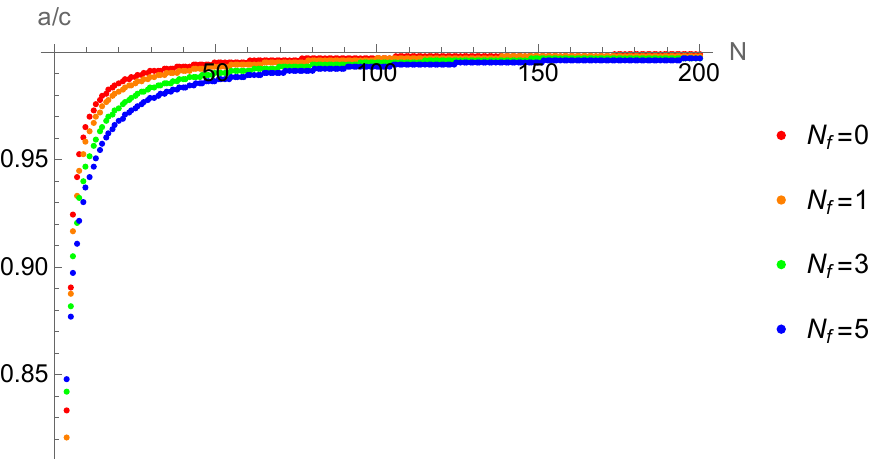}
    \caption{
    The central charge ratio $a/c$ versus $N$ for $SU(N)$ theory with 3 $\antisym$ + 3 $\overline{\antisym}$ + $N_f$ ( $\fund$ + $\overline{\fund}$ ).
    \label{fig:a3A3ratio}}
  \end{figure}

\paragraph{Conformal window}
  The asymptotic freedom bounds the upper limit of the conformal windows to be $N_f<6$. For small $N_f$, the $a$-maximization always yields a unique solution, and the resulting central charges lie within the Hofman-Maldacena bound. We therefore conjecture that this theory flows to an interacting SCFT for $0\leq N_f <6$.

\paragraph{Relevant operators}
  For generic $N$ and $N_f$, except in cases where decoupled operators are present, there exist the following relevant operators:
  \begin{itemize}
    \item nine operators of the form $\Tr A_{M_1}\widetilde{A}_{\widetilde{M}_2}$ with a dimension $1.2641\lesssim \D <2\,,$
    \item $N_f^2$ operators of the form $Q_I\widetilde{Q}_{\tilde{J}}$ with a dimension $1.4157\lesssim\D<2\,,$
    \item $\frac{3N_f(N_f-1)}{2}$ operators of the form $Q_I\widetilde{A}_MQ_J$ and $\widetilde{Q}_{\tilde{I}}A_M\widetilde{Q}_{\tilde{J}}$ with a dimension $2.3209\lesssim \D <3\,,$
  \end{itemize}
  The number of relevant operators does not depend on $N$ for a fixed $N_f$. The low-lying operator spectrum is sparse in the large $N$ limit.

  It was proposed in \cite{Abel:2009ty} that with the deformation
  \begin{align}\label{eq:a3A3 duality}
  \begin{split}
      W&=M_1\Tr (A_1\widetilde{A}_1)^{k_1+1}+M_2\Tr(A_2\widetilde{A}_2)^{k_2+1}+M_3\Tr(A_3\widetilde{A}_3)^{k_3+1}\\
      &\quad+M_4(\Tr A_1\widetilde{A}_2 + \Tr A_2\widetilde{A}_1+\Tr A_2\widetilde{A}_3+\Tr\widetilde{A}_2A_3)\,.
  \end{split}
  \end{align}
  with flip fields (gauge-singlets) $M_i$, the theory admits a dual description as an $SU(\widetilde{N})$ gauge theory with a certain superpotential, where $\widetilde{N}$ is given by
  \begin{align}
      \widetilde{N}=(2k^*+1)N_f-4k^*-N\,,\qquad k^*=\half\left[(2k_1+1)(2k_2+1)(2k_3+1)-1\right]\,.
  \end{align}
  However, we note that the operator of the form $M (A \tilde{A})^{k+1}$ is irrelevant both at the $W=0$ fixed point and at the fixed point where the last term in \eqref{eq:a3A3 duality} is turned on for $k \ge 1$. Therefore, the validity of this proposed duality remains doubtful.
  
\paragraph{Conformal manifold}
  There are no marginal operators for generic $N_f$ and $N$ in the absence of a superpotential. However, upon a suitable superpotential deformation, the theory flows to a superconformal fixed point with a non-trivial conformal manifold.

  For the $N_f=0$ case, we find that upon the following superpotential deformation, a non-trivial conformal manifold emerges at the IR fixed point:
  \begin{align}
    W=M_1\Tr A_1\widetilde{A}_1 + M_2 \Tr A_2\widetilde{A}_2 + M_3\Tr A_3\widetilde{A}_3\,.
  \end{align}
  Here, $M_i$'s are flip fields. 
  We can test this by computing the superconformal index. For example, the reduced superconformal index for the $SU(5)$ gauge theory with this superpotential is given by
  \begin{align}
    \CI_{\text{red}}=6 t^{2.67} + 3 t^{3.33} + 57 t^{5.33} + t^6 + \cdots \,.
  \end{align}
  The positivity of the coefficient at the $t^6$ term indicates the existence of a non-trivial conformal manifold. 
  The marginal operators take the form $M_1 \Tr A_3\widetilde{A}_2$, $M_1 \Tr A_2\widetilde{A}_3$, $M_2 \Tr A_3\widetilde{A}_1$, $M_2 \Tr A_1\widetilde{A}_3$, $M_3 \Tr A_2\widetilde{A}_1$, and $M_3 \Tr A_1\widetilde{A}_2$. 
  Such a deformation exists for general $N$ with a fixed $N_f=0$.

  For the $N_f=1$ case, consider the following superpotential deformation:
  \begin{align}
    W=M Q \widetilde{Q} + M^2\,.
  \end{align}
  Here, $M$ is a flip field, which is a gauge-singlet chiral superfield. At the IR fixed point, this theory possesses a non-trivial conformal manifold. We find that the reduced index for the $SU(5)$ gauge theory with this superpotential has a positive coefficient at the $t^6$ term:
  \begin{align}
    \CI_{\text{red}}=10 t^3 + 12 t^{4.5} + 97 t^6 + \cdots \,.
  \end{align}
  The $R$-charges are given by $R_Q=R_{\widetilde{Q}}=R_A=R_{\widetilde{A}}=1/2$, so the marginal operators are of the quartic form $\Tr A_{M_1}\widetilde{A}_{\widetilde{M}_2} A_{M_3} \widetilde{A}_{\widetilde{M}_4}$, $Q\widetilde{Q}\Tr A_{M_1}\widetilde{A}_{M_2}$, and similar operators. 
  Such a deformation exists for general $N$ with a fixed value of $N_f=1$.

\paragraph{Weak Gravity Conjecture}
  
  \begin{figure}[t]
    \centering
    \begin{subfigure}[b]{0.45\textwidth}
      \centering
      \includegraphics[width=\linewidth]{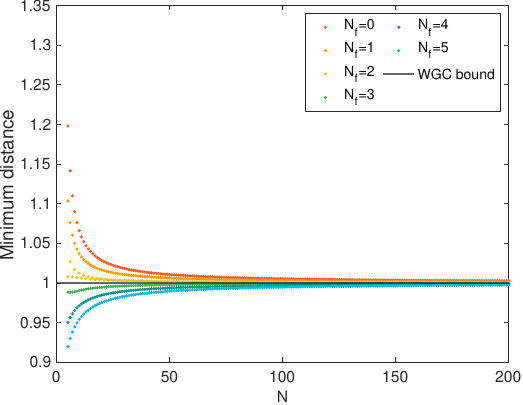}
      \caption{original WGC}
    \end{subfigure}
    \hspace{4mm}
    \begin{subfigure}[b]{0.45\textwidth}
      \centering
      \includegraphics[width=\linewidth]{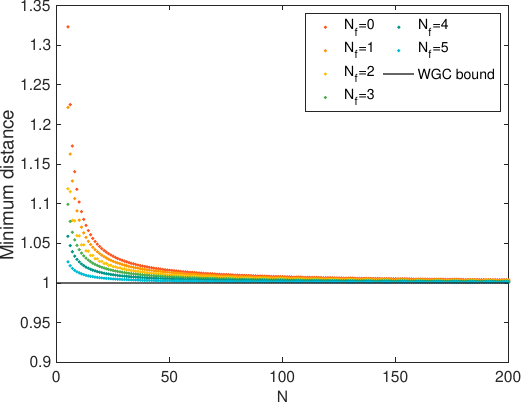}
      \caption{modified WGC}
    \end{subfigure}
    \hfill
    \caption{Testing AdS WGC for $SU(N)$ theory with 3 $\antisym$ + 3 $\overline{\antisym}$ + $N_f$ ( $\fund$ + $\overline{\fund}$ ).
    The minimum distance from the origin to the convex hull with a fixed $N_f$. Theories below the solid line at the minimum distance 1 do not satisfy the WGC.\label{fig:wgc_a3A3ven}}
  \end{figure}
We examine the AdS WGC using the gauge-invariant operators and $U(1)$ flavor charges identified at the beginning of this section. We find that, for $N_f=3,4,5$, this theory does not satisfy the NN-WGC for any $N$, whereas the modified WGC always holds. The result is shown in Figure \ref{fig:wgc_a3A3ven}.

\subsection{\texorpdfstring{1 \textbf{Adj} + 2 $\antisym$ + 2 $\overline{\antisym}$ + $N_f$ ( $\fund$ + $\overline{\fund}$ )}{1 Adj + 2 A + 2 At + Nf ( Q + Qt )}}
\paragraph{Matter content and symmetry charges}
  The next example of the Type III theory is an $SU(N)$ gauge theory with an adjoint and a pair of rank-2 anti-symmetric and its conjugate, and $N_f$ pairs of fundamental and anti-fundamental chiral multiplets. The matter fields and their $U(1)$ global charges are listed in Table \ref{tab:adj1a2A2}.
  {\renewcommand\arraystretch{1.6}
  \begin{table}[h]
    \centering
    \begin{tabular}{|c|c||c|c|c|c|c|c|c|}
    \hline
    \# & Fields & $SU(N)$ & $U(1)_1$ & $U(1)_2$ & $U(1)_3$ & $ U(1)_4$  & $U(1)_R$ \\\hline
    
     $N_f$ & $Q$ & $\fund$  & $N$ & $2N-4$ & 1  & 0 & $R_Q$\\
     
     $N_f$ & $\widetilde{Q}$ & $\overline{\fund}$ & $N$ & $2N-4$ & -1 & 0 & $R_{\widetilde{Q}}$ \\
     
     1 & $\Phi$ & \textbf{Adj} & $-N_f$ & 0 & 0 & 0  & $R_\Phi$ \\
     
     2 & $A$  &$\antisym$& 0  & $-N_f$& 0 & 1  & $R_A$ \\
     
     2 & $\widetilde{A}$  &$\overline{\antisym}$& 0 & $-N_f$ & 0  & -1 & $R_{\widetilde{A}}$\\\hline
    \end{tabular}
    \caption{The matter contents and their corresponding charges in $SU(N)$ gauge theory with 1 \textbf{Adj} + 2 $\antisym$ + 2 $\overline{\antisym}$ + $N_f$ ( $\fund$ + $\overline{\fund}$ ).\label{tab:adj1a2A2}}
  \end{table}}

 When $N=3$, the anti-symmetric representation is identical to the anti-fundamental representation. Therefore, we consider only $N>3$ cases here.

\paragraph{Gauge-invariant operators}
  Let $I$ and $J$ denote the flavor indices for $Q$, and $M$ denote the flavor indices for $A$. We present a sample of single-trace gauge-invariant operators in schematic form as follows:
  \begin{enumerate}
    \item $\Tr\Phi^n\,,\quad n = 2,3,\dots, N\,.$
    \item $\Tr\Phi^n(A_{M_1}\widetilde{A}_{\widetilde{M}_1}\cdots A_{M_m}\widetilde{A}_{\widetilde{M}_m})\,,\quad n=0,1,\dots,N\,,\quad m=1,2,\dots\,.$
    \item $Q_{I}\Phi^n(\widetilde{A}_{\widetilde{M}_1}A_{M_1}\cdots\widetilde{A}_{\widetilde{M}_m}A_{M_m})\widetilde{Q}_{\tilde{J}}\,,\quad n,m=0,1,\dots\,.$
    \item $Q_{I}\widetilde{A}_{\widetilde{M}_0}\Phi^n(\widetilde{A}_{\widetilde{M}_1}A_{M_1}\cdots\widetilde{A}_{\widetilde{M}_m}A_{M_m}) Q_{J},\quad n,m=0,1,\dots\,.$
    \item $\e\, A_{M_1}\cdots A_{M_i} \CQ_{I_1}^{n_1}\cdots\CQ_{I_{N-2i}}^{n_{N-2i}}\,.$
    \item $\e\,\e\, A_{M_1}^{n_1}A_{M_2}^{n_2}\Phi\,,\quad n_1+n_2=N\,,\quad n_1,n_2=1,2,\dots\quad$ if $N$ is odd.
    \item The conjugates of the above-listed operators.
    
    $\vdots$
  \end{enumerate}
  The ellipsis indicates that only the low-lying operators have been listed. This subset is sufficient to identify relevant operators or to test the Weak Gravity Conjecture.
  Here, the dressed quarks are defined as
  \begin{align}
    \CQ_I^n=\begin{cases}
        (A_{M_1}\widetilde{A}_{\widetilde{M}_1}\cdots A_{M_{n/2}}\widetilde{A}_{\widetilde{M}_{n/2}})Q_I & n=0,2,4,\dots\,,\\
        (A_{M_1}\widetilde{A}_{\widetilde{M}_1}\cdots A_{M_{(n-1)/2}}\widetilde{A}_{\widetilde{M}_{(n-1)/2}})A\widetilde{Q}_{\tilde{I}} & n=1,3,5,\dots\,.
    \end{cases}
  \end{align}

\paragraph{$R$-charges and central charges}
  
  We perform the $a$-maximization to compute the $R$-charges of the matter fields and central charges. In the large $N$ limit with a fixed $N_f$, they are given by
  \begin{align}
  \begin{split}
    R_Q=R_{\widetilde{Q}}&\sim \frac{2}{3} + \frac{N_f-4}{18N}+O(N^{-2}) \,,\\
    R_\Phi&=\frac{2}{3}+\frac{N_f-4}{9N} + O(N^{-2})\,,\\
    R_A=R_{\widetilde{A}}&\sim \frac{2}{3}+\frac{N_f-4}{9N}+O(N^{-2})\,,\\
    a&\sim \frac{1}{4}N^2+ \frac{N_f-1}{24}N+O(N^0)\,,\\
    c&\sim \frac{1}{4}N^2 + \frac{N_f}{16}N + O(N^0)\,,\\
    a/c&\sim 1-\frac{N_f+2}{12N}+O(N^{-2})\,.
  \end{split}
  \end{align}
  We find that no gauge-invariant operators decouple along the RG flow. The leading-order behavior of the $R$-charges and central charges is universal across Type III theories.

  We observe that the ratio of central charges $a/c$ is always less than one, taking a value within the range $0.8879 \simeq 103/116 \leq  a / c < 1$. The minimum value of $a/c$ arises when $(N_f,N)=(4,4)$. The maximum value of $a/c$ can be obtained in the large $N$ limit with $N_f=0$. Figure \ref{fig:adj1a2A2ratio} illustrates the behavior of the ratio $a/c$.
  \begin{figure}[t]
    \centering
    \includegraphics[width=.5\linewidth]{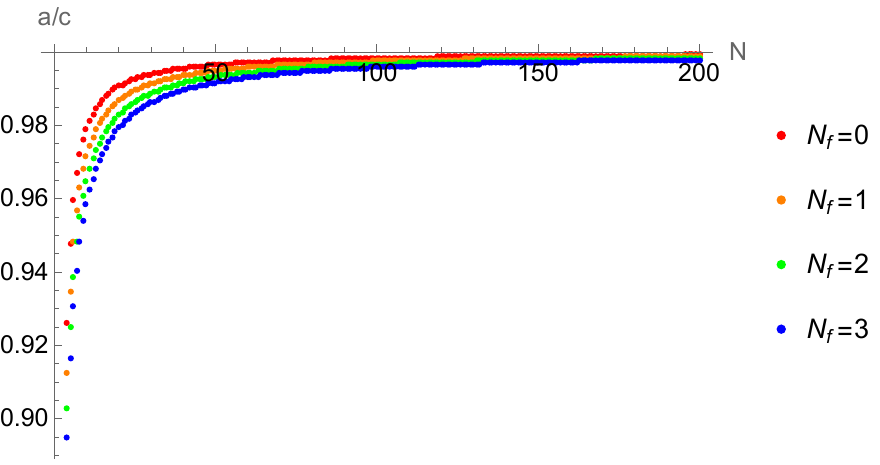}
    \caption{The central charge ratio $a/c$ versus $N$ for $SU(N)$ theory with 1 \textbf{Adj} + 2 $\antisym$ + 2 $\overline{\antisym}$ + $N_f$ ( $\fund$ + $\overline{\fund}$ ).\label{fig:adj1a2A2ratio}}
  \end{figure}

\paragraph{Conformal window}
  The asymptotic freedom bounds the upper limit of the conformal windows to be $N_f \le 4$. When $N_f=4$, there is a non-trivial conformal manifold \cite{Razamat:2020pra}. For small $N_f$, the $a$-maximization always yields a unique solution, and the resulting central charges lie within the Hofman-Maldacena bound. We therefore conjecture that this theory describes an interacting SCFT for $0\leq N_f \le 4$.

\paragraph{Relevant operators}
  For generic $N$ and $N_f$, there exist the following relevant operators:
  \begin{itemize}
    \item an operator of the form $\Tr \Phi^2$ with a dimension $1.4031\lesssim \D < 2\,,$
    \item four operators of the form $\Tr A_{M_1}\widetilde{A}_{\widetilde{M}_2}$ with a dimension $1.5969\lesssim \D < 2\,,$
    \item an operator of the form $\Tr \Phi^3$ with a dimension $2.1047\lesssim \D <3\,,$
    \item four operators of the form $\Tr A_{M_1}\widetilde{A}_{\widetilde{M}_2}\Phi$ with a dimension $2.2984\lesssim \D < 3\,,$
    \item $N_f^2$ operators of the form $Q_I\widetilde{Q}_{\tilde{J}}$ with a dimension $1.7222 \lesssim \D <2\,,$
    \item $N_f^2$ operators of the form $Q_I\Phi \widetilde{Q}_{\tilde{J}}$ with a dimension $2.5764\lesssim \D <3 \,,$
    \item $N_f(N_f-1)$ operators of the form $Q_I\widetilde{A}_{\widetilde{M}}Q_J$ and $\widetilde{Q}_{\tilde{I}}A_M\widetilde{Q}_{\tilde{J}}$ with a dimension $2.7833\lesssim\D<3\,,$
  \end{itemize}
  The number of relevant operators does not depend on $N$ for a fixed $N_f$. The low-lying operator spectrum is sparse in the large $N$ limit.

\paragraph{Conformal manifold}
  When $N_f=4$, which lies at the upper bound of the conformal window, the theory possesses a non-trivial conformal manifold that contains a one-dimensional subspace that preserves $\CN=2$ supersymmetry \cite{Bhardwaj:2013qia, Razamat:2020pra}. At this value, the theory is a conformal gauge theory with no running coupling.
  
  For $N_f< 4$, there are no marginal operators in the absence of a superpotential. However, upon a suitable superpotential deformation, this theory flows to a superconformal fixed point with a non-trivial conformal manifold.
  
  For instance, by turning on the superpotential $W=M_1\Tr\Phi^2 + M_1^2$, where $M_1$ is a flip field, the theory admits an exactly marginal operator given as $\Tr\Phi^4$, which is neutral under the remaining flavor symmetry after the deformation. Such a deformation (and also the marginal operator) exists for general $N$ with a fixed $N_f$.

\paragraph{Weak Gravity Conjecture}
 
  \begin{figure}[t]
    \centering
    \begin{subfigure}[b]{0.45\textwidth}
      \centering
      \includegraphics[width=\linewidth]{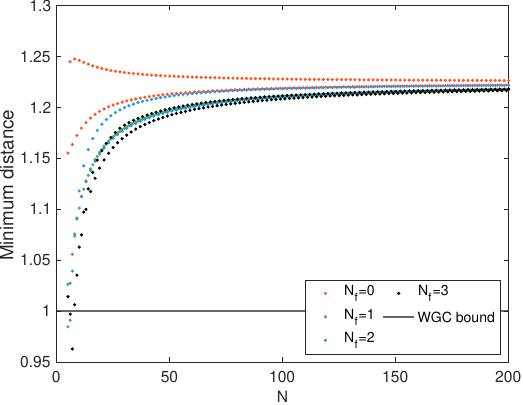}
      \caption{NN-WGC}
    \end{subfigure}
    \hspace{4mm}
    \begin{subfigure}[b]{0.45\textwidth}
      \centering
      \includegraphics[width=\linewidth]{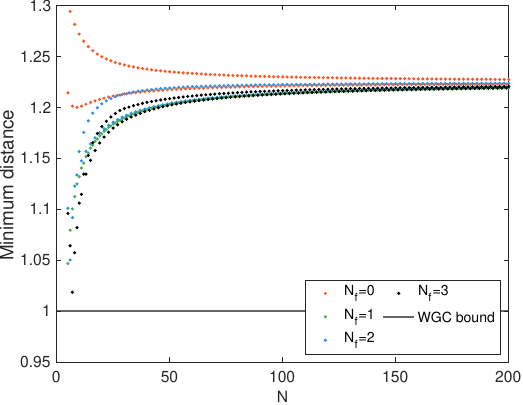}
      \caption{modified WGC}
    \end{subfigure}
    \hfill
    \caption{Testing AdS WGC for $SU(N)$ theory with 1 \textbf{Adj} + 2 $\antisym$ + 2 $\overline{\antisym}$ + $N_f$ ( $\fund$ + $\overline{\fund}$ ).
    The minimum distance from the origin to the convex hull with a fixed $N_f$. Theories below the solid line at the minimum distance 1 do not satisfy the WGC.\label{fig:wgc_adj1a2A2ven}}
  \end{figure}
 We test the AdS WGC using the gauge-invariant operators and $U(1)$ flavor charges identified at the beginning of this section. We find that, when $(N_f,N)=(1,5),(2,6),(3,6),(3,7)$, this theory does not satisfy the NN-WGC, whereas the modified WGC always holds. The result is shown in Figure \ref{fig:wgc_adj1a2A2ven}. 

\subsection{\texorpdfstring{1 \textbf{Adj} + 1 $\sym$ + 1 $\overline{\sym}$ + 1 $\antisym$ + 1 $\overline{\antisym}$}{1 Adj + 1 S + 1 St + 1 A + 1 At}}
\paragraph{Matter content and symmetry charges}
  The next example of the Type III theory is an $SU(N)$ gauge theory with an adjoint, a pair of rank-2 symmetric and its conjugate, and a pair of rank-2 anti-symmetric and its conjugate chiral multiplets. This theory is a conformal gauge theory with no running coupling. The matter fields and their $U(1)$ global charges are listed in Table \ref{tab:adj1s1S1a1A1}.
  {\renewcommand\arraystretch{1.6}
  \begin{table}[h]
    \centering
    \begin{tabular}{|c|c||c|c|c|c|c|c|c|}
    \hline
    \# & Fields & $SU(N)$ & $U(1)_1$ & $U(1)_2$  & $ U(1)_3$ & $U(1)_4$ & $U(1)_R$ \\\hline
     
     1 & \textbf{Adj} & \textbf{Adj} & $N+2$ & $N-2$ & 0 & 0   & $R_\Phi$\\
     
     1 & $S$ &$\sym$ & $-N$ & 0 & 1 & 0  & $R_S$ \\
     
     1 & $\widetilde{S}$ & $\overline{\sym}$ & $-N$ & 0 & -1 & 0 &  $R_{S}$\\
     
     1 & $A$ &$\antisym$ & 0 & $-N$& 0 & 1  & $R_A$ \\
     
     1 & $\widetilde{A}$ &$\overline{\antisym}$ & 0 & $-N$ & 0 & -1  & $R_{{A}}$\\\hline
    \end{tabular}
    \caption{The matter contents and their corresponding charges in $SU(N)$ gauge theory with 1 \textbf{Adj} + 1 $\sym$ + 1 $\overline{\sym}$ + 1 $\antisym$ + 1 $\overline{\antisym}$.\label{tab:adj1s1S1a1A1}}
  \end{table}}
For $N=3$, the rank-2 anti-symmetric tensor is the same as the anti-fundamental representations. Thus, we only consider cases with $N>3$ here.

\paragraph{Gauge-invariant operators}
  Let $P$ denote the flavor indices for $\Phi$. We present a sample of single-trace gauge-invariant operators in schematic form as follows: 
  \begin{enumerate}
    \item $\Tr\Phi^n\,,\quad n=2,\dots,N\,.$
    \item $\Tr (S\widetilde{S})^n\,,\quad \Tr (A\widetilde{A})^m\,,\quad \Tr (S\widetilde{A})^{2m}\,,\quad n=1,2,\dots,N-1\,,\quad m=1,2,\dots, \lfloor \frac{N-1}{2}\rfloor\,.$
    \item $\Tr \Phi^i(S\widetilde{S})^j(A\widetilde{A})^k(S\widetilde{A})^{l}(A\widetilde{S})^{m}\,,\quad i,j,k,l,m=1,2,\dots$.
    \item $Q_I\Phi^i(S\widetilde{S})^j(A\widetilde{A})^k(S\widetilde{A})^{l}(A\widetilde{S})^{m}\widetilde{Q}_{\tilde{J}}\,,\quad i,j,k,l,m=0,1,\dots$.
    \item $Q_I\widetilde{S}\Phi^i(S\widetilde{S})^j(A\widetilde{A})^k(S\widetilde{A})^{l}(A\widetilde{S})^{m}Q_J,\quad i,j,k,l,m=0,1,\dots$.
    \item $Q_I\widetilde{A}\Phi^i(S\widetilde{S})^j(A\widetilde{A})^k(S\widetilde{A})^{l}(A\widetilde{S})^{m}Q_J,\quad i,j,k,l,m=0,1,\dots$.
    \item $\e A^i \CQ_{{I}_1}^{n_1}\cdots \CQ_{{I}_{N-2i}}^{n_{N-2i}}\,,\quad \e \widetilde{A}^i \widetilde{\CQ}_{I_1}^{n_1}\cdots \widetilde{\CQ}_{I_{N-2i}}^{n_{N-2i}}$.
    \item $\e\, A^N\Phi^3\quad$ if $N$ is odd.
    \item $\e\,\e\, S^i A^{2j}  (\CQ_{I_1}\CQ_{J_1})\cdots (\CQ_{I_k}\CQ_{J_k}),\quad i+2j+k=N$.
    \item The conjugates of the above-listed operators.

    $\vdots$
  \end{enumerate}
  The ellipsis indicates that only the low-lying operators have been listed. This subset is sufficient to identify relevant operators or to test the Weak Gravity Conjecture.
  Here, the dressed quarks are defined as:
  \begin{align}
    \CQ^n_I=\begin{cases}
        \Phi^i(S\widetilde{S})^j(S\widetilde{A})^k(A\widetilde{S})^{n/2-i-j-k}Q_I & n=0,2,4,\dots\,,\\
        \Phi^i(S\widetilde{S})^j(S\widetilde{A})^k(A\widetilde{S})^{(n-1)/2-i-j-k} S\widetilde{Q}_{\tilde{I}} & n=1,3,5,\dots\,.
    \end{cases}
  \end{align}

\paragraph{$R$-charges and central charges}
   \begin{figure}[t]
    \centering
    \includegraphics[width=.5\linewidth]{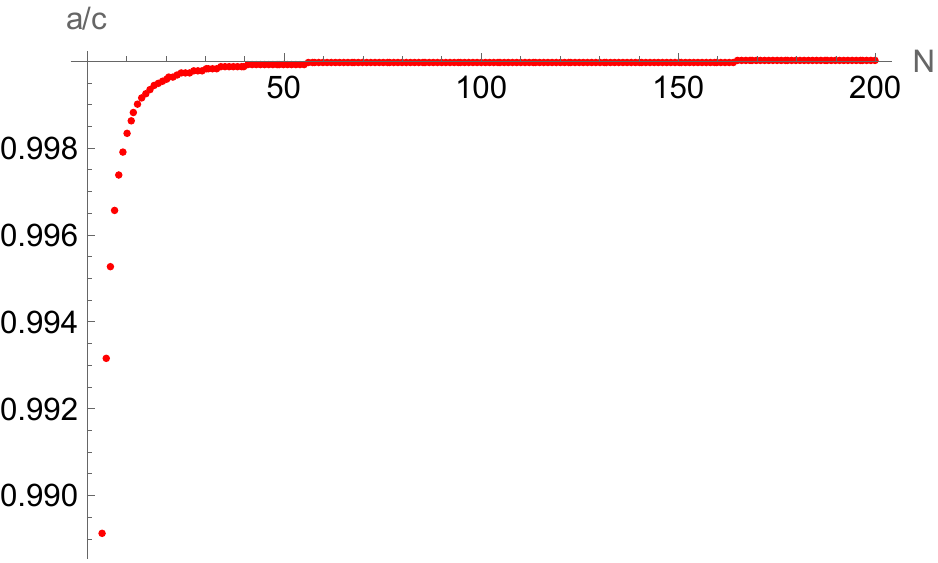}
    \caption{The central charge ratio $a/c$ versus $N$ for $SU(N)$ theory with 1 \textbf{Adj} + 1 $\sym$ + 1 $\overline{\sym}$ + 1 $\antisym$ + 1 $\overline{\antisym}$.\label{fig:adj1s1S1a1A1ratio}}
  \end{figure}
  
  We perform the $a$-maximization to obtain the $R$-charges and central charges to be
  \begin{align}
  \begin{split}
    R_\Phi&=R_S=R_{\widetilde{S}}=R_A=R_{\widetilde{A}}=\frac{2}{3} \,,\\
    a&=\frac{1}{4}N^2 - \frac{5}{24}\,,\quad c= \frac{1}{4}N^2  -\frac{1}{6}\,,\quad a/c= 1-\frac{1}{6N^2-4}\,.
  \end{split}
  \end{align}
  We find that no gauge-invariant operators decouple along the RG flow. The leading-order behavior of the $R$-charges and central charges is universal across Type III theories.

  We observe that the ratio of central charges $a/c$ is always less than one, taking a value within the range $\frac{91}{92}\simeq 0.9891 \leq  a / c < 1$. The minimum value of $a/c$ arises when $N=4$. The maximum value of $a/c$ can be obtained in the large $N$ limit. Figure \ref{fig:adj1s1S1a1A1ratio} illustrates the behavior of the ratio $a/c$.

\paragraph{Relevant operators}
  For generic $N$, there exist the following relevant operators:
  \begin{itemize}
    \item an operator of the form $\Tr \Phi^2$ with a dimension $\D=2\,,$
    \item an operators of the form $\Tr S\widetilde{S}$ with a dimension $\D=2\,,$
    \item an operators of the form $\Tr A\widetilde{A}$ with a dimension $\D=2\,.$
  \end{itemize}

\paragraph{Conformal manifold}
  This is a conformal gauge theory whose coupling does not run. 
  There exist five marginal operators of the form $\Tr\Phi^3$, $\Tr \Phi S\widetilde{S}$, $\Tr \Phi A\widetilde{A}$, $\Tr \Phi S\widetilde{A}$, and $\Tr \Phi A\widetilde{S}$. The conformal manifold has dimension two for $N\neq6$ and three for $N=6$ and contains a one-dimensional subspace which preserves $\CN=2$ supersymmetry \cite{Bhardwaj:2013qia, Razamat:2020pra}.

\paragraph{Weak Gravity Conjecture}
  \begin{figure}[t]
    \centering
    \begin{subfigure}[b]{0.45\textwidth}
      \centering
      \includegraphics[width=\linewidth]{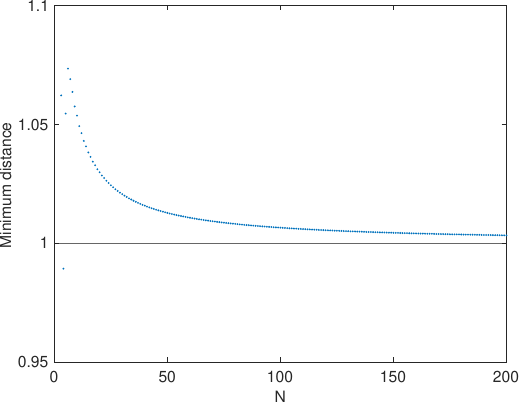}
      \caption{NN-WGC}
    \end{subfigure}
    \hspace{4mm}
    \begin{subfigure}[b]{0.45\textwidth}
      \centering
      \includegraphics[width=\linewidth]{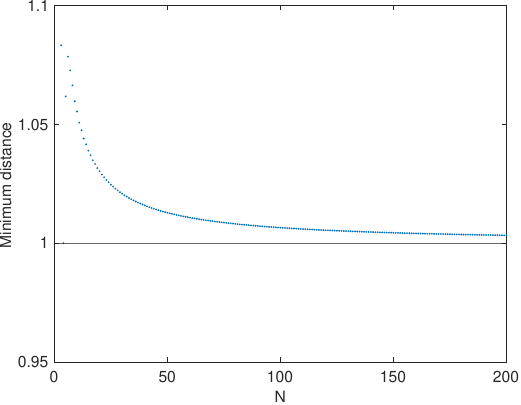}
      \caption{modified WGC}
    \end{subfigure}
    \hfill
    \caption{Testing AdS WGC for $SU(N)$ theory with 1 \textbf{Adj} + 1 $\sym$ + 1 $\overline{\sym}$ + 1 $\antisym$ + 1 $\overline{\antisym}$.
    The minimum distance from the origin to the convex hull. \label{fig:wgc_adj1s1S1a1A1}}
  \end{figure}
  
We examine the AdS WGC using the gauge-invariant operators and $U(1)$ flavor charges identified at the beginning of this section. We find that, when $N=4$, this theory does not satisfy the NN-WGC, whereas the modified WGC always holds. The result is shown in Figure \ref{fig:wgc_adj1s1S1a1A1}.

\subsection{\texorpdfstring{2 \textbf{Adj} + 1 $\antisym$ + 1 $\overline{\antisym}$ + $N_f$ ( $\fund$ + $\overline{\fund}$ )}{2 Adj + 1 A + 1 At + Nf ( Q + Qt )}}

\paragraph{Matter content and symmetry charges}
  The next example of the Type III theory is an $SU(N)$ gauge theory with two adjoints and a pair of rank-2 anti-symmetric and its conjugate chiral multiplets. The matter fields and their $U(1)$ global charges are listed in Table \ref{tab:adj2a1A1}.
  For $N=3$, the rank-2 anti-symmetric tensor is the same as the anti-fundamental representations. Thus, we only consider cases with $N>3$ here.
  {\renewcommand\arraystretch{1.6}
  \begin{table}[h]
    \centering
    \begin{tabular}{|c|c||c|c|c|c|c|c|c|}
    \hline
    \# & Fields & $SU(N)$ & $U(1)_1$ & $U(1)_2$  & $ U(1)_3$ & $U(1)_4$ & $U(1)_R$ \\\hline

     $N_f$ & $Q$ & $\fund$ & $N$ & 0 & 1 & 0  & $R_Q$ \\
     
     $N_f$ & $\widetilde{Q}$ & $\overline{\fund}$ & $N$ & 0 & -1 & 0 &  $R_{\widetilde{Q}}$\\
     
     2 & \textbf{Adj} & \textbf{Adj} & $-N_f$ & $N-2$ & 0 & 0   & $R_\Phi$\\
     
     1 & $A$ &$\antisym$ & 0 & $-2N$& 0 & 1  & $R_A$ \\
     
     1 & $\widetilde{A}$ &$\overline{\antisym}$ & 0 & $-2N$ & 0 & -1  & $R_{\widetilde{A}}$\\\hline
    \end{tabular}
    \caption{The matter contents and their corresponding charges in $SU(N)$ gauge theory with 2 \textbf{Adj} + 1 $\antisym$ + 1 $\overline{\antisym}$ + $N_f$ ( $\fund$ + $\overline{\fund}$ ).\label{tab:adj2a1A1}}
  \end{table}}

\paragraph{Gauge-invariant operators}
  Let $I$ and $J$ denote the flavor indices for $Q$, and $P$ denote the flavor indices for $\Phi$. We present a sample of single-trace gauge-invariant operators in schematic form as follows:
  \begin{enumerate}
      \item $\Tr\Phi_{P_1}\cdots\Phi_{P_n}\,,\quad \Tr (A\widetilde{A})^m\,,\quad n=2,3,\dots\,,\quad m=1,2,\dots,\lfloor\frac{N-1}{2}\rfloor\,.$
      \item $\Tr(\Phi_{P_1}\cdots\Phi_{P_n})(A\widetilde{A})^m\,,\quad n=0,1,\dots\,,\quad m=1,2,\dots,\lfloor\frac{N-1}{2}\rfloor\,.$
      \item $Q_I(\Phi_{P_1}\cdots\Phi_{P_n})(\widetilde{A}A)^m \widetilde{Q}_{\tilde{I}}\,,\quad n,m=0,1,\dots\,.$
      \item $Q_I\widetilde{A} (\Phi_{P_1}\cdots\Phi_{P_n})(A\widetilde{A})^m Q_J\,,\widetilde{Q}_{\tilde{I}}A (\Phi_{P_1}\cdots\Phi_{P_n})(A\widetilde{A})^m \widetilde{Q}_{\tilde{J}}\,,\quad n,m=0,1,\dots\,.$
      \item $\e\, A^n \CQ_{I_1}\cdots \CQ_{I_{N-2n}}\,,\quad \e\, \widetilde{A}^n \widetilde{\CQ}_{\tilde{I}_1}\cdots \widetilde{\CQ}_{\tilde{I}_{N-2n}}\,.$
  \end{enumerate}
  The ellipsis indicates that only the low-lying operators have been listed. This subset is sufficient to identify relevant operators or to test the Weak Gravity Conjecture.
  Here, we define the dressed quarks as
  \begin{align}
      \CQ_I^n=\begin{cases}
          \Phi_{P_1}\cdots \Phi_{P_{n/2}} Q_I &  n=0,2,4,\dots\,,\\
          \Phi_{P_1}\cdots\Phi_{P_{(n+1)/2}}A\widetilde{Q}_{\tilde{I}} & n=1,3,5,\dots\,.
      \end{cases}
  \end{align}

\paragraph{$R$-charges and central charges}
  
  We perform the $a$-maximization to compute the $R$-charges of the matter fields and central charges. In the large $N$ limit with a fixed $N_f$, they are given by
  \begin{align}
  \begin{split}
    R_Q = R_{\widetilde{Q}} & \sim \frac{2}{3} + \frac{N_f-2}{18N}+O(N^{-2}) \,,\\
    R_\Phi & \sim \frac{2}{3}+\frac{N_f-2}{9N}  + O(N^{-2}) \,,\\
    R_A = R_{\widetilde{A}} & \sim \frac{2}{3} + \frac{N_f-4}{9N} + O(N^{-2}) \,,\\
    a & \sim\frac{1}{4}N^2 + \frac{2N_f-1}{48}N + O(N^0) \,,\\
    c & \sim \frac{1}{4}N^2  + \frac{N_f}{16}N +O(N^0) \,,\\
    a/c & \sim 1-\frac{N_f+1}{12N} + O(N^{-2}) \,.
  \end{split}
  \end{align}
  We find that no gauge-invariant operators decouple along the RG flow. The leading-order behavior of the $R$-charges and central charges is universal across Type III theories.

  We observe that the ratio of central charges $a/c$ is always less than one, taking a value within the range $0.9369\simeq \frac{193}{206} \leq  a / c < 1$. The minimum value of $a/c$ arises when $(N_f,N)=(2,4)$. The maximum value of $a/c$ can be obtained in the large $N$ limit with $N_f=0$. Figure \ref{fig:adj2a1A1ratio} illustrates the behavior of the ratio $a/c$.
  \begin{figure}[t]
    \centering
    \includegraphics[width=.5\linewidth]{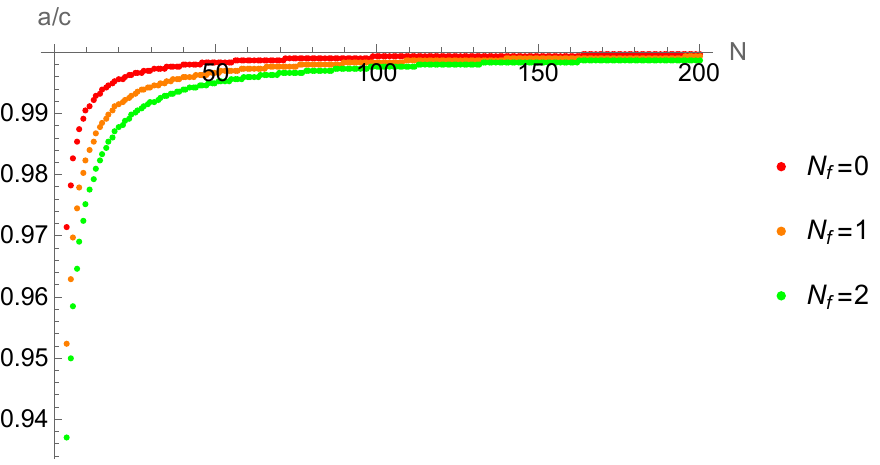}
    \caption{The central charge ratio $a/c$ versus $N$ for $SU(N)$ theory with 2 \textbf{Adj} + 1 $\antisym$ + 1 $\overline{\antisym}$ + $N_f$ ( $\fund$ + $\overline{\fund}$ ).\label{fig:adj2a1A1ratio}}
  \end{figure}

\paragraph{Conformal window}
  The upper bound of the conformal window comes from the asymptotic freedom. When $N_f=2$, there is a non-trivial conformal manifold so that it is an interacting SCFT \cite{Razamat:2020pra}. For small $N_f$, the $a$-maximization always yields a unique solution, and the resulting central charges lie within the Hofman-Maldacena bound. We therefore conjecture that this theory describes an interacting SCFT for $0\leq N_f \leq 2$.

\paragraph{Relevant operators}
  For generic $N$ and $N_f$, there exist the following relevant operators:
  \begin{itemize}
    \item three operators of the form $\Tr \Phi_{P_1}\Phi_{P_2}$ with a dimension $1.7848\lesssim \D \leq 2\,,$
    \item an operator of the form $\Tr A\widetilde{A}$ with a dimension $1.8607\lesssim \D \leq 2\,,$
    \item four operators of the form $\Tr \Phi_{P_1}\Phi_{P_2}\Phi_{P_3}$ with a dimension $2.6772\lesssim \D \leq 3\,,$
    \item two operators of the form $\Tr A\widetilde{A}\Phi_P$ with a dimension $2.7531\lesssim \D \leq 3\,,$
    \item $N_f^2$ operators of the form $Q_I\widetilde{Q}_{\tilde{J}}$ with a dimension $1.9507 \lesssim \D <2\,,$
    \item $2N_f^2$ operators of the form $Q_I\Phi_P \widetilde{Q}_{\tilde{J}}$ with a dimension $2.8994\lesssim \D <3 $.
  \end{itemize}
  The number of relevant operators does not depend on $N$ for a fixed $N_f$. The low-lying operator spectrum is sparse in the large $N$ limit.
  
\paragraph{Conformal manifold}
  When $N_f=2$, the one-loop beta function for the gauge coupling vanishes. At this value, the theory possesses a non-trivial conformal manifold \cite{Razamat:2020pra}.
  
  For $N_f< 2$, there are no marginal operators in the absence of a superpotential. However, upon a suitable superpotential deformation, this theory flows to a superconformal fixed point with a non-trivial conformal manifold.

  For the $N_f=0$ case, upon the following superpotential deformation, a non-trivial conformal manifold emerges at the IR fixed point:
  \begin{align}
    W=M_1\Tr\Phi_1\Phi_2 + M_2\Tr A\widetilde{A} + M_1M_2\,.
  \end{align}
  Here, $M_i$'s are flip fields. 
  We can test this by computing the superconformal index. For example, the reduced superconformal index for the $SU(5)$ gauge theory with this superpotential is given by
  \begin{align}
    \CI_{\text{red}}=t^{3.43} + 2t^{3.86} + 4t^{4.29} - 2 t^{5.14}\left(y+\frac{1}{y}\right) + 2t^6 + \cdots\,.
  \end{align}
  The positivity of the coefficient at the $t^6$ term indicates the existence of a non-trivial conformal manifold. The marginal operators includes $\Tr \Phi_{P_1}\Phi_{P_2}A\widetilde{A}$ and $\Tr \Phi_{P_1}A\Phi_{P_2}\widetilde{A}$. This deformation--along with possible flipping of decoupled operators--and the presence of marginal operators are universal for general $N$ with a fixed $N_f=0$.

  For the $N_f=1$, the following superpotential deformation produces a non-trivial conformal manifold in the IR:
  \begin{align}
    W= \Tr\Phi_1 A\widetilde{A} + \Phi_1\Phi_2^2\,.
  \end{align}
  The theory admits exactly marginal operators given as $Q\Phi_1^2\widetilde{Q}$ and $(\Tr\Phi_1^2)(Q\widetilde{Q})$, which are neutral under the remaining flavor symmetry after the deformation. Such a deformation (and also the marginal operator) exists for general $N$ with a fixed $N_f=1$.

\paragraph{Weak Gravity Conjecture}
  We examine the AdS WGC using the gauge-invariant operators and $U(1)$ flavor charges identified at the beginning of this section. This theory satisfies both versions of the WGC. The result is shown in Figure \ref{fig:wgc_adj2a1A1}.
  \begin{figure}[t]
    \centering
    \begin{subfigure}[b]{0.45\textwidth}
      \centering
      \includegraphics[width=\linewidth]{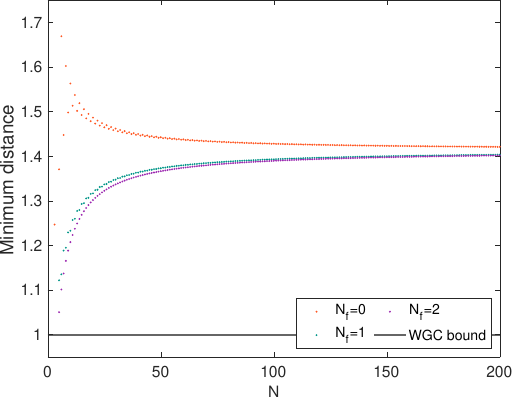}
      \caption{NN-WGC}
    \end{subfigure}
    \hspace{4mm}
    \begin{subfigure}[b]{0.45\textwidth}
      \centering
      \includegraphics[width=\linewidth]{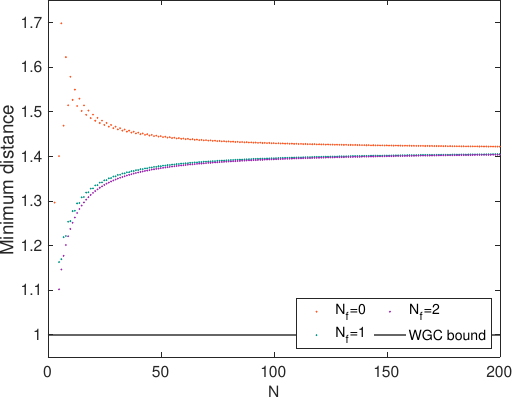}
      \caption{modified WGC}
    \end{subfigure}
    \hfill
    \caption{Testing AdS WGC for $SU(N)$ theory with 2 \textbf{Adj} + 1 $\antisym$ + 1 $\overline{\antisym}$ + $N_f$ ( $\fund$ + $\overline{\fund}$ ).
    The minimum distance from the origin to the convex hull with a fixed $N_f$. Theories below the solid line at the minimum distance 1 do not satisfy the WGC.\label{fig:wgc_adj2a1A1}}
  \end{figure}

\subsection{\texorpdfstring{3 \textbf{Adj}}{3 Adj}}
  The last case of the Type III theory is an $SU(N)$ gauge theory with three adjoint chiral multiplets.
  This theory is a conformal gauge theory with no running coupling. The $R$-charge and the central charges are given by
  \begin{align}
    R_\Phi=\frac{2}{3},\quad a=c=\frac{1}{4} \left(N^2 - 1 \right).
  \end{align}
  The leading-order behavior of the $R$-charges and central charges is universal across Type III theories.

  The single-trace gauge-invariant operators are of the form
  \begin{align}
      \Tr\Phi_{P_1}\cdots\Phi_{P_n}\,,\quad n=2,3,\dots\,.
  \end{align}
  There exist six relevant operators of the form
  \begin{align}
      \Tr\Phi_{P_1}\Phi_{P_2}\,.
  \end{align}
  For $N>2$, there are eleven marginal operators of the form
  \begin{align}
      \Tr \Phi_{P_1}\Phi_{P_2}\Phi_{P_3}\,,
  \end{align}
  which decompose into the $\mathbf{1}\oplus\mathbf{10}$ representations of the $SU(3)$ flavor symmetry.
  For $N=2$, however, only a single marginal operator exists, corresponding to the singlet of the $SU(3)$ flavor symmetry.
  For $N>2$, the dimension of the conformal manifold is three, and a one-dimensional subspace, parametrized by the singlet of $SU(3)$, preserves $\CN=4$ supersymmetry. Along another direction, one can instead preserve a $U(1)\times U(1)$ flavor symmetry, corresponding to the so-called $\b$-deformation \cite{Leigh:1995ep}. A non-Lagrangian dual description for this theory is obtained in \cite{Kang:2023dsa}.


\section{Superconformal SO(N) gauge theories with large N limit} \label{sec:SO}
  In this section, we classify and study all possible superconformal $SO(N)$ gauge theories that flow to SCFTs and admit a large $N$ limit. As discussed in Section \ref{sec:classification}, only the representations in Table \ref{tab:SOindex} are allowed. 
  {\renewcommand\arraystretch{1.4}
  \begin{table}[t]
    \centering
    \begin{tabular}{|c||c|c|c|}
    \hline
      Fields & $SO(N)$ & $T(\mathbf{R})$ & $|\mathbf{R}|$ \\\hline

      $Q$ &$\fund$& $1$ & $N$\\

      $\Phi$ & \textbf{Adj} & $N-2$ & $\frac{N(N-1)}{2}$\\
     
      $S$ &$\sym$ & $N+2$  & $\frac{N(N+1)}{2}-1$ \\
    
      $A$ & $\antisym$ & $N-2$ & $\frac{N(N-1)}{2}$ \\\hline
    \end{tabular}
    \caption{The Dynkin indices and dimensions of irreducible representations \textbf{R} in $SO(N)$.\label{tab:SOindex}}
  \end{table}}
  The Young diagrams $\fund\,$, $\sym\,$, and $\antisym\,$ denote the fundamental, rank-2 traceless symmetric, and rank-2 anti-symmetric representations, respectively. Since these representations are real, our $SO(N)$ theories considered here are free from gauge anomalies.

  From the asymptotic freedom, we require the one-loop beta function to be negative. This implies
  \begin{align}
    b_0=3(N-2)-N_Q-(N+2)N_S-(N-2)N_A\geq0\,,
  \end{align}
  where $N_Q$, $N_S$, $N_A$ denotes the number of fundamental (vector), symmetric and anti-symmetric tensor chiral multiplets respectively. 
  This condition constrains the set of admissible theories to a finite number of families. The result is listed in Table \ref{tab:SOlist}. We also summarize the range of the $a/c$ ratio and the condition for the existence of a non-trivial conformal manifold in Table \ref{tab:SOratio}.
  {\renewcommand\arraystretch{1.6}
  \begin{table}[t]
    \centering
    \begin{tabular}{|c|c|c|c|}
      \hline
     Type & $SO(N)$ theory &  NN-WGC & Conformal window  \\\hline \hline 
      
\multirow{2}{*}{I} &     1 $\sym$ + $N_f$ $\fund$ &  $\begin{aligned}[c]&(N_f,N)\neq(10,9),(11,10),\\&(12,10),(12,11),(13,11),\\&(14,11),(13,12),(14,12),\\&(15,12),(14,13),(15,13),\\&(15,14),(16,15) \end{aligned}$  & $0 \leq N_f\leq 2N-8^*$\\ \cline{2-4}
      
 &     1 $\antisym$\, + $N_f$ {$\fund$} & Always & $1\leq N_f \leq 2N-4^*$  {\rule[-2ex]{0pt}{-3.0ex}} \\ \hline\hline
      
  \multirow{3}{*}{II} &   2 $\sym$ + $N_f$ $\fund$  & Always & $0\le N_f\leq N-10^*$  \\ \cline{2-4}
      
&      1 $\sym$ + 1 $\antisym$ + $N_f$ $\fund$ & $(N_f,N)\neq (4,10)$ & $0\le N_f\leq N-6^*$  \\ \cline{2-4}
      
&      2 $\antisym$ + $N_f$ $\fund$ & Always & $0\le N_f < N-2$  \\ \hline\hline
      
 III &     3 $\antisym$ & Always  & $*$  \\ \hline
    \end{tabular}
    \caption{List of all possible superconformal $SO(N)$ theories with large $N$ limit in Type I,II,III. The third column lists the condition for each theory to satisfy the NN-WGC in the Veneziano limit. The last column specifies the condition for each theory to flow to a superconformal fixed point. The entries with $*$ (if $N_f$ saturates the upper bound) do not flow but have non-trivial conformal manifolds \cite{Razamat:2020pra, Bhardwaj:2013qia}.\label{tab:SOlist}}
  \end{table}}

  {\renewcommand\arraystretch{1.6}
  \begin{table}[t]\small
    \centering
    \begin{tabular}{|c|c|c|c|}
      \hline
     Type & $SO(N)$ theory &  Range of $a/c$ & \makecell{Conformal manifold}  \\\hline \hline 
      
\multirow{2}{*}{I} &     1 $\sym$ + $N_f$ $\fund$ & $0.875=\frac{7}{8}<a/c\leq \frac{417}{368}\simeq 1.1332$ & \makecell{$N_f=0\,,$\\$N\in 2\IZ$ or\\$N_f=2N-8$} \\ \cline{2-4}
      
&      1 $\antisym$\, + $N_f$ $\fund$ & $0.875=\frac{7}{8}<a/c\leq\frac{257+77\sqrt{985}}{2668}\simeq1.0021$ & $(2N-4)/N_f\in\IZ$  {\rule[-2ex]{0pt}{-3.0ex}} \\ \hline\hline
      
 \multirow{3}{*}{II} &    2 $\sym$ + $N_f$ $\fund$  & $0.9286\simeq \frac{13}{14}<a/c\leq \frac{19}{18}\simeq 1.0556$ & $N_f=N-10$  \\ \cline{2-4}
      
&      1 $\sym$ + 1 $\antisym$ + $N_f$ $\fund$ & $0.9286\simeq \frac{13}{14}<a/c\leq \frac{17}{16}\simeq 1.0625$ & $N_f=N-6$ \\ \cline{2-4}
      
&      2 $\antisym$ + $N_f$ $\fund$ & $0.9286\simeq\frac{13}{14} < a/c\leq 1$ & $N_f=0$  \\ \hline\hline
      
 III &    3 $\antisym$ & $a/c=1$  & Always  \\ \hline
    \end{tabular}
    \caption{Central charge ratio and the conformal manifold of $SO(N)$ theories. The third column lists the range of the ratio $a/c$ for general $N$ and $N_f$ within the conformal window. The last column denotes the condition for the theory to have a non-trivial conformal manifold (with $W=0$). 
    \label{tab:SOratio}}
  \end{table}}

  In this section, we describe each family of theories, parametrized by the number of pairs of fundamental (vector) chiral multiplets. We determine their $R$-symmetry, central charges, and analyze the spectra of relevant and marginal operators. We then identify the conditions on $N_f$ for the theory to be an interacting SCFT. We also test two versions of the AdS Weak Gravity Conjecture and find that NN-WGC is sometimes violated, whereas our modified WGC always holds, even for a finite $N$.

\subsection{\texorpdfstring{1 $\sym$ + $N_f$ $\fund$}{1 S + Nf Q}}\label{sec:SOs1}

\paragraph{Matter content and symmetry charges}
  The first entry of the Type I theories is $SO(N)$ gauge theory with a rank-2 traceless symmetric tensor and $N_f$ vector (fundamental) chiral multiplets. The matter fields and their $U(1)$ global charges are listed in Table \ref{tab:SOs1}.
  {\renewcommand\arraystretch{1.4}
  \begin{table}[h]
    \centering
    \begin{tabular}{|c|c||c|c|c|}
    \hline
    \# & Fields & $SO(N)$ & $U(1)_F$  & $U(1)_R$ \\\hline
     $N_f$ & $Q$ & $\fund$ &  $N+2$ & $R_Q$ \\
     1 & $S$ & $\sym$ &  $-N_f$  & $R_S$  \\\hline
    \end{tabular}
    \caption{The matter contents and their corresponding charges in $SO(N)$ gauge theory with 1 $\sym$ + $N_f$ $\fund$.\label{tab:SOs1}}
  \end{table}}

\paragraph{Gauge-invariant operators}
  Let $I$ and $J$ denote the flavor indices for $Q$. We present a sample of single-trace gauge-invariant operators in schematic form as follows:
  \begin{enumerate}
      \item $\Tr S^n\,,\quad n=1,2,\dots,N\,.$
      \item $Q_I S^n Q_J\,,\quad n=0,1,\dots,N-1\,.$
      \item $\e\, Q_{I_1}\cdots Q_{I_N}\,.$
      
        $\vdots$
  \end{enumerate}
  This subset is sufficient to identify relevant operators or to test the Weak Gravity Conjecture.

\paragraph{$R$-charges and central charges}
  We perform the $a$-maximization to compute the $R$-charges of the matter fields and central charges. In the large $N$ limit with a fixed $N_f$, they are given by
  \begin{align}
      \begin{split}\label{eq:SOs1 charges}
          R_Q&\sim \frac{3-\sqrt{5}}{3} + \frac{8\sqrt{5}+ 3 N_f}{6N} + O(N^{-2})\,,\\
          R_S&\sim \frac{12 + \sqrt{5}N_f}{3N} - \frac{N_f^2 + 4\sqrt{5} N_f + 16}{2N^2}+ O(N^{-3})\,,\\
          a & \sim \frac{63+5\sqrt{5}N_f}{48} N -\frac{15 N_f^2 + 80 \sqrt{5} N_f + 516}{64} + O(N^{-1})\,,\\
          c & \sim \frac{120+11\sqrt{5}N_f}{96} N -\frac{8N_f^2 + 41\sqrt{5}N_f + 256}{32} + O(N^{-1})\,,\\
          a/c &\sim \frac{126+10\sqrt{5}N_f}{120+11\sqrt{5}N_f}-\frac{3(5\sqrt{5}N_f^3 + 84 N_f^2 -176\sqrt{5}N_f -2592)}{2(120+11\sqrt{5}N_f)^2N} + O(N^{-2})\,.
      \end{split}
  \end{align}
  For a fixed $N_f$, the $R$-charge of a rank-2 symmetric chiral multiplet scales as $1/N$. As a result, the spectrum of gauge-invariant operators becomes dense, with $O(N)$ operators decoupling, and hence this theory belongs to the Type I class.

  In the Veneziano limit $N,N_f\goto\infty$ with a $\a=N_f/N$ fixed, the result of $a$-maximization is as follows:
  \begin{align}
      \begin{split}\label{eq:SOs1ven charges}
          R_Q&\sim \frac{6-\sqrt{20-\a^2}+3\a-3\a^2}{3(2-\a^2)} + O(N^{-1}) \,,\\
          R_S &\sim \frac{\a\sqrt{20-\a^2}-3\a^2}{3(2-\a^2)} + O(N^{-1})\,,\\
          a & \sim \frac{-18\a^2(5-\a^2)+\a(20-\a^2)^{3/2}}{48(2-\a^2)^2} \times \half N^2 + O(N)\,,\\
          c & \sim \frac{-96\a^2+21\a^4+(22\a-2\a^3)\sqrt{20-\a^2}}{48(2-\a^2)^2} \times \half  N^2 + O(N)\,,\\
          a/c &\sim \frac{440-76\a^2-3\a\sqrt{20-\a^2}}{484-89\a^2}+  O(N^{-1})\,.
      \end{split}
  \end{align}  
  The leading-order terms of $R$-charges, $a/\dim(G)$, $c/\dim(G)$, and $a/c$ are the same as those given in equation (\ref{eq:adj1ven charges}) for the $SU(N)$ Type I theories. This leading behavior of the $R$-charges and central charges is universal across Type I theories. However, since the set of decoupled operators differs, the $R$-charges have to be corrected. As discussed in Section \ref{sec:SUclass}, we obtain a universal result if we take the additional limit $1\ll N_f\ll N$.

  In the presence of decoupled operators, the naive results of $a$-maximization are no longer valid, and we have to perform $a$-maximization again after flipping decoupled operators to obtain the correct answer.\footnote{When $N_f=0$, we do not have to invoke $a$-maximization. Anomaly constraint determines the $R$-charge.} 
  In the case of $N_f=0$, upon removing the decoupled operators, we obtain the following result: 
  \begin{align}
  \begin{split}
      R_S &=\frac{4}{N+2}\,, \\
      a &\sim\frac{3059}{2304}N-\frac{18725}{2304}+O(N^{-1})\,,\\
      c &\sim\frac{965}{768}N - \frac{18593}{2304}+O(N^{-1})\\
    a/c &\sim \frac{3059}{2895} + \frac{2667112}{8381025N}+O(N^{-2})\,.
  \end{split}
  \end{align}
  
  We find that the ratio of central charges $a/c$ lies within the range $0.875=7/8< a/c \leq 417/368 \simeq  1.1332$. The ratio $a/c$ is greater than one when $N_f\leq 3$. The minimum value of $a/c$ can be obtained in the Veneziano limit with $\a\goto 2$. The maximum values arises when $(N_f,N)=(0,5)$. The plot of the ratio $a/c$ is shown in Figure \ref{fig:SOs1ratio}.
  \begin{figure}[t]
    \centering
    \begin{subfigure}[b]{0.45\textwidth}
        \includegraphics[width=\linewidth]{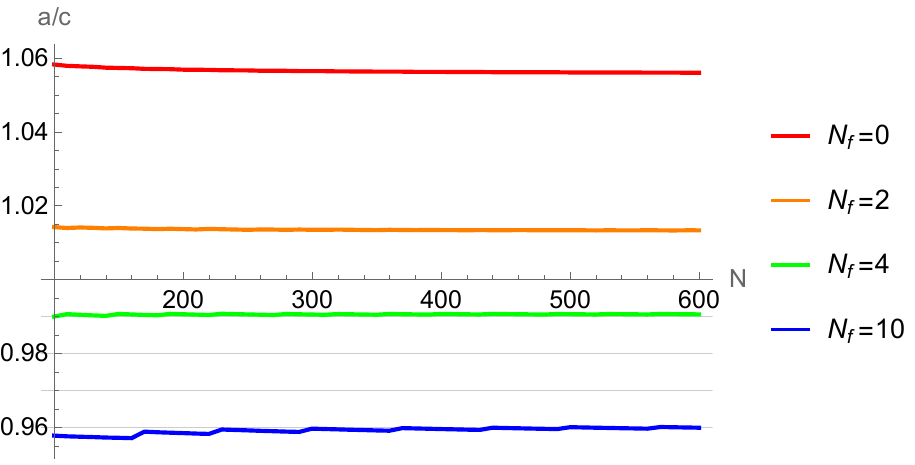}
    \end{subfigure}
    \hspace{4mm}
    \begin{subfigure}[b]{0.45\textwidth}
        \includegraphics[width=\linewidth]{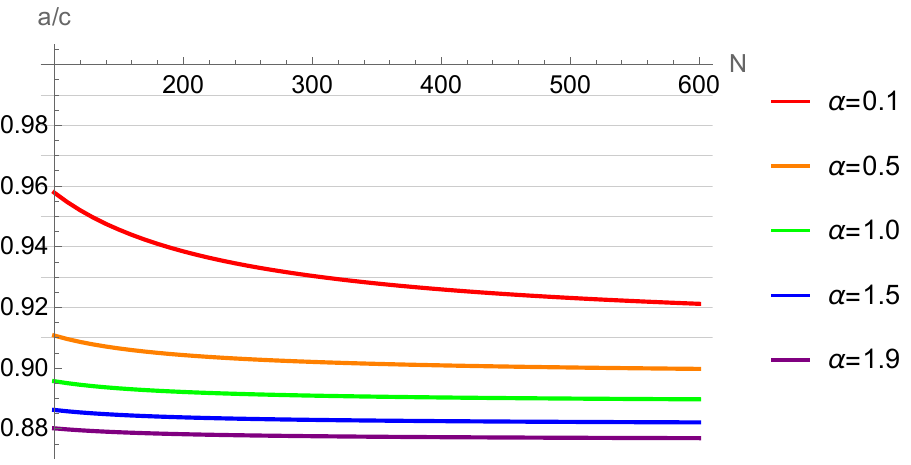}
    \end{subfigure}
    \hfill
     \caption{The central charge ratio for $SO(N)$ theory with 1 $\sym$ + $N_f$ $\fund$. Left: $a/c$ versus $N$ with a fixed $N_f$. Right: $a/c$ versus $N$ with a fixed $\a=N_f/N$.}
    \label{fig:SOs1ratio}   
  \end{figure}

\paragraph{Conformal window}
  The upper bound comes from the asymptotic freedom. For this theory, when it is at the top of the conformal window $N_f=2N-8$, there is a non-trivial conformal manifold \cite{Razamat:2020pra} so that it is an interacting SCFT. For small $N_f$, we find that the $a$-maximization always yields a unique solution, and the resulting central charges lie within the Hofman-Maldacena bound. We therefore conjecture that this theory describes an interacting SCFT for $0\leq N_f \leq  2N-8$.

\paragraph{Relevant operators}
  For a fixed $N_f$, since the $R$-charge of a rank-2 symmetric chiral multiplet scales as $1/N$, there are $O(N)$ relevant operators of the form $\Tr S^n$ and $Q_I S^n Q_J$. The $O(N)$ amount of them get decoupled along the RG flow.

  One can consider the deformation of the form $W=\Tr S^{2(k+1)}$. This operator is indeed relevant for $N$ large enough so that the theory flows to a good fixed point (with some operators decoupled). 
  In this case, the theory has a dual description given by $SO\left((2k+1)N_f-N+4\right)$ gauge theory \cite{Intriligator:1995ff} with a number of operators flipped, similar to Kutasov-Schwimmer duality. 

\paragraph{Conformal manifold}
  In the case of $N_f=0$, the $U(1)_R$ symmetry is uniquely determined by the anomaly-free condition as
  \begin{align}\label{eq:R_S}
      R_S=\frac{4}{N+2}\,.
  \end{align}
  For even $N$ (except for $N=2$), the equation \eqref{eq:R_S} implies that the operators of the form
  \begin{align}
      \Tr S^{(N+2)/2}\,,\quad (\Tr S^n)(\Tr S^m)\,,\quad n+m=(N+2)/2\in\IZ\,.
  \end{align}
  are marginal. Here, the operators of the form $\Tr S^p$ with $p\leq \lfloor(N+2)/6\rfloor$ are decoupled and are not present in the interacting SCFT. Nevertheless, there are $O(N)$ possible ways to partition $n$ and $m$ such that $n+m=(N+2)/2$ while satisfying $n,m>\lfloor (N+2)/6\rfloor$. Thus, the theory contains $O(N)$ marginal operators, leading to an $O(N)$-dimensional conformal manifold. 
  For odd $N$, there is no operator of this form as $(N+2)/2$ is not an integer. Empirically, we have not identified any marginal operators for odd $N$ with $N_f=0$.
  
  Let us consider the case where $N_f > 0$. As in the $SU(N)$ gauge theories, the anomaly-free condition for the $U(1)_R$ symmetry allows us to identify the possible combinations of elementary fields producing marginal operators. From the $R$-anomaly cancellation condition, we have
  \begin{align}
  \begin{split}
      &N-2 + (N+2)(R_S - 1) + N_f(R_Q-1)=0\\
      &\quad \quad \implies \quad \frac{2(N+2)}{N_f+4}R_S + \frac{2N_f}{N_f+4}R_Q =2\,.
  \end{split}
  \end{align}
  This equation suggests that an operator of the form
  \begin{align}
      S^{2(N+2)/(N_f+4)}Q^{2N_f/(N_f+4)}\,,\quad \frac{2(N+2)}{N_f+4}\,, \,\frac{2N_f}{N_f+4}\in\IZ\,.
  \end{align}
  could be marginal. $2N_f/(N_f+4)$ becomes an integer only for $N_f=4$, where it takes the value of 1. However, even in this case, there is no gauge-invariant operator composed only of a single $Q$. Thus, no marginal operators of this form can exist. Empirically, we have not identified any marginal operators for generic $N$ and $0<N_f<2N-8$. When $N_f=2N-8$, a non-trivial conformal manifold exists \cite{Razamat:2020pra}.
  
\paragraph{Weak Gravity Conjecture}
  We examine the AdS WGC using the gauge-invariant operators and $U(1)$ flavor charges identified at the beginning of this section. We find that this theory does not satisfy the NN-WGC for the following values of $N_f$ and $N$:
  \begin{align}
  \begin{split}
      (N_f,N) &=(10,9),(11,10),(12,10),(12,11),(13,11),(14,11),(13,12),(14,12)\,,\\
      &\quad (15,12),(14,13),(15,13),(15,14),(16,15)\,,
  \end{split}
  \end{align}
  On the other hand, the modified WGC always holds. 
  The result is shown in Figure \ref{fig:wgc_SOs1ven}.

\begin{figure}[t]
    \centering
     \begin{subfigure}[b]{0.45\textwidth}
     \centering
    \includegraphics[width=\linewidth]{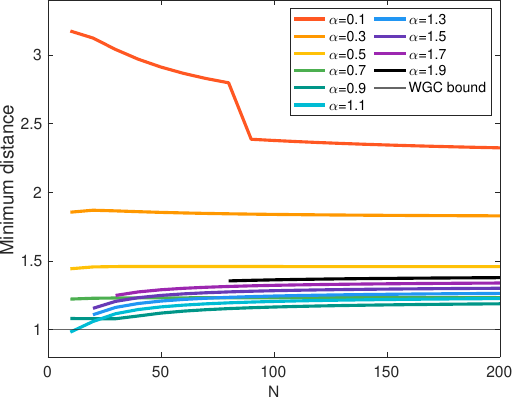}
    \caption{NN-WGC}
     \end{subfigure}
     \hspace{4mm}
     \begin{subfigure}[b]{0.45\textwidth}
     \centering
        \includegraphics[width=\linewidth]{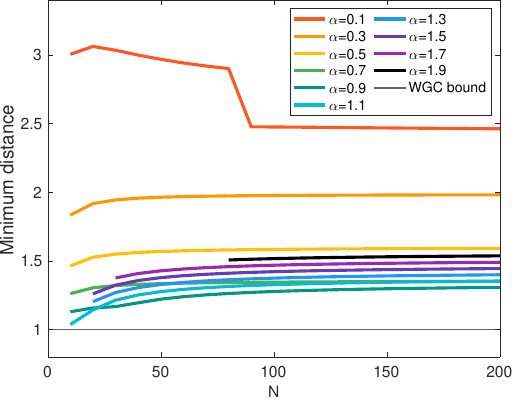}
        \caption{modified WGC}
     \end{subfigure}
     \hfill
        \caption{Testing AdS WGC for $SO(N)$ theory with 1 $\sym$ + $N_f$ $\fund$. The minimum distance from the origin to the convex hull with a fixed $\a=N_f/N$. Theories below the solid line at the minimum distance 1 do not satisfy the WGC.\label{fig:wgc_SOs1ven}}
 \end{figure}

\subsection{\texorpdfstring{1 $\antisym$ + $N_f$ $\fund$}{1 A + Nf Q}}\label{sec:SOadjSQCD}

\paragraph{Matter content and symmetry charges}
  Next case of the Type I theory is the $SO(N)$ gauge theory with a rank-2 anti-symmetric tensor and $N_f$ vector (fundamental) chiral multiplets. Notice that the rank-2 anti-symmetric tensor is the adjoint representation of $SO(N)$. 
  The matter fields and their $U(1)$ global charges are listed in Table \ref{tab:SOa1}.
  {\renewcommand\arraystretch{1.4}
  \begin{table}[h]
    \centering
    \begin{tabular}{|c|c||c|c|c|}
    \hline
    \# & Fields & $SO(N)$ & $U(1)_F$  & $U(1)_R$ \\\hline
     $N_f$ & $Q$ & $\fund$ &  $N-2$ & $R_Q$ \\
     1 & $A$ & $\antisym$ &  $-N_f$  & $R_A$  \\\hline
    \end{tabular}
    \caption{The matter contents and their corresponding charges in $SO(N)$ gauge theory with 1 $\antisym$ + $N_f$ $\fund$.\label{tab:SOa1}}
  \end{table}}
  For $N=3$, this theory is the same as the $SU(2)$ gauge theory with an adjoint chiral multiplet, which is already discussed in Section \ref{sec:adjSQCD}. Moreover, since $SO(4)$ is not a simple group, we restrict our analysis to the cases with $SO(N>4)$ gauge group.

\paragraph{Gauge-invariant operators}
  Let $I$ and $J$ denote the flavor indices for $Q$. We present a sample of single-trace gauge-invariant operators in schematic form as follows:
  \begin{enumerate}
      \item $\Tr A^{2n}\,,\quad n=1,2,\dots,\lfloor \frac{N-1}{2}\rfloor \,.$
      \item $Q_I A^{n} Q_J\,,\quad 0,1,\dots,N \,.$
      \item $\e\, A^n Q_{I_1}\cdots Q_{I_{N-2n}}\,,\quad n=\lceil \frac{N-N_f}{2}\rceil,\dots, \lfloor \frac{N}{2}\rfloor\,.$

      $\vdots$
  \end{enumerate}
  This subset is sufficient to identify relevant operators or to test the Weak Gravity Conjecture.

\paragraph{$R$-charges and central charges}
  
  We perform the $a$-maximization to compute the $R$-charges of the matter fields and central charges. 
  In the large $N$ limit with a fixed $N_f$, they are given by
  \begin{align}
      \begin{split}
          R_Q&\sim \frac{3-\sqrt{5}}{3} + \frac{-4 \sqrt{5}+ 15 N_f}{30N} + O(N^{-2})\,,\\
          R_A&\sim \frac{\sqrt{5}N_f}{3N} - \frac{5N_f^2 - 8\sqrt{5} N_f}{10N^2}+ O(N^{-3})\,,\\
          a & \sim \frac{5\sqrt{5}N_f}{48} N -\frac{15 N_f^2 - 8 \sqrt{5} N_f}{64} + O(N^{-1})\,,\\
          c & \sim \frac{11\sqrt{5}N_f}{96} N -\frac{40N_f^2 - 19\sqrt{5}N_f}{160} + O(N^{-1})\,,\\
          a/c &\sim \frac{10}{11}-\frac{3\sqrt{5}N_f-36}{242N} + O(N^{-2})\,.
      \end{split}
  \end{align}
  For a fixed $N_f$, the $R$-charge of a rank-2 anti-symmetric chiral multiplet scales as $1/N$. As a result, the spectrum of gauge-invariant operators becomes dense, with $O(N)$ operators decoupling, and hence this theory belongs to the Type I class.

  In the Veneziano limit $N,N_f\goto\infty$ with $\a=N_f/N$ fixed, the leading-order terms of $R$-charges, $a/\dim(G)$, $c/\dim(G)$, and $a/c$ are the same as those given in equation (\ref{eq:adj1ven charges}) for $SU(N)$ Type I theories. This behavior is universal across Type I theories. 
  However, since the decoupled operators are distinct across the Type I theories, the final corrected $R$-charges and central charges are not universal. As shown in Section \ref{sec:SUclass}, we do obtain a universal result if we take the double-scaling limit of $1\ll N_f\ll N$.

  In the presence of decoupled operators, $a$-maximization requires correction. 
  In this theory, operators of the form $\Tr A^{2n}$ and $Q_I A^n Q_J$ get decoupled along the RG flow.
  After flipping decoupled operators, we perform the $a$-maximization again. For example, in the case of $N_f=1$, after flipping all decoupled operators, we get the central charge ratio $a/c$ given as $a/c\sim 0.9948 - 0.0120/N$.

  We find that the ratio $a/c$ lies within in the range $0.875 =7/8 < a/c \leq (257+77\sqrt{985})/2668\simeq 1.0021$. The minimum value of $a/c$ can be obtained in the Veneziano limit with $\a\goto 2$. The maximum value arises when $(N_f,N)=(1,6)$. Figure \ref{fig:SOa1ratio} illustrates the behavior of the ratio $a/c$.
  \begin{figure}[t]
    \centering
    \begin{subfigure}[b]{0.45\textwidth}
        \includegraphics[width=\linewidth]{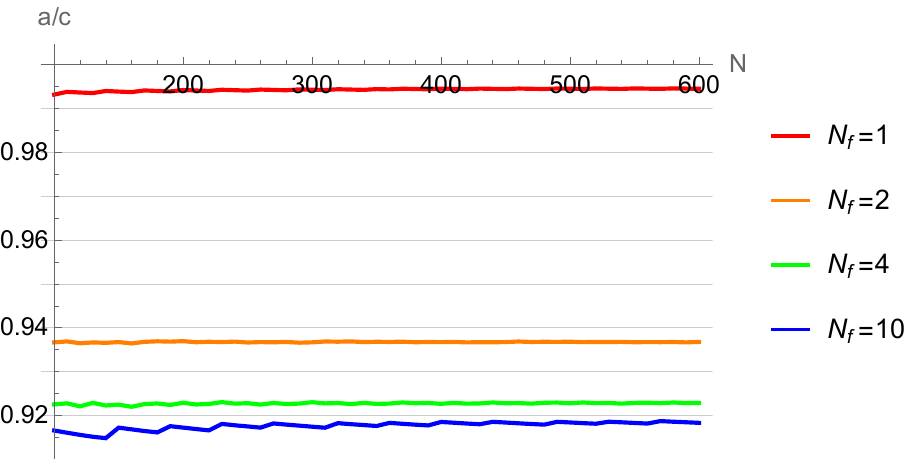}
    \end{subfigure}
    \hspace{4mm}
    \begin{subfigure}[b]{0.45\textwidth}
        \includegraphics[width=\linewidth]{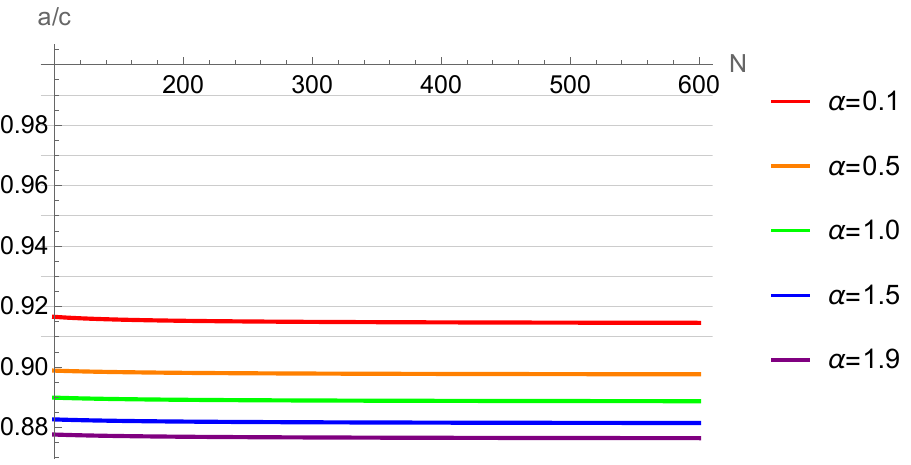}
    \end{subfigure}
    \hfill
     \caption{The central charge ratio for $SO(N)$ theory with 1 $\antisym$ + $N_f$ $\fund$. Left: $a/c$ versus $N$ with a fixed $N_f$. Right: $a/c$ versus $N$ with a fixed $\a=N_f/N$.}
    \label{fig:SOa1ratio}   
  \end{figure}

\paragraph{Conformal window}
  The upper bound of the conformal window comes from the asymptotic freedom. 
  When $N_f = 2N-4$, the gauge coupling does not run, and the theory possesses a conformal manifold, whose one-dimensional locus has $\CN=2$ supersymmetry. 
  When $N_f=0$, it is identical to $\CN=2$ SYM theory and does not flow to an interacting superconformal fixed point. Instead, the IR theory is described by an abelian gauge theory with massive charged particles \cite{Seiberg:1994rs}. 
  Otherwise, the $a$-maximization procedure poses no issues: it always yields a unique solution, and the resulting central charges fall within the Hofman-Maldacena bound. We therefore conjecture that this theory flows to an interacting SCFT for $1\leq N_f \leq 2N-4$.

\paragraph{Relevant operators}
  For a fixed $N_f$, since the $R$-charge of a rank-2 anti-symmetric chiral multiplet scales as $1/N$, there are $O(N)$ relevant operators of the form $\Tr A^{2n}$ and $Q_I A^n Q_J$. The $O(N)$ amount of them get decoupled along the RG flow.

  One can consider the deformation of the form $W=\Tr A^{2(k+1)}$, which is relevant for $N$ large enough. In this case, the theory has a dual description given by $SO\left((2k+1)N_f+4-N\right)$ gauge theory \cite{Leigh:1995qp} with a set of gauge-invariant operators flipped. This is nothing but $SO$ version of the Kutasov-Schwimmer duality. 

\paragraph{Conformal manifold}
  From the anomaly-free condition for the $U(1)_R$ symmetry, we obtain
  \begin{align}\label{eq:SOa1 marginal}
  \begin{split}
      &N-2 + (N-2)(R_A - 1) + N_f(R_Q-1)=0\\
      &\quad \quad \implies \quad \frac{2(N-2)}{N_f}R_A + 2R_Q =2\,.
  \end{split}
  \end{align}
  This implies that the operators of the form
  \begin{align}
      (\Tr A^{2n})(QA^{m}Q)\,,\quad 2n+m=\frac{2(N-2)}{N_f}\,,
  \end{align}
  are marginal for $2(N-2)/N_f\in\IZ$.
  
  For $N_f=1$, the operators of the form
  \begin{align}
      (\Tr A^{2n})(QA^{m}Q)\,,\quad 2n+m=2(N-2)\,,
  \end{align}
  are marginal. Here, the operators of the form $\Tr A^{2p}$ with $p\lesssim 0.4658 N$ and $QA^{2q}Q$ with $q \lesssim 0.0684 N $ are decoupled and not present in the interacting SCFT. 
  Nevertheless, there are $O(N)$ possible ways to partition $n$, $m$ such that $2n+m=2(N-2)$ while satisfying $n>0.4658 N$ and $m > 0.0684 N$. Thus, the theory contains $O(N)$ marginal operators, leading to an $O(N)$-dimensional conformal manifold.
  
  Similarly, for $N_f=2$, there exist $O(N)$ marginal operators of the form
  \begin{align}
      (\Tr A^{2n})(QA^mQ)\,,\quad 2n+m=N-2\,,
  \end{align}
  leading to an $O(N)$-dimensional conformal manifold.

  For $2<N_f<2(N-2)$, there are no marginal operators for general $N$, except for values of $N$ satisfying $2(N-2)/N_f\in\IZ$. For such an $N$ with a fixed $N_f$, there are again $O(N)$ marginal operators, resulting in an $O(N)$-dimensional conformal manifold.

  For the case of $N_f=2N-4$, there is a one-dimensional conformal manifold that preserves $\CN=2$ supersymmetry \cite{Bhardwaj:2013qia,Razamat:2020pra}, which is nothing but $\CN=2$ $SO(N)$ SQCD with $N-2$ fundamental hypermultiplets. 
  
\paragraph{Weak Gravity Conjecture}

  \begin{figure}[t]
    \centering
    \begin{subfigure}[b]{0.45\textwidth}
      \centering
      \includegraphics[width=\linewidth]{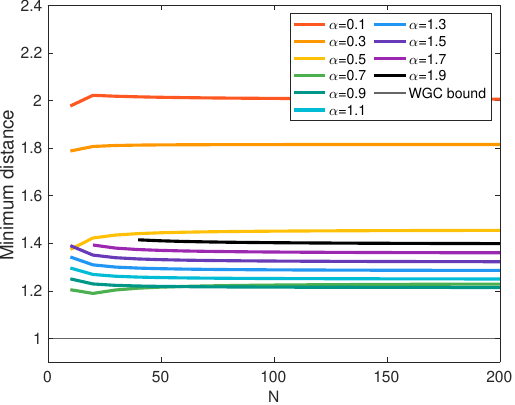}
      \caption{NN-WGC}
    \end{subfigure}
    \hspace{4mm}
    \begin{subfigure}[b]{0.45\textwidth}
      \centering
      \includegraphics[width=\linewidth]{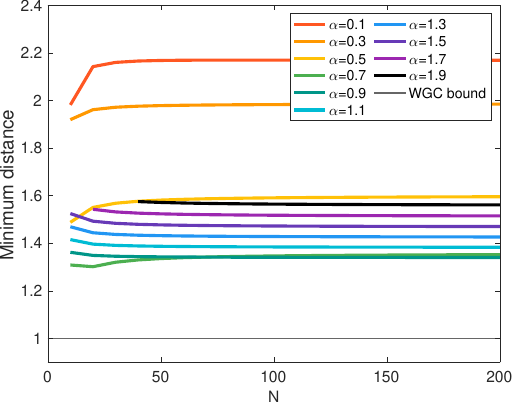}
      \caption{modified WGC}
    \end{subfigure}
    \hfill
    \caption{
    Testing AdS WGC for $SO(N)$ theory with 1 $\antisym$ + $N_f$ $\fund$. 
    The minimum distance from the origin to the convex hull with a fixed $\a=N_f/N$. \label{fig:wgc_SOa1ven}}
  \end{figure}
  We examine the AdS WGC using the gauge-invariant operators and $U(1)$ flavor charges identified at the beginning of this section. We find that this theory satisfies both versions of the WGC. The result is shown in Figure \ref{fig:wgc_SOa1ven}.

\subsection{\texorpdfstring{2 $\sym$ + $N_f$ $\fund$}{2 S + Nf Q}}\label{sec:SOs2}

\paragraph{Matter content and symmetry charges}
  The first example of the Type II theory is $SO(N)$ gauge theory with two rank-2 traceless symmetric tensors and $N_f$ vector (fundamental) chiral multiplets. The matter fields and their $U(1)$ global charges are listed in Table \ref{tab:s2}.
  {\renewcommand\arraystretch{1.6}
  \begin{table}[h]
    \centering
    \begin{tabular}{|c|c||c|c|c|}
    \hline
    \# & Fields & $SO(N)$ &$U(1)_F$ & $U(1)_R$ \\
    \hline
     $N_f$ & $Q$ &$\fund$& $2N+4$ & $R_Q$\\
     2 & $S$ &$\sym$ & $-N_f$  & $R_S$ \\
     \hline
    \end{tabular}
    \caption{
    The matter contents and their corresponding charges in $SO(N)$ gauge theory with 2 $\sym$ + $N_f$ $\fund$.\label{tab:s2}}
  \end{table}}

\paragraph{Gauge-invariant operators}
  Let $I$ and $J$ denote the flavor indices for $Q$, and $K$ denote the flavor indices for $S$. We present a sample of single-trace gauge-invariant operators in schematic form as follows:
  \begin{enumerate}
    \item $\Tr S_{K_1}\cdots S_{K_n}\,,\quad n=2,3,\dots\,.$
    \item $Q_I S_{K_1}\cdots S_{K_n} Q_J,\quad n=0,1,\dots\,.$
    \item $\e\, (S_{(1}S_{2)})^n Q_{I_1}\cdots Q_{I_{N-2n}}\,.$

    $\vdots$
  \end{enumerate}
  The ellipsis indicates that only the low-lying operators have been listed. This subset is sufficient to identify relevant operators or to test the Weak Gravity Conjecture.

\paragraph{$R$-charges and central charges}
  We perform the $a$-maximization to compute the $R$-charges of the matter fields and central charges. In the large $N$ limit with a fixed $N_f$, they are given by
  \begin{align}
  \begin{split}
    R_Q & \sim \frac{12-\sqrt{26}}{12} + \frac{39N_f + 77\sqrt{26}}{312N}+ O(N^{-2})\,, \\
    R_S & \sim  \half + \frac{48 + N_f \sqrt{26} }{24N} + O(N^{-2})\,,\\
    a & \sim \frac{27}{128}\times \half N^2 + \frac{13(18 + N_f\sqrt{26})}{1536}N+O(N^0)\,,\\
    c & \sim \frac{27}{128}\times \half N^2 + \frac{138 + 17N_f\sqrt{26}}{1536}N+O(N^0)\,,\\
    a/c & \sim 1 - \frac{2 ( -24 + N_f \sqrt{26} )}{81N} + O(N^{-2})\,.
  \end{split}
  \end{align}
  We find that no gauge-invariant operators decouple along the RG flow. The leading-order behavior of the $R$-charges, $a/\dim(G)$, and $c/\dim(G)$ is universal across Type II theories, matching the results in equations (\ref{eq:s2S2 rcharges}) and (\ref{eq:s2S2 central charges}).
  
  In the Veneziano limit with a fixed $\a=N_f/N$, the $R$-charges of the matter fields and the central charges are given by
  \begin{align}\label{eq:SOs2S2ven rcharges}
  \begin{split}
    R_Q  & \sim \frac{24-2\sqrt{26-\a^2}+3\a-3\a^2}{3(8-\a^2)}+O(N^{-1})\,,\\
    R_S  & \sim \frac{12-3\a^2+\a\sqrt{26-\a^2}}{3(8-\a^2)}+O(N^{-1})\,,\\
    a & \sim \frac{648-279\a^2 +18\a^4+2\a(26-\a^2)^{3/2}}{48(8-\a^2)^2}\times\half N^2+O(N)\,,\\
    c & \sim \frac{648-303\a^2 +21\a^4+4\a(17-\a^2)\sqrt{26-\a^2}}{48(8-\a^2)^2}\times\half N^2+O(N)\,,\\
    a/c & \sim 1-\frac{13\a(8-\a^2)}{162\sqrt{26-\a^2}+\a (685-71\a^2-30\a \sqrt{26-\a^2})}+O(N^{-1})\,.
  \end{split}
  \end{align}
  The leading-order terms of the $R$-charges, $a/\dim(G)$, and $c/\dim(G)$ are the same as those, (\ref{eq:s2S2ven rcharges}) and (\ref{eq:s2S2ven central charges}), in $SU(N)$ Type II theories.

  We observe that the ratio of central charges $a/c$ is always less than one, taking a value within the range $ 0.9286\simeq 13/14<a/c\leq19/18\simeq 1.0556$. The minimum value of $a/c$ can be obtained in the Veneziano limit with $\a\goto 1$. The maximum value of $a/c$ arises when $(N_f,N)=(0,10)$. Figure \ref{fig:SOs2ratio} illustrates the behavior of the ratio $a/c$.
  \begin{figure}[h]
    \centering
    \begin{subfigure}[b]{0.45\textwidth}
        \includegraphics[width=\linewidth]{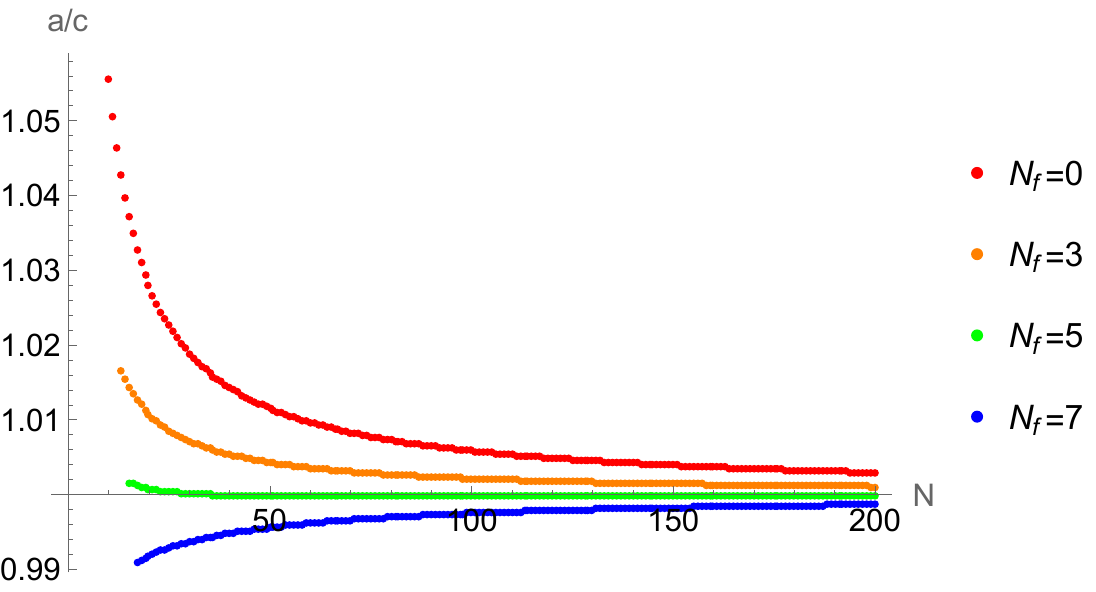}
    \end{subfigure}
    \hspace{4mm}
    \begin{subfigure}[b]{0.45\textwidth}
        \includegraphics[width=\linewidth]{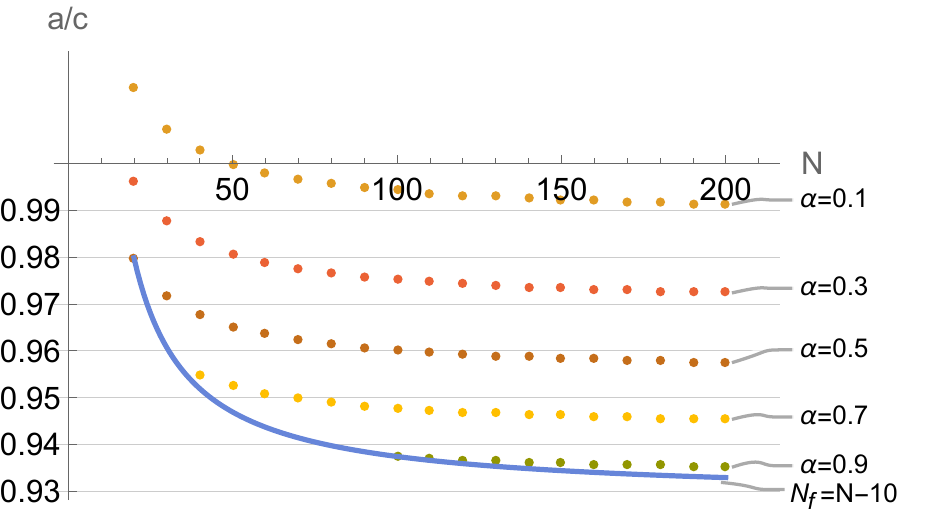}
    \end{subfigure}
    \hfill
    \caption{
    The central charge ratio for $SO(N)$ theory with 2 $\sym$ + $N_f$ $\fund$. 
    Left: $a/c$ versus $N$ with a fixed $N_f$. Right: $a/c$ versus $N$ with a fixed $\a=N_f/N$.\label{fig:SOs2ratio}}
  \end{figure}

\paragraph{Conformal window}
  The upper bound of the conformal window is determined by asymptotic freedom. For this theory, when it is at the top of the conformal window $N_f=N-10$, there is a non-trivial conformal manifold \cite{Razamat:2020pra} so that it is an interacting SCFT. For small $N_f$, the $a$-maximization procedure always yields a unique solution, and the resulting central charges lie within the Hofman-Maldacena bound. We therefore conjecture that this theory describes an interacting SCFT for $0\leq N_f\leq N-10$.

\paragraph{Relevant operators}
  For generic $N$ and $N_f$, there exist the following relevant operators:
  \begin{itemize}
    \item three operators of the form $\Tr S_{K_1}S_{K_2}$ with a dimension $\frac{3}{2}<\D \leq  2\,,$
    \item four operators of the form $\Tr S_{K_1}S_{K_2}S_{K_3}$ with a dimension $\frac{9}{4}<\D\leq 3\,,$
    \item $\frac{N_f(N_f+1)}{2}$ operators of the form $Q_I Q_J$ with a dimension $1.7253\simeq \frac{12-\sqrt{26}}{4}< \D \leq 2\,,$
    \item $N_f(N_f+1)$ operators of the form $Q_I S_K Q_J$ with a dimension $2.4753\simeq \frac{15-\sqrt{26}}{4}< \D \leq 3\,,$
  \end{itemize}
  where the upper bound arises when $N_f=N-10$. The number of relevant operators does not depend on $N$ for a fixed $N_f$. The low-lying operator spectrum is sparse in the large $N$ limit.

  Upon deforming the theory by the superpotential $W=\Tr S_1^{k+1} + \Tr S_1 S_2^2$, a dual description is proposed in \cite{Brodie:1996xm}, which is given by $SO(3kN_f+8k+4-N)$ gauge theory with a certain superpotential and flip fields. We find that the first term in the superpotential for $k>2$ is not relevant near $W=0$, but at the fixed point of $W=\Tr S_1 S_2^2$, it is relevant for large enough $N$ with a fixed $N_f$.
  
\paragraph{Conformal manifold}
  When $N_f=N-10$, the one-loop beta function for the gauge coupling vanishes. At this value, the theory possesses a non-trivial conformal manifold \cite{Razamat:2020pra}. 
  
  When $N_f<N-10$, there are no marginal operators for generic $N$ in the absence of a superpotential. However, upon a suitable superpotential deformation, the theory flows to a superconformal fixed point with a non-trivial conformal manifold.

  For the $N_f=0$ case, upon the following superpotential deformation, a non-trivial conformal manifold emerges at the IR fixed point:
  \begin{align}
      W=M_1\Tr S_1S_2 + M_2 \Tr S_1^2 + M_2^2\,.
  \end{align}
  Here, $M_i$'s are flip fields, which are gauge-singlet chiral superfields. We can test this by computing the superconformal index. For example, the reduced superconformal index for the $SO(11)$ gauge theory with this superpotential is given by
  \begin{align}
      \CI_{\text{red}} = t^{2.08}+t^{3}+t^{4.15}+t^{4.5}+t^{4.85}+t^{5.42}+t^{6} + \cdots \,.
  \end{align}
  The positivity of the coefficient at the $t^6$ term indicates a non-trivial conformal manifold to exist. 
  The marginal operator takes the form $\Tr S_1^4$. Such a deformation and exists for an arbitrary $N$ with a fixed $N_f=0$.

  For the $N_f=1$ case, consider the following superpotential deformation:
  \begin{align}
      W=M_1\Tr S_1S_2 + M_1^2\,.
  \end{align}
  At the IR fixed point, this theory possesses a non-trivial conformal manifold. We can verify this using the superconformal index. For instance, the reduced index for the $SO(12)$ gauge theory with this superpotential has a positive coefficient at the $t^6$ term:
  \begin{align}
      \CI_{\text{red}} = 3 t^{3}+4 t^{4.5}+9 t^{6}+ \cdots \,.
  \end{align}
  The marginal operator includes, for example, $\Tr S_1^4$, $(\Tr S_1^2)^2$, $\Tr(S_1S_2)^2$, and others. Such a deformation and marginal operators exist for general $N$ with a fixed $N_f=1$.

\paragraph{Weak Gravity Conjecture}
  We test the AdS WGC using the gauge-invariant operators and $U(1)$ flavor charges identified at the beginning of this section. We find that this theory always satisfies both versions of the WGC. The result is shown in Figure \ref{fig:wgc_SOs2ven}.
  \begin{figure}[t]
    \centering
    \begin{subfigure}[b]{0.45\textwidth}
      \centering
      \includegraphics[width=\linewidth]{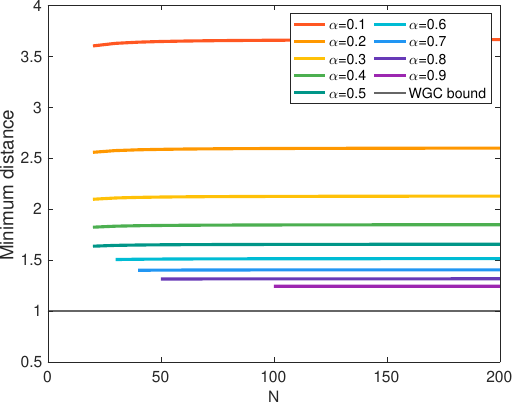}
      \caption{original WGC}
    \end{subfigure}
    \hspace{4mm}
    \begin{subfigure}[b]{0.45\textwidth}
      \centering
      \includegraphics[width=\linewidth]{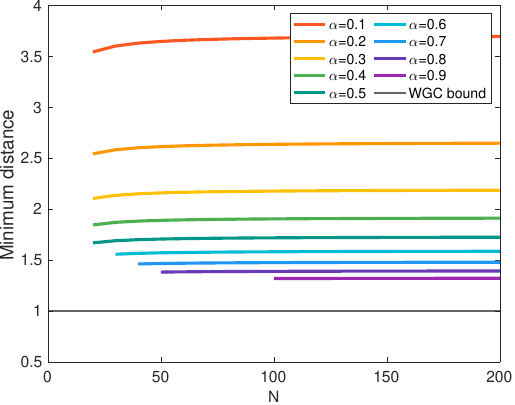}
      \caption{modified WGC}
    \end{subfigure}
    \hfill
    \caption{
    Testing AdS WGC for $SO(N)$ theory with 2 $\sym$ + $N_f$ $\fund$. The minimum distance from the origin to the convex hull with a fixed $\a=N_f/N$. \label{fig:wgc_SOs2ven}}
  \end{figure}

\subsection{\texorpdfstring{1 $\sym$ + 1 $\antisym$ + $N_f$ $\fund$}{1 S + 1 A + Nf Q}}\label{sec:SOs1a1}

\paragraph{Matter content and symmetry charges}
  Next entry of the Type II theories is $SO(N)$ gauge theory with a rank-2 traceless symmetric tensor, a rank-2 anti-symmetric tensor, and $N_f$ vector (fundamental) chiral multiplets. The matter fields and their $U(1)$ global charges are listed in Table \ref{tab:SOs1a1}.
  {\renewcommand\arraystretch{1.6}
    \begin{table}[h]
    \centering
    \begin{tabular}{|c|c||c|c|c|c|}
    \hline
    \# & Fields & $SO(N)$ &$U(1)_1$& $U(1)_2$ & $U(1)_R$ \\\hline

    $N_f$ & $Q$ & $\fund$  &  $N+2$ & $N-2$ & $R_Q$\\
    
     1 & $S$ &$\sym$ & $-N_f$ & 0 &$R_S$ \\
     
     1 & $A$ & $\antisym$ & 0 & $-N_f$  & $R_A$\\\hline
    \end{tabular}
    \caption{The matter contents and their corresponding charges in $SO(N)$ gauge theory with 1 $\sym$ + 1 $\antisym$ + $N_f$ $\fund$.\label{tab:SOs1a1}}
  \end{table}}

\paragraph{Gauge-invariant operators}
  Let $I$ and $J$ denote the flavor indices for $Q$. We present a set of single-trace gauge-invariant operators in schematic form:
  \begin{enumerate}
    \item $\Tr S^n\,,\quad \Tr A^{2m}\,,\quad  n=2,3,\dots,N\,,\quad m=1,2,\dots,\lfloor\frac{N-1}{2}\rfloor\,.$
    \item $\Tr A^nS^m\,,\quad n=2,3,\dots\,,\quad m=1,2,\dots\,.$
    \item $Q_I S^n A^m Q_J\,,\quad  n=0,1,\dots\,,\quad m=0,1,\dots\,.$
    \item $\e\,A^n  Q_{I_1}\cdots Q_{I_{N-2n}}\,.$
  \end{enumerate}

\paragraph{$R$-charges and central charges}
  We perform $a$-maximization procedure to compute the $R$-charges of the matter fields and central charges. In the large $N$ limit with a fixed $N_f$, they are given by
  \begin{align}
  \begin{split}
    R_Q & \sim \frac{12-\sqrt{26}}{12} + \frac{13N_f + 12\sqrt{26}}{104N}+ O(N^{-2})\,, \\
    R_S & \sim  \half + \frac{62 + 3N_f \sqrt{26} }{72N} + O(N^{-2})\,,\\
    R_A & \sim \half + \frac{82 + 3N_f \sqrt{26}}{72N} + O(N^{-2})\,,\\
    a & \sim \frac{27}{128}\times \half N^2 + \frac{36 + 13N_f\sqrt{26}}{1536}N+O(N^0)\,,\\
    c & \sim \frac{27}{128}\times \half N^2 + \frac{-12 + 17N_f\sqrt{26}}{1536}N+O(N^0)\,,\\
    a/c & \sim 1 - \frac{ -24 + 2N_f \sqrt{26}}{81N} + O(N^{-2})\,.
  \end{split}
  \end{align}
  We find no gauge-invariant operators decouple along the RG flow. 
  The leading-order behavior of the $R$-charges, $a/\dim(G)$, and $c/\dim(G)$ is universal across Type II theories. Similarly, in the Veneziano limit with a fixed $\a=N_f/N$, the leading terms of $R$-charges and central charges are also universal across Type II theories.
  
  We observe that the ratio of central charges $a/c$ lies within the range $0.9286 \simeq 13/14< a/c\leq 17/16=1.0625$. The ratio $a/c$ is greater than one when $N_f\leq 2$. The minimum value of $a/c$ can be obtained in the Veneziano limit with $\a \goto 1$. The maximum value of $a/c$ arises when $(N_f,N)=(0,6)$. Figure \ref{fig:SOs1a1ratio} illustrates the behavior of the ratio $a/c$.
  \begin{figure}[t]
    \centering
    \begin{subfigure}[b]{0.45\textwidth}
        \includegraphics[width=\linewidth]{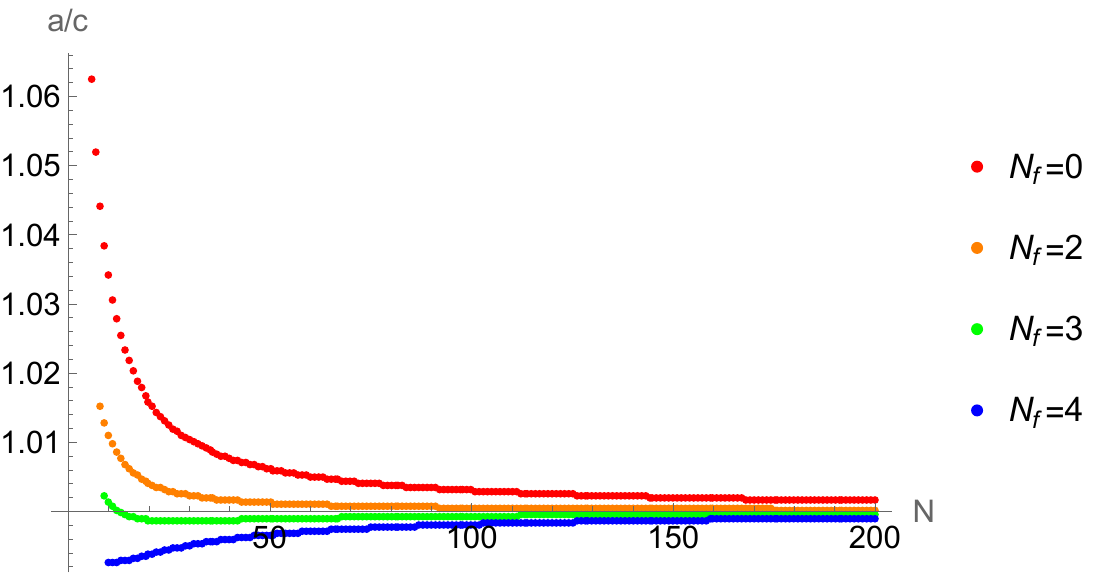}
    \end{subfigure}
    \hspace{4mm}
    \begin{subfigure}[b]{0.45\textwidth}
        \includegraphics[width=\linewidth]{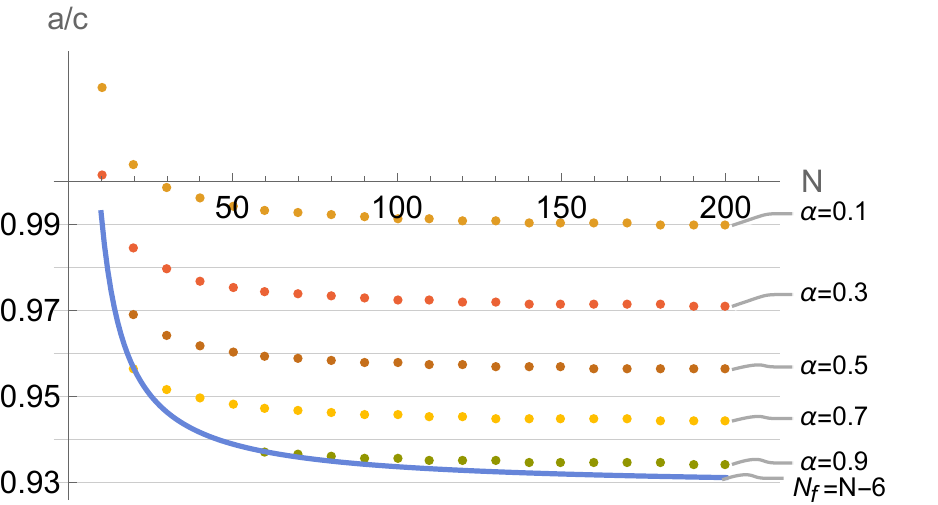}
    \end{subfigure}
    \hfill
    \caption{
    The central charge ratio for $SO(N)$ theory with 1 $\sym$ + 1 $\antisym$ + $N_f$ $\fund$. Left: $a/c$ versus $N$ with a fixed $N_f$. Right: $a/c$ versus $N$ with a fixed $\a=N_f/N$.\label{fig:SOs1a1ratio}}
  \end{figure}

\paragraph{Conformal window}
  The upper bound of the conformal window comes from the asymptotic freedom. When it is at the top of the conformal window $N_f=N-6$, there is a non-trivial conformal manifold \cite{Razamat:2020pra} so that it is an interacting SCFT. For small $N_f$, the $a$-maximization always yields a unique solution, and the resulting central charges lie within the Hofman-Maldacena bound. We therefore conjecture that this theory describes an interacting SCFT for $0\leq N_f\leq N-6$.

\paragraph{Relevant operators}
  For generic $N$ and $N_f$, there exist the following relevant operators:
  \begin{itemize}
    \item an operators of the form $\Tr S^2$ with a dimension $\frac{3}{2}<\D \leq  2\,,$
    \item an operators of the form $\Tr A^2$ with a dimension $\frac{3}{2}<\D \leq 2\,,$
    \item $\frac{N_f(N_f+1)}{2}$ operators of the form $Q_I Q_J$ with a dimension $1.7253\simeq \frac{12-\sqrt{26}}{4}< \D \leq 2\,,$
    \item an operators of the form $\Tr S^3$ with a dimension $\frac{9}{4}<\D \leq 3\,,$
    \item an operators of the form $\Tr A^2S$ with a dimension $\frac{9}{4}<\D \leq 3\,,$
    \item $\frac{N_f(N_f+1)}{2}$  operators of the form $Q_I S Q_J$ with a dimension $2.4753 \simeq\frac{15-\sqrt{26}}{4}< \D \leq 3\,,$
    \item $\frac{N_f(N_f-1)}{2}$  operators of the form $Q_I A Q_J$ with a dimension $2.4753 \simeq\frac{15-\sqrt{26}}{4} < \D \leq 3\,.$
  \end{itemize}
  The number of relevant operators does not depend on $N$ for a fixed $N_f$. The low-lying operator spectrum is sparse in the large $N$ limit.

  Upon deforming the theory by the superpotential $W=\Tr S^{k+1} + \Tr SA^2$, a dual description is proposed in \cite{Brodie:1996xm}, which is given by $SO(3kN_f+8k-4-N)$ gauge theory with a certain superpotential and flip fields. We find that the first term in the superpotential for $k>2$ is not relevant near $W=0$, but at the fixed point of $W=\Tr S A^2$, it is relevant for large enough $N$ with a fixed $N_f$.
  
\paragraph{Conformal manifold}
  When $N_f=N-6$, the one-loop beta function for the gauge coupling vanishes. At this value, the theory possesses a non-trivial conformal manifold \cite{Razamat:2020pra}.
  
  When $N_f<N-6$, there are no marginal operators for generic $N$ in the absence of a superpotential. However, upon a suitable superpotential deformation, the theory flows to a superconformal fixed point with a non-trivial conformal manifold.

  For the $N_f=0$ case, upon the following superpotential deformation, a non-trivial conformal manifold emerges at the IR fixed point:
  \begin{align}
      W=M_1\Tr A^2 + M_1^2 \,.
  \end{align}
  Here, $M_1$ is a flip field, which is a gauge-singlet chiral superfield. We can test this by computing the superconformal index. For example, the reduced superconformal index for the $SO(7)$ gauge theory with this superpotential is given by
  \begin{align}
      \CI_{\text{red}} = t^{3}+t^{13/3}-t^{9/2}\left(y+\frac{1}{y}\right)+t^{31/6}+t^{6} + \cdots \,.
  \end{align}
  The positivity of the coefficient at the $t^6$ term indicates the existence of a non-trivial conformal manifold. The marginal operator takes the form $\Tr A^4$. This deformation works for general $N$ with a fixed $N_f=0$.

  For the $N_f=1$ case, consider the following superpotential deformation:
  \begin{align}
      W=M_1\Tr A^2 + M_1^2 \,.
  \end{align}
  At the IR fixed point, this theory possesses a non-trivial conformal manifold. We can verify this using the superconformal index. For instance, the reduced index for the $SO(8)$ gauge theory with this superpotential has a positive coefficient at the $t^6$ term:
  \begin{align}
      \CI_{\text{red}} = t^{3}+t^{4.16}+t^{4.38}-t^{4.5}\left(y+\frac{1}{y}\right)+t^{5.19}+t^{6}+ \cdots \,.
  \end{align}
  The marginal operator takes the form $\Tr A^4$. Such a deformation exists for general $N$ with a fixed $N_f=1$.

\paragraph{Weak Gravity Conjecture}
  \begin{figure}[t]
    \centering
    \begin{subfigure}[b]{0.45\textwidth}
      \centering
      \includegraphics[width=\linewidth]{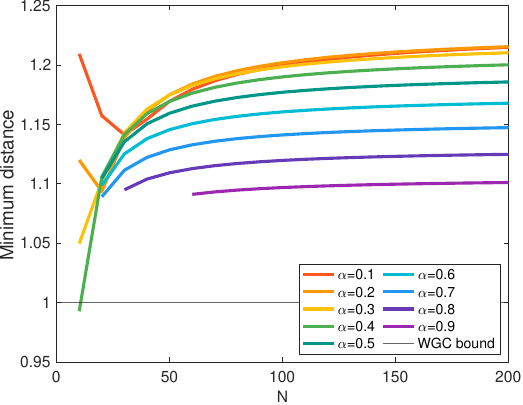}
      \caption{original WGC}
    \end{subfigure}
    \hspace{4mm}
    \begin{subfigure}[b]{0.45\textwidth}
      \centering
      \includegraphics[width=\linewidth]{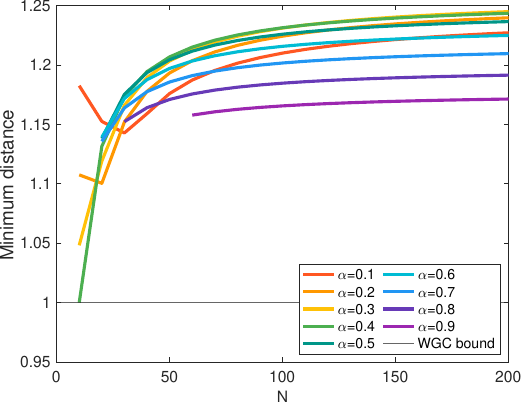}
      \caption{modified WGC}
    \end{subfigure}
    \hfill
    \caption{
    Testing AdS WGC for $SO(N)$ theory with 1 $\sym$ + 1 $\antisym$ + $N_f$ $\fund$. The minimum distance from the origin to the convex hull with a fixed $\a=N_f/N$. Theories below the solid line at the minimum distance 1 do not satisfy the WGC.\label{fig:wgc_SOs1a1ven}}
  \end{figure}
We examine the AdS WGC using the gauge-invariant operators and $U(1)$ flavor charges identified at the beginning of this section. We find that this theory does not satisfy the NN-WGC for $(N_f,N)=(4,10)$, whereas the modified WGC always holds. The result is shown in Figure \ref{fig:wgc_SOs1a1ven}. The detailed calculation is presented in \cite{Cho:2023koe}.

\subsection{\texorpdfstring{2 $\antisym$ + $N_f$ $\fund$}{2 A + Nf Q}}

\paragraph{Matter content and symmetry charges}
  The last entry of the Type II theories is $SO(N)$ gauge theory with two rank-2 anti-symmetric tensors and $N_f$ vector (fundamental) chiral multiplets. The matter fields and their $U(1)$ global charges are listed in Table \ref{tab:SOa2}.
  {\renewcommand\arraystretch{1.6}
  \begin{table}[t]
    \centering
    \begin{tabular}{|c|c||c|c|c|}
    \hline
    \# & Fields & $SO(N)$ &$U(1)_F$& $U(1)_R$ \\\hline

    $N_f$ & $Q$ & $\fund$ & $2N-4$  & $R_Q$\\
     
     2 & $A$ &$\antisym$& $-N_f$  &$R_A$\\\hline
     
    \end{tabular}
    \caption{The matter contents and their corresponding charges in $SO(N)$ gauge theory in 2 $\antisym$ + $N_f$ $\fund$.\label{tab:SOa2}}
  \end{table}}

\paragraph{Gauge-invariant operators}
  Let $I$ and $J$ denote the flavor indices for $Q$, and $M$ denote the flavor indices for $A$. We present a sample of single-trace gauge-invariant operators in schematic form as follows:
  \begin{enumerate}
    \item $\Tr A_{M_1}\cdots A_{M_n}\,,\quad n=2,3,\dots\,.$
    \item $Q_I A_{M_1}\cdots A_{M_n} Q_J,\quad n=0,1,\dots\,.$
    \item $\e\, A_{M_1}\cdots A_{M_i} \CQ_{I_1}^{n_1}\cdots \CQ_{I_{N-2i}}^{n_{N-2i}}\,.$
  \end{enumerate}
  The ellipsis indicates that only the low-lying operators have been listed. This subset is sufficient to identify relevant operators or to test the Weak Gravity Conjecture. Here, we define the dressed quarks as
  \begin{align}
      \CQ_I^n=\begin{cases}
          Q_I\,, & n=0\,,\\
          A_{M_n}Q_I\,, & n=1,2\,.
      \end{cases}
  \end{align}

\paragraph{$R$-charges and central charges}
  We perform the $a$-maximization to compute the $R$-charges of the matter fields and central charges. In the large $N$ limit with a fixed $N_f$, they are given by
  \begin{align}
  \begin{split}
    R_Q & \sim \frac{12-\sqrt{26}}{12} + \frac{39N_f -5\sqrt{26}}{312 N}+ O(N^{-2})\,, \\
    R_A & \sim \half + \frac{N_f\sqrt{26}}{24 N} + O(N^{-2})\,,\\
    a & \sim \frac{27}{128}\times \half N^2 + \frac{-162 + 13N_f\sqrt{26}}{1536}N+O(N^0)\,,\\
    c & \sim \frac{27}{128}\times \half N^2 + \frac{-162 + 17N_f\sqrt{26}}{1536}N+O(N^0)\,,\\
    a/c & \sim 1 - \frac{2 N_f \sqrt{26}}{81 N} + O(N^{-2})\,.
  \end{split}
  \end{align}
  We find that no gauge-invariant operators decouple along the RG flow. 
  The leading-order term of the $R$-charges, $a/\dim(G)$, and $c/\dim(G)$ are universal across Type II theories. Similarly, in the Veneziano limit with a fixed $\a=N_f/N$, the leading terms of $R$-charges and central charges are also universal across Type II theories.
  
  We observe that the ratio of central charges $a/c$ lies within the range $13/14\simeq 0.9286 < a/c\leq 1$. The minimum value of $a/c$ can be obtained in the Veneziano limit with $\a\goto 1$. The maximum value of $a/c$ arises when $N_f=0$. Figure \ref{fig:SOa2ratio} illustrates the behavior of the ratio $a/c$.
  \begin{figure}[t]
    \centering
    \begin{subfigure}[b]{0.45\textwidth}
        \includegraphics[width=\linewidth]{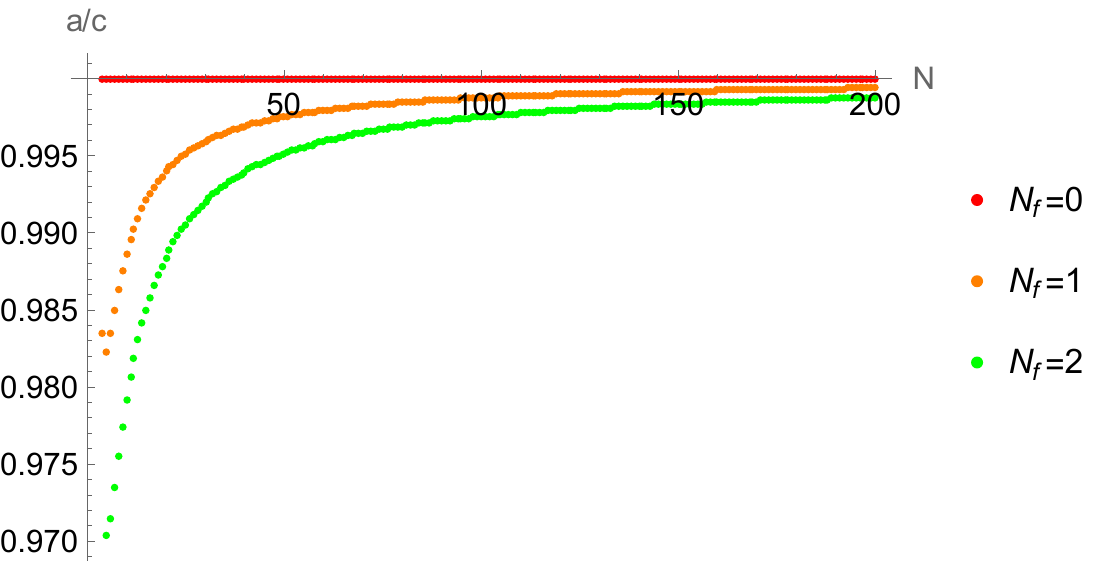}
    \end{subfigure}
    \hspace{4mm}
    \begin{subfigure}[b]{0.45\textwidth}
        \includegraphics[width=\linewidth]{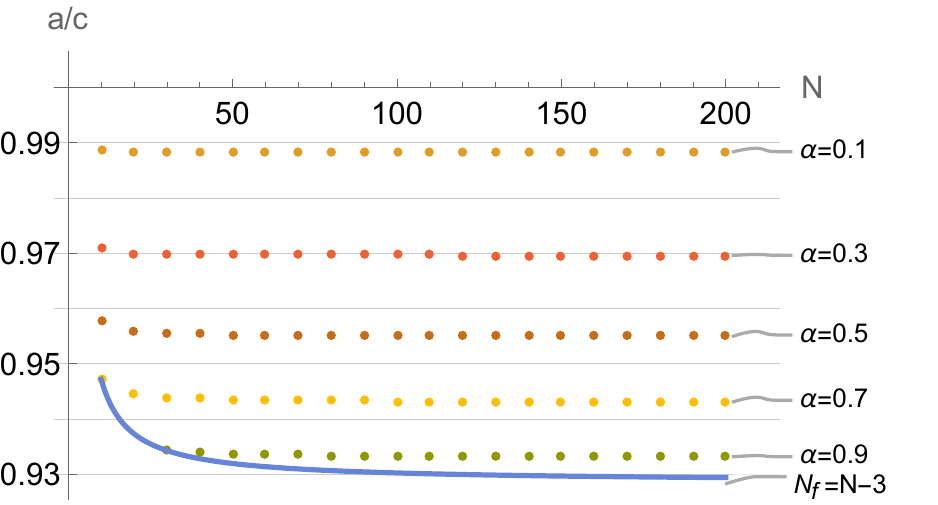}
    \end{subfigure}
    \hfill
    \caption{The central charge ratio for $SO(N)$ theory with 2 $\antisym$ + $N_f$ $\fund$. Left: $a/c$ versus $N$ with a fixed $N_f$. Right: $a/c$ versus $N$ with a fixed $\a=N_f/N$.\label{fig:SOa2ratio}}
  \end{figure}

\paragraph{Conformal window}
  The upper bound of the conformal window comes from the asymptotic freedom. For small $N_f$, the $a$-maximization procedure always yields a unique solution, and the resulting central charges lie within the Hofman-Maldacena bound. We therefore conjecture that this theory flows to an interacting SCFT for $0\leq N_f < N-2$.

\paragraph{Relevant operators}
  For generic $N$ and $N_f$, there exist the following relevant operators:
  \begin{itemize}
    \item three operators of the form $\Tr A_{M_1}A_{M_2}$ with a dimension $\frac{3}{2}\leq \D < 2\,,$
    \item $\frac{N_f(N_f+1)}{2}$ operators of the form $Q_I Q_J$ with a dimension $1.7253\simeq \frac{12-\sqrt{26}}{4}< \D < 2\,,$
    \item $N_f(N_f-1)$  operators of the form $Q_I A_M Q_J$ with a dimension $2.4753 \simeq\frac{15-\sqrt{26}}{4} < \D < 3\,.$
  \end{itemize}
  The number of relevant operators does not depend on $N$ for a fixed $N_f$. The low-lying operator spectrum is sparse in the large $N$ limit.

\paragraph{Conformal manifold}
  When $N_f=0$, there are twelve marginal operators: six are single-trace operators of the form
  \begin{align}
       \Tr A_{M_1}A_{M_2}A_{M_3}A_{M_4}\,,
  \end{align}
  and the remaining six are multi-trace operators of the form
  \begin{align}
      (\Tr A_{M_1}A_{M_2})(\Tr A_{M_3}A_{M_4})\,.
  \end{align}
  These marginal operators lead to a non-trivial conformal manifold.
  
  When $N_f>0$, there are no marginal operators for generic $N$ in the absence of a superpotential. However, upon a suitable superpotential deformation, the theory flows to a superconformal fixed point with a non-trivial conformal manifold.

  For the $N_f=1$ case, the following superpotential deformation gives rise a non-trivial conformal manifold in the IR:
  \begin{align}
      W= M_1\Tr A_1 A_2 + M_1^2\,.
  \end{align}
  The theory admits exactly marginal operators of the form $\Tr A_1^2A_2^2$, $\Tr (A_1A_2)^2$, and $\Tr A_1^2\Tr A_2^2$, which are neutral under the remaining flavor symmetry after the deformation. This deformation works for general $N$ with a fixed $N_f=1$.

\paragraph{Weak Gravity Conjecture}
  We examine the AdS WGC using the gauge-invariant operators and $U(1)$ flavor charges identified at the beginning of this section. We find that this theory satisfies both versions of the WGC. The result is shown in Figure \ref{fig:wgc_SOa2ven}.
  \begin{figure}[t]
    \centering
    \begin{subfigure}[b]{0.45\textwidth}
      \centering
      \includegraphics[width=\linewidth]{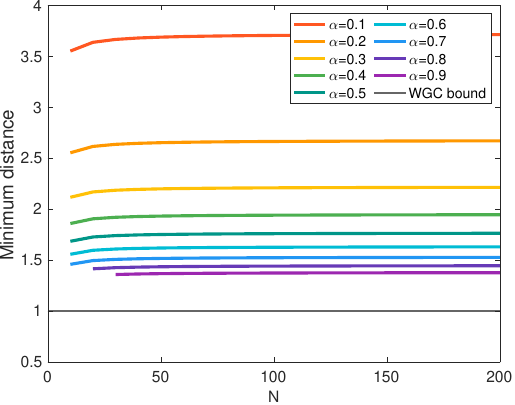}
      \caption{original WGC}
    \end{subfigure}
    \hspace{4mm}
    \begin{subfigure}[b]{0.45\textwidth}
      \centering
      \includegraphics[width=\linewidth]{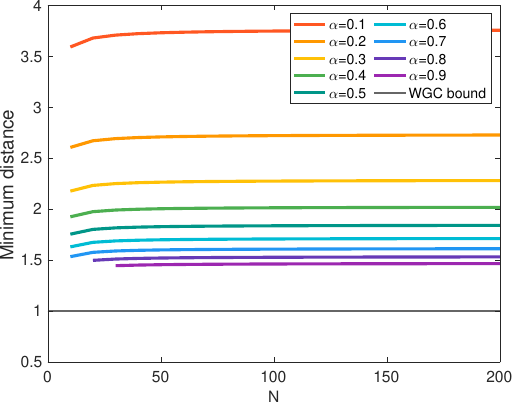}
      \caption{modified WGC}
    \end{subfigure}
    \hfill
    \caption{
    Testing AdS WGC for $SO(N)$ theory with 2 $\antisym$ + $N_f$ $\fund$. The minimum distance from the origin to the convex hull with a fixed $\a=N_f/N$. \label{fig:wgc_SOa2ven}}
  \end{figure}

\subsection{\texorpdfstring{3 $\antisym$}{}}
  The only Type III theory with $SO(N)$ gauge group is the one with three rank-2 anti-symmetric tensor (adjoint) chiral multiplets. It has the same matter content as $\CN=4$ SYM theory. 
  This theory is a conformal gauge theory with no running coupling. The $R$-charges and the central charges are given by
  \begin{align}
      R_A=\frac{2}{3}\,,\quad a=c = \frac{1}{4} \dim G=\frac{1}{4}\cdot \half N(N-1)\,.
  \end{align}
  The leading-order behavior of the $R$-charges and central charges is universal across the Type III theories.

  The single-trace gauge-invariant operators are of the form
  \begin{align}
      \Tr A_{M_1}\cdots A_{M_{n}}\,,\quad n=1,2,\dots\,.
  \end{align}
  There are six relevant operators of the form $\Tr A_{M_1}A_{M_2}$ and a marginal operator of the form $\Tr A_1 A_2 A_3$. 
  The theory possesses a one-dimensional subspace in the conformal manifold that preserves $\CN = 4$ supersymmetry.

\section{Superconformal Sp(N) gauge theories with large N limit} \label{sec:Sp}
  In this section, we analyze all possible superconformal $Sp(N)$ gauge theories that can have a large $N$ limit. As discussed in Section \ref{sec:classification}, only the representations in Table \ref{tab:Spindex} are allowed. The Young diagrams $\fund\,$, $\sym\,$, and $\antisym\,$ denote the fundamental, rank-2 symmetric tensor, and rank-2 traceless anti-symmetric tensor representations, respectively. Since these representations are (pseudo-)real, our $Sp(N)$ theories considered here are free from gauge anomalies. The Witten's global anomaly \cite{Witten:1982df} restricts the number of chiral multiplets in the fundamental representation to be even.
  {\renewcommand\arraystretch{1.4}
  \begin{table}[t]
    \centering
    \begin{tabular}{|c||c|c|c|}
    \hline
    Fields & $Sp(N)$ & $T(\textbf{R})$ & $|\mathbf{R}|$ \\\hline

    $Q$ & $\fund$& $\frac{1}{2}$ & $2N$\\

    $\Phi$ & \textbf{Adj} & $N+1$ & $N(2N+1)$\\
     
    $S$ & $\sym$ & $N+1$  & $N(2N+1)$ \\
    
    $A$ & $\antisym$ & $N-1$ & $N(2N-1)-1$ \\\hline
    \end{tabular}
    \caption{The Dynkin indices and dimensions of irreducible representations $\mathbf{R}$ in $Sp(N)$.\label{tab:Spindex}}
  \end{table}}

  The one-loop beta function is required to be negative:
  \begin{align}
    b_0=3(N+1)-\half N_{Q}-(N+1)N_{S}-(N-1)N_{A}\geq0\,,
  \end{align}
  where $N_Q$, $N_S$, and $N_A$ denote the number of fundamentals, symmetric tensors, and anti-symmetric tensors, respectively. 
  Note that $N_{Q}$ must be even due to the Witten anomaly. 
  When the one-loop beta function becomes zero, the coupling is generally marginally irrelevant, unless there is a non-trivial conformal manifold. These two conditions constrain the set of admissible theories to a finite number of families. The result is listed in Table \ref{tab:Splist}. 
  We also summarize the range of the central charge ratio $a/c$ and the condition for a non-trivial conformal manifold to exist in Table \ref{tab:SOratio}.

  {\renewcommand\arraystretch{1.6}
  \begin{table}[t]
    \centering
    \begin{tabular}{|c|c|c|c|}
      \hline
     Type & $Sp(N)$ theory &  NN-WGC & Conformal window  \\\hline \hline

\multirow{2}{*}{I} &     1 $\sym$\, + $2N_f$ $\fund$ & Always & $1\le N_f \leq 2N+2^*$  {\rule[-2ex]{0pt}{-3.0ex}} \\ \cline{2-4}
      
&      1 $\antisym$\, + $2N_f$ $\fund$  & Always & $4 \le N_f < 2N+4$  \\ \hline\hline
      
 \multirow{4}{*}{II} &    2 $\sym$\, + $2N_f$ $\fund$ & Always & $0\le N_f < N+1$  \\ \cline{2-4}
      
  &    1 $\sym$\, + 1 $\antisym$\, + $2N_f$ $\fund$ & Always & $0\le N_f \leq N+3^*$  \\ \cline{2-4}
      
&      2 $\antisym$\, + $2N_f$ $\fund$ & Always & $0\le N_f \leq N+5^*$\tablefootnote{Except for the cases of $(N_f,N)=(0,2), (0,3), (1,2)$. When $(N_f,N)=(0,2)$, the theory is equivalent to $SO(5)$ gauge theory with two vectors, which does not flow to an interacting SCFT. For $(N_f,N)=(0,3),(1,2)$, the $a$-maximization procedure yields negative central charges, so we conjecture that these theories do not flow to unitary interacting SCFTs.}  \\ \cline{2-4}
      
  &    2 $\sym$\, + 1 $\antisym$\, + $2N_f$ $\fund$ & Always & $0\le N_f < 2$  \\ \hline\hline
      
   \multirow{3}{*}{III} &  1 $\sym$\, + 2 $\antisym$\, + $2N_f$ $\fund$ & $\begin{aligned}[c]&(N_f,N)\neq(2,5),(3,6),\\&(4,2),(4,6),(4,7),(4,8) \end{aligned}$  & $0\le N_f \leq 4^*$  \\ \cline{2-4}
      
&      3 $\antisym$\, + $2N_f$ $\fund$ & $(N_f,N)\neq (5,3),(6,3)$ & $0\le N_f \leq  6^*$\tablefootnote{Except for the $(N_f,N)=(0,2)$ case, in which the theory is equivalent to $SO(5)$ gauge theory with three vectors and does not flow to an interacting SCFT.}  \\ \cline{2-4}
      
&      3 \textbf{Adj} & Always  & $*$ \\ \hline
	\end{tabular}
	\caption{List of all possible superconformal $Sp(N)$ theories with large $N$ limit in Type I,II,III. The third column lists the condition for each theory to satisfy the NN-WGC in the Veneziano limit. The last column specifies the condition for each theory to flow to a superconformal fixed point. The entries with $*$ (if $N_f$ saturates the upper bound) do not flow but have non-trivial conformal manifolds \cite{Razamat:2020pra, Bhardwaj:2013qia}.}
	\label{tab:Splist}
\end{table}}

  {\renewcommand\arraystretch{1.6}
  \begin{table}[t]\small
    \centering
    \begin{tabular}{|c|c|c|c|}
      \hline
Type &      $Sp(N)$ theory &  Range of $a/c$ & \makecell{Conformal manifold}  \\\hline \hline 
      
\multirow{2}{*}{I}  &    1 $\sym$\, + $2N_f$ $\fund$ & $0.8409\simeq \frac{37}{44}< a/c \lesssim 0.9367$ & $2(N+1)/N_f\in\IZ$  {\rule[-2ex]{0pt}{-3.0ex}} \\ \cline{2-4}
      
&      1 $\antisym$\, + $2N_f$ $\fund$  & $0.7028\simeq \frac{473}{673} \leq a/c \lesssim 0.9159$  & $N_f=4$  \\ \hline\hline
      
 \multirow{4}{*}{II} &    2 $\sym$\, + $2N_f$ $\fund$ & $0.9261\simeq \frac{213779-812\sqrt{2753}}{184826}\leq a/c \leq 1$ & $N_f=0$ \\ \cline{2-4}
      
&      1 $\sym$\, + 1 $\antisym$\, + $2N_f$ $\fund$ & $0.8529\simeq \frac{29}{34}<a/c < 1$ & $N_f=N+3$ \\ \cline{2-4}
      
&      2 $\antisym$\, + $2N_f$ $\fund$ & $0.6788\simeq \frac{2(148+13\sqrt{19})}{603}\leq a/c < 1$ & $N_f=N+5$ \\ \cline{2-4}
      
&      2 $\sym$\, + 1 $\antisym$\, + $2N_f$ $\fund$ & $0.9475\lesssim a/c < 1$ & *  \\ \hline\hline
      
\multirow{3}{*}{III}  &    1 $\sym$\, + 2 $\antisym$\, + $2N_f$ $\fund$ & $0.8659\simeq \frac{71}{82}\leq a/c < 1$ & $N_f=4$  \\ \cline{2-4}
     
&      3 $\antisym$\, + $2N_f$ $\fund$ & $0.7742\simeq \frac{3(243842-985\sqrt{849})}{833662}\leq a/c < 1$ &  $N_f=6$ \\ \cline{2-4}
      
 &     3 \textbf{Adj} & $a/c=1$  &  Always \\ \hline
	\end{tabular}
	\caption{Central charge ratio and the condition for the conformal manifold to exist in $Sp(N)$ theories. The third column lists the range of the ratio $a/c$ for general $N$ and $N_f$ within the conformal window. The last column denotes the condition for the theory to have a non-trivial conformal manifold. The entries with $*$ do not have non-trivial conformal manifolds. \label{tab:Spratio}}
\end{table}}

\subsection{\texorpdfstring{1 $\sym$ + 2 $N_f$ $\fund$}{1 S + 2 Nf Q}}\label{sec:SpadjSQCD}
\paragraph{Matter content and symmetry charges}
  The first entry of the Type I theories is $Sp(N)$ gauge theory with a rank-2 symmetric and $2N_f$ fundamental chiral multiplets. The matter fields and their $U(1)$ global charges are listed in Table \ref{tab:Sps1}.
 {\renewcommand\arraystretch{1.4}
  \begin{table}[t]
    \centering
    \begin{tabular}{|c|c||c|c|c|}
    \hline
    \# & Fields & $Sp(N)$ & $U(1)_F$  & $U(1)_R$ \\\hline
     $2N_f$ & $Q$ & $\fund$ &  $N+1$ & $R_Q$ \\
     1 & $S$ & $\sym$ &  $-N_f$  & $R_S$  \\\hline
    \end{tabular}
    \caption{The matter contents and their corresponding charges in $Sp(N)$ gauge theory with 1 $\sym$ + 2 $N_f$ $\fund$.\label{tab:Sps1}}
  \end{table}}

\paragraph{Gauge-invariant operators}
  Let $I$ and $J$ denote the flavor indices for $Q$. Let by $\O$ be the invariant anti-symmetric matrix of the $Sp(N)$ group, defined as
  \begin{align}
      \O=\begin{pmatrix}
          0 & - I_N \\ I_N & 0
      \end{pmatrix}\,,
  \end{align}
  where $I_N$ is the $N\times N$ identity matrix. We present a set of single-trace gauge-invariant operators in schematic form as follows:
  \begin{enumerate}
      \item $\Tr (\O S)^{2n}\,,\quad n=1,2,\dots,N\,.$
      \item $Q_I (\O S)^n \O Q_J\,,\quad 0,1,\dots,2N-1\,.$
  \end{enumerate}

\paragraph{$R$-charges and central charges}
  We perform the $a$-maximization to compute the $R$-charges of the matter fields and central charges. In the large $N$ limit with a fixed $N_f$, they are given by
  \begin{align}
      \begin{split}\label{eq:Sps1 charges}
          R_Q&\sim \frac{3-\sqrt{5}}{3} + \frac{2\sqrt{5}+ 15 N_f}{30N} + O(N^{-2})\,,\\
          R_S&\sim \frac{ \sqrt{5}N_f}{3N} - \frac{5N_f^2 + 4\sqrt{5} N_f + 16}{10N^2}+ O(N^{-3})\,,\\
          a & \sim \frac{5\sqrt{5}N_f}{12} N -\frac{15 N_f^2 + 4 \sqrt{5} N_f }{16} + O(N^{-1})\,,\\
          c & \sim \frac{11\sqrt{5}N_f}{24} N -\frac{80N_f^2 + 19\sqrt{5}N_f }{80} + O(N^{-1})\,,\\
          a/c &\sim \frac{10}{11}-\frac{3\sqrt{5}N_f + 18}{242N} + O(N^{-2})\,.
      \end{split}
  \end{align}
  For a fixed $N_f$, the $R$-charge of a rank-2 symmetric chiral multiplet scales as $1/N$. As a result, the spectrum of gauge-invariant operators becomes dense, with $O(N)$ operators decoupling, and therefore this theory belongs to the Type I class.

  In the Veneziano limit $N,N_f\goto\infty$ with a fixed $\a=N_f/N$, the result of $a$-maximization is as follows:
  \begin{align}
      \begin{split}\label{eq:Sps1ven charges}
          R_Q&\sim \frac{6-\sqrt{20-\a^2}+3\a-3\a^2}{3(2-\a^2)} + O(N^{-1}) \,,\\
          R_S &\sim \frac{\a\sqrt{20-\a^2}-3\a^2}{3(2-\a^2)} + O(N^{-1})\,,\\
          a & \sim \frac{-18\a^2(5-\a^2)+\a(20-\a^2)^{3/2}}{48(2-\a^2)^2} \times 2 N^2 + O(N)\,,\\
          c & \sim \frac{-96\a^2+21\a^4+(22\a-2\a^3)\sqrt{20-\a^2}}{48(2-\a^2)^2} \times 2  N^2 + O(N)\,,\\
          a/c &\sim \frac{440-76\a^2-3\a\sqrt{20-\a^2}}{484-89\a^2}+  O(N^{-1})\,.
      \end{split}
  \end{align}
  The leading-order terms of $R$-charges, $a/\dim(G)$, $c/\dim(G)$, and $a/c$ are the same as those given in equation (\ref{eq:adj1ven charges}) for $SU(N)$ Type I theories and equation (\ref{eq:SOs1ven charges}) for $SO(N)$ Type I theories. This leading behavior is universal across Type I theories. However, since the decoupled operators are distinct across the Type I theories, the resulting corrected $R$-charges and central charges are not universal. As discussed in Section \ref{sec:SUclass}, we obtain a universal result if we take the additional limit $1\ll N_f\ll N$.
  
  Since there exist decoupled operators, the naive results of the $a$-maximization are no longer valid. We must correct the $a$-maximization by flipping decoupled operators.  For example, in the $N_f=1$ case, upon flipping all decoupled operators, the central charge ratio $a/c$ is given by $0.9367 - 0.1261/N$.

  We find that the ratio $a/c$ lies within the range $0.8409\simeq 37/44 < a/c \lesssim 0.9367$. The minimum value of $a/c$ arises when $(N_f,N)=(6,2)$. The maximum value can be obtained in the large $N$ limit with $N_f=1$. Figure \ref{fig:Sps1ratio} illustrates the behavior of the ratio $a/c$.
  \begin{figure}[t]
    \centering
    \begin{subfigure}[b]{0.45\textwidth}
        \includegraphics[width=\linewidth]{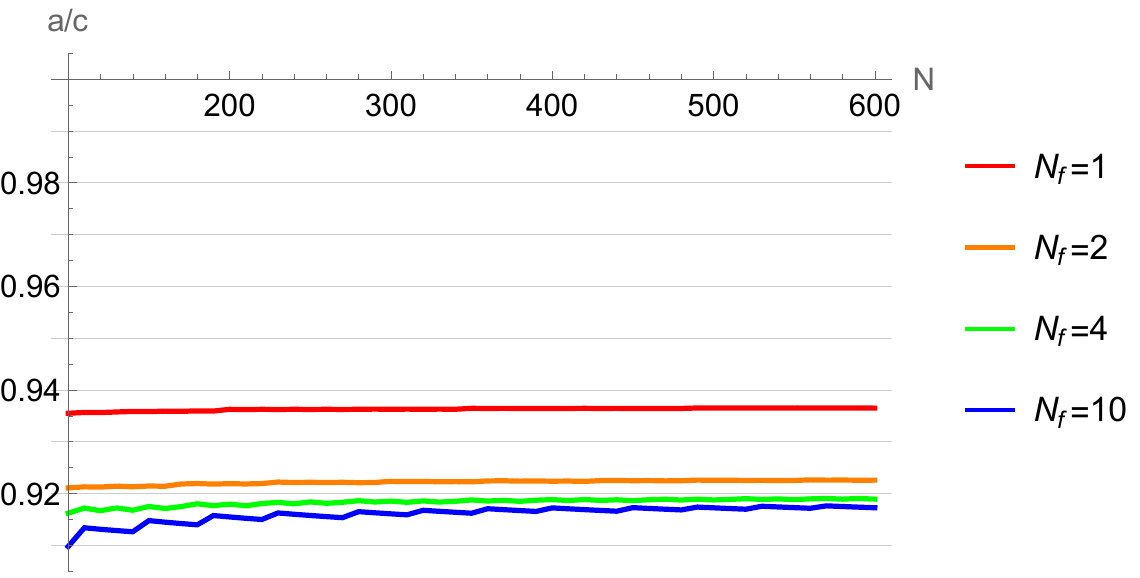}
    \end{subfigure}
    \hspace{4mm}
    \begin{subfigure}[b]{0.45\textwidth}
        \includegraphics[width=\linewidth]{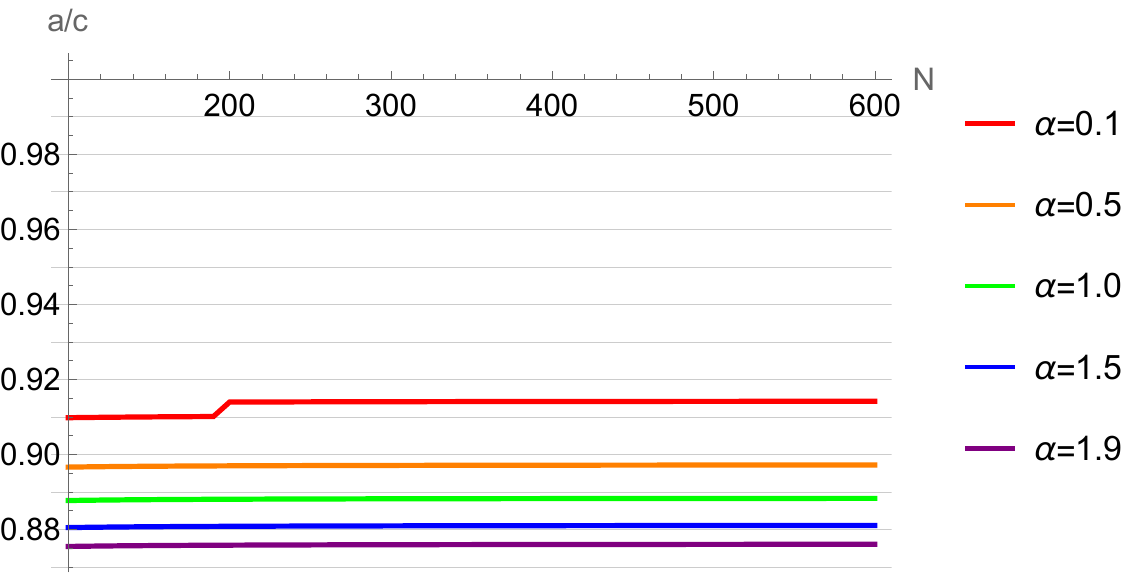}
    \end{subfigure}
    \hfill
     \caption{The central charge ratio for $Sp(N)$ theory with 1 $\sym$ + 2 $N_f$ $\fund$. Left: $a/c$ versus $N$ with a fixed $N_f$. Right: $a/c$ versus $N$ with a fixed $\a=N_f/N$.}
    \label{fig:Sps1ratio}   
  \end{figure}

\paragraph{Conformal window}
  The upper bound is determined by the asymptotic freedom. When it is at the top of the conformal window $N_f=2N+2$, there is a non-trivial conformal manifold \cite{Razamat:2020pra} so that it is an interacting SCFT. When $N_f=0$, it is identical to $\CN=2$ SYM and does not flow to an interacting superconformal fixed point. Instead, the IR theory is described by an abelian gauge theory with massive charged particles \cite{Seiberg:1994rs}. For small $N_f$, the $a$-maximization procedure always yields a unique solution, and the resulting central charges lie within the Hofman-Maldacena bound. We therefore conjecture that this theory flows to an interacting SCFT for $1\leq N_f \leq  2N+2$.

\paragraph{Relevant operators}
  For a fixed $N_f$, since the $R$-charge of a rank-2 symmetric chiral multiplet scales as $1/N$, there are $O(N)$ relevant operators of the form $\Tr S^{2n}$ and $Q_I S^n Q_J$. The $O(N)$ amount of them get decoupled along the RG flow.

  One can consider the deformation of the form $W=\Tr (\O S)^{2k+2}$, which is relevant for sufficiently large $N$. In this case, the theory has a dual description given by $Sp\left((2k+1)N_f-N-2\right)$ gauge theory \cite{Leigh:1995qp}, with a set of gauge-invariant operators flipped. This is the $Sp$ version of the Kutasov-Schwimmer duality.

  By considering a flip deformation of all the `Coulomb branch operators' of the form $\Tr (\Omega S)^{2k}$ and `generalized mesons' of the form $Q S^n Q$, we obtain the $(A_1, A_{2N})$ Argyres-Douglas theory as its fixed point with emergent $\CN=2$ supersymmetry \cite{Maruyoshi:2016aim}. 

\paragraph{Conformal manifold}
  From the anomaly-free condition for $U(1)_R$ symmetry, we obtain
  \begin{align}
  \begin{split}
      &N+1 + (N+1)(R_S - 1) + N_f(R_Q-1)=0\\
      &\quad \quad \implies \quad \frac{2(N+1)}{N_f}R_S + 2R_Q =2\,.
  \end{split}
  \end{align}
  This implies that the operators of the form
  \begin{align}
      \left(\Tr (\O S)^{2n}\right)\left(Q_I(\O S)^mQ_J\right)\,,\quad 2n+m = \frac{2(N+1)}{N_f}
  \end{align}
  are marginal whenever $2(N+1)/N_f\in\IZ$.
  For such an $N$ satisfying $2(N+1)/N_f \in \IZ$ with a fixed $N_f$, the theory contains $O(N)$ marginal operators, leading to an $O(N)$-dimensional conformal manifold. 
  
  When $N_f=2N+2$, there is a one-dimensional conformal manifold that preserves $\CN=2$ supersymmetry \cite{Bhardwaj:2013qia, Razamat:2020pra}, which is nothing but $\CN=2$ $Sp(N)$ conformal SQCD with $N+1$ fundamental hypermultiplets.
  
\paragraph{Weak Gravity Conjecture}
  We test the AdS WGC using the gauge-invariant operators and $U(1)$ flavor charges identified at the beginning of this section. We find that this theory satisfies both versions of the WGC. The result is shown in Figure \ref{fig:wgc_Sps1ven}.

\begin{figure}[t]
    \centering
     \begin{subfigure}[b]{0.45\textwidth}
     \centering
    \includegraphics[width=\linewidth]{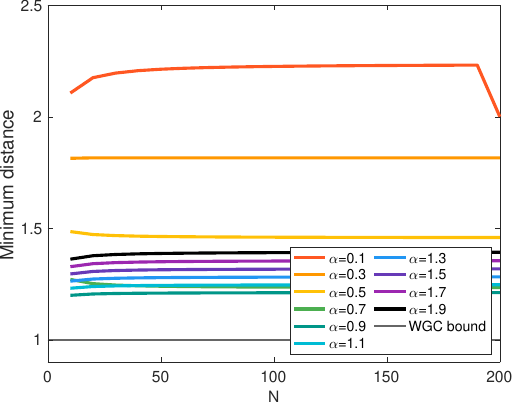}
    \caption{NN-WGC}
     \end{subfigure}
     \hspace{4mm}
     \begin{subfigure}[b]{0.45\textwidth}
     \centering
        \includegraphics[width=\linewidth]{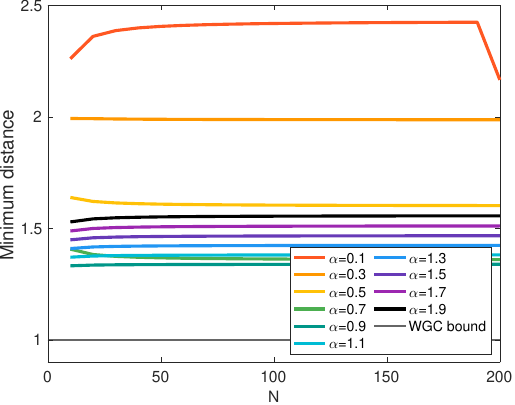}
        \caption{modified WGC}
     \end{subfigure}
     \hfill
        \caption{
        Testing AdS WGC for $Sp(N)$ theory with 1 $\sym$ + 2 $N_f$ $\fund$. The minimum distance from the origin to the convex hull with a fixed $\a=N_f/N$.\label{fig:wgc_Sps1ven}}
 \end{figure}

\subsection{\texorpdfstring{1 $\antisym$ + 2 $N_f$ $\fund$}{1 A + 2 Nf Q}} \label{sec:sp1antisym}
\paragraph{Matter content and symmetry charges}
  The next case of the Type I theory is the $Sp(N)$ gauge theories with a rank-2 anti-symmetric tensor and $2N_f$ fundamental chiral multiplets. The matter fields and their $U(1)$ global charges are listed in Table \ref{tab:Spa1}.
 {\renewcommand\arraystretch{1.4}
  \begin{table}[h]
    \centering
    \begin{tabular}{|c|c||c|c|c|}
    \hline
    \# & Fields & $Sp(N)$ & $U(1)_F$  & $U(1)_R$ \\\hline
     $2N_f$ & $Q$ & $\fund$ &  $N-1$ & $R_Q$ \\
     1 & $A$ & $\antisym$ &  $-N_f$  & $R_A$  \\\hline
    \end{tabular}
    \caption{The matter contents and their corresponding charges in $Sp(N)$ gauge theory with 1 $\antisym$ + 2 $N_f$ $\fund$.\label{tab:Spa1}}
  \end{table}}

\paragraph{Gauge-invariant operators}
  Let $I$ and $J$ denote the flavor indices for $Q$. We present a set of single-trace gauge-invariant operators in schematic form as follows:
  \begin{enumerate}
      \item $\Tr (\O A)^{n}\,,\quad n=2,3,\dots,N\,.$
      \item $Q_I (\O A)^n \O Q_J\,,\quad 0,1,\dots,N-1\,.$
  \end{enumerate}

\paragraph{$R$-charges and central charges}
  We perform the $a$-maximization to compute the $R$-charges of the matter fields and central charges. In the large $N$ limit with a fixed $N_f$, they are given by
  \begin{align}
      \begin{split}\label{eq:Spa1 charges}
          R_Q&\sim \frac{3-\sqrt{5}}{3} + \frac{-4\sqrt{5}+ 3 N_f}{6N} + O(N^{-2})\,,\\
          R_A&\sim \frac{ -6 + \sqrt{5}N_f}{3N} - \frac{N_f^2 - 2\sqrt{5} N_f + 4}{2N^2}+ O(N^{-3})\,,\\
          a & \sim \frac{-63+10\sqrt{5}N_f}{24} N -\frac{15 N_f^2 - 40 \sqrt{5} N_f +129}{16} + O(N^{-1})\,,\\
          c & \sim \frac{-60+11\sqrt{5}N_f}{24} N -\frac{16N_f^2 - 41\sqrt{5}N_f + 128}{16} + O(N^{-1})\,,\\
          a/c &\sim \frac{-63+10\sqrt{5}N_f}{-60+11\sqrt{5}N_f} - \frac{3(5\sqrt{5}N_f^3 -42N_f^2-44\sqrt{5}N_f+ 324)}{2(-60+11\sqrt{5}N_f)^2N} + O(N^{-2})\,.
      \end{split}
  \end{align}
  For a fixed $N_f$, the $R$-charge of a rank-2 anti-symmetric chiral multiplet scales as $1/N$. As a result, the spectrum of gauge-invariant operators becomes dense, with $O(N)$ decoupled operators, and hence this theory belongs to the Type I class.

  In the Veneziano limit with a fixed $\a=N_f/N$, the leading-order terms of $R$-charges, $a/\dim(G)$, $c/\dim(G)$, and $a/c$ are universally across Type I theories. As discussed in Section \ref{sec:SUclass}, we obtain a universal result if we take the additional limit $1\ll N_f\ll N$.
  
  In the presence of the decoupled operators, the results of $a$-maximization above are no longer valid since there exists accidental symmetry acting on decoupled operators. To fix this, we flip the decoupled operators and perform $a$-maximization again. For example, in the case of $N_f=4$, upon flipping all decoupled operators, we obtain the central charge ratio $a/c$ to be $0.7540 - 0.2410/N$.
  
  We find that the ratio lies within the range $0.7028\simeq \frac{473}{673} \leq a/c \lesssim 0.9159 $. The minimum value of $a/c$ arises when $(N_f,N)=(4,3)$. The maximum value can be obtained in the Veneziano limit with $\a\sim 0.1$. The plot of the ratio $a/c$ is shown in Figure \ref{fig:Spa1ratio}. 
  \begin{figure}[t]
    \centering
    \begin{subfigure}[b]{0.45\textwidth}
        \includegraphics[width=\linewidth]{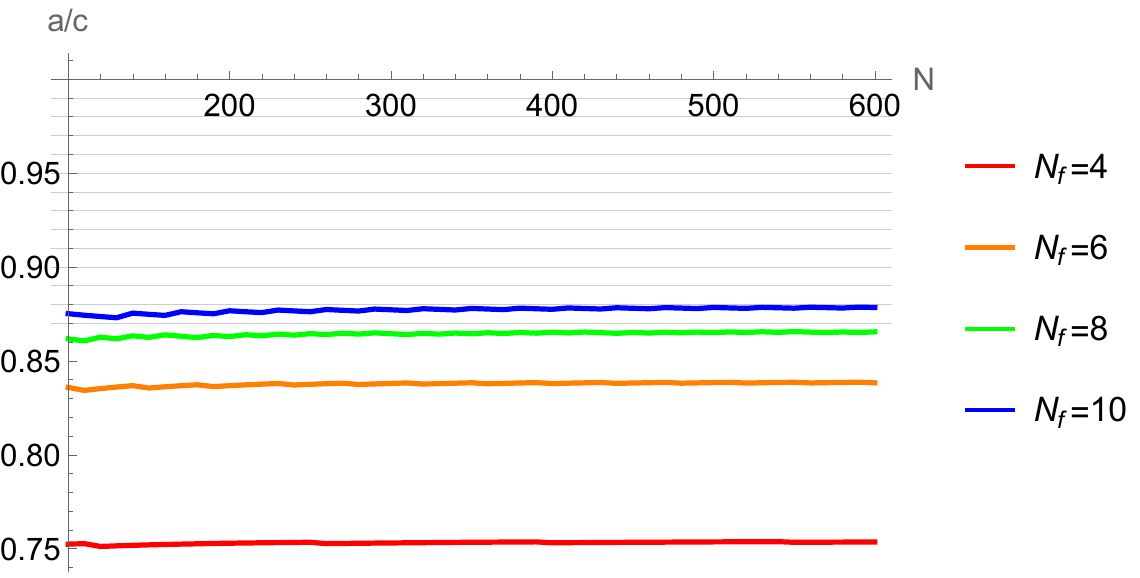}
    \end{subfigure}
    \hspace{4mm}
    \begin{subfigure}[b]{0.45\textwidth}
        \includegraphics[width=\linewidth]{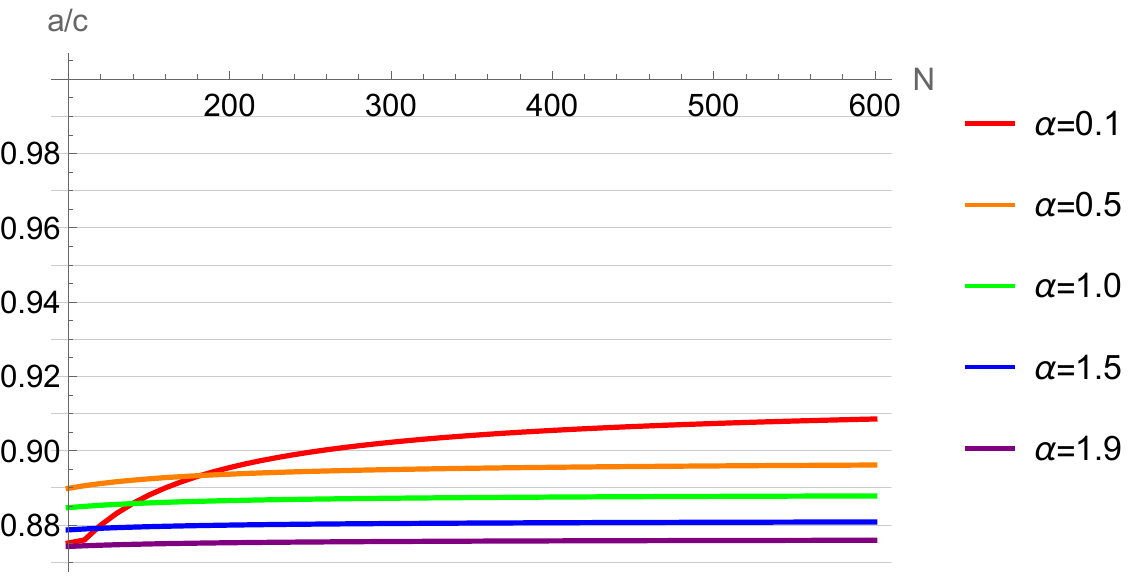}
    \end{subfigure}
    \hfill
     \caption{The central charge ratio for $Sp(N)$ theory with 1 $\antisym$ + 2 $N_f$ $\fund$. Left: $a/c$ versus $N$ with a fixed $N_f$. Right: $a/c$ versus $N$ with a fixed $\a=N_f/N$.}
    \label{fig:Spa1ratio}   
  \end{figure}

\paragraph{Conformal window}
  The upper bound of the conformal window comes from the asymptotic bound. For $N_f<3$, the $a$-maximization procedure yields negative central charges, so these theories cannot flow to unitary interacting SCFTs. In fact, we expect a runaway superpotential to be dynamically generated \cite{Csaki:1996eu} as in the case of $SU(N)$ SQCD with $N_f < N$ \cite{Affleck:1983mk}. 
  When $N_f=3$, the theory confines without chiral symmetry breaking \cite{Csaki:1996eu}. The theory at the IR fixed point is described by massless free fields. This can also be seen from our analysis via $a$-maximization, as all the gauge-invariant operators decouple and the central charges of the would-be interacting sector become zero \cite{Agarwal:2020pol}. For all other cases, the $a$-maximization procedure always yields a unique solution, and the resulting central charges lie within the Hofman-Maldacena bound. We therefore conjecture that the theory flows to an interacting SCFT for $4\leq N_f <  2N+4$.

\paragraph{Relevant operators}
  For a fixed $N_f$, since the $R$-charge of a rank-2 symmetric chiral multiplet scales as $1/N$, there are $O(N)$ relevant operators of the form $\Tr (\O A)^{n}$ and $Q_I (\O A)^m \O Q_J$. The $O(N)$ number of them get decoupled along the RG flow.

  One can consider the deformation of the form $W=\Tr (\O A)^{k+1}$, which is relevant for sufficiently large $N$. In this case, the theory has a dual description given by $Sp\left(k(N_f-2)-N\right)$ gauge theory \cite{Intriligator:1995ff}, with a set of gauge-invariant operators flipped. 
  When $N_f=4$, a dual description for the case without superpotential is proposed in \cite{Csaki:1996eu}. 

\paragraph{Conformal manifold}  
  From the anomaly-free condition for $U(1)_R$ symmetry, we obtain
  \begin{align}\label{eq:spa1marginal}
  \begin{split}
      &N+1 + (N-1)(R_A - 1) + N_f(R_Q-1)=0\\
      &\quad \quad \implies \quad \frac{2(N-1)}{N_f-2}R_A + \frac{2N_f}{N_f-2}R_Q =2\,.
  \end{split}
  \end{align}
  For example, when $N_f=4$, this equation \eqref{eq:spa1marginal} suggests that an operator of the form
  \begin{align}
      (Q_{I_1}A^nQ_{J_1})(Q_{I_2}A^m Q_{J_2})\,,\quad n+m=N-1\,,
  \end{align}
  to be marginal. This gives $O(N)$ amount of marginal operators, leading to an $O(N)$-dimensional conformal manifold.
  
  For $N_f>4$, the expression $2N_f/(N_f-2)$ in \eqref{eq:spa1marginal} cannot be an even integer. Since the gauge-invariant operators always contain an even number of $Q$'s, marginal operators of the above form cannot exist. Empirically, we have not identified any marginal operators for generic $N$ and $N_f>4$.

\paragraph{Weak Gravity Conjecture}
  We examine the AdS WGC using the gauge-invariant operators and $U(1)$ flavor charges identified at the beginning of this section. We find that this theory satisfies both versions of the WGC. The result is shown in Figure \ref{fig:wgc_Spa1ven}.

\begin{figure}[t]
    \centering
     \begin{subfigure}[b]{0.45\textwidth}
     \centering
    \includegraphics[width=\linewidth]{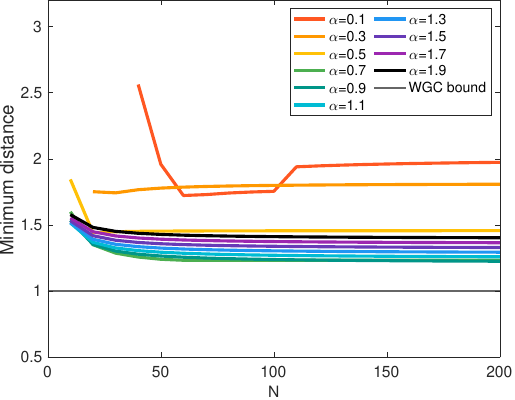}
    \caption{NN-WGC}
     \end{subfigure}
     \hspace{4mm}
     \begin{subfigure}[b]{0.45\textwidth}
     \centering
        \includegraphics[width=\linewidth]{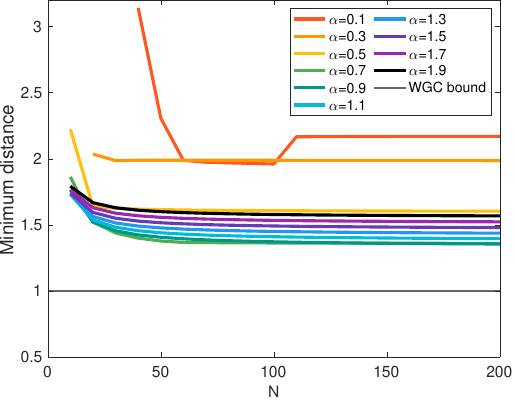}
        \caption{modified WGC}
     \end{subfigure}
     \hfill
        \caption{Testing AdS WGC for $Sp(N)$ theory with 1 $\antisym$ + 2 $N_f$ $\fund$. The minimum distance from the origin to the convex hull with a fixed $\a=N_f/N$.}
        \label{fig:wgc_Spa1ven}
 \end{figure}

\subsection{\texorpdfstring{2 $\sym$ + 2 $N_f$ $\fund$}{2 S + 2 Nf Q}}

\paragraph{Matter content and symmetry charges}
  The first entry of the Type II theories is $Sp(N)$ gauge theory with two rank-2 symmetric and $2N_f$ fundamental chiral multiplets. The matter fields and their $U(1)$ global charges are listed in Table \ref{tab:sps2}.
  {\renewcommand\arraystretch{1.6}
  \begin{table}[h]
    \centering
    \begin{tabular}{|c|c||c|c|c|}
    \hline
    \# & Fields & $Sp(N)$ &$U(1)_F$ & $U(1)_R$ \\\hline

     $2N_f$ & $Q$ &$\fund$& $2N+2$ & $R_Q$\\
     
     2 & $S$ &$\sym$ & $-N_f$  & $R_S$ \\\hline
    \end{tabular}
    \caption{The matter contents and their corresponding charges in $Sp(N)$ gauge theory with 2 $\sym$ + 2 $N_f$ $\fund$.\label{tab:sps2}}
  \end{table}}

\paragraph{Gauge-invariant operators}
  Let $I$ and $J$ denote the flavor indices for $Q$, and $K$ denote the flavor indices for $S$. We present a sample of single-trace gauge-invariant operators in schematic form as follows: 
  \begin{enumerate}
    \item $\Tr (\O S_{K_1})\cdots (\O S_{K_{2n}})\,,\quad n=1,2,\dots\,.$
    \item $Q_I (\O S_{K_1})\cdots (\O S_{K_n}) \O Q_J,\quad n=0,1,\dots\,.$

    $\vdots$
  \end{enumerate}
  The ellipsis indicates that only the low-lying operators have been listed. This subset is sufficient to identify relevant operators or to test the Weak Gravity Conjecture.

\paragraph{$R$-charges and central charges}
  We perform the $a$-maximization to compute the $R$-charges of matter fields and central charges. In the large $N$ limit with a fixed $N_f$, they are given by
  \begin{align}
  \begin{split}
    R_Q & \sim \frac{12-\sqrt{26}}{12} + \frac{78N_f + 5\sqrt{26}}{624N}+ O(N^{-2})\,, \\
    R_S & \sim  \half + \frac{N_f \sqrt{26} }{24N} + O(N^{-2})\,,\\
    a & \sim \frac{27}{128}\times 2 N^2 + \frac{81 + 13N_f\sqrt{26}}{384}N+O(N^0)\,,\\
    c & \sim \frac{27}{128}\times 2 N^2 + \frac{81 + 17N_f\sqrt{26}}{384}N+O(N^0)\,,\\
    a/c & \sim 1 - \frac{2  N_f \sqrt{26}}{81N} + O(N^{-2})\,.
  \end{split}
  \end{align}
  We find that no gauge-invariant operators decouple along the RG flow. The leading-order behavior of the $R$-charges, $a/\dim(G)$, and $c/\dim(G)$ is universal across Type II theories.
  
  In the Veneziano limit with a fixed $\a=N_f/N$, the $R$-charges of the matter fields and the central charges are given by
  \begin{align}
  \begin{split}
    R_Q  & \sim \frac{24-2\sqrt{26-\a^2}+3\a-3\a^2}{3(8-\a^2)}+O(N^{-1})\,,\\
    R_S  & \sim \frac{12-3\a^2+\a\sqrt{26-\a^2}}{3(8-\a^2)}+O(N^{-1})\,,\\
    a & \sim \frac{648-279\a^2 +18\a^4+2\a(26-\a^2)^{3/2}}{48(8-\a^2)^2}\times\half N^2+O(N)\,,\\
    c & \sim \frac{648-303\a^2 +21\a^4+4\a(17-\a^2)\sqrt{26-\a^2}}{48(8-\a^2)^2}\times\half N^2+O(N)\,,\\
    a/c & \sim 1-\frac{13\a(8-\a^2)}{162\sqrt{26-\a^2}+\a (685-71\a^2-30\a \sqrt{26-\a^2})}+O(N^{-1})\,.
  \end{split}
  \end{align}
  The leading-order terms of the $R$-charges, $a/\dim(G)$, and $c/\dim(G)$ are the same as those in $SU(N)$ and $SO(N)$ Type II theories.
  
  We find that the ratio of central charges $a/c$ lies within the range $0.9261\simeq (213779-812\sqrt{2753})/184826\leq a/c\leq 1$. The minimum value of $a/c$ arises when $(N_f,N)=(3,3)$. If $N_f=0$, $a/c$ is exactly one. Figure \ref{fig:Sps2ratio} illustrates the behavior of the ratio $a/c$.
  \begin{figure}[t]
    \centering
    \begin{subfigure}[b]{0.45\textwidth}
        \includegraphics[width=\linewidth]{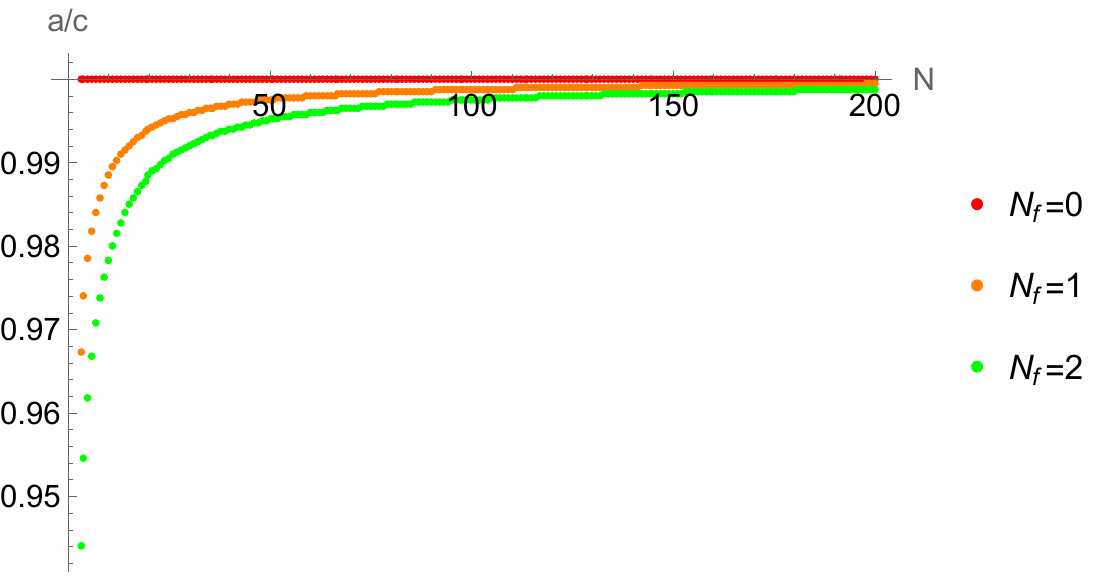}
    \end{subfigure}
    \hspace{4mm}
    \begin{subfigure}[b]{0.45\textwidth}
        \includegraphics[width=\linewidth]{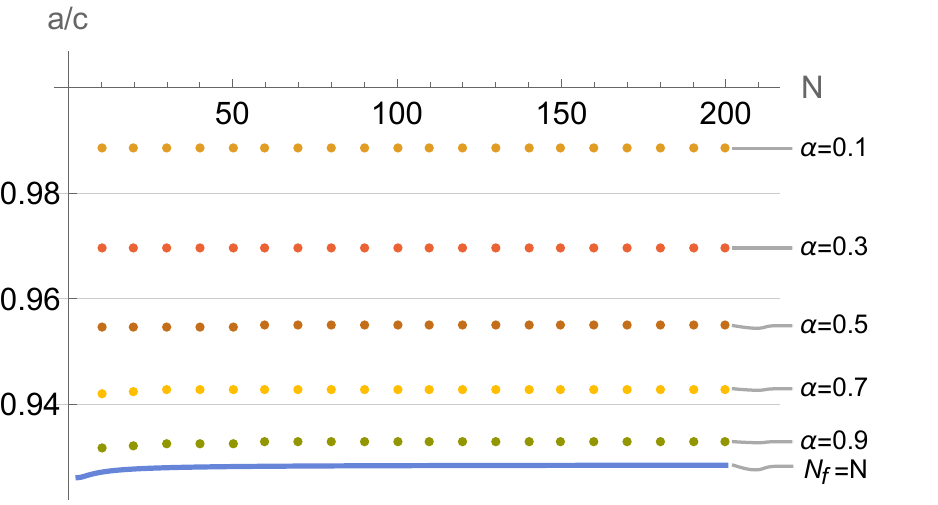}
    \end{subfigure}
    \hfill
    \caption{The central charge ratio for $Sp(N)$ theory with 2 $\sym$ + 2 $N_f$ $\fund$. Left: $a/c$ versus $N$ with a fixed $N_f$. Right: $a/c$ versus $N$ with a fixed $\a=N_f/N$.\label{fig:Sps2ratio}}
  \end{figure}

\paragraph{Conformal window}
  The upper bound of the conformal window comes from the asymptotic freedom. For small $N_f$, the $a$-maximization procedure always yields a unique solution, and the resulting central charges lie within the Hofman-Maldacena bound. 
  Especially when $N_f=0$, it is in the same conformal manifold as the mass-deformed $\CN=4$ SYM theory obtained by integrating out one of the adjoint chiral multiplets. 
  We therefore conjecture that this theory flows to an interacting SCFT for $0\leq N_f < N+1$.

\paragraph{Relevant operators}
  For generic $N$ and $N_f$, there exist the following relevant operators:
  \begin{itemize}
    \item three operators of the form $\Tr (\O S_{K_1})(\O S_{K_2})$ with a dimension $\frac{3}{2}<\D \leq  2\,,$
    \item $N_f(2N_f-1)$ operators of the form $Q_I \O Q_J$ with a dimension $1.7253\simeq \frac{12-\sqrt{26}}{4}< \D \leq 2\,,$
    \item $2N_f(2N_f+1)$ operators of the form $Q_I (\O S_K) \O Q_J$ with a dimension $2.4753 \simeq\frac{15-\sqrt{26}}{4} < \D \leq 3\,.$
  \end{itemize}
  The number of relevant operators does not depend on $N$ for a fixed $N_f$. The low-lying operator spectrum is sparse in the large $N$ limit.

\paragraph{Conformal manifold}
  When $N_f=0$, there exist twelve marginal operators of the form
  \begin{align}\label{eq:spadj2 marginal}
      \Tr(\O S_{K_1})(\O S_{K_2})\Tr(\O S_{K_3})(\O S_{K_4})\,,\quad \Tr(\O S_{K_1})(\O S_{K_2})(\O S_{K_3})(\O S_{K_4})
  \end{align}
  and the theory possesses a non-trivial conformal manifold.
  
  For $N_f>0$, there are no marginal operators for generic $N$ in the absence of a superpotential. However, upon a suitable superpotential deformation, we find that the theory flows to a superconformal fixed point with a non-trivial conformal manifold.

  For the $N_f=1$ case, we find that upon the following superpotential deformation, a non-trivial conformal manifold emerges at the IR fixed point:
  \begin{align}
      W= Q_1 (\O S_1) Q_1 + Q_1\O Q_2 \,.
  \end{align}
  We can test this by computing the superconformal index. For example, the reduced superconformal index for the $Sp(2)$ gauge theory with this superpotential is given by
  \begin{align}
      \CI_{\text{red}} = 3 t^{3}-2t^{9/2}\left(y+\frac{1}{y}\right)+9 t^{6}+ \cdots \,.
  \end{align}
  The positivity of the coefficient at the $t^6$ term indicates the existence of a non-trivial conformal manifold. The marginal operator takes the same form as (\ref{eq:spadj2 marginal}). This deformation is valid for any $N$ with a fixed $N_f=1$.

\paragraph{Weak Gravity Conjecture}
  We examine the AdS WGC using the gauge-invariant operators and $U(1)$ flavor charges identified at the beginning of this section. We find that this theory always satisfies both versions of the WGC. The result is shown in Figure \ref{fig:wgc_sps2ven}.
  \begin{figure}[t]
    \centering
    \begin{subfigure}[b]{0.45\textwidth}
      \centering
      \includegraphics[width=\linewidth]{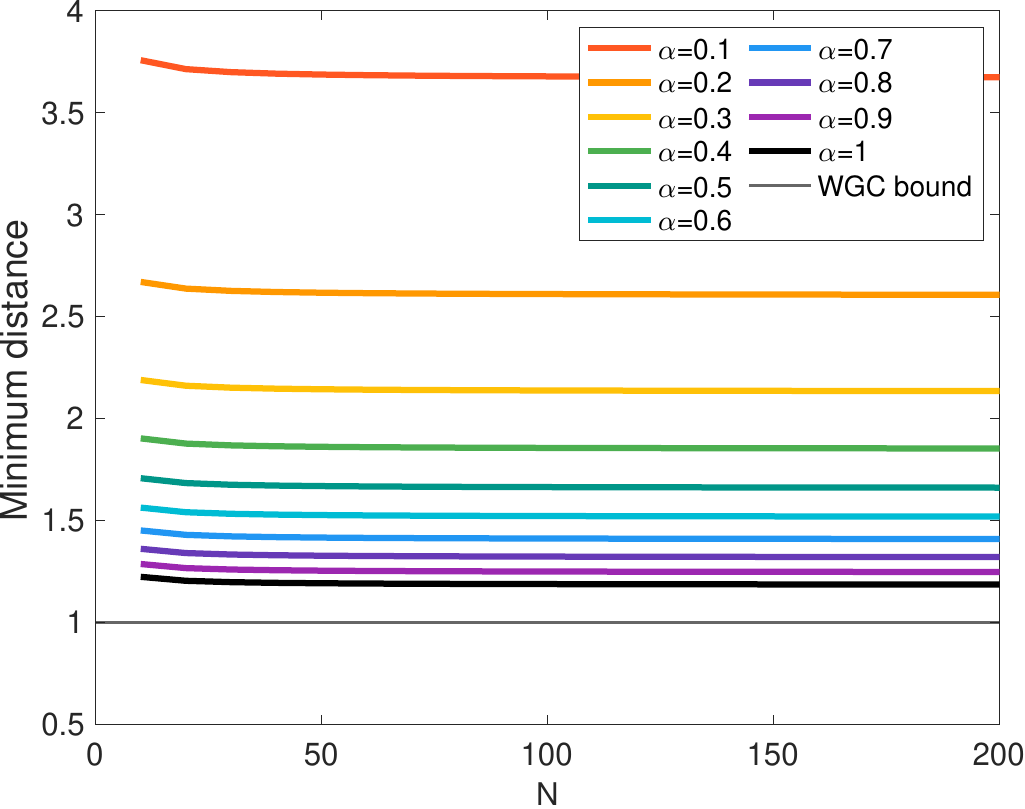}
      \caption{original WGC}
    \end{subfigure}
    \hspace{4mm}
    \begin{subfigure}[b]{0.45\textwidth}
      \centering
      \includegraphics[width=\linewidth]{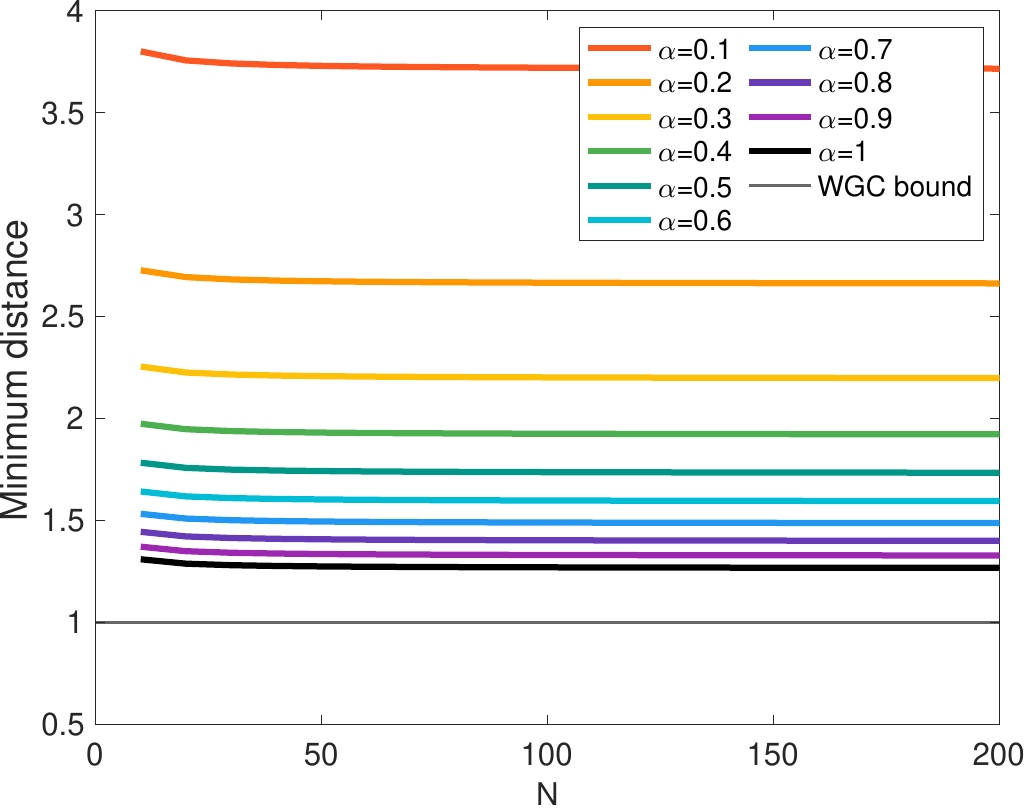}
      \caption{modified WGC}
    \end{subfigure}
    \hfill
    \caption{Testing AdS WGC for $Sp(N)$ theory with 2 $\sym$ + 2 $N_f$ $\fund$. The minimum distance from the origin to the convex hull with a fixed $\a=N_f/N$. \label{fig:wgc_sps2ven}}
   \end{figure}

\subsection{\texorpdfstring{1 $\sym$ + 1 $\antisym$ + 2 $N_f$ $\fund$}{1 S + 1 A + 2 Nf Q}}

\paragraph{Matter content and symmetry charges}
  Next case of the Type II theories is $Sp(N)$ gauge theory with a rank-2 symmetric, a rank-2 anti-symmetric, and $2N_f$ fundamental chiral multiplets. The matter fields and their $U(1)$ global charges are listed in Table \ref{tab:sps1a1}.
  {\renewcommand\arraystretch{1.6}
  \begin{table}[h]
    \centering
    \begin{tabular}{|c|c||c|c|c|c|}
    \hline
    \# & Fields & $Sp(N)$ &$U(1)_1$ & $U(1)_2$ &  $U(1)_R$ \\\hline

    $2 N_f$ & $Q$ & $\fund$ & $N+1$ &  $N-1$  & $R_Q$ \\

    1 & $S$ & $\sym$ & $-N_f$  &  0  &$R_S$\\
     
    1 & $A$ &$\antisym$& 0 &  $-N_f$  &$R_A$\\\hline
    \end{tabular}
    \caption{The matter contents and their corresponding charges in $Sp(N)$ gauge theory with 1 $\sym$ + 1 $\antisym$ + 2 $N_f$ $\fund$.\label{tab:sps1a1}}
  \end{table}}

\paragraph{Gauge-invariant operators}
  Let $I$ and $J$ denote the flavor indices for $Q$. We present a sample of single-trace gauge-invariant operators in schematic form as follows:
  \begin{enumerate}
    \item $\Tr (\O A)^{n}\,,\quad\Tr (\O S)^{2m}\,,\quad  n=2,3,\dots,N\,,\quad m=1,2,\dots, N\,.$
    \item $\Tr (\O A)^n (\O S)^{2m}\,,\quad n,m=1,2,\dots\,.$
    \item $Q_I (\O A)^n (\O S)^{2m}\O Q_J\,,\quad n,m = 0,1,\dots\,.$

    $\vdots$
  \end{enumerate}
  The ellipsis indicates that only the low-lying operators have been listed. This subset is sufficient to identify relevant operators or to test the Weak Gravity Conjecture.

\paragraph{$R$-charges and central charges}
  We perform $a$-maximization to compute the $R$-charges of the matter fields and central charges. We find that for small values of $N$, there are special cases in which a gauge-invariant operator decouples along the RG flow. We must flip this decoupled operator to obtain the correct answer from the $a$-maximization.
  
  When $(N_f,N)=(0,2),(0,3)$, we find that the operator of the form $\Tr (\O S)^2$ is decoupled. Upon flipping this operator, we obtain the correct result of the $a$-maximization given by
  \begin{align}
  \begin{split}
      R_S &= \frac{-18+\sqrt{10831}}{399}\simeq 0.2157\,,\quad R_A = \frac{151-\sqrt{10831}}{133}\simeq 0.3528\,,\quad\\
      a&=\frac{-130320+10831\sqrt{10831}}{849072}\simeq 1.1741\,,\quad c=\frac{11762 \sqrt{10831}-200145}{849072}\simeq 1.2060
  \end{split}
  \end{align}
  for $(N_f,N)=(0,2)$, and 
  \begin{align}
      \begin{split}
          R_S &= \frac{-17+\sqrt{2269}}{99}\simeq 0.3094\,,\quad R_A = \frac{1}{99} \left(133-2 \sqrt{2269}\right)\simeq 0.3811\,,\quad\\
      a&=\frac{2269 \sqrt{2269}-52136}{17424}\simeq 3.2108\,,\quad c=\frac{37 \left(16 \sqrt{2269}-371\right)}{4356}\simeq 3.3224
      \end{split}
  \end{align}
  for $(N_f,N)=(0,3)$.
  
  For all other cases, none of the gauge-invariant operators decouple along the RG flow. The $R$-charges of the matter fields and central charges, in the large $N$ limit with a fixed $N_f$, are given by
  \begin{align}
  \begin{split}
    R_Q & \sim \frac{12-\sqrt{26}}{12} + \frac{13 N_f -6\sqrt{26}}{104 N}+ O(N^{-2})\,, \\
    R_S & \sim  \half + \frac{ -41 + 3 N_f \sqrt{26} }{72 N} + O(N^{-2})\,,\\
    R_A & \sim  \half + \frac{ -31 + 3 N_f \sqrt{26} }{72 N} + O(N^{-2})\,,\\
    a & \sim \frac{27}{128}\times 2 N^2 + \frac{-18 + 13 N_f \sqrt{26}}{384}N + O(N^0)\,,\\
    c & \sim \frac{27}{128}\times 2 N^2 + \frac{6 + 17N_f\sqrt{26}}{384}N + O(N^0)\,,\\
    a/c & \sim 1 - \frac{12 + 2  N_f \sqrt{26}}{81N} + O(N^{-2})\,.
  \end{split}
  \end{align}
  The leading-order behavior of the $R$-charges, $a/\dim(G)$, and $c/\dim(G)$ is universal across Type II theories. Similarly, in the Veneziano limit with a fixed $\a=N_f/N$, the leading terms of $R$-charges and central charges are also universal across Type II theories.
  
  We find that the ratio of central charges $a/c$ is always less than one, taking a value within the range $0.8529\simeq \frac{29}{34}<a/c < 1$. The minimum value of $a/c$ arises when $(N_f,N)=(5,2)$. The maximum value of $a/c$ can be obtained in the large $N$ limit with $N_f=0$. Figure \ref{fig:Sps1a1ratio} illustrates the behavior of the ratio $a/c$.
  \begin{figure}[t]
    \centering
    \begin{subfigure}[b]{0.45\textwidth}
        \includegraphics[width=\linewidth]{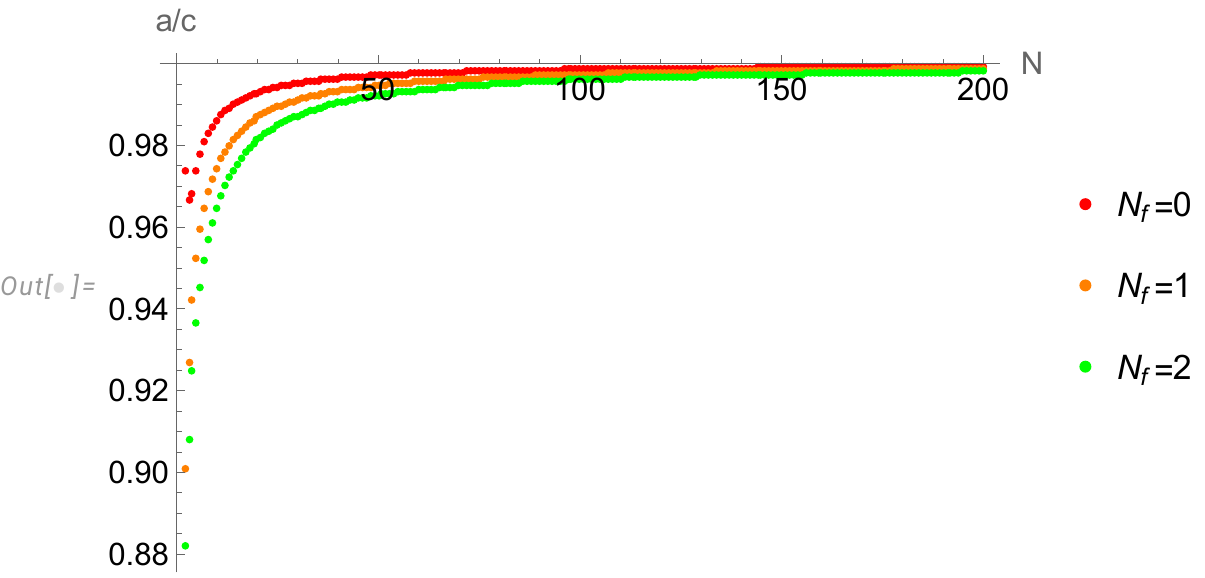}
    \end{subfigure}
    \hspace{4mm}
    \begin{subfigure}[b]{0.45\textwidth}
        \includegraphics[width=\linewidth]{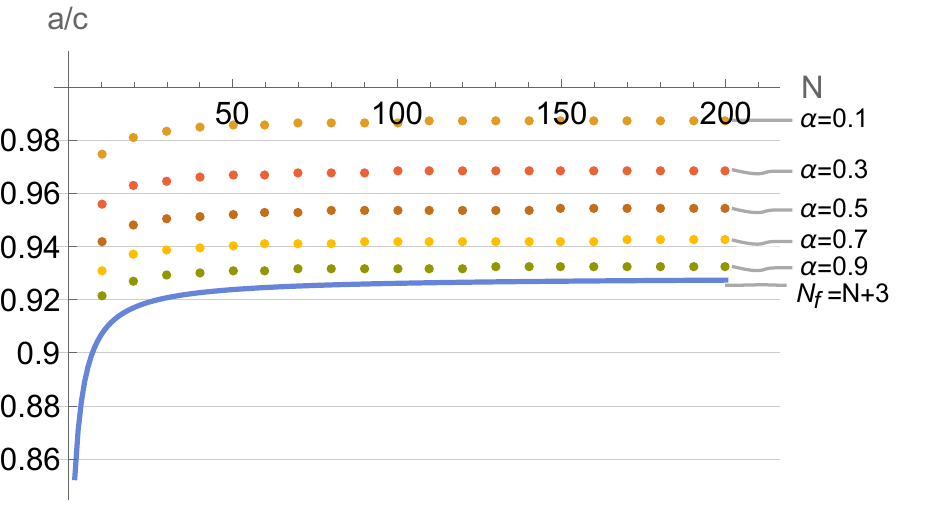}
    \end{subfigure}
    \hfill
    \caption{The central charge ratio for $Sp(N)$ theory with 1 $\sym$ + 1 $\antisym$ + 2 $N_f$ $\fund$. Left: $a/c$ versus $N$ with a fixed $N_f$. Right: $a/c$ versus $N$ with a fixed $\a=N_f/N$.\label{fig:Sps1a1ratio}}
  \end{figure}

\paragraph{Conformal window}
  The upper bound of the conformal window comes from the asymptotic freedom. When it is at the top of the conformal window $N_f=N+3$, there is a non-trivial conformal manifold \cite{Razamat:2020pra} so that it is an interacting SCFT. For $N_f<N+3$, the $a$-maximization procedure always yields a unique solution, and the resulting central charges lie within the Hofman-Maldacena bound. Thus, we conjecture that the theory flows to an interacting SCFT for $0\leq N_f \leq N+3$.

\paragraph{Relevant operators}
  For generic $N$ and $N_f$, except for the cases with small $N$ where some operators decouple, there exist the following relevant operators:
  \begin{itemize}
    \item an operator of the form $\Tr (\O S)^2$ with a dimension $1.0537\lesssim\D \leq  2\,,$
    \item $N_f(2N_f-1)$ operators of the form $Q_I \O Q_J$ with a dimension $1.5297 \lesssim \D \leq 2\,,$
    \item $N_f(2N_f+1)$ operators of the form $Q_I (\O S) \O Q_J$ with a dimension $ 2.0566 \lesssim \D \leq 3\,,$
    \item $N_f(2N_f-1)$ operators of the form $Q_I (\O A) \O Q_J$ with a dimension $ 2.1843 \lesssim \D \leq 3\,,$
    \item (if $N_f\leq 2$) an operator of the form $\Tr (\O S)^4$ and $(\Tr(\O S)^2)^2$ with a dimension $2.1074 \lesssim \D \leq 3\,,$
    \item an operator of the form $\Tr (\O A)^2$ with a dimension $1.2167\lesssim \D\leq 2$
    \item an operator of the form $\Tr (\O A)(\O S)^2$ with a dimension $1.6783 \lesssim \D \leq 3$
    \item an operator of the form $\Tr (\O A)^3$ with a dimension $1.8251\lesssim \D \leq 3\,,$
    \item (if $N_f\leq 2$) an operator of the form $\Tr (\O A)^4$ and $(\Tr(\O A)^2)^2$ with a dimension $2.4335 \lesssim \D \leq 3\,,$
  \end{itemize}
  The number of relevant operators does not depend on $N$ for a fixed $N_f$. The low-lying operator spectrum is sparse in the large $N$ limit.

  It was proposed in \cite{Brodie:1996xm} that upon deforming the theory by the superpotential $W=\Tr(\O A)^{k+1}+\Tr A S^2$, the theory admits a dual description given by $Sp(3kN_f-4k+2-N)$ gauge theory with a set of gauge-invariant operators flipped. The first term in the superpotential for $k>2$ is irrelevant near $W=0$, but at the fixed point of $W=\Tr AS^2$, it is relevant for sufficiently large $N$ with a fixed $N_f$. Therefore, this superpotential indeed describes a good fixed point. 

\paragraph{Conformal manifold}
  When $N_f=N+3$, the one-loop beta function for the gauge coupling vanishes. At this value, the theory possesses a non-trivial conformal manifold \cite{Razamat:2020pra}.
  
  When $N_f<N+3$, there are no marginal operators for generic $N$ in the absence of a superpotential. However, we find that upon a suitable superpotential deformation, the theory flows to a superconformal fixed point with a non-trivial conformal manifold.

  For the $N_f=0$ case, upon the following superpotential deformation, a non-trivial conformal manifold emerges at the IR fixed point:
  \begin{align}
      W= \Tr(\O A)^4 + X_1 \Tr (\O S)^2\,.
  \end{align}
  Here, $X_1$ is a flip field, which is a gauge-singlet chiral superfield.
  We can test this by computing the superconformal index. For example, the reduced superconformal index for $Sp(4)$ gauge theory with this superpotential is given by
  \begin{align}
      \CI_{\text{red}} = t^3 + t^{3.3} + \cdots + t^6 + \cdots \,.
  \end{align}
  The positivity of the coefficient at the $t^6$ term indicates the existence of a non-trivial conformal manifold. 
  The marginal operator takes the form of $(\Tr(\O A)^2)^2$. Such a deformation exists for general $N$ with a fixed $N_f=0$.

  For the $N_f=1$ case, consider the following superpotential deformation:
  \begin{align}
      W= Q_1 (\O A) \O Q_2 + \Tr(\O S)^2(\O A)^2 + X_1\Tr (\O S)^2\,.
  \end{align}
  The marginal operators include, for example, $Q_1(\O S)^4 \O Q_2$ and $\Tr(\O A)(\O S)^6$. There is only one $U(1)$ flavor symmetry, under which $Q_1$ carries charge $+1$ and $Q_2$ carries charge $-1$. Thus, these marginal operators are exactly marginal. This deformation works for general $N$ with a fixed $N_f=1$.

\paragraph{Weak Gravity Conjecture}
 
  \begin{figure}[t]
    \centering
    \begin{subfigure}[b]{0.45\textwidth}
      \centering
      \includegraphics[width=\linewidth]{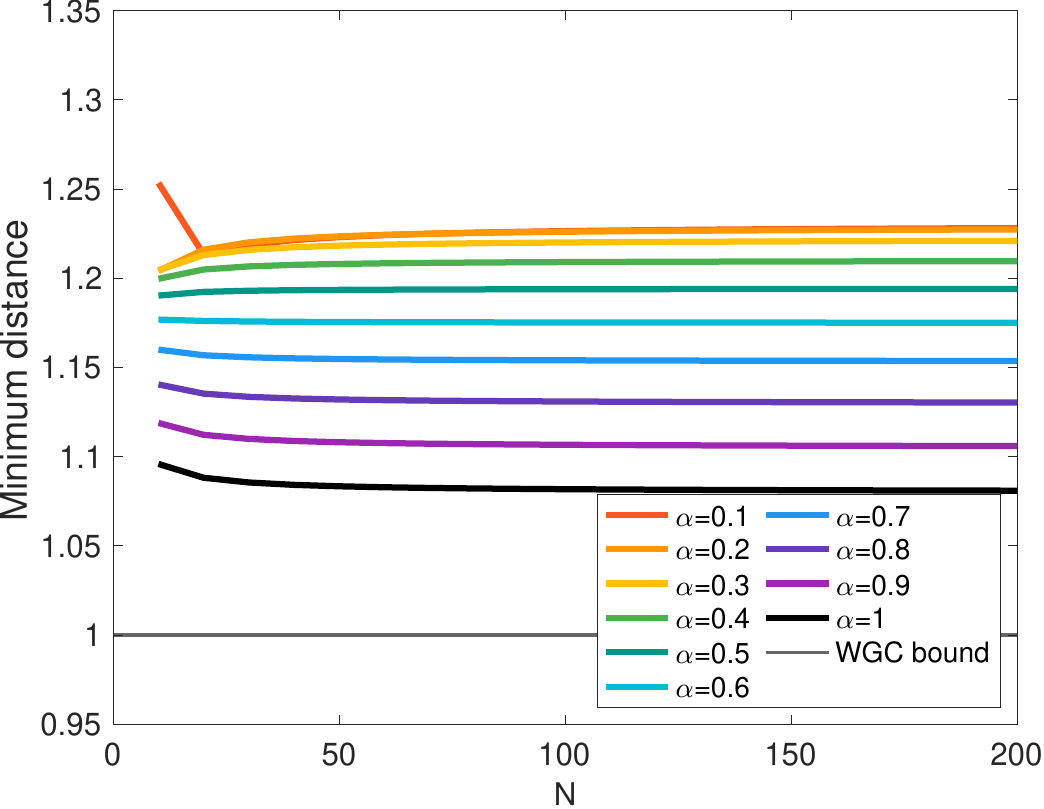}
      \caption{original WGC}
    \end{subfigure}
    \hspace{4mm}
    \begin{subfigure}[b]{0.45\textwidth}
      \centering
      \includegraphics[width=\linewidth]{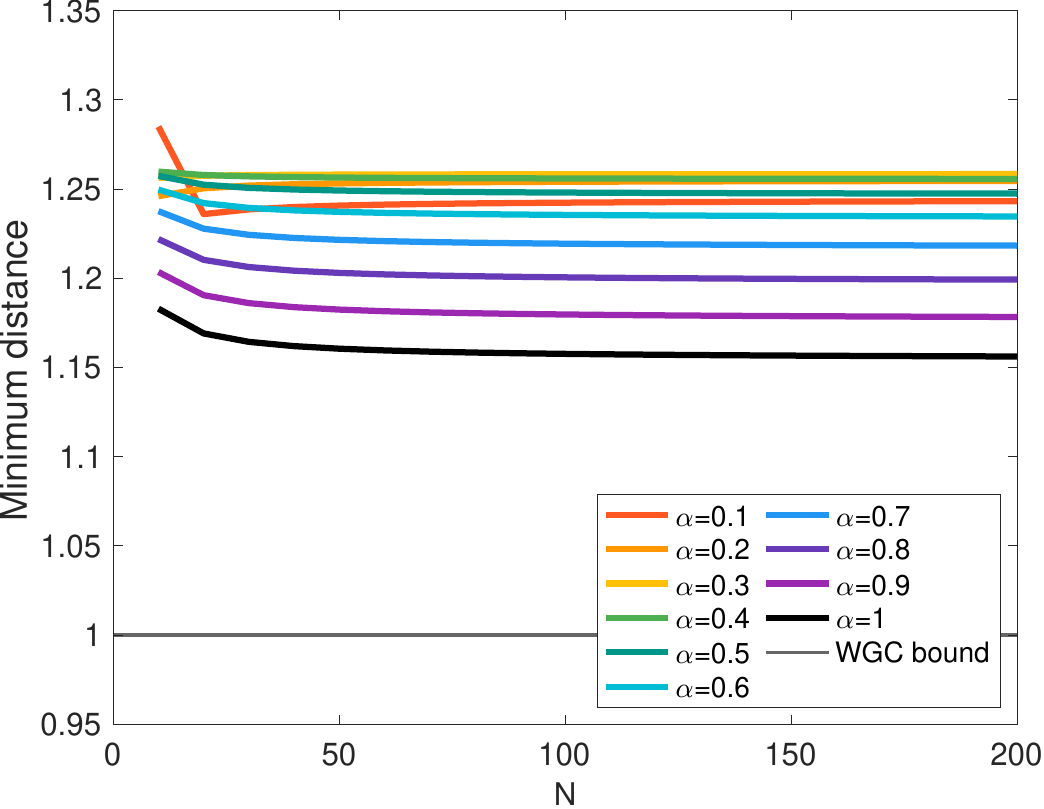}
      \caption{modified WGC}
    \end{subfigure}
    \hfill
    \caption{Testing AdS WGC for $Sp(N)$ theory with 1 $\sym$ + 1 $\antisym$ + 2 $N_f$ $\fund$. The minimum distance from the origin to the convex hull with a fixed $\a=N_f/N$.\label{fig:wgc_sps1a1ven}}
   \end{figure}

 We test the AdS WGC using the gauge-invariant operators and $U(1)$ flavor charges identified at the beginning of this section. We find that this theory always satisfies both versions of the WGC. The result is shown in Figure \ref{fig:wgc_sps1a1ven}.
 
\subsection{\texorpdfstring{2 $\antisym$ + 2 $N_f$ $\fund$}{2 A + 2 Nf Q}}\label{sec:Spa2}

\paragraph{Matter content and symmetry charges}
  The last entry of the Type II theories is $Sp(N)$ gauge theory with two rank-2 anti-symmetric tensors and $2N_f$ fundamental chiral multiplets. The matter fields and their $U(1)$ global charges are listed in Table \ref{tab:spa2}.
  {\renewcommand\arraystretch{1.6}
  \begin{table}[h]
    \centering
    \begin{tabular}{|c|c||c|c|c|}
    \hline
    \# & Fields & $Sp(N)$ &$U(1)_F$ &  $U(1)_R$ \\\hline

    2 $N_f$ & $Q$ & $\fund$ & $2N-2$  & $R_Q$\\
     
    2 & $A$ & $\antisym$& $-N_f$  &$R_A$\\\hline
    \end{tabular}
    \caption{The matter contents and their corresponding charges in $Sp(N)$ gauge theory with 2 $\antisym$ + 2 $N_f$ $\fund$.\label{tab:spa2}}
  \end{table}}

\paragraph{Gauge-invariant operators}
  Let $I$ and $J$ denote the flavor indices for $Q$, and $M$ denote the flavor indices for $A$. We present a sample of single-trace gauge-invariant operators in schematic form as follows: 
  \begin{enumerate}
    \item $\Tr (\O A_{M_1})\cdots(\O A_{M_n})\,,\quad  n=2,3,\dots\,.$
    \item $Q_I (\O A_{M_1})\cdots(\O A_{M_n})Q_J\,,\quad n= 0,1,\dots\,.$

    $\vdots$
  \end{enumerate}
  The ellipsis indicates that only the low-lying operators have been listed. This subset is sufficient to identify relevant operators or to test the Weak Gravity Conjecture.

\paragraph{$R$-charges and central charges}
  We perform $a$-maximization to compute the $R$-charges of the matter fields and central charges. We find that for small values of $N$, there are special cases that require separate discussion. We first spell these special cases out and then describe the generic case. 
  
  For the $Sp(2)$ theory, the rank-2 anti-symmetric tensor is nothing but the vector representation of $SO(5)$. When $N_f=0$, this theory is identical to the $SO(5)$ theory with two vectors, which does not flow to an interacting SCFT, rather, confines without a chiral symmetry breaking \cite{Intriligator:1995id}.
  
  When $(N_f,N)=(0,3),(1,2)$, the $a$-maximization procedure yields negative central charges, so these theories do not flow to unitary interacting SCFTs. In fact, it generates a runaway-type dynamical superpotential as in the case of $Sp(N)$ with 1 anti-symmetric tensor, as we discussed in Section \ref{sec:sp1antisym}. 
  
  When $(N_f,N)=(0,4)$, we find the gauge-invariant operators of the form
  \begin{align}
      \Tr (\O A_{M_1})(\O A_{M_2}),\ \Tr (\O A_{M_1})(\O A_{M_2})(\O A_{M_3}),\ \Tr (\O A_{M_1})(\O A_{M_2})(\O A_{M_3})(\O A_{M_4})
  \end{align}
  hit the unitarity bound along the RG flow and decouple from the rest of the system. Therefore, we must flip these decoupled operators. This can be done by introducing the following superpotential:
  \begin{align}
  \begin{split}
      W&=X_1\Tr(\O A_1)^2+X_2\Tr(\O A_2)^2+X_3\Tr(\O A_1)(\O A_2)+ X_4 \Tr(\O A_1)^3+ X_5 \Tr(\O A_2)^3  \\
      &~~+ X_6\Tr(\O A_1)^2(\O A_2) + X_7 \Tr(\O A_1)(\O A_2)^2 + X_8 \Tr(\O A_1)^4 + X_9 \Tr(\O A_2)^4  \\
      &~~+ X_{10} \Tr(\O A_1)^3(\O A_2) + X_{11} \Tr(\O A_1)(\O A_2)^3 + X_{12} \Tr(\O A_1)^2(\O A_2)^2 \\
      &~~+ X_{13} \Tr((\O A_1)(\O A_2))^2(\O A_1)(\O A_2) \,,
  \end{split}
  \end{align}
  where $X_i$'s are gauge-singlet chiral superfields. Upon flipping decoupled operators, the $R$-charges of the matter fields and central charges are given by
  \begin{align}
      R_A=\frac{1}{6}\simeq 0.1667\,,\quad a= \frac{265}{128} \simeq 2.0703 \,,\quad c=\frac{289}{128}\simeq 2.2578\,,\quad a/c=\frac{265}{289}\simeq0.9170\,,
  \end{align}
  which has no unitarity-violating gauge-invariant operators in the chiral ring. 

  When $(N_f,N)=(1,3)$, the gauge-invariant operators of the form $\Tr (\O A_{M_1})(\O A_{M_2})$ and $\Tr (\O A_{M_1})(\O A_{M_2})(\O A_{M_3})$ decouple. In the cases of $(N_f,N)=(0,5)$, $(0,6)$, $(0,7)$, $(1,4)$, $(1,5)$, $(2,2)$, and $(2,3)$, only operators of the form $\Tr (\O A_{M_1})(\O A_{M_2})$ decouple.

  For all other cases, none of the gauge-invariant operators decouple along the RG flow. The $R$-charges of the matter fields and central charges, in the large $N$ limit with a fixed $N_f$, are given by
  \begin{align}
  \begin{split}
    R_Q & \sim \frac{12-\sqrt{26}}{12} + \frac{78 N_f - 77\sqrt{26}}{624 N}+ O(N^{-2})\,, \\
    R_A & \sim  \half + \frac{ -24 +  N_f \sqrt{26} }{24 N} + O(N^{-2})\,,\\
    a & \sim \frac{27}{128}\times 2 N^2 + \frac{13(-9 +  N_f \sqrt{26})}{384}N + O(N^0)\,,\\
    c & \sim \frac{27}{128}\times 2 N^2 + \frac{-69 + 17N_f\sqrt{26}}{384}N + O(N^0)\,,\\
    a/c & \sim 1 - \frac{24 + 2  N_f \sqrt{26}}{81N} + O(N^{-2})\,.
  \end{split}
  \end{align}
  The leading-order behavior of the $R$-charges, $a/\dim(G)$, and $c/\dim(G)$ is universal across Type II theories. Similarly, in the Veneziano limit with a fixed $\a=N_f/N$, the leading terms of $R$-charges and central charges are also universal across Type II theories.
  
  We find that the ratio of central charges $a/c$ is always less than one, taking a value within the range $0.6788\simeq 2(148+13\sqrt{19})/603\leq a/c < 1$. The minimum value of $a/c$ arises when $(N_f,N)=(2,2)$. The maximum value of $a/c$ can be obtained in the large $N$ limit with $N_f=0$. Figure \ref{fig:Spa2ratio} illustrates the behavior of the ratio $a/c$.
  \begin{figure}[t]
    \centering
    \begin{subfigure}[b]{0.45\textwidth}
        \includegraphics[width=\linewidth]{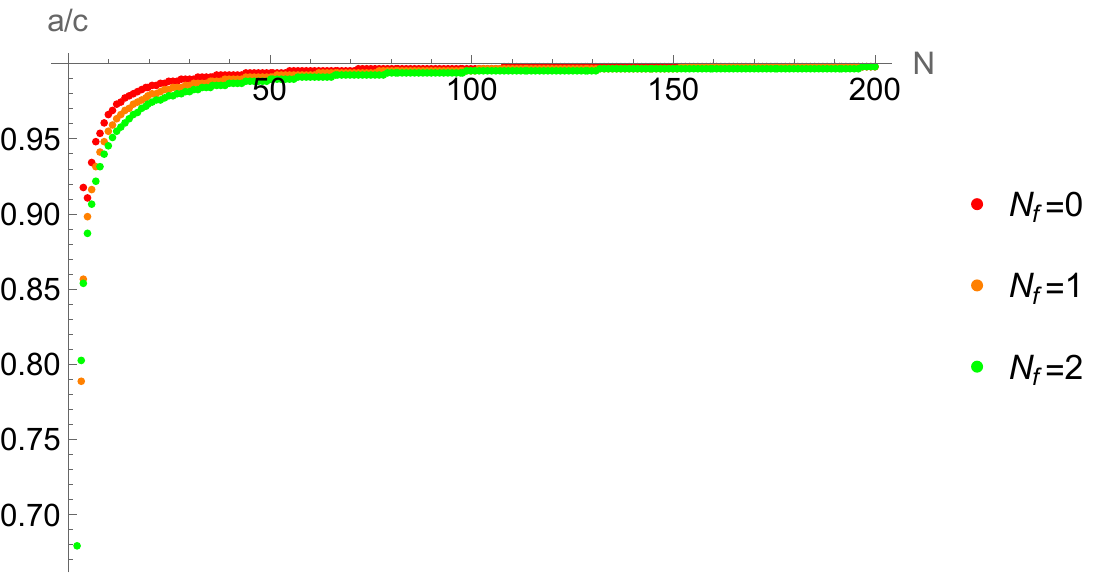}
    \end{subfigure}
    \hspace{4mm}
    \begin{subfigure}[b]{0.45\textwidth}
        \includegraphics[width=\linewidth]{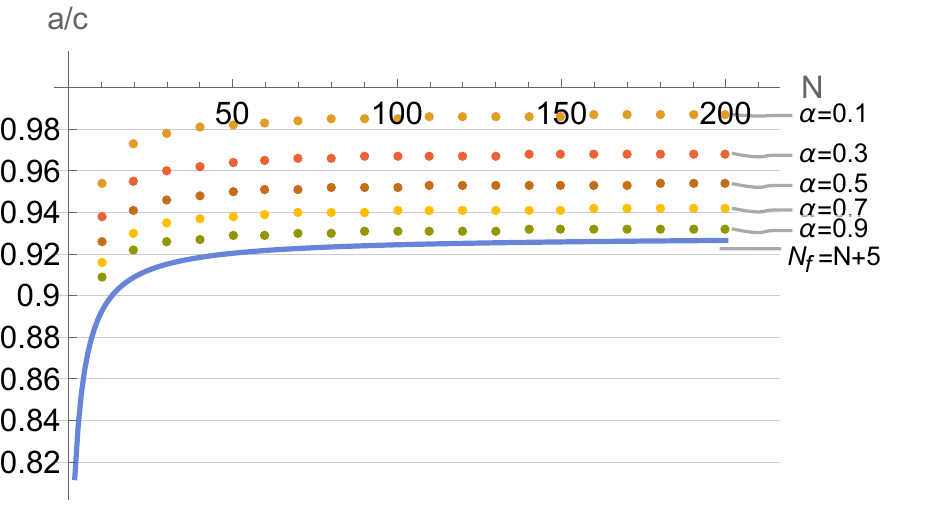}
    \end{subfigure}
    \hfill
    \caption{The central charge ratio for $Sp(N)$ with 2 $\antisym$ + 2 $N_f$ $\fund$. Left: $a/c$ versus $N$ with a fixed $N_f$. Right: $a/c$ versus $N$ with a fixed $\a=N_f/N$.\label{fig:Spa2ratio}}
  \end{figure}

\paragraph{Conformal window}
  The upper bound of the conformal window is determined by the asymptotic freedom. At the top of the conformal window $N_f=N+5$, the theory possesses a non-trivial conformal manifold \cite{Razamat:2020pra}, and is therefore an interacting SCFT. However, there are special cases for small $N_f$. As discussed earlier, the theory does not flow to an interacting SCFT for $(N_f,N)=(0,2)$, $(0,3)$, and $(1,2)$. In all other cases, the $a$-maximization procedure is well-behaved: it always yields a unique solution, and the resulting central charges lie within the Hofman-Maldacena bound. We therefore conjecture that this theory flows to an interacting SCFT for $0\leq N_f\leq N+5$, with the exception of $(N_f,N)=(0,2),(0,3),(1,2)$.

\paragraph{Relevant operators}
  For generic $N$ and $N_f$, except in cases where decoupled operators are present, there exist the following relevant operators:
  \begin{itemize}
    \item three operators of the form $\Tr (\O A_{M_1})(\O A_{M_2})$ with a dimension $1.0206 \lesssim \D \leq  2\,,$
    \item $N_f(2N_f-1)$ operators of the form $Q_I \O Q_J$ with a dimension $1.3196 \lesssim \D \leq 2\,,$
    \item $2N_f(2N_f-1)$ operators of the form $Q_I (\O A_{M}) \O Q_J$ with a dimension $1.8299 \lesssim \D \leq 3\,,$
    \item four operators of the form $\Tr (\O A_{M_1})(\O A_{M_2})(\O A_{M_3})$ with a dimension $1.5567 \lesssim \D \leq  2\,,$
    \item (if $N_f\leq 4$) six operators of the form $\Tr (\O A_{M_1})(\O A_{M_2})(\O A_{M_3})(\O A_{M_4})$ with a dimension $2.0756\lesssim\D\leq 3\,,$
  \end{itemize}
  The number of relevant operators does not depend on $N$ for a fixed $N_f$. The low-lying operator spectrum is sparse in the large $N$ limit.

  It was proposed in \cite{Brodie:1996xm} that upon deforming the theory by the superpotential $W=\Tr(\O A_1)^{k+1}+\Tr A_1 A_2^2$, the theory admits a dual description given by $Sp(3kN_f-4k-2-N)$ gauge theory with a set of gauge-invariant operators flipped. We find that the first term in the superpotential for $k>2$ is not relevant near $W=0$, but at the fixed point of $W=\Tr A_1A_2^2$, it is relevant for sufficiently large $N$ with a fixed $N_f$. Hence, the IR theory described by this superpotential is a good interacting SCFT.

\paragraph{Conformal manifold}
  When $N_f=N+5$, the one-loop beta function for the gauge coupling vanishes. At this value, the theory possesses a non-trivial conformal manifold \cite{Razamat:2020pra}.
  
  When $N_f<N+5$, there are no marginal operators for generic $N$ in the absence of a superpotential. However, it turns out that upon a suitable superpotential deformation, the theory flows to a superconformal fixed point with a non-trivial conformal manifold.

  For the $N_f=0$ case, upon the following superpotential deformation, we find a non-trivial conformal manifold to emerge at the IR fixed point:
  \begin{align}
      W= \Tr (\O A_1)(\O A_2)^3\,.
  \end{align}
  Under this superpotential deformation, no $U(1)$ flavor symmetry survives. Therefore, we get an exactly marginal operator given as $(\Tr A_1 A_2)(\Tr A_2^2)$. Such a deformation (and also the marginal operator) exists for general $N \geq 8$ with a fixed $N_f=0$.

  For the $N_f=1$ case, consider the following superpotential deformation:
  \begin{align}
      W= Q_1 (\O A_1)(\O A_2)\O Q_2\,.
  \end{align}
  In the deformed theory, the marginal operators take the form of
  \begin{align}
      Q_1(\O A_1)^2 Q_2\,,\quad Q_1(\O A_2)^2 Q_2\,,\quad \Tr (\O A_1 \O A_2)(Q_1\O Q_2)\,.
  \end{align}
  There are two $U(1)$ flavor symmetries: under the first $U(1)$, the fundamental and anti-fundamental chiral multiplets are charged to $+1$ and $-1$, respectively; under the second, the rank-2 anti-symmetric fields and their conjugate are charged to $+1$ and $-1$, respectively. There exist exactly marginal operators that are neutral under those flavor symmetries, so that we have a non-trivial conformal manifold. 
  Such a deformation exists for general $N\geq 7$ with a fixed $N_f=1$. 
  For the case of $N=6$, the marginal operators are of the form $\Tr(\O A_{M_1})\cdots (\O A_{M_6})$.

\paragraph{Weak Gravity Conjecture}
  
  \begin{figure}[t]
    \centering
    \begin{subfigure}[b]{0.45\textwidth}
      \centering
      \includegraphics[width=\linewidth]{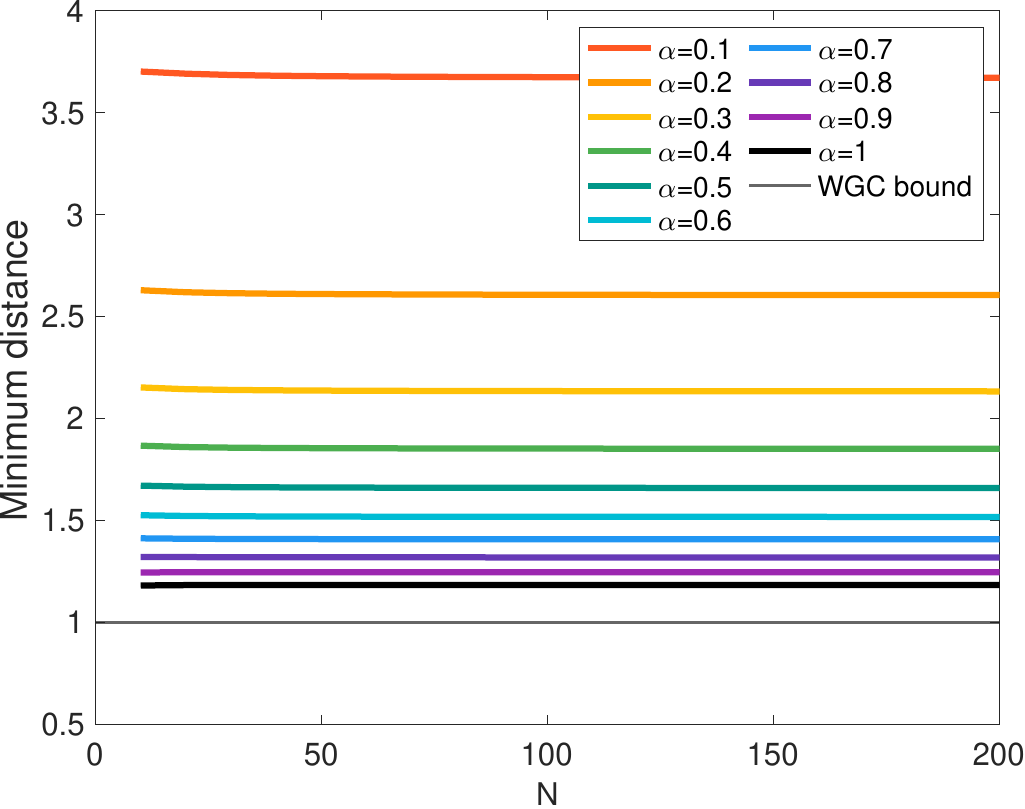}
      \caption{original WGC}
    \end{subfigure}
    \hspace{4mm}
    \begin{subfigure}[b]{0.45\textwidth}
      \centering
      \includegraphics[width=\linewidth]{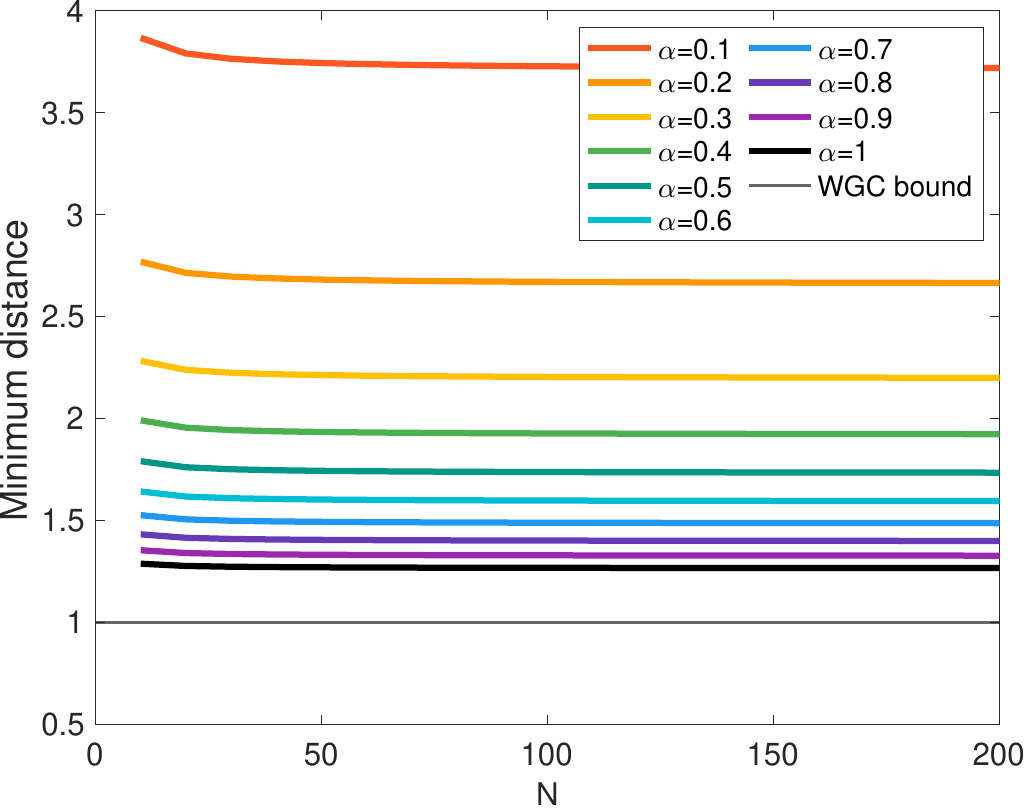}
      \caption{modified WGC}
    \end{subfigure}
    \hfill
    \caption{Testing AdS WGC for $Sp(N)$ theory with 2 $\antisym$ + 2 $N_f$ $\fund$. We plot the minimum distance from the origin to the convex hull with a fixed $\a=N_f/N$. \label{fig:wgc_spa2ven}}
   \end{figure}
   
We examine the AdS WGC using the gauge-invariant operators and $U(1)$ flavor charges identified at the beginning of this section. We find that this theory always satisfies both versions of the WGC. The result is shown in Figure \ref{fig:wgc_spa2ven}.

\subsection{\texorpdfstring{2 $\sym$ + 1 $\antisym$ + 2 $N_f$ $\fund$}{2 S + 1 A + 2 Nf Q}}

\paragraph{Matter content and symmetry charges}
  The first entry of the Type III theories is $Sp(N)$ gauge theories with two rank-2 symmetric tensors, a rank-2 anti-symmetric tensor, and $2N_f$ fundamental chiral multiplets. The matter fields and their $U(1)$ global charges are listed in Table \ref{tab:sps2a1}.
  {\renewcommand\arraystretch{1.6}
  \begin{table}[h]
    \centering
    \begin{tabular}{|c|c||c|c|c|c|}
    \hline
    \# & Fields & $Sp(N)$ &$U(1)_1$& $U(1)_2$ & $U(1)_R$ \\\hline

    $2 N_f$ & $Q$ & $\fund$ & $2N+2$ &  $N-1$  & $R_Q$ \\

    2 & $S$ & $\sym$ & $-N_f$  &  0  &$R_S$\\
     
    1 & $A$ &$\antisym$& 0 &  $-N_f$  &$R_A$\\\hline
    \end{tabular}
    \caption{The matter contents and their corresponding charges in $Sp(N)$ gauge theory with 2 $\sym$ + 1 $\antisym$ + 2 $N_f$ $\fund$.\label{tab:sps2a1}}
  \end{table}}

\paragraph{Gauge-invariant operators}
  Let $I$ and $J$ denote the flavor indices for $Q$, and $K$ denote the flavor indices for $S$. We present a sample of single-trace gauge-invariant operators in schematic form as follows:
  \begin{enumerate}
    \item $\Tr (\O A)^n\,,\quad \Tr (\O S_{K_1})\cdots (\O S_{K_{2m}}) \,,\quad n=2,3,\dots,N\,,\quad m=1,2,\dots\,.$
    \item $\Tr (\O A)^n (\O S_{K_1})\cdots (\O S_{K_{2m}}) \,,\quad n,m=1,2,\dots\,.$
    \item $Q_I (\O A)^n(\O S_{K_1})\cdots (\O S_{K_m}) \O Q_J,\quad n,m=0,1,\dots\,.$

    $\vdots$
  \end{enumerate}
  The ellipsis indicates that only the low-lying operators have been listed. This subset is sufficient to identify relevant operators or to test the Weak Gravity Conjecture.

\paragraph{$R$-charges and central charges}
  We perform the $a$-maximization to compute the $R$-charges of the matter fields and central charges. In the large $N$ limit with a fixed $N_f$, they are given by
  \begin{align}
  \begin{split}
    R_Q & \sim \frac{2}{3} + \frac{N_f-2}{18N}+ O(N^{-2})\,, \\
    R_S & \sim \frac{2}{3} + \frac{N_f-2}{9N} + O(N^{-2})\,,\\
    R_A & \sim \frac{2}{3}+ \frac{N_f-2}{9N}\,,\\
    a & \sim \frac{1}{4}\times 2 N^2 + \frac{2N_f+5}{24} N + O(N^0)\,,\\
    c & \sim \frac{1}{4}\times 2 N^2 + \frac{N_f+2}{8} N + O(N^0)\,,\\
    a/c & \sim 1 - \frac{N_f+1}{12N} + O(N^{-2})\,.
  \end{split}
  \end{align}
  We find that no gauge-invariant operators decouple along the RG flow. 
  The leading-order behavior of the $R$-charges, $a/\dim(G)$, and $c/\dim(G)$ is universal across Type III theories.
  
  We find that the ratio of central charge $a/c$ is always less than one, taking a value within the range $0.9475\lesssim a/c < 1$. The minimum value of $a/c$ arises when $(N_f,N)=(1,2)$. The maximum value of $a/c$ can be obtained in the large $N$ limit with $N_f=0$. Figure \ref{fig:Sps2a1ratio} illustrates the behavior of the ratio $a/c$.
  \begin{figure}[t]
    \centering
    \begin{subfigure}[b]{0.45\textwidth}
        \includegraphics[width=\linewidth]{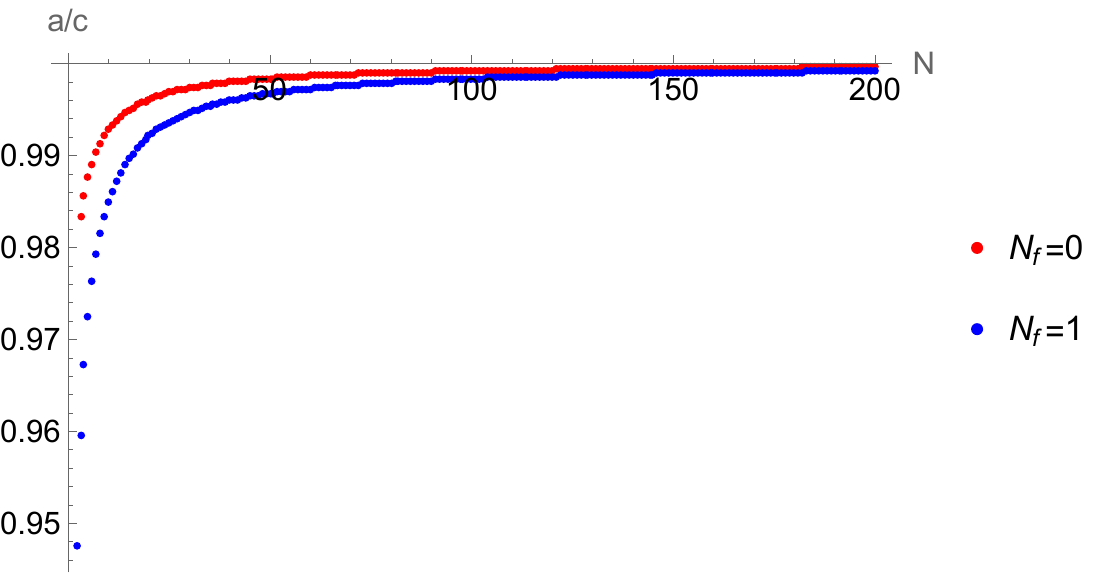}
    \end{subfigure}
    \hfill
    \caption{The central charge ratio for $Sp(N)$ theory with 2 $\sym$ + 1 $\antisym$ + 2 $N_f$ $\fund$. \label{fig:Sps2a1ratio}}
  \end{figure}

\paragraph{Conformal window}
  The upper bound of the conformal window comes from the asymptotic freedom. The $a$-maximization procedure always yields a unique solution, and the resulting central charges lie within the Hofman-Maldacena bound. Thus, we conjecture that the theory flows to an interacting SCFT for $0\leq N_f < 2$. We see that there cannot be a Veneziano-type limit for this theory.

\paragraph{Relevant operators}
  For generic $N$ and $N_f$, there exist the following relevant operators:
  \begin{itemize}
    \item three operators of the form $\Tr (\O S_{K_1})(\O S_{K_2})$ with a dimension $1.7902\lesssim\D <  2\,,$
    \item an operator of the form $\Tr (\O A)^2$ with a dimension $1.8391\lesssim\D <  2\,,$
    \item three operators of the form $\Tr (\O S_{M_1})(\O S_{M_2})(\O A)$ with a dimension $2.7098\lesssim\D <  3\,,$
    \item an operator of the form $\Tr (\O A)^3$ with a dimension $2.7587\lesssim\D <  3\,,$
    \item $N_f(2N_f-1)$ operators of the form $Q_I \O Q_J$ with a dimension $1.9392\lesssim \D < 2\,,$
    \item $2N_f(2N_f+1)$ operators of the form $Q_I (\O S_K) \O Q_J$ with a dimension $2.8689\lesssim \D < 3\,,$
    \item $N_f(2N_f-1)$ operators of the form $Q_I (\O A) \O Q_J$ with a dimension $2.8913 \lesssim \D < 3\,,$
  \end{itemize}
  The number of relevant operators does not depend on $N$ for a fixed $N_f$. The low-lying operator spectrum is sparse in the large $N$ limit.

\paragraph{Conformal manifold}
  There are no marginal operators for generic $N_f$ and $N$ in the absence of a superpotential. However, upon a suitable superpotential deformation, we find that the theory flows to a superconformal fixed point with a non-trivial conformal manifold.

  For the $N_f=0$ case, upon the following superpotential deformation, we find a non-trivial conformal manifold to emerge at the IR fixed point:
  \begin{align}
      W= \Tr (\O A)(\O S_1)^2\,.
  \end{align}
  Upon this deformation, a single $U(1)$ flavor symmetry remains, under which $S_1$ carries charge $+1$, $A$ carries charge $-2$, and $S_2$ is neutral. Therefore, there exist exactly marginal operators given as $\Tr (\O S_2)^4$ and $(\Tr \O S_2^2)^2$, both of which are neutral under this symmetry. Such a deformation (and also the marginal operators) exists for general $N$ with a fixed $N_f=0$.

  For the $N_f=1$ case, consider the following superpotential deformation:
  \begin{align}
      W= \Tr ((\O S_1)^2 \O A_1) + \Tr (Q_1 \O S_1 \O Q_1 ) + \Tr ((\O S_2)^2 \O A_1) \,.
  \end{align}
  We can verify this using the superconformal index. For instance, the reduced index for the $Sp(2)$ gauge theory with this superpotential has a positive coefficient at the $t^6$ term:
  \begin{align}
      \CI_{\text{red}} = 3 t^{3.43}+t^{4.29}-2 t^{4.71}\left(y+\frac{1}{y}\right)+t^{5.14}+t^{6} + \cdots \,.
  \end{align}
  The marginal operators take the form of $Q_1 \O S_2 \O Q_1$, $Q_1 \O S_2\O Q_2$, $Q_2 \O S_2 Q_2$. Such a deformation exists for general $N$ with a fixed $N_f=1$.

\paragraph{Weak Gravity Conjecture}
 
  \begin{figure}[t]
    \centering
    \begin{subfigure}[b]{0.45\textwidth}
      \centering
      \includegraphics[width=\linewidth]{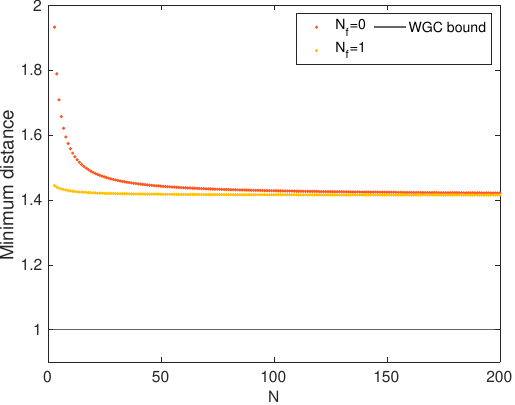}
      \caption{original WGC}
    \end{subfigure}
    \hspace{4mm}
    \begin{subfigure}[b]{0.45\textwidth}
      \centering
      \includegraphics[width=\linewidth]{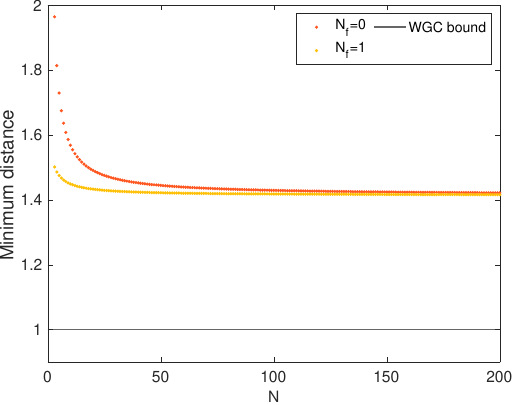}
      \caption{modified WGC}
    \end{subfigure}
    \hfill
    \caption{Testing AdS WGC for $Sp(N)$ theory with 2 $\sym$ + 1 $\antisym$ + 2 $N_f$ $\fund$. We plot the minimum distance from the origin to the convex hull for $N_f=0, 1$. \label{fig:wgc_sps2a1ven}}
   \end{figure}

 We examine the AdS WGC using the gauge-invariant operators and $U(1)$ flavor charges identified at the beginning of this section. We find that this theory always satisfies both versions of the WGC. The result is shown in Figure \ref{fig:wgc_sps2a1ven}.

\subsection{\texorpdfstring{1 $\sym$ + 2 $\antisym$ + 2 $N_f$ $\fund$}{1 S + 2 A + 2 Nf Q}}

\paragraph{Matter content and symmetry charges}
  Next case of the Type III theories is $Sp(N)$ gauge theory with a rank-2 symmetric tensor, two rank-2 anti-symmetric tensors, and $2N_f$ fundamental chiral multiplets. The matter fields and their $U(1)$ global charges are listed in Table \ref{tab:sps1a2}.
  {\renewcommand\arraystretch{1.6}
  \begin{table}[h]
    \centering
    \begin{tabular}{|c|c||c|c|c|c|}
    \hline
    \# & Fields & $Sp(N)$ &$U(1)_1$& $U(1)_2$ & $U(1)_R$ \\\hline

    $2 N_f$ & $Q$ & $\fund$ & $N+1$ &  $2N-2$  & $R_Q$ \\

    1 & $S$ & $\sym$ & $-N_f$  &  0  &$R_S$\\
     
    2 & $A$ &$\antisym$& 0 &  $-N_f$  &$R_A$\\\hline
    \end{tabular}
    \caption{The matter contents and their corresponding charges in $Sp(N)$ gauge theory with 1 $\sym$ + 2 $\antisym$ + 2 $N_f$ $\fund$.\label{tab:sps1a2}}
  \end{table}}

\paragraph{Gauge-invariant operators}
  Let $I$ and $J$ denote the flavor indices for $Q$, and $M$ denote the flavor indices for $A$. We present a sample of single-trace gauge-invariant operators in schematic form as follows:
  \begin{enumerate}
    \item $\Tr (\O A_{M_1})\cdots(\O A_{M_n})\,,\quad \Tr (\O S)^{2m} \,,\quad n=2,3,\dots,N\,,\quad m=1,2,\dots\,.$
    \item $\Tr (\O A_{M_1})\cdots(\O A_{M_n}) (\O S)^{2m}\,,\quad n,m=1,2,\dots\,.$
    \item $Q_I (\O A_{M_1})\cdots(\O A_{M_n})(\O S)^m \O Q_J,\quad n,m=0,1,\dots\,.$

    $\vdots$
  \end{enumerate}
  The ellipsis indicates that only the low-lying operators have been listed. This subset is sufficient to identify relevant operators or to test the Weak Gravity Conjecture.

\paragraph{$R$-charges and central charges}
  We perform the $a$-maximization to compute the $R$-charges of the matter fields and central charges. In the large $N$ limit with a fixed $N_f$, they are given by
  \begin{align}
  \begin{split}
    R_Q & \sim \frac{2}{3} + \frac{N_f-4}{18N}+ O(N^{-2})\,, \\
    R_S & \sim \frac{2}{3} + \frac{N_f-4}{9N} + O(N^{-2})\,,\\
    R_A & \sim \frac{2}{3}+ \frac{N_f-4}{9N}\,,\\
    a & \sim \frac{1}{4}\times 2 N^2 + \frac{N_f+2}{12} N + O(N^0)\,,\\
    c & \sim \frac{1}{4}\times 2 N^2 + \frac{N_f+2}{8} N + O(N^0)\,,\\
    a/c & \sim 1 - \frac{N_f+2}{12N} + O(N^{-2})\,.
  \end{split}
  \end{align}
  We find that none of the gauge-invariant operators decouple along the RG flow. 
  The leading-order behavior of the $R$-charges, $a/\dim(G)$, and $c/\dim(G)$ is universal across Type III theories.
  
  We observe that the ratio of central charges $a/c$ is always less than one, taking a value within the range $0.8659\simeq 71/82 \leq  a/c < 1$. The minimum value of $a/c$ arises when $(N_f,N)=(4,2)$. The maximum value of $a/c$ can be obtained in the large $N$ limit with $N_f=0$. Figure \ref{fig:Sps1a2ratio} illustrates the behavior of the ratio $a/c$.
  \begin{figure}[t]
    \centering
    \begin{subfigure}[b]{0.6\textwidth}
        \includegraphics[width=\linewidth]{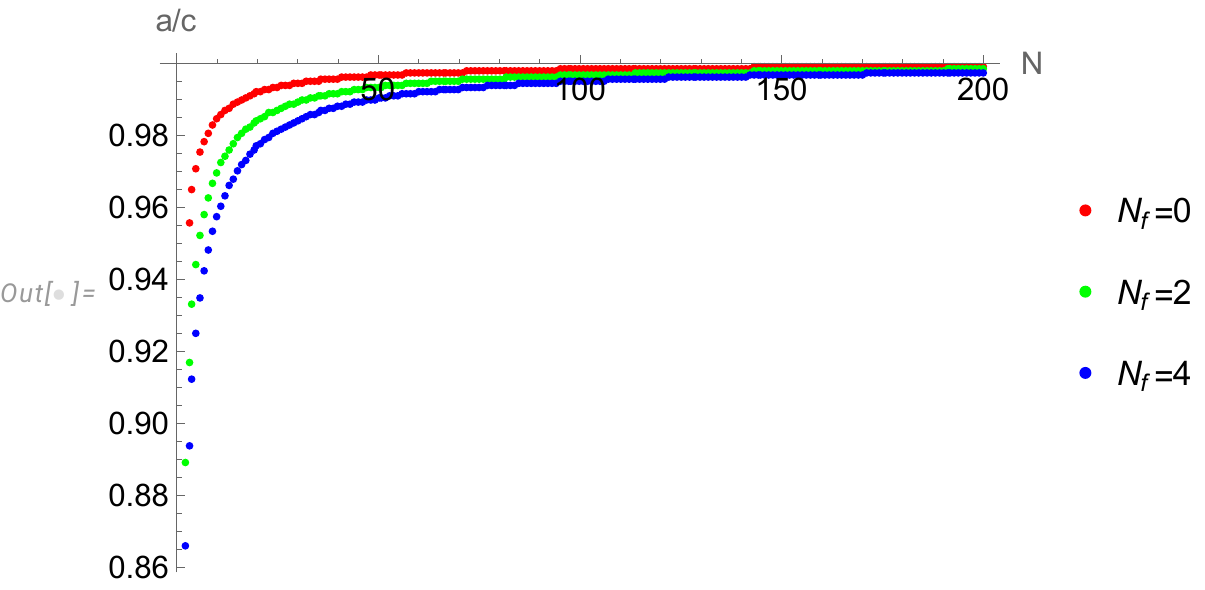}
    \end{subfigure}
    \hfill
    \caption{The central charge ratio for $Sp(N)$ theory with 1 $\sym$ + 2 $\antisym$ + 2 $N_f$ $\fund$. \label{fig:Sps1a2ratio}}
  \end{figure}

\paragraph{Conformal window}
  The upper bound of the conformal window is determined by the asymptotic freedom. At the top of the conformal window $N_f=4$, the theory possesses a non-trivial conformal manifold \cite{Razamat:2020pra}, and is therefore an interacting SCFT. For small $N_f$, the $a$-maximization procedure always yields a unique solution, and the resulting central charges lie within the Hofman-Maldacena bound. Thus, we conjecture that the theory flows to an interacting SCFT for $0\leq N_f \leq 4$.

\paragraph{Relevant operators}
  For generic $N$ and $N_f$, there exist the following relevant operators:
  \begin{itemize}
    \item three operators of the form $\Tr (\O A_{M_1})(\O A_{M_2})$ with a dimension $1.5597\lesssim\D \leq  2\,,$
    \item four operators of the form $\Tr (\O A_{M_1})(\O A_{M_2})(\O A_{M_3})$ with a dimension $2.3395\lesssim\D \leq  3\,,$
   \item an operator of the form $\Tr (\O S)^2$ with a dimension $1.3883\lesssim\D \leq  2\,,$
    \item $N_f(2N_f-1)$ operators of the form $Q_I \O Q_J$ with a dimension $1.7094\lesssim \D \leq 2\,,$
    \item $N_f(2N_f+1)$ operators of the form $Q_I (\O S) \O Q_J$ with a dimension $2.4035\lesssim \D \leq 3\,,$
    \item $2N_f(2N_f-1)$ operators of the form $Q_I (\O A_M) \O Q_J$ with a dimension $2.4908 \lesssim \D \leq 3\,,$
  \end{itemize}
  The number of relevant operators does not depend on $N$ for a fixed $N_f$. The low-lying operator spectrum is sparse in the large $N$ limit.

\paragraph{Conformal manifold}
  When $N_f=4$, the one-loop beta function for the gauge coupling vanishes. At this value, the theory possesses a non-trivial conformal manifold \cite{Razamat:2020pra}.
  
  When $N_f<4$, there is no marginal operator for generic $N$ in the absence of a superpotential. However, upon a suitable superpotential deformation, we find that the theory flows to a superconformal fixed point with a non-trivial conformal manifold.

  For the $N_f=0$ case, upon the following superpotential deformation, a non-trivial conformal manifold emerges at the IR fixed point:
  \begin{align}
      W= M_1\Tr (\O A_1 \O A_2) + M_1\Tr(\O S)^2 + M_2 \Tr(\O S)^2\,.
  \end{align}
  Here, $M_i$'s are flip fields, which are gauge-singlet chiral superfields. In the IR, the marginal operators take the form of $M_2 \Tr (\O A_1)^2$ and $M_2(\O A_2)^2$. Since these operators carry opposite charges under the flavor symmetry, the combination of them generates a non-trivial conformal manifold \cite{Green:2010da}. This deformation works for general $N$ with a fixed $N_f=0$.

  For the $N_f=1$ case, consider the following superpotential deformation:
  \begin{align}
      W= Q_1 \O A_1 Q_2 + \Tr (\O A_1)^2 + \Tr(\O A_1 \O A_2 \O S) + X_1 \Tr (\O S)^2\,.
  \end{align}
  Here, $X_1$ is a flip field, which is a gauge-singlet chiral superfield. 
  We can verify the existence of the conformal manifold from the superconformal index. For instance, the reduced superconformal index for $Sp(2)$ gauge theory with this superpotential is given by
  \begin{align}
      \CI_{\text{red}} = 2t^3 + 4t^{15/4} + \cdots + 3 t^6 +\cdots \,.
  \end{align}
  The positivity of the coefficient at the $t^6$ term indicates a non-trivial conformal manifold to exist. 
  The marginal operators take the form of $Q_I \O S \O A_2 \O Q_J$, $\Tr (\O A_2)^2 (\O S)^2$, $Q_1 (\O S)^4\O Q_1$, and $\Tr (\O S)^8$. Such a deformation exists for general $N$ with a fixed $N_f=1$.

\paragraph{Weak Gravity Conjecture}
  We examine the AdS WGC using the gauge-invariant operators and $U(1)$ flavor charges identified at the beginning of this section. We find that, when $(N_f,N)=(2,5),(3,6),(4,2),(4,6),(4,7),(4,8)$, this theory does not satisfy the NN-WGC, whereas the modified WGC always holds. The result is shown in Figure \ref{fig:wgc_sps1a2ven}.
  \begin{figure}[t]
    \centering
    \begin{subfigure}[b]{0.45\textwidth}
      \centering
      \includegraphics[width=\linewidth]{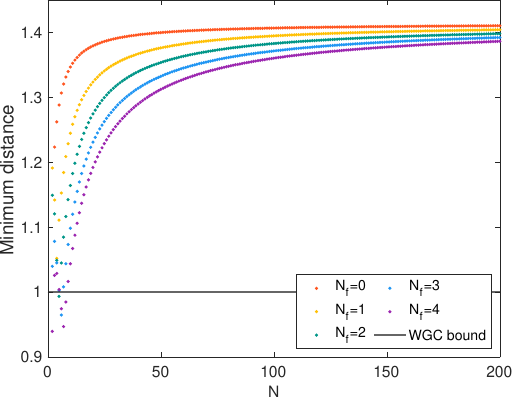}
      \caption{original WGC}
    \end{subfigure}
    \hspace{4mm}
    \begin{subfigure}[b]{0.45\textwidth}
      \centering
      \includegraphics[width=\linewidth]{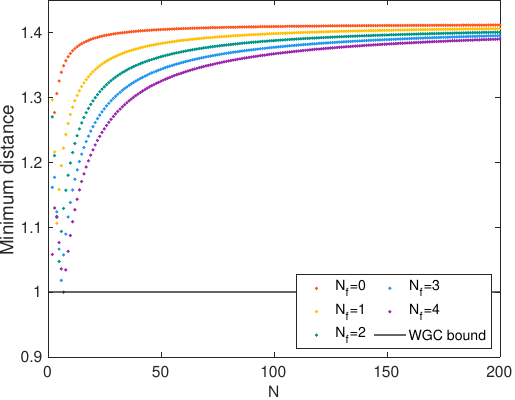}
      \caption{modified WGC}
    \end{subfigure}
    \hfill
    \caption{Testing AdS WGC for $Sp(N)$ theory with 1 $\sym$ + 2 $\antisym$ + 2 $N_f$ $\fund$. The minimum distance from the origin to the convex hull with a fixed $\a=N_f/N$. Theories below the solid line at the minimum distance 1 do not satisfy the WGC.\label{fig:wgc_sps1a2ven}}
   \end{figure}

\subsection{\texorpdfstring{3 $\antisym$ + 2 $N_f$ $\fund$}{3 A + 2 Nf Q}}
\paragraph{Matter content and symmetry charges}
  The next entry of the Type III theory is $Sp(N)$ gauge theory with three rank-2 anti-symmetric tensors and $2N_f$ fundamental chiral multiplets. The matter fields and their $U(1)$ global charges are listed in Table \ref{tab:spa3}.
  {\renewcommand\arraystretch{1.6}
  \begin{table}[h]
    \centering
    \begin{tabular}{|c|c||c|c|c|}
    \hline
    \# & Fields & $Sp(N)$ &$U(1)_F$& $U(1)_R$ \\\hline

    $2N_f$ & $Q$ & $\fund$ & $3N-3$  & $R_Q$\\
     
     3 & $A$ &$\antisym$& $-N_f$  &$R_A$\\\hline
    \end{tabular}
    \caption{The matter contents and their corresponding charges in $Sp(N)$ gauge theory with 3 $\antisym$ + 2 $N_f$ $\fund$.\label{tab:spa3}}
  \end{table}}

\paragraph{Gauge-invariant operators}
  Let $I$ and $J$ denote the flavor indices for $Q$, and $M$ denote the flavor indices for $A$. We present a sample of single-trace gauge-invariant operators in schematic form as follows:
  \begin{enumerate}
    \item $\Tr (\O A_{M_1})\cdots(\O A_{M_n})\,,\quad  n=2,3,\dots\,.$
    \item $Q_I (\O A_{M_1})\cdots(\O A_{M_n})Q_J\,,\quad n= 0,1,\dots\,.$

    $\vdots$
  \end{enumerate}
  The ellipsis indicates that only the low-lying operators have been listed. This subset is sufficient to identify relevant operators or to test the Weak Gravity Conjecture.

\paragraph{$R$-charges and central charges}
  We perform the $a$-maximization to compute the $R$-charges of the matter fields and central charges. However, for small values of $N$, we identify special cases that must be discussed separately. For the $Sp(2)$ theory, the rank-2 anti-symmetric tensor is the same as the vector representation of $SO(5)$. When $N_f=0$, this theory is identical to the $SO(5)$ theory with three vectors, which does not flow to an interacting SCFT. Instead, it is in a Coulomb phase with a massless photon and charged particles \cite{Intriligator:1995id}.
  
  When $(N_f,N)=(0,3),(1,2)$, the gauge-invariant operators of the form $\Tr (\O A_{M_1})(\O A_{M_2})$ hit the unitarity bound along the RG flow and decouple from the rest of the system. Therefore, we must flip these decoupled operators. This can be done by introducing the following superpotential:
  \begin{align}
  \begin{split}
      W &= X_1\Tr(\O A_1)^2+X_2\Tr(\O A_2)^2+X_3\Tr(\O A_3)^2 \\
      &\quad + X_4 \Tr(\O A_1)(\O A_2)+ X_5 \Tr(\O A_1)(\O A_3)+ X_6\Tr(\O A_2)(\O A_3)
  \end{split}
  \end{align}
  where $X_i$'s are flip fields. Upon flipping decoupled operators, the $R$-charges of the matter fields and central charges are given by
  \begin{align}
      R_A=\frac{1}{3}\simeq 0.3333\,,\quad a=\frac{47}{16}\simeq2.9375 \,,\quad c=\frac{13}{4}=3.25 \,,\quad a/c=\frac{47}{52}\simeq0.9038\,,
  \end{align}
  for $(N_f,N)=(0,3)$, and
  \begin{align}
      R_Q&=\frac{74-\sqrt{1409}}{83}\simeq 0.4393\,,\quad R_A=\frac{9+\sqrt{1409}}{249}\simeq 0.1869\,,\\ a&=\frac{39324+1409\sqrt{1409}}{110224}\simeq0.8366 \,,\quad c=\frac{18832+995\sqrt{1409}}{55112}\simeq1.0194\,, \\
      a/c&=\frac{179243+1828\sqrt{1409}}{302018}\simeq0.8207\,,
  \end{align}
  for $(N_f,N)=(1,2)$.

  For all other cases, none of the gauge-invariant operators decouple along the RG flow. The $R$-charges of the matter fields and central charges, in the large $N$ limit with a fixed $N_f$, are given by
  \begin{align}
  \begin{split}
    R_Q & \sim \frac{2}{3} + \frac{N_f-6}{18 N}+ O(N^{-2})\,, \\
    R_A & \sim  \frac{2}{3} + \frac{ N_f-6 }{9 N} + O(N^{-2})\,,\\
    a & \sim \frac{1}{4}\times 2 N^2 + \frac{2N_f+3}{24}N + O(N^0)\,,\\
    c & \sim \frac{1}{4}\times 2 N^2 + \frac{N_f+2}{8}N + O(N^0)\,,\\
    a/c & \sim 1 - \frac{N_f+3}{12N} + O(N^{-2})\,.
  \end{split}
  \end{align}
  The leading-order behavior of the $R$-charges, $a/\dim(G)$, and $c/\dim(G)$ is universal across Type III theories. 
  
  We find that the ratio of central charges $a/c$ is always less than one, taking a value within the range $0.7742\simeq 3(243842-985\sqrt{849})/833662\leq a/c < 1$. The minimum value of $a/c$ arises when $(N_f,N)=(2,2)$. The maximum value of $a/c$ can be obtained in the large $N$ limit with $N_f=0$. Figure \ref{fig:Spa3ratio} illustrates the behavior of the ratio $a/c$.
  \begin{figure}[t]
    \centering
    \begin{subfigure}[b]{0.45\textwidth}
        \includegraphics[width=\linewidth]{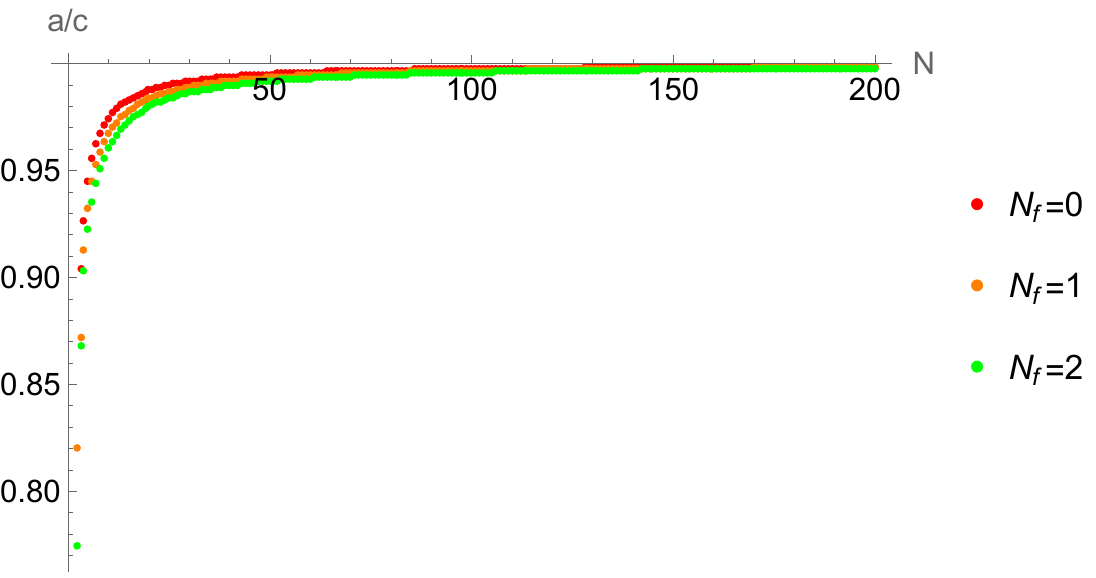}
    \end{subfigure}
    \hfill
    \caption{The central charge ratio for $Sp(N)$ theory with 3 $\antisym$ + 2 $N_f$ $\fund$. \label{fig:Spa3ratio}}
  \end{figure}

\paragraph{Conformal window}
  The upper bound of the conformal window is determined by the asymptotic freedom. At the top of the conformal window $N_f=6$, the theory possesses a non-trivial conformal manifold \cite{Razamat:2020pra}, and is therefore an interacting SCFT.
  
  As discussed before, the theory does not flow to an interacting SCFT for $(N_f,N)=(0,2)$. In all other cases, the $a$-maximization procedure is well-behaved: it always yields a unique solution, and the resulting central charges lie within the Hofman-Maldacena bound. We therefore conjecture that this theory flows to an interacting SCFT for $0\leq N_f\leq 6$, with the exception of $(N_f, N)=(0,2)$.

\paragraph{Relevant operators}
  For generic $N$ and $N_f$, except in cases where decoupled operators are present, there exist the following relevant operators:
  \begin{itemize}
    \item six operators of the form $\Tr (\O A_{M_1})(\O A_{M_2})$ with a dimension $1.0875 \lesssim \D \leq  2\,,$
    \item $N_f(2N_f-1)$ operators of the form $Q_I \O Q_J$ with a dimension $1.3687 \lesssim \D \leq 2\,,$
    \item $3N_f(2N_f-1)$ operators of the form $Q_I (\O A_{M}) \O Q_J$ with a dimension $1.9125 \lesssim \D \leq 3\,,$
    \item ten operators of the form $\Tr (\O A_{M_1})(\O A_{M_2})(\O A_{M_3})$ with a dimension $1.8715 \lesssim \D \leq  3\,,$
  \end{itemize}
  The number of relevant operators does not depend on $N$ for a fixed $N_f$. The low-lying operator spectrum is sparse in the large $N$ limit.

\paragraph{Conformal manifold}
  When $N_f=6$, the one-loop beta function for the gauge coupling vanishes. At this value, the theory possesses a non-trivial conformal manifold \cite{Razamat:2020pra}.
  
  When $N_f<6$, there are no marginal operators for generic $N$ in the absence of a superpotential. However, upon a suitable superpotential deformation, the theory flows to a superconformal fixed point with a non-trivial conformal manifold. 
  
  For instance, by turning on the superpotential $W=M_1\Tr(\O A_1)^2+M_2\Tr(\O A_2)^2+M_3\Tr(\O A_3)^2$, where $M_i$'s are flip fields, the theory admits a non-trivial conformal manifold. We can verify the existence of the conformal manifold from the superconformal index. For instance, the reduced superconformal index for the $(N_f,N)=(0,4)$ case with this superpotential is given by
  \begin{align}
    \CI_{\text{red}}=3 t^{8/3}+3t^{10/3}+10t^4+27 t^{16/3}+ t^6+\cdots\,.
  \end{align}
  The positive of the coefficient at the $t^6$ term indicates a non-trivial conformal manifold to exist. The marginal operators take the form of $M_1\Tr(\O A_2)(\O A_3)$, $M_2\Tr((\O A_3)(\O A_1))$, and $M_3(\Tr(\O A_1)(\O A_2))$. This deformation works for general $N$ with a fixed $N_f$.

\paragraph{Weak Gravity Conjecture}
  We examine the AdS WGC using the gauge-invariant operators and $U(1)$ flavor charges identified at the beginning of this section. We find that, when $(N_f,N)=(5,3),(6,3)$, this theory does not satisfy the NN-WGC, whereas the modified WGC always holds. The result is shown in Figure \ref{fig:wgc_spa3ven}.
  \begin{figure}[t]
    \centering
    \begin{subfigure}[b]{0.45\textwidth}
      \centering
      \includegraphics[width=\linewidth]{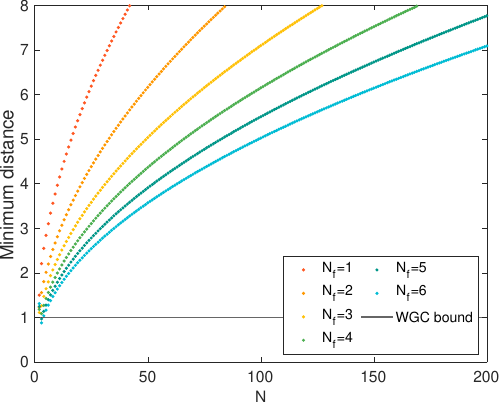}
      \caption{original WGC}
    \end{subfigure}
    \hspace{4mm}
    \begin{subfigure}[b]{0.45\textwidth}
      \centering
      \includegraphics[width=\linewidth]{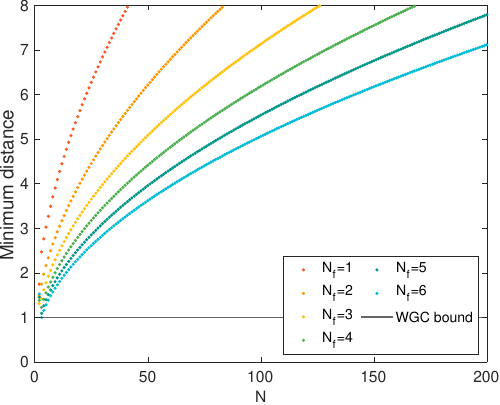}
      \caption{modified WGC}
    \end{subfigure}
    \hfill
    \caption{Testing AdS WGC for $Sp(N)$ theory with 3 $\antisym$ + 2 $N_f$ $\fund$. We plot the minimum distance from the origin to the convex hull. Theories below the solid line at the minimum distance 1 do not satisfy the WGC.\label{fig:wgc_spa3ven}}
   \end{figure}

\subsection{\texorpdfstring{3 $\sym$}{3 S}}
  The last case of the Type III theories is $Sp(N)$ gauge theory with three rank-2 symmetric chiral multiplets. It has the same matter content as the $\CN=4$ SYM theory. This theory is a conformal gauge theory with no running coupling. The $R$-charges and the central charges are given by
  \begin{align}
      R_A=\frac{2}{3}\,,\quad a=c=\frac{1}{4}N(2N+1)\,.
  \end{align}
  The leading-order behavior of the $R$-charges and central charges is universal across the Type III theories.

  The single-trace gauge-invariant operators are of the form
  \begin{align}
      \Tr (\O S_{M_1})\cdots (\O S_{M_n})\,,\quad  n=1,2,\dots\,.
  \end{align}
  There exist six relevant operators of the form $\Tr (\O S_{M_1})(\O S_{M_2})$ and a single marginal operators of the form $\Tr(\O S_{1})(\O S_{2})(\O S_{3})$. The theory possesses a one-dimensional subspace in the conformal manifold that preserves $\CN=4$ supersymmetry.

\section{Discussion} \label{sec:discussion}

In this paper, we classified all supersymmetric simple gauge theories that 1) flow to interacting SCFTs in the IR, and 2) admit a large $N$ limit while flavor symmetry is fixed, 3) without a superpotential. 
We find that such theories fall into one of three classes, which we call Type I, II, and III, by the number of rank-2 tensor matters. 
We find surprising universality across these theories. 
Each of Type II and III shares universal $R$-charges of matters, and it turns out that all universal $R$-charges and consequent spectral behaviors are solely determined by $N_{\text{rank-}2}$ regardless of the gauge group and the number of matters. 
For the Type I theories, upon taking the double scaling limit $N\gg N_f\gg 1$, we find universal $R$-charges. 
We also find that once we take $N, N_f$ to be large with $\alpha = N_f/N$ kept fixed, the Veneziano limit, in the Type I and II theories, they again share universal $R$-charges determined by two parameters $N_{\text{rank-}2}$ and $\alpha$. 
We have studied various aspects of each theory in detail: central charges, the spectrum of relevant gauge-invariant operators, the conformal window, the conformal manifold, and the Weak Gravity Conjecture. 
In particular, we find that every single theory (even for a finite $N$) satisfies the modified WGC, even though some of them do not satisfy the original version of \cite{Nakayama:2015hga}. It suggests that WGC is a generic property of 4d (S)CFT. 

The Type I theories are rather unusual in many ways. The UV description is still given by a conventional gauge theory, whose degrees of freedom are naturally described by matrices. Therefore, the linear scaling of central charges in $N$ is very surprising. There is a very strong quantum effect, yielding large anomalous dimensions. This phenomenon is also in conjunction with the dense spectrum, meaning that the gap on the operator dimensions scale as $1/N$. This implies that Type I theories cannot be holographic. Nevertheless, we find they all satisfy the (modified) WGC. It is interesting to look for a bulk spacetime interpretation of the Type I theory, which has a dense spectrum of low-lying operators.  

In many respects, Type II and III theories are good candidates to be holographic. 
They have $a=c$ in the large $N$ limit, a sparse spectrum at low energy, and exhibit a Hawking-Page-like phase transition of the superconformal index. 
Moreover, they all can be deformed via appropriate superpotentials (and a number of flip fields), whose fixed points are equipped with non-trivial conformal manifolds. This implies that the fixed point theory is not an isolated SCFT without any tunable coupling; rather, it has the potential to be deformed to one described by a weakly-coupled gravity in the bulk. It would be interesting to find a way to further test if these theories are holographic, or directly construct supergravity backgrounds giving the dual theory. 

Let us also mention that almost none of our theories have exactly marginal couplings connected to the free theory. The only such case is when we are at the upper end of the conformal window, at which the theory does not flow. Such examples are already classified in \cite{Razamat:2020pra}. If we invoke the CFT version of the Swampland Distance Conjecture (SDC) \cite{Ooguri:2006in, Perlmutter:2020buo}, this implies that all of our theories (except the ones that do not flow) should have a compact conformal manifold. See \cite{Calderon-Infante:2024oed} for the study of 4d conformal gauge theories whose conformal manifold is non-compact. Unlike these previous studies, our SCFTs are strongly-coupled with no obvious cusps at which we get free gauge fields. Another possibility is that the conformal manifold remains non-compact. At the same time, at an infinite distance limit, there exists a dual description involving the coupling of strongly-coupled components that disintegrate with a free vector or chiral multiplet. We cannot rule this out in an obvious way, since we know that in many $\CN=2$ theories at infinite coupling, a strongly coupled SCFT emerges that is weakly gauged \cite{Argyres:2007cn, Argyres:2007tq, Gaiotto:2009we}. 
For the Type III theories, it may be possible to have a non-compact direction in the conformal manifold at large $N$, given that the $R$-charges of the fields asymptote to that of the free value and get closer and closer to conformal gauge theories in the limit. However, since the theory is not strictly connected to the free theory, it may mean that the distance in the conformal manifold grows with $N$. It would be interesting to find a method to estimate the size of a conformal manifold and test the AdS SDC for the strongly-coupled theories.

In this paper, we classified all possible supersymmetric simple gauge theories whose large $N$ limit can be taken while the number of flavors is kept constant. We restricted to such case since we wanted to keep the bulk gauge group to be fixed.
However, if we relax this assumption, we obtain many SCFTs whose rank of the flavor symmetry increases in the large $N$ limit. 
For example, the $SU(N)$ SQCD in the conformal window should have $\frac{3}{2}N <N_f < 3N$. 
More generally, $SU(N)$ gauge theory with $N_{\text{rank-}2}=0, 1/2, 3/2, 5/2$ should contain $N_f\propto N$ to cancel the gauge anomaly while keeping asymptotic freedom in the large $N$ limit. 
Except for the case of $N_{\text{rank-}2}=1/2$, the spectrum of low-lying operators is expected to be sparse, and at most $O(1)$ number of operators could decouple. 
See Table \ref{table:type0} for the full list. The detailed aspects of these theories will appear elsewhere.

{\renewcommand\arraystretch{1.3}
  \begin{table}[htbp] 
    \footnotesize
    \centering
    \begin{tabular}{|c|c|c|}
      \hline
  $N_{\text{rank-}2}$   & Theories & Conformal window  \\\hline \hline 
    
   0 &  $N_f$  ( $\fund$ + $\overline{\fund}$ )  & $\frac{3}{2}N < N_f<3N$ \\\hline\hline

\multirow{2}{*}{1/2}   &     1 $\sym$ + $(N+4)$ $\overline{\fund}$ + $N_f$ ( $\fund$ + $\overline{\fund}$ ) & $0.1705 N - 2.2853 \lesssim N_f < 2N-3$ \\\cline{2-3}

&    1 $\antisym$ + $(N-4)$ $\overline{\fund}$ + $N_f$ ( $\fund$ + $\overline{\fund}$ ) & $ 0.1711N + 3.8517 \lesssim N_f < 2N+3$ \\\hline\hline

\multirow{3}{*}{1} &     2 $\sym$ + $2(N+4)$ $\overline{\fund}$ + $N_f$ ( $\fund$ + $\overline{\fund}$ ) & $0 \leq N_f < N-6$ \\\cline{2-3}

&      1 $\sym$ + 1 $\antisym$ + $2N$ $\overline{\fund}$ + $N_f$ ( $\fund$ + $\overline{\fund}$ ) & $0 \leq N_f < N$ \\\cline{2-3}
      
&      2 $\antisym$ + $2(N-4)$ $\overline{\fund}$ + $N_f$ ( $\fund$ + $\overline{\fund}$ ) & $ 0
\footnotemark \leq  N_f < N+6$ \\\hline\hline
      
\multirow{10}{*}{3/2} &      2 $\sym$ + 1 $\overline{\sym}$ + $(N+4)$ $\overline{\fund}$ + $N_f$ ( $\fund$ + $\overline{\fund}$ ) & $0 \leq N_f < N-5$ \\\cline{2-3}

&      2 $\sym$ + 1 $\overline{\antisym}$ + $(N+12)$ $\overline{\fund}$ + $N_f$ ( $\fund$ + $\overline{\fund}$ ) & $0 \leq N_f < N-7$ \\\cline{2-3}

&      1 $\sym$ + 1 $\overline{\sym}$ + 1 $\antisym$ + $(N-4)$ $\overline{\fund}$ + $N_f$ ( $\fund$ + $\overline{\fund}$ ) & $0 \leq N_f < N+1$ \\\cline{2-3}
      
&      1 $\sym$ + 1 $\antisym$ + 1 $\overline{\antisym}$ + $(N+4)$ $\overline{\fund}$ + $N_f$ ( $\fund$ + $\overline{\fund}$ ) & $0 \leq N_f < N-1$ \\\cline{2-3}

&      1 $\sym$ + 2 $\overline{\antisym}$ + $(N+4)$ $\fund$ + 16 $\overline{\fund}$ + $N_f$ ( $\fund$ + $\overline{\fund}$ ) & $ 0 \leq N_f < N-9$ \\\cline{2-3}

&      2 $\antisym$ + 1 $\overline{\antisym}$ + $(N-4)$ $\overline{\fund}$ + $N_f$ ( $\fund$ + $\overline{\fund}$ ) & $0
\footnotemark\leq N_f < N+5$ 
\\\cline{2-3}

&      1 \textbf{Adj} + 1 $\sym$ + $(N+4)$ $\overline{\fund}$ + $N_f$ ( $\fund$ + $\overline{\fund}$ ) & $0 \leq N_f < N-3$ \\\cline{2-3}
      
&      1 \textbf{Adj} + 1 $\antisym$ + $(N-4)$ $\overline{\fund}$ + $N_f$ ( $\fund$ + $\overline{\fund}$ ) & $0 \leq N_f < N+3$ \\\cline{2-3}

&      1 $\sym$ + 2 $\antisym$ + $(3N-4)$ $\overline{\fund}$ & $0\leq N_f < 3$ \\\cline{2-3}

&      3 $\antisym$ + $3(N-4)$ $\overline{\fund}$ + $N_f$ ( $\fund$ + $\overline{\fund}$ ) & $0
\footnotemark\leq N_f < 9$ \\\hline\hline

\multirow{6}{*}{2}  &    1 $\sym$ + 1 $\overline{\sym}$ + 2 $\antisym$ + $2(N-4)$ $\overline{\fund}$ & $0 \leq N_f < 4$ \\\cline{2-3}
      
&      1 $\sym$ + 2 $\antisym$ + 1 $\overline{\antisym}$ + $2N$ $\overline{\fund}$ & $0 \leq N_f < 2$ \\\cline{2-3}

&      3 $\antisym$ + 1 $\overline{\sym}$ + $2(N-8)$ $\overline{\fund}$ & $0\leq N_f < 10$\\\cline{2-3}
      
&      3 $\antisym$ + 1 $\overline{\antisym}$ + $2(N-4)$ $\overline{\fund}$ + $N_f$ ( $\fund$ + $\overline{\fund}$ ) & $0\leq N_f < 8$ \\\cline{2-3}

&      1 \textbf{Adj} + 1 $\sym$ + 1 $\antisym$ + $2N$ $\overline{\fund}$ & $*$ \\\cline{2-3}

&      1 \textbf{Adj} + 2 $\antisym$ + $2(N-4)$ $\overline{\fund}$ + $N_f$ ( $\fund$ + $\overline{\fund}$ ) & $0\leq N_f < 6$ \\\hline\hline
      
\multirow{10}{*}{5/2}  &    1 $\sym$ + 2 $\overline{\sym}$ + 2 $\antisym$ + $(N-12)$ $\overline{\fund}$ + $N_f$ ( $\fund$ + $\overline{\fund}$ ) & $0 \leq N_f < 5$ \\\cline{2-3}

&      1 $\sym$ + 1 $\overline{\sym}$ + 2 $\antisym$ + 1 $\overline{\antisym}$ + $(N-4)$ $\overline{\fund}$ & $0 \leq N_f < 3$ \\\cline{2-3}

&      1 $\sym$ + 2 $\antisym$ + 2 $\overline{\antisym}$ + $(N+4)$ $\overline{\fund}$ & $0 \leq N_f < 1$ \\\cline{2-3}

&      3 $\antisym$ + 2 $\overline{\sym}$ + $(N-20)$ $\overline{\fund}$ + $N_f$ ( $\fund$ + $\overline{\fund}$ ) & $0\leq N_f < 11$ \\\cline{2-3}

&      3 $\antisym$ + 1 $\overline{\sym}$ + 1 $\overline{\antisym}$ + $(N-12)$ $\overline{\fund}$ + $N_f$ ( $\fund$ + $\overline{\fund}$ ) & $0\leq N_f < 9$ \\\cline{2-3}

&      3 $\antisym$ + 2 $\overline{\antisym}$ + $(N-4)$ $\overline{\fund}$ + $N_f$ ( $\fund$ + $\overline{\fund}$ ) & $0\leq N_f < 7$ \\\cline{2-3}

&      1 \textbf{Adj} + 1 $\sym$ + 1 $\overline{\sym}$ + 1 $\antisym$ + $(N-4)$ $\overline{\fund}$ & $0\leq N_f < 1$ \\\cline{2-3}

&      1 \textbf{Adj} + 1 $\sym$ + 2 $\overline{\antisym}$ + $(N-12)$ $\fund$ + $N_f$ ( $\fund$ + $\overline{\fund}$ ) & $0\leq N_f < 7$ \\\cline{2-3}

&      1 \textbf{Adj} + 2 $\antisym$ + 1 $\overline{\antisym}$ + $(N-4)$ $\overline{\fund}$ + $N_f$ ( $\fund$ + $\overline{\fund}$ ) & $0\leq N_f < 5$ \\\cline{2-3}

&      2 \textbf{Adj} + 1 $\antisym$ + $(N-4)$ $\overline{\fund}$  + $N_f$ ( $\fund$ + $\overline{\fund}$ ) & $0\leq N_f < 3$ \\\hline
  
    \end{tabular}
    \caption{List of all possible superconformal $SU(N)$ theories where only the Veneziano limit is allowed. We also specify the condition for the gauge theory to flow an interacting SCFT.}
    \label{table:type0}
  \end{table}
  }

In addition, we only considered gauge theories whose gauge groups are simple Lie groups. On the other hand, most of the known good holographic conformal gauge theories are of quiver type \cite{Klebanov:1998hh, Martelli:2004wu, Benvenuti:2004dy}. 
Once we allow the gauge group to be semi-simple, we get many more possibilities. 
Moreover, in the current paper we restricted ourselves to the case without superpotential $W=0$. It would also be interesting to investigate these theories in the presence of superpotentials. For instance, one encounters a rich variety of generalized Kutasov dualities \cite{Kutasov:1995ve,Kutasov:1995np}. Such dualities for all Type I theories were studied in the mid-1990s \cite{Kutasov:1995ve,Kutasov:1995np,Aharony:1995ne,Leigh:1995qp,Intriligator:1995ax,Intriligator:1995ff}. Kutasov-type dualities are also known for Type II $SU(N)$ theories which include adjoint matter fields, as well as certain Type II $SO(N)$ and $Sp(N)$ theories \cite{Brodie:1996vx,Brodie:1996xm}. In addition, certain Type II $SU(N)$ theories with no adjoint matter fields but with additional gauge singlets are explored in \cite{Abel:2009ty}. 
One can also combine flip deformation along with ordinary relevant deformations to obtain a vast landscape of SCFTs \cite{Maruyoshi:2018nod, Cho:2024civ}. It may be possible to find new dualities and universal features in such SCFT landscape.

    \footnotetext[12]{Except for the cases $(N_f,N)=(0,4),(1,4),(2,4),(3,4),(0,5),(1,5),(2,5),(0,6),(1,6),(0,7)$.}
    \footnotetext[13]{Except for the cases $(N_f,N)=(0,4),(1,4),(0,5)$.}
    \footnotetext[14]{Except for the cases $(N_f,N)=(0,4),(1,4),(0,5)$.}

\begin{acknowledgments}
We thank Prarit Agarwal, Sunjin Choi, Monica Kang, Seungyu Kim, Craig Lawrie, and Matteo Lotito for discussions and related collaborations. 
The work of M.C. and J.S. is supported by the National Research Foundation of Korea (NRF) Grants Grant RS-2023-00208602 and RS-2024-00405629, and also by the KAIST-KIAS collaboration program. 
K.H.L. is supported by the Israel Science Foundation under grant no. 759/23. K.H.L thanks the hospitality of KAIST during the collaboration. J.S. is also supported by the Walter Burke Institute for Theoretical Physics at Caltech. 

\end{acknowledgments}

\bibliographystyle{JHEP}
\bibliography{refs}

\end{document}